\documentclass[twocolumn,showpacs,showkeys,preprintnumbers,amsmath,amssymb]{revtex4-1} 

\newcommand{\m}[1]{\macro{#1}}

\usepackage{amsmath,epsfig}

\newcommand{\be}{\begin{equation}} 
\newcommand{\ee}{\end{equation}} 
\newcommand{\bea}{\begin{eqnarray}} 
\newcommand{\eea}{\end{eqnarray}} 
\newcommand{\eml}{\end{mathletters}} 
\newcommand{\nn}{\nonumber\\} 
\newcommand{\oh}{\frac{1}{2}}

\newcommand{\pa}{\partial} 
\newcommand{\la}{\langle} 
\newcommand{\ra}{\rangle} 
 
\newcommand{\ov}{\overline} 
\newcommand{\tl}{\tilde} 
\newcommand{\wtl}{\widetilde}

\newcommand{\ba}{\begin{eqnarray}}
\newcommand{\ea}{\end{eqnarray}}
\newcommand{\sba}{\begin{subequations}}
\newcommand{\sea}{\end{subequations}}
\newcommand{\barr}{\begin{array}}
\newcommand{\earr}{\end{array}}
\def\A {{{\cal A}}}
\def\B {{{\cal B}}}

\def\N {{{\cal N}}}
\def\X {{{\cal X}}}
\def\Y {{{\cal Y}}}
\def\Z {{{\cal Z}}}
\def\W {{{\cal W}}}
\def\a{\alpha}
\def\b{\beta}
\def\c{\gamma}
\def\d{\delta}
\def\D{\Delta}
\def\e{\epsilon}

\def\m{\mu}
\def\n{\nu}
\def\r{\rho}

\def\t{\tau}

\def\w{\omega}
\def\W{\Omega}
\def\vec#1{\mbox{\boldmath $#1$}}

\begin{document} 

\title{ Equation of Motion Method to strongly correlated Fermi systems \\ and Extended RPA approaches}

\author{P. Schuck}

\affiliation{Universit\'e Paris-Saclay, CNRS-IN2P3, IJCLab, 91405 Orsay, France and \\
Universit\'e Grenoble Alpes, CNRS, LPMMC, 38000 Grenoble, France}

\author{D.S. Delion}

\affiliation{"Horia Hulubei" National Institute of Physics and
Nuclear Engineering, 407 Atomistilor, RO-077125 Bucharest-Magurele, Rom\^ania and \\
Academy of Romanian Scientists, 3 Ilfov RO-050044,
Bucharest, Rom\^ania}

\author{J. Dukelsky}

\affiliation{Instituto de Estructura de la Materia, CSIC,
Serrano 123, E-28006 Madrid, Spain}

\author{M. Jemai}

\affiliation{Laboratory of Advanced Materials and Quantum Phenomena, Physics Department, FST, El-Manar University, 2092 Tunis,Tunisia
  \\
  ISSATM, Carthage University, Avenue de la R\'epublique P.O. Box 77-1054
  Amilcar, Tunis, Tunisia}

\author{E. Litvinova}

\affiliation{Department of Physics, Western Michigan University, Kalamazoo, MI 49008, USA}
\affiliation{National Superconducting Cyclotron Laboratory, Michigan State University, East Lansing, MI 48824, USA}
\affiliation{GANIL, CEA/DRF-CNRS/IN2P3, F-14076 Caen, France}

\author{G. R\"opke}

\affiliation{Universit\"at Rostock, FB Physik, Universit\"atplatz,
D-18051 Rostock, Germany}

\author{M. Tohyama}

\affiliation{Kyorin University School of Medicine, Mitaka, Tokyo 181-8611, Japan}

\date{September 1, 2020}

\begin{abstract}
{The status of different extensions of the
Random Phase Approximation (RPA) is reviewed.
The general framework is given within the Equation of Motion Method
and the equivalent Green's function approach for the so-called
Self-Consistent RPA (SCRPA). The role of the Pauli principle is analyzed. 
A comparison among various approaches to include Pauli correlations,
in particular, renormalized RPA (r-RPA), is performed. 
The thermodynamic properties of nuclear matter are studied with several
cluster approximations for the self-energy of the single-particle Dyson equation.
More particle RPA's are shortly discussed with a particular
attention to the $\alpha$-particle condensate.
Results obtained concerning the Three-level Lipkin, Hubbard and Picket Fence Models, 
respectively, are outlined.
Extended second RPA (ESRPA) is presented.}
\end{abstract}


\vfill\eject
\maketitle

\newpage
\begin{widetext}

\begin{tabular}{rrl}
     &   & ~~~~~~~~~~~~~~~~~~~~~~~~~~~~~~~~~~~~~{\bf CONTENT} \cr
     &   & \cr
I    &   & Introduction \cr
     &   & \cr
II   &   & The Equation of Motion Method \cr
     & A & Rowe's Equation of Motion method, self-consistent RPA (SCRPA), \cr 
     &   & and connection with Coupled Cluster theory \cr
     & B & Renormalized RPA \cr
     & C & The correlation energy and the boson aspect of the \cr
     &   & Self-Consistent Random Phase Approximation (SCRPA) \cr
     & D & SCRPA in the particle-particle channel \cr
     & E & Self-Consistent Quasi-particle RPA (SCQRPA) \cr
     & F & Number conserving ph-RPA in superfluid nuclei \cr
     & G & Odd-particle number random phase approximation \cr
     &   & \cr
III  &   & Applications of SCRPA \cr
     & A & Picket Fence (pairing) Model \cr
     & B & Three level Lipkin model \cr
     & C & Hubbard model \cr
     & D & Various applications and extensions of the renormalized RPA \cr
     &   & \cr
IV   &   & The Green's Function Formalism \cr
     & A &  Static part of the BSE kernel \cr
     & B & Dynamic part of the BSE kernel \cr
     & C & The ph-channel \cr
     & D & The particle-vibration-coupling (PVC) approach  \cr
     & E & Application of the Green's Function Approach \cr
     &   & to the pairing model at finite temperature \cr
     &   & \cr
V    &   & Single-particle Green's function, Dyson equation, \cr
     &   & and applications to thermodynamic properties of nuclear matter \cr
     & A & Relation of the s.p. Green's function to the ground state energy. \cr
     &   & The tadpole, perturbative particle-vibration coupling, and the spurious mode \cr
     & B & An application to the Lipkin model of the coupling constant integration with SCRPA \cr
     & C & Inclusion of particle-particle RPA correlations into the self-energy. \cr
     &   & The T-matrix approximation \cr 
     & D & Single-particle Green's function from Coupled Cluster Doubles (CCD) wave function. \cr
     &   & Even-odd self-consistent RPA. Application to the Lipkin model \cr
     & E & Cluster expansion of the single-particle self-energy \cr 
     &   & and applications to infinite nuclear matter problems \cr
     & F & Applications of the in-medium two nucleon problem \cr
     &   & and the T-matrix approximation for the s.p. self-energy \cr
     &   & \cr
VI   &   & Quartetting and $\alpha$-particle condensation \cr
     & A & Critical temperature for $\alpha$ condensation \cr
     & B & 'Gap' equation for quartet order parameter \cr
     &   & \cr
VII  &   & Second RPA and extensions \cr
     & A & Extended RPA (ERPA) equation \cr
     & B & Hermiticity of ERPA matrix \cr
     & C & Orthonormal condition \cr
     & D & Energy-weighted sum rule \cr
     & E & Spurious modes in ERPA and SCRPA \cr
     & F & Approximate forms of ERPA \cr
     & G & particle-particle ERPA equation \cr
     &   & \cr
VIII &   & Applications \cr
     & A & Selfconsistent second RPA in the exactly solvable single shell pairing model \cr
     & B & Lipkin model \cr
     & C & Hubbard model \cr
     & D & Damping of giant resonances \cr
     &   & \cr
IX   &   & Discussion and Conclusions \cr
     &   & References \cr
\end{tabular}
\end{widetext}

\newpage
\section{Introduction}
\label{sec:intro}
\setcounter{equation}{0}
\renewcommand{\theequation}{1.\arabic{equation}}

The solution of the many body problem of quantum gases or quantum fluids
is a formidable challenge. In spite of considerable progress and tremendous
effort in the past fifty years, we still have no general theory at hand
which allows to accurately calculate many properties of strongly correlated
many body quantum systems. Of course the Hartree-Fock (HF) or effective mean field
approaches \cite{Bla86,Mah81,Neg88,Fet71,Rin80} 
are well accepted in almost every branch of many body physics as the
first basic and necessary step. Many qualitative features can be explained by
this method and, if one takes the case of Bardeen-Cooper-Schrieffer (BCS)
theory \cite{Bla86,Mah81,Neg88,Fet71,Rin80} as an example for the description of superconducting 
or superfluid Fermi systems, sometimes even very accurate 
predictions of the phenomena
can be obtained. These one body mean field approaches are in general
non perturbative and in the case of the pure HF theory this
corresponds to a Rayleigh-Ritz variational principle yielding an upper
bound to the true ground state energy which is, of course, a very desirable
feature. However, effective mean field theories based on density functionals
or effective forces, like they are in use for band structure calculations in
condensed matter or for ground state energies of atomic nuclei, usually
cannot assure such an upper bound limit of the energy. The consensus
which prevails on the level of one body mean field theory, unfortunately,
is already lost on the next level of sophistication, when it comes to 
two body correlations or quantum fluctuations. Indeed, quite a variety of
formalisms exist to deal with correlation functions beyond the mean
field level. Of course the most ambitious attempt is to calculate two
body correlations also from a Rayleigh-Ritz variational principle.
Since mean field theory corresponds to a variational wave function of the
coherent state type with a one body operator in the exponent,
it is natural to extend this to include also a two body operator
for two body correlations \cite{Bla86,Bar07,Ful}. 
However, a most general two-body operator in the exponent is
by far too complicated for practical purposes, so that various restrictions
on the two-body term have been imposed in the past \cite{Bla86}. 
A most natural choice is a {\it local} two body operator leading to
the famous Jastrow or Gutzwiller type of variational wave functions \cite{Bla86,Ful}
together with Quantum Monte Carlo (QMC) methods \cite{Wag16,Car15}.

However, even these restricted variational ground state wave functions
are extremely complicated to be put in full operation. The method of
correlated basis functions \cite{Cla66}, the hypernetted chain expansion
\cite{Bla86,Fab02} and renormalisation group methods (RGM) \cite{Bau19}
are, besides QMC, ways of how to treat this problem. 
Once the ground state problem is solved, there remains the question
of how to obtain the excited states.
For this, separate developments based on the previously obtained ground state
wave functions are necessary. Though a Rayleigh-Ritz variational method
seems conceptually the cleanest way to treat correlations with its
nonperturbative and well controlled aspects, because of its high complexity
and numerical difficulties in practical applications, quite a variety of
other methods is in use. The oldest but because of its simplicity still very
much employed consists of partial resummation of bubbles (Random Phase
Approximation, RPA) in the particle-hole (ph) channel \cite{Bla86,Mah81,Neg88,Fet71,Rin80} or of ladders
(Bethe-Goldstone equation, Brueckner-Hartree-Fock or Galitskii-Feynman
T-matrix equation) in the particle-particle (pp) channel \cite{Fet71,Rin80}.

In spite of the general usefulness of these approaches, they suffer from
obvious short comings, like violation of the Pauli principle, uncontrolled
{\it (e.g. non-conserving)} approximations, self-energy but not vertex 
corrections, etc.
Therefore, in order to correct for these short comings, at least partially,
more sophisticated approaches have been invented correcting one or
several of these deficiencies, but in general not all of them.
For example Coupled Cluster Theory (CCT) \cite{Bla86,Bar07,Bis91} also starts from an exponential
with, as a first correction to HF, a two body operator in the exponent.
However, it is not used as a variational wave function but the Schr\"odinger
equation is closed by its projection on the basis of uncorrelated HF states.
This leads to a non-Hermitian problem, which lacks the upper bound theorem
of the Rayleigh-Ritz variation, but which is otherwise quite general and
has been successfully applied to a variety of physics problems \cite{Bar07}.
It contains RPA as a limiting case, but in general for excited states and
also for finite temperature extra ingredients have to be and have been
invented \cite{Bar07}.
Usually what  makes the problem with a two-body operator in
the exponent difficult, is the fact that the corresponding wave function  
does not correspond to a unitary
transformation of some reference state and then the norm of the correlated
wave function is very difficult to evaluate. The method of flow
equations \cite{Keh} just tries to establish  a unitary
transformation going beyond HF in a systematic way. This is a relatively new
approach, which is quite general. It seems, however, that correlation
functions are very difficult to obtain from this theory. As mentioned, other well established
methods are the (Quantum) Monte Carlo, or Path Integral Approaches 
\cite{Neg88,Cep10,Rub17}.
For Bose systems they are quite efficient approaching the exact solutions
of various quantum many body systems quite accurately, but for
correlated Fermi systems the so called sign problem has so far prevented
from a real break through and mostly the method is restricted to
separable interactions.The methods described above being
quite general and applicable practically to any system of interacting fermions
or bosons, there also exist numerous methods more or less tailored to
specific problems. The Gutzwiller ansatz for the ground state wave function
of the Hubbard model is a famous example but again the ansatz can in general
not be carried through and is accompanied by the so-called Gutzwiller
approximation \cite{Ful}.
Other methods try to attack the many-body fermion problem by
diagonalizing huge matrices with more or less sophisticated algorithms
like, e.g. the one by Lanczos or by different renormalisation 
group methods \cite{Sch05}.

It is, however, not our intention here to be exhaustive in the description
of all existing theories. We rather will now give the motivation and
a basic outline of the many body formalism which shall be the subject
of the present article. Roughly speaking our approach can be characterised
by the Equation of Motion (EOM) method in conjunction with extended RPA theories. EOM has, of course,
been applied to the many body problem since its early days.
However, we believe that the potential of this method has never been fully
exploited. In the last couple of years we have developed this formalism
and applied it with very good success to various physical problems.
In spite of the fact that the theory still can certainly be developed
further, we believe that we have explored it sufficiently far by now to
present a quite coherent and self contained frame on this subject in this
report.

Let us start explaining the physical idea behind our approach. Standard
single-particle mean field or HF theory aims at finding the best possible
single-particle description of the system. This leads to the well known
self consistent HF mean field Hamiltonian, where the two body interaction
is averaged over the single-particle density. The idea is now that
a many body quantum system not only consists out of a gas of independent
mean field quasiparticles but also, in a further step, out of a gas of
quantum fluctuations, built out of fermion or boson pairs.
These quantum fluctuations then make up their own mean field, in spirit
very similar to the ordinary single-particle mean field. 
As an example, if bound states are formed, they may be considered as 
new entities producing their own mean field. The formulation of this Cluster 
Mean-Field (CMF) or Self-Consistent RPA (SCRPA) approach 
\cite{Row68,Rop80,Rop95,Rop09,Her16,Duk90,Duk98,Duk99}  
will be given below in Sect. II and IV. 
If the quantum
fluctuations can be represented by bosons and the fermion Hamiltonian is
mapped into one of interacting bosons, then the concept of a mean field
for these bosons can be easily accepted. The difficulty comes from the fact
that we want to avoid as far as possible bosonisation and always stay within
the original fermion description and then the concept of the
mean field for quantum fluctuations (correlated fermion)
becomes less evident. In the main text we, however, will show how
this concept can be worked out quite rigorously starting from different
initial descriptions of the many body system leading, however, to the same
final result.\\

In this review, we will concentrate on interacting Fermi systems while our approach can rather straightforwardly also be applied to Bose systems or to mixed Bose-Fermi ones.
Let us here give a short outline of the main ingredients of our approach 
based on the Equation of Motion   method.
One particularly simple way to introduce the generalised mean field equations
via the EOM is given by the minimisation of the energy weighted sum
rules \cite{Bar70}.
For pedagogical reasons we  want to start out with the
rederivation of a well known example which are the Hartree-Fock-Bogoliubov
(HFB) equations for interacting fields of bosons 
$b^{\dag}_{\alpha}, b_{\alpha}$ 
\cite{Bla86,Rin80}. 
The Bogoliubov transformation among these operators reads
\bea
\label{Bog}
q^{\dag}_{\nu}=\sum_{\alpha}\left[
U_{\nu\alpha}b^{\dag}_{\alpha}-V_{\nu\alpha}b_{\alpha}\right]~.
\eea
The transformation shall be unitary and therefore the amplitudes
$U$ and $V$ obey the usual orthonormality and completeness relations 
\cite{Bla86,Rin80}.

The coefficients $U$ and $V$ will be determined from extremum of the
following energy weighted sum rule \cite{Bar70} 
\bea
\label{sumrule}
e_{\nu}=\frac{1}{2}\frac{\la 0|[q_{\nu},[H,q^{\dag}_{\nu}]]|0\ra}
{\la 0|[q_{\nu},q^{\dag}_{\nu}]|0\ra}~,
\eea
where the ground state $|0\rangle$ is defined below.
Schematically the minimisation leads to the following set of equations
\bea
\label{HFB}
\left(\begin{matrix} h & \Delta \cr -\Delta^* & -h^*\end{matrix}\right)
\left(\begin{matrix} U \cr V\end{matrix}\right)=E
\left(\begin{matrix} U \cr V\end{matrix}\right)~,
\eea
with $h=\la 0|[b,[H,b^{\dag}]]|0\ra$ and $\Delta=\la 0|[b,[H,b]]|0\ra$.
With $H$ containing a two body boson interaction of the form
$\sum v b^{\dag}b^{\dag}bb$ we easily verify that $h$ and $\Delta$
are given in terms of single-particle densities 
$\la 0|b^{\dag}b|0\ra$ and $\la 0|b^{\dag}b^{\dag}|0\ra$, respectively.
In the EOM one always assumes the existence of a ground state
$|0\ra$, which is the well known  vacuum of the new quasiparticle operators
$q_{\nu}|0\ra=0$, for all $\nu$, see, e.g., \cite{Bla86,Rin80} . The states $|\nu\ra=q^{\dag}_{\nu}|0\ra$
are then the excited states of the system.
Either now one constructs the ground state from the vacuum condition
and one evaluates the single-particle densities in terms of the
amplitudes $U$ and $V$, or one demands that the transformation (\ref{Bog})
be unitary in which case this relation can be inverted and the operators
$b^{\dag},b$ can be expressed in terms of $q^{\dag},q$.
Inserting this into the expression for the densities, moving the destruction
operators to the right and exploiting the above mentioned vacuum condition,
again one evaluates the densities in terms of the amplitudes $U$ and $V$.
The resulting nonlinear and self-consistent equations are, of course,
identical with the original HFB equations for bosons
\cite{Bla86,Rin80}.
In a very similar way one can derive the HFB equations
for fermions. 

Let us now indicate how in complete analogy to the HFB equations one
derives self consistent equations for e.g. fermion pair operators,
or any other clusters of fermion or boson operators, or a mixture of both.
As a definite case let us consider the well known example of density
fluctuations in a Fermi system. We start with the definition of an RPA-type of
excitation operator in the particle-hole channel, i.e. describing 
density excitations
\bea
\label{Qdag}
Q^{\dag}_{\nu}=\sum_{ph}\left[
  X^{\nu}_{ph}a^{\dag}_pa_h-Y^{\nu}_{ph}a^{\dag}_ha_p\right]~,
\label{s-RPA-op}
\eea
where $a^{\dag},a$ are fermion creation/destruction operators and the
indices $p,h$ stand for "particle" and "hole" states of a yet to be
defined "optimal" single-particle basis. It is recognized that the operators
$Q^{\dag}_{\nu}$ of (\ref{Qdag}) contain a Bogoliubov transformation
of fermion pair operators $a^{\dag}_pa_h$.
If they are approximated by ideal Bose operators 
$a^{\dag}_pa_h\rightarrow B^{\dag}_{ph}$, as in standard RPA 
\cite{Rin80}, (\ref{Qdag})
constitutes a Bogoliubov transformation among bosons quite analogous to
(\ref{Bog}). We, however, want to stress the point that we will avoid
"bosonisation" as far as possible and stay with the fermion pair operators,
as in (\ref{Qdag}).

Furthermore, the operator of (\ref{Qdag}), as in standard HFB, should have
the properties
\bea
\label{creat}
Q^{\dag}_{\nu}|0\ra=\nu,
\eea
\bea
\label{anih}
Q_{\nu}|0\ra=0~,
\eea
that is the application of $Q^{\dag}_{\nu}$ on the ground state $|0\ra$
of the system creates an excited state and at the same time the
ground state should be the "vacuum" to the destructors $Q_{\nu}$.
In order to determine the amplitudes $X,Y$ of (\ref{Qdag}) we
use in analogy with (\ref{sumrule}) a generalised sum rule
\bea
\label{gensumrule}
\Omega_{\nu}=\frac{1}{2}\frac{\la 0|[Q_{\nu},[H,Q^{\dag}_{\nu}]]|0\ra}
{\la 0|[Q_{\nu},Q^{\dag}_{\nu}]|0\ra}~,
\eea
which we make stationary with respect to $X,Y$.
This leads to the RPA-type of equations of the form
\bea
\label{RPA}
\left(\begin{matrix} {\cal A} & {\cal B} \cr -{\cal B}^* & -{\cal A}^*
\end{matrix}\right)
\left(\begin{matrix} X \cr Y\end{matrix}\right)=\Omega
\left(\begin{matrix} {\cal N} & 0 \cr 0 & -{\cal N}\end{matrix}\right)
\left(\begin{matrix} X \cr Y\end{matrix}\right)~,
\eea
which are the counterpart of the HFB equations for bosons described above.
The matrices ${\cal A}$ and ${ \cal B}$ contain corresponding double commutators
involving the fermion pair operators and the matrix ${\cal N}$ stems from the
fact that the fermion pair operators do not have ideal Bose commutation
relations. With a Hamiltonian containing  a two body interaction,
one easily convinces oneself that the matrices ${\cal A,B,N}$ contain
no more than single-particle and two particle densities
of the schematic form $\la 0|a^{\dag}a|0\ra$ and
$\la 0|a^{\dag}a^{\dag}aa|0\ra$. Evaluating these expectation
values with the HF ground state leads to the standard HF-RPA
equations (as obtained from Time Dependent Hartree-Fock (TDHF) in the small amplitude limit, that is with exchange) . However, in general (\ref{anih}) is not fulfilled with a HF state
but leads to a correlated state containing the $X$ and $Y$ amplitudes.
Evaluating the one and two particle densities with such correlated 
ground state leads to matrices ${\cal A,B,N}$ which depend on the amplitudes
$X,Y$ and therefore a selfconsistency problem is established quite
analogous to the HFB problem for bosons described above.
We call these generalised RPA equations the Self-Consistent RPA (SCRPA)
equations.

Contrary to the original HFB approach for bosons, the determination
of functionals ${\cal A}[X,Y],{\cal B}[X,Y],{\cal N}[X,Y]$ is, in general,
not possible without some approximation.
This stems from the fact that Eq. (\ref{anih})
can, besides in exceptional model cases, not be solved exatly for
the ground state $|0\ra$. However, as we will show in the main text, it is possible to solve (\ref{anih}) with a somewhat extended RPA operator and the corresponding ground state wave function will be the well known Coupled-Cluster Doubles (CCD) wave function. We will explain this in detail in Section II. On the other hand, if one sticks to the usual RPA ph-operator (\ref{s-RPA-op}), in general the condition (\ref{anih}) will only be approximately fulfilled. Essentially two strategies are then possible:
either one evaluates the one- and two-body densities with  an approximate ground state 
as, e.g. the HF one, or, in the case of a broken symmetry, projected HF, etc.
Or one inverts relation (\ref{Qdag}), inserts the  $ph$
pair operators into the densities, commutes the destructors $Q_{\nu}$
to the right and  uses (\ref{anih}). We will show in the main text
that the second method, i.e. the one using the inversion
of (\ref{Qdag}), leads mostly to much better results. Details of the method and applications also will be given. 

We should stress at this point that the above  mentioned
necessary approximations again lead to certain violation of the Pauli
principle. However, as we will show in our examples, SCRPA often quite
dramatically improves over standard RPA. Naturally this occurs, for instance,
in situations where standard RPA breaks down, i.e close to a phase
transition point or for finite systems with very few number of particles.
Let us point out here again that (\ref{anih}) can be solved for the ground state if 
an extension of the operator (\ref{Qdag}) including some specific two body terms
is used. We will present this extended approach in section II.B.

The above summary describes the essentials  of our method on the example
of density fluctuations. However, EOM is not at all restricted to this
case. One can in the same way treat pair-fluctuations involving
fermion pairs $a^{\dag}a^{\dag}$ and $aa$.
Formally there is no restriction in the choice of the composite operators.
To describe quartetting, quadruple operators like 
$a^{\dag}a^{\dag}a^{\dag}a^{\dag}$ shall be used. One can consider second
order density fluctuations with $a^{\dag}aa^{\dag}a$, odd numbers
of operators as $a^{\dag}a^{\dag}a$ ond so on. The same can be repeated
for Bose systems using clusters of Bose operators \cite{Duk91} 
Also mixtures of bosonic and fermionic operators can be treated in an analogous way \cite{Sog13,Wat08,Sto05}. 

The above formalism can also be derived using many body Green's 
functions \cite{Rop80,Rop09,Duk98,Kru94}.
This has the important advantage that generalisation to finite temperature
is straigthforward and we will give an example where SCRPA at finite 
temperature is solved. SCRPA equations can numerically be solved
for pairs of fermion operators $a^{\dag}a$ or $aa$, since the equations are of the Schroedinger type. They are  not more complicated as, e.g., self-consistent Bruckner-HF equations \cite{Rin80}.  However, in general, for higher clusters this is not possible at present without drastic approximations. We will further point out that SCRPA is a conserving approach with all the appreciable properties of standard RPA, as, e.g., Ward identities, maintained.
We want to point out that the Green's function formalism used, is the one 
based on so-called two times Green's functions $-i\langle T A(t) A^+(t')\rangle$
where the operators $A$ may be clusters of single fermion (or boson) operators. 
Quite naturally this then leads to Dyson type of equations for those 
'cluster' Green's functions which, at equilibrium, depend only on one energy 
variable. This is contrary to the usual where many body Green's functions 
depend on as many times (energies) as there are single-particle operators 
involved \cite{Bla86}. It has, however, become evident that equations for those 
many time 
Green's functions, involving parquet diagram techniques \cite{Bla86}, are extremely difficult to solve numerically (besides lowest order equations, this was not achieved) and, therefore, 
we stick to the above type of propagators depending on only one energy 
variable. This then leads to Schr\"odinger type of equations which are 
much more accessible for a numerical treatment. In this vain we will introduce a Dyson-Bethe-Salpeter Equation (Dyson-BSE) for fermion pairs with an integral kernel which, at equilibrium, depends only on a time difference as  the initial pair propagator or, after Fourier transform, this kernel depends only on one frequency, that is, in the case of the response function on the frequency of the external field. The kernel can be expressed by higher correlation functions and, thus, has a definite form ready for well chosen approximations. This one frequency Dyson-BSE is formally as exact as is the usual multi-time BSE. 

The review is organized as follows. In Sect. II we will explain the EOM in detail for the example of the response function leading to the self-consistent RPA (SCRPA). A sub product is the renormalized RPA (r-RPA) presented in Sect. II.A. The boson aspect of SCRPA and the SCRPA correlation energy is discussed in Sect. II.B. In Sect. II.C we show how an extended RPA operator can annihilate the CCD wave function. The SCRPA in the particle-particle channel and the self-consistent quasiparticle RPA are outlined in Sects. II.D and II.E, respectively. The very interesting number conserving ph-RPA in superfluid nuclei is presented in Sect. II.F. For odd particle numbers we derive an odd-RPA (o-RPA) in Sect. II.G. In Sections III.A,B,C, we give examples, where SCRPA is applied to the pairing model, the three-level Lipkin model, the Hubbard model, respectively. In Sect. III.D applications of the r-RPA are discussed. In Sect. IV the Green's function formalism with the EOM method is shown to be equivalent to SCRPA with, however, extensions to higher correlations leading to a formally exact Bethe-Salpeter equation of the Dyson form (Dyson-BSE) with an integral kernel depending only on one frequency. The static and dynamic parts of the kernel are presented in Sects. IV.A and IV.B and in Sect. IV.C special attention is payed to the ph-channel. In Sect. IV.D the particle-vibration coupling model and its applications to nuclear structure are presented. In Sect. IV.E an application to the pairing model at finite temperature is given. In Sect. V we discuss the single-particle Green's function and its self-energy, also at finite temperature. In Sects. V.A,B,C the self-energy is presented in various approximate forms including ph-correlation and pp-ones and in general a cluster expansion of the self-energy is discussed. Sect. V.D is devoted to the cluster expansion of the single-particle self-energy and applications to infinite nuclear matter problems In Sect. V.E applications of the so-called T-matrix approximation of the self-energy is applied to several problems of nuclear matter. In Sect. VI we discuss quartet ($\alpha$ particle) condensation also based on the EOM method. In Sects. VI.A,B the critical temperature and the four-nucleon order parameter are calculated in infinite nuclear matter. In Sect. VII the so-called second RPA with extensions (ERPA) is introduced with a discussion of several interesting properties of this scheme and in Sect. VIII some applications of ERPA are given. Finally, in Sect. IX we present our conclusions and perspectives.

\section{The Equation of Motion Method}
\label{sec:EMM}
\setcounter{equation}{0}
\renewcommand{\theequation}{2.\arabic{equation}}

In this section we want to present the details of the Equation of Motion (EOM)
method. As in the introduction, we will consider as a specific first example
the density excitations of a many-body fermion system (later, we also will consider the-two particle, that is the pairing channel). In particular,
we want to derive details of the Self-Consistent RPA (SCRPA) scheme.
Pioneering work in this direction has been performed about half a century ago
by D. Rowe (see e.g. the review article \cite{Row68}). Numerous other
studies have followed \cite{Rop80,Rop95,Rop09,Duk90,Duk98,Duk99,Sch73,Ada89,Sch00}. 
But extensions of RPA have also spread into other fields like chemical physics 
\cite{Cha12,Esh12,Per14,Per18}
and electronic, that is condensed matter systems 
\cite{Shi70,Shi73,Las77}.
Let us now set the detailed frame of the EOM method following D. Rowe and also give 
a connection with the Coupled Cluster Doubles  wave function. 

\subsection{Rowe's Equation of Motion method, self-consistent RPA (SCRPA), and connection with Coupled Cluster theory}

The basic observation of D. Rowe \cite{Row68} was that, given the exact non-degenerate ground state $|0\ra$ of a
many-body system with $N$ particles, an excited state of the system can be
obtained in applying a creation operator $Q^{\dag}$ on this  ground state,
which at the same time is the vacuum to the corresponding destruction
operator, that is 
\bea
\label{creat1}
Q^{\dag}_{\nu}|0\ra=|\nu\ra~,
\eea
with
\bea
\label{anih1}
Q_{\nu}|0\ra=0~.
\eea
Given that $|0\ra$ and $|\nu\ra$ are, respectively, exact ground state and
excited states of the many body Hamiltonian, i.e.
$H|\nu\ra=E_{\nu}|\nu\ra$ and $H|0\ra=E_{0}|0\ra$, one easily can write
down such an excitation operator. With $\la\nu|0 \ra=0$ the solution
to (\ref{creat1}) and (\ref{anih1}) is \cite{Row68}
\bea
\label{nu0}
Q^{\dag}_{\nu}=|\nu\ra\la 0|~.
\eea
With the help of the Schr\"odinger equation we then obtain
\bea
\label{HQ}
[H,Q^{\dag}_{\nu}]|0\ra=\Omega_{\nu}Q^{\dag}_{\nu}|0\ra~,
\eea
with $\Omega_{\nu}=E_{\nu}-E_0$ the excitation energy.
Multiplying from the left with an arbitrary variation of the form
$\la 0|\delta Q$ we obtain
\bea
\label{QHQ}
\la 0|[\delta Q,[H,Q^{\dag}_{\nu}]]|0\ra=\Omega_{\nu}
\la 0|[\delta Q,Q^{\dag}_{\nu}]|0\ra~.
\label{double-comm-eq}
\eea
In the remainder of the review we will use a two body Hamiltonian of the form

\bea
H &=& \sum_{k_1k_2}H_{0,k_1k_2}a^{\dag}_{k_1}a_{k_2}+ \frac{1}{4}
\sum_{k_1k_2k_3k_4}\bar v_{k_1k_2k_3k_4}a^{\dag}_{k_1}a^{\dag}_{k_2}
a_{k_4}a_{k_3} \nn
&\equiv& H_0 + V~,
\label{H-0-V}
\eea
with the antisymmetrised matrix element of the two body force  $\bar v_{k_1k_2k_3k_4} = \langle k_1k_2|v|k_3k_4\rangle -\langle k_1k_2|v|k_4k_3\rangle$.
The use of a three-body force is in principle feasible, but would unnecessarily
complicate all formulas. So, we refrain from this.
In (\ref{QHQ}) we can use the double commutator because 
$\la 0|Q^{\dag}_{\nu}=\la 0|HQ^{\dag}_{\nu}=0$ in the exact case.
The variation $\delta Q^{\dag}|0\ra$, exhausting the complete Hilbert
space (\ref{QHQ}), is equivalent to consider the extremum of the mean excitation
energy given by an energy weighted sum rule
\bea
\label{gensumrule1}
\Omega_{\nu}=\frac{1}{2} \frac{\la 0|[Q_{\nu},[H,Q^{\dag}_{\nu}]]|0\ra}
{\la 0|[Q_{\nu},Q^{\dag}_{\nu}]|0\ra}~.
\eea
With the exact operator (\ref{nu0}), (\ref{gensumrule1}) is equal
to exact excitation energy of the state $|\nu\ra$, i.e.
$\Omega_{\nu}=E_{\nu}-E_0$. However, for restricted operators
$Q^{\dag}_{\nu}$ the minimisation of (\ref{gensumrule1}) with variations
$\delta Q,\delta Q^{\dag}$ (both are independent), one sees that this
corresponds to minimise the energy weighted sum rule with respect to the
trial operator $Q^{\dag}$. One directly verifies that this again leads
to (\ref{QHQ}). An obvious but important observation is that the creation
operator (\ref{creat1}) is an $N$-body operator.
It is therefore  a natural idea to develop this operator in a series
of one, two, ..., $N$-body operators as follows
\bea
\label{Qexp}
Q^{\dag}_{\nu}&=&\sum_{k_1k_2}\chi^{\nu}_{k_1k_2}a^{\dag}_{k_1}a_{k_2}
\nn&+& \frac{1}{4}
\sum_{k_1k_2k_3k_4}\chi^{\nu}_{k_1k_2k_3k_4}:a^{\dag}_{k_1}a^{\dag}_{k_2}
a_{k_3}a_{k_4}:+ ... ~,
\label{gen-Q}
\eea
where :....: means that no contractions of fermion operators are allowed within the double dots. 
If there are $N$ particles in the system and one pushes above expansion up to the $Np-Nh$ configuration, the exact result will be recovered. A demonstration of this is given in \cite{Ter17}. Of course, the more terms are kept in the expansion (\ref{gen-Q}), the more difficult it will become to solve the ensuing equations (\ref{double-comm-eq}), for instance, from the numerical point of view. So in the course of this review, we will restrict ourselves to the one-body and two-body terms shown in (\ref{gen-Q}).\\
  
Before entering the details, it may, however, be instructive to present the theory from a slightly different point of view.
From the Thouless theorem, see, e.g., \cite{Bla86,Rin80} we know that a general Slater determinant and, 
in particular, the HF determinant can be written as

\be
|\Phi_1\rangle \propto \exp\left(\sum_{ph}z_{ph} K^{\dag}_{ph}\right) |\Phi_0\rangle~,
\label{gen-SL}
\ee
with $K^{\dag}_{ph} = c^{\dag}_pc_h$
and $|\Phi_1\rangle$ not orthogonal to $|\Phi_0\rangle$. Obtaining the $z_{ph}$ from the minimisation of the energy, one arrives at the HF Slater-determinant

\be
|{\rm HF}\rangle = \Pi_ha^{\dag}_h|\rm vac\rangle~,
\label{HF-orbits}
\ee
where the $a^{\dag}_k,a_k$ represent orthonormalised creators and destructors of the HF-orbitals. As is well known, the standard RPA is based on the HF Slater determinant as ground state \cite{Rin80}. The $ph$ annihilator in standard RPA is then given by, see (\ref{s-RPA-op})

\be
a^{\dag}_ha_p|{\rm HF}\rangle = 0
\label{HF-killer}.
\ee



For a theory which goes beyond mean-field approximation like RPA with extensions, it is then natural to consider the following wave function

\be
|Z\rangle = e^{\hat Z}|\mbox{HF}\rangle~,
\label{Z}
\ee
with
\be
\hat Z = \frac{1}{4}\sum_{p_1p_2h_1h_2}z_{p_1p_2h_1h_2}\tilde K^{\dag}_{p_1h_1} \tilde K^{\dag}_{p_2h_2} ~.
\label{Z-op}
\ee
with $\tilde K^{\dag}_{ph}= a^{\dag}_pa_h$
where, instead of a single $ph$ operator in the exponent, there is in addition a quadratic one.
\noindent
It can be shown that this so-called Coupled Cluster Doubles  wave function  is the vacuum to the following generalized RPA operator \cite{Jem11,Jem13}.

\begin{eqnarray}
\tilde Q^+_{\nu} &=& \sum_{ph}[\tilde X^{\nu}_{ph}\tilde K^{\dag}_{ph} -\tilde Y^{\nu}_{ph}\tilde K_{ph}
\nn&+& \frac{1}{2}\sum_{php_1p_2}\eta_{php_1p_2}a^+_{p_1}a_{p_2}\tilde K^{\dag}_{ph} 
\nn&-& \frac{1}{2}\sum_{phh_1h_2}\eta_{h_1h_2ph}a^+_{h_1}a_{h_2}\tilde K_{ph},
\label{Q+}
\end{eqnarray}

\noindent
that is there exists the annihilating condition

\begin{equation}
\tilde Q_{\nu}|Z\rangle = 0~,
\label{killing}
\end{equation}

\noindent
with the following relations between the various amplitudes 

\begin{eqnarray}
 \tilde Y^{\nu}_{ph} &=& \sum_{p'h'}z_{pp'hh'}\tilde X^{\nu}_{p'h'}\nonumber\\ 
z_{pp'hh'} &=& \sum_{\nu}\tilde Y^{\nu}_{ph}(\tilde X^{-1})^{\nu}_{p'h'}\nonumber\\
\eta^{\nu}_{p_1p_2ph} &=& \sum_{h_1}z_{pp_2hh_1}\tilde X^{\nu}_{p_1h_1}\nonumber\\
\eta^{\nu}_{h_1h_2ph}&=&\sum_{p_1}z_{pp_1hh_2}\tilde X^{\nu}_{p_1h_1}~.
\label{amplitudes}
\end{eqnarray}

The amplitudes $z_{pp'hh'}$ are antisymmetric in $pp'$ and $hh'$. With the above relations, the vacuum state is entirely expressed by the RPA amplitudes $\tilde X, \tilde Y$. As mentioned, this vacuum state is exactly the one of coupled cluster theory (CCT) truncated at the-two body level which is called CCD \cite{Bla86,Bar07}.
However, the use we will make of this vacuum is very different from CCT. Of course, for the moment, all remains formal because this generalized RPA operator contains, besides the standard one-body terms, also specific two-body terms, which cannot be handled in a straightforward way. For instance, this non-linear transformation among fermion operators cannot be inverted in a simple manner as this is the case for HF or BCS quasiparticle destructors, which are annihilators of their respective wave functions. And, thus, despite being the vacuum of a annihilating operator, it is not immediately clear how to make calculations with this wave function. However, the mere existence of an exact annihilator of the CCD wave function is quite remarkable and we will see later in Sect. V.C, how this CCD with the generalized RPA may be handled in an approximate but efficient way. One may also notice that the operator (\ref{Q+}) is  part of the extended RPA operator considered in (\ref{gen-Q}).

On the other hand, there exists a very suggestive and eventually very valid approximation, which replaces in (\ref{Q+}) in the $\eta$ terms the density operators $a^{\dag}_{p_1}a_{p_2}$ and $a^{\dag}_{h_1}a_{h_2}$ by their expectation values

\[ a^{\dag}_{p_1}a_{p_2} \rightarrow  \langle a^{\dag}_{p_1}a_{p_2}\rangle \simeq \delta_{p_1p_2}n_{p_1}\]
    and
\[a^{\dag}_{h_1}a_{h_2} \rightarrow  \langle a^{\dag}_{h_1}a_{h_2}\rangle \simeq \delta_{h_1h_2}n_{h_1} \]
with $n_k = \langle a^{\dag}_ka_k\rangle $ being the single-particle (s.p.) occupation numbers. Of course, replacing operators by c-numbers implies to violate the Pauli principle. There exists, unfortunately, no simple measure which tells in general how severe this violation is. However, in some non-trivial models, where this approximation could be tested, it turned out that the violation stays quite mild \cite{Jem13}.
This is, for instance, the case in the Richardson pairing model, where the respect of the Pauli principle is extremely important \cite{Hir02}, because the s.p. levels are only two-fold degenerate. In any case, adopting above approximation leads us immediately to the usual ansatz for the RPA creation operator, which is

\be
Q^{\dag}_{\nu} = \sum_{ph}\bigg [X^{\nu}_{ph}a^{\dag}_pa_h - Y^{\nu}_{ph}a^{\dag}_ha_p\bigg ]~,
\label{rpa-Q}
\ee
and which has already been presented in the Introduction (\ref{s-RPA-op}). Besides the hypothesis that the replacements of density operators by their expectation values,  leading to (\ref{rpa-Q}), is in general a good approximation, we can 
now also give all the well-known arguments under which the ansatz (\ref{rpa-Q}) should yield a good description of excited states of a Fermi system. As we know, this is usually the case for collective excitations of the system. For instance, the plasma oscillation in electronic systems or Giant Resonances (GR) in nuclei are, among many other examples, of this kind. Of course, in considering finite systems like finite electronic devices and nuclei the size of those systems also plays a role: the number of particles should be large in order that collectivity can develop.


As we already mentioned, we make the reasonable hypothesis that, considering the reduced RPA operator (\ref{rpa-Q}), does not violate the Pauli-principle strongly. We, thus, can suppose that the annihilating condition (\ref{anih1}) is also still valid and Eq. (\ref{double-comm-eq}) can be used to calculate excited states. Before giving the details of the equations, we, however, want to proceed to a generalisation. Since Eq. (\ref{double-comm-eq}) implies ground state correlations, the s.p. occupation numbers $n_k$ will not be any longer of the step function form like with the HF approach but will be rounded close to the Fermi surface. Then, there is no need any longer to restrict the summation in the RPA operator to the $ph$ domain, but the amplitudes $X, Y $ can also contain $hh'$ and $pp'$ configurations. 
Consequently, we  will choose the amplitudes in (\ref{gen-Q})
$\chi^{\nu}_{mi}\equiv\tl{X}^{\nu}_{mi}$ with $m>i$ different
from the amplitudes $\chi^{\nu}_{im}\equiv-\tl{Y}^{\nu}_{im}$ with
$i<m$ and all $\chi^{\nu}_{kk}\equiv 0$.
We then write for the one
body part of (\ref{Qexp}) (unless otherwise stated, we will hitherto make the convention that indices $m,n > i,j$)
\bea
\label{Qdag1}
Q^{\dag}_{\nu}=
\sum_{m>i}\left[\tl{X}^{\nu}_{mi}a^{\dag}_{m}a_{i}-
\tl{Y}^{\nu}_{mi}a^{\dag}_{m}a_{i}\right]~.
\eea
It is, of course, evident, that the operator (\ref{Qdag1}) depends very
much on the  single-particle basis, since any change of the basis
will again create a hermitian part $\chi_{kk}a^{\dag}_ka_k$.
Therefore, it is very important to write down  the operator $Q^{\dag}_{\nu}$
of (\ref{Qdag1}) in a single-particle basis, which is optimal.
As usual, we will choose  the one which minimises  the ground state energy. 
It turns out that the ensuing equation is given by 
$\langle 0|[H,Q_{\nu}]|0\rangle = 0$.
How this goes in detail will be demonstrated  below. It is, however, 
clear that this relation is just another equation of motion, fullfilled in 
the exact case. This single-particle
basis will be given by a generalised single-particle mean-field Hamiltonian. It may be instructive to
divide for a moment the space into occupied levels ($h$: holes) and
unoccupied levels ($p$: particles). To be definite let us consider 4 levels
with the Fermi energy in the middle. We then order the states according
to this energy $p_4>p_3>h_2>h_1$. We thus have six $X^{\nu}$ amplitudes:
$X^{\nu}_{p_4p_3},~X^{\nu}_{h_2h_1},~X^{\nu}_{p_4h_2},~X^{\nu}_{p_4h_1},
~X^{\nu}_{p_3h_2},~X^{\nu}_{p_3h_1}$ and coresponding six $Y^{\nu}$ amplitudes.
We anticipate that in the standard RPA \cite{Bla86,Mah81,Neg88,Fet71,Rin80}
only the $ph$ amplitudes  survive.
However, as we will see, in the more general approach
of SCRPA also all other amplitudes can, in principle, be included, which 
may give non-negligible contributions.
This will, for instance, become important later, when we shall discuss
conservation laws and the Goldstone theorem in the case of spontaneously broken symmetries. 

From (\ref{Qdag1}) we see that this leads to an excited state
$|\nu\ra=Q^{\dag}_{\nu}|0\ra$, which is not normalised, i.e.
$\la\nu|\nu\ra=\la 0|[Q_{\nu},Q^{\dag}_{\nu}]|0\ra\neq 1$.
We therefore introduce slightly modified amplitudes and write
\bea
\label{Qdag2}
Q^{\dag}_{\nu}=
\sum_{m>i}\left(X^{\nu}_{mi}\delta Q^{\dag}_{mi}-
Y^{\nu}_{mi}\delta Q_{mi}\right)~,
\eea
where
\bea
\label{QQ}
\delta Q^{\dag}_{mi}&=&\frac{A_{mi}}{\sqrt{n_{i}-n_{m}}}~,~~~
A_{mi}=a^{\dag}_{m}a_{i}~,
\eea
are the normalised pair creation operators and
\bea
\label{norm}
n_i=\la 0|a^{\dag}_ia_i|0\ra~,
\eea
are the single-particle occupation numbers. With this choice
one immediately verifies that with
\bea
\label{XYnorm}
\sum_{m>i}\left(
\left|X^{\nu}_{mi}\right|^2-
\left|Y^{\nu}_{mi}\right|^2\right)=1~,
\eea
the excited states $|\nu\ra$ are normalised under the assumption that
the single-particle density matrix only has diagonal elements that is
$\rho_{kk'}=\la 0|a^{\dag}_ka_{k'}|0\ra=n_k\delta_{kk'}$, a fact
which will become clear in a moment, see after Eq. (\ref{aa}). With this we finally can write
for Eq. (\ref{QHQ})
\bea
\label{RPAeq}
\left(\begin{matrix} {\cal A}_{mim'i'} & {\cal B}_{mim'i'} \cr
 -{\cal B}^*_{mim'i'} & -{\cal A}^*_{mim'i'}\end{matrix} \right)
\left(\begin{matrix}X^{\nu}_{m'i'}\cr Y^{\nu}_{m'i'}\end{matrix}\right)=
\Omega_{\nu}
\left(\begin{matrix}X^{\nu}_{mi}\cr Y^{\nu}_{mi}\end{matrix}\right)~,
\nn
\label{SCRPA-gen}
\eea
where
\bea
\label{A}
{\cal A}_{mim'i'}=\la 0|\left[\delta Q_{mi}
\left[H,\delta Q^{\dag}_{m'i'}\right]\right]|0\ra~,
\eea
and
\bea
\label{B}
{\cal B}_{mim'i'}=-\la 0|\left[\delta Q^{\dag}_{mi}
\left[H,\delta Q^{\dag}_{m'i'}\right]\right]|0\ra~.
\eea
We realise that (\ref{RPAeq}) has exactly the same mathematical structure
as the standard RPA equations 
(see e.g. \cite{Bla86,Mah81,Neg88,Fet71,Rin80}). 
Therefore in this respect all standard
RPA properties are preserved \cite{Bla86,Mah81,Neg88,Fet71,Rin80}.
For instance we see that the eigenvectors
$\left(\begin{matrix}X^{\nu}\cr Y^{\nu}\end{matrix}\right)$
form a complete orthonormal set.
It is useful to introduce the matrices
\bea
{\cal X}=\left(\begin{matrix}X & Y^* \cr Y & X^* \end{matrix}\right)~,~~~
{\cal N}=\left(\begin{matrix}1 & 0 \cr 0 & -1 \end{matrix}\right)~.
\eea
Equation (\ref{RPAeq}) can then be written as
\bea
    {\cal S X}= {\cal N X}\Omega~,
    \label{short-RPA}
\eea
where
$
{\cal S}=\left(\begin{matrix}{\cal A} & {\cal B} \cr {\cal B}^* & {\cal A}^*
\end{matrix}\right)~,
$
and the diagonal matrix $\Omega$ contains the eigenvalues
$
\left(\begin{matrix}\Omega_{\nu}\cr-\Omega_{\nu}\end{matrix}\right)~,
$
if ${\cal S}$ is a positive definite matrix.
Simple matrix algebra shows that
\bea
\left[\Omega,{\cal X}^{\dag}{\cal N X}\right]&=&
\left({\cal N X}\Omega\right)^{\dag}{\cal X}-
{\cal X}^{\dag}\left({\cal N X}\Omega\right)
\nn&=&
{\cal X}^{\dag}\left({\cal S}^{\dag}-{\cal S}\right){\cal X}=0~,
\eea
that is, $\Omega$ commutes with ${\cal X}^{\dag}{\cal N X}$, and thus
${\cal X}^{\dag}{\cal N X}$ is diagonal together with $\Omega$.
The normalisation (\ref{XYnorm}) corresponds to the more general
orthogonality relations
\bea
\label{orth}
{\cal X}^{\dag}{\cal NX}={\cal N}~.
\eea
This closure condition  is obtained by multiplying (\ref{orth}) with
${\cal N}$, which shows that ${\cal NXN}$ is the inverse of ${\cal X}^{\dag}$,
or
\bea
\label{clos}
{\cal XNX}^{\dag}={\cal N}~,
\eea
which gives explicitly
\bea
\label{clos1}
\sum_{\nu}\left(
X^{\nu}_{mi}X^{\nu*}_{m'i'}-
Y^{\nu*}_{mi}Y^{\nu}_{m'i'}\right)=
\delta_{mm'}\delta_{ii'}~.
\eea
These orthonormality relations allow us to invert the operator (\ref{Qdag2})
\bea
\label{aa}
a^{\dag}_{m}a_{i}=\sqrt{n_{i}-n_{m}}
\sum_{\nu}\left(
X^{\nu*}_{mi}Q^{\dag}_{\nu}+Y^{\nu*}_{mi}Q_{\nu}\right)~.
\label{inv}
\eea
With (\ref{anih1}), it then follows that the density matrix 
$\la 0|a^{\dag}_ka_{k'}|0\ra$
only has diagonal elements, as postulated after eq.(\ref{XYnorm}).\\


The matrix ${\mathcal S}$ in (\ref{short-RPA}) can be written in the following way \cite{Sch16}.
\bea
&&\tilde {\mathcal S}_{minj} = \sqrt{n_{i}-n_{m}}{\mathcal S}_{minj}\sqrt{n_{j}-n_{n}}
\nn
&&= (\epsilon_m-\epsilon_i)N_{mi}\delta_{ij}\delta_{mn} + N_{mi}\bar v_{mjin}N_{nj} +\nonumber\\
&&\bigg [-\frac{1}{2}\sum_{ll'l''}(\delta_{ij}\bar 
v_{mll'l''}C_{ll''nl}+\delta_{mn}\bar v_{ll'il''}C_{jl''ll'})\nonumber\\
&+&\sum_{ll'}(\bar v_{mlnl'}C_{jl'il}+\bar v_{jlil'}C_{ml'nl})
\nonumber\\
&-&\frac{1}{2}\sum_{ll'}(\bar v_{mjll'}C_{ll'in}+\bar v_{ll'in}C_{mjll'})\bigg ]~,
\label{cal-S}
\eea
where $\epsilon_k$ are the HF s.p. energies, $N_{mi} = n_{i}-n_{m}$, and

\be
C_{mim'i'} = \langle a^{\dag}_{m'}a^{\dag}_{i'}a_{i}a_{m}\rangle -n_{m}n_{i}\delta_{mi,m'i'}~.
\label{C-dm}
\ee
With the inversion (\ref{inv}) and the annihilating condition (\ref{anih1}) the RPA matrix can entirely be expressed by the $X,Y$ amplitudes which then will depend in a very non-linear way of those amplitudes. This then constitutes the most general SCRPA scheme.\\
It can immediately be verified that, if all expectation values
in (\ref{RPAeq}) are evaluated with the HF ground state, then the standard
RPA equations are recovered with, in particular, only $X_{ph}$ and $Y_{ph}$
amplitudes surviving.

Before we come to the explicit evaluation of the matrix elements 
${\cal A},{\cal B}$
in (\ref{RPAeq}) in terms of $X,Y$
we first shall deal with the already mentioned and very important question 
of the optimal single-particle basis. This basis
is to be determined from the minimisation of the ground state energy.
However, as shown in \cite{Duk99,Del05}, there exists a very elegant but 
equivalent way which we now will explain. If, istead of closing the 
EOM (\ref{HQ}) from the left with a variation, we project from the left
with the ground state, we obtain with (\ref{anih1})
\bea
\label{HQ1}
\la 0|[H,Q^{\dag}_{\nu}]|0\ra=
\la 0|[H,Q_{\nu}]|0\ra=0~.
\eea
Because there are as many operators $Q^{\dag}_{\nu},Q_{\nu}$
as there are components $a^{\dag}_{m}a_{i},~a^{\dag}_{i}a_{m}$
we also can write for (\ref{HQ1})
\bea
\label{Haa}
\la 0|[H,a^{\dag}_{m}a_{i}]|0\ra=
\la 0|[H,a^{\dag}_{i}a_{m}]|0\ra=0~,
\label{gen-mf}
\eea
where we again recall our convention $m>i$. One also checks that with these relations the eventual non-hermiticity of the off-diagonal matrices in the RPA matrix (\ref{SCRPA-gen}) disappears. It also implies that the time derivative of the single-particle  density matrix is zero at equilibrium, that is, it is stationary. 

Equations (\ref{Haa}) are of the one-body type and one can directly verify
that with a Slater determinant as a ground state they reduce to the HF
equations. However, with the RPA ground state the single-particle basis
becomes coupled to the two-body RPA correlations as follows
\be
\label{HFeq}
\sum_{m'} H_{mm'}C_{m'\a} = \epsilon_{\a} n_{\a} C_{m\a}~,
\ee
where $C_{m\a}$ are the transformation coefficients defining the basis in which the density matrix is diagonal, the so-called canonical basis, that is
\bea
\label{ac0}
a^{\dag}_{k\mu}=\sum_{\a}C_{k\a}c^{\dag}_{\a\mu}~.
\eea
We also introduced as short-hand notation
\bea
\label{Hmm}
&&H_{mm'}
\equiv n_m\sum_{\mu}\epsilon_{\mu}C_{m\mu}C_{m'\mu}
\nn
&+&\frac{1}{2}\sum_{jkl}\sum_{\mu\b\c\d}
[\langle mjkl\rangle V_{\a\b\c\d}+
 \langle jmkl\rangle V_{\b\a\c\d}
\nn&+&
 \langle kjml\rangle V_{\c\b\a\d}+
 \langle ljkm\rangle V_{\d\b\c\a}]
C_{m'\mu}C_{j\b}C_{k\c}C_{l\d}~,
\nn
\eea
where $\la ijkl\ra\equiv\la a^{\dag}_ia_ja^{\dag}_ka_l\ra$ 
are the two-body densities which,  together with occupation numbers $n_{m}$,
depend on the RPA amplitudes.\\

So this is the outline of the most general RPA scheme with a correlated ground state based on a one-body operator to generate excited states. We now will pass to some useful and simplifying approximations.

 \subsection{Renormalized RPA}

There exists a first relatively easy to handle approximation of the SCRPA equations which is usually called the renormalized RPA. Due to its simplicity for numerical realisation with existing standard RPA-codes, it has been applied in the past quite frequently. We, therefore, will give in Sect. II.D a summary of applications and possible properties and here we will only present the basics.
 The so-called renormalized RPA (r-RPA) is a particular version of 
SCRPA, defined by the factorisation of two-body densities.
\bea
\label{HFfact}
\la  a^{\dag}_ma^{\dag}_na_ja_i\ra \simeq 
\la a^{\dag}_m a_i\ra \la a^{\dag}_na_j \ra - \la a^{\dag}_m a_j\ra \la a^{\dag}_na_i \ra~.
\eea
It was introduced by Hara \cite{Har64}, but it became 
popular after the paper of Catara {\it et al.} \cite{Cat96},
introducing a simple boson mapping method to estimate one-body
densities in terms of RPA amplitudes (the so-called Catara method).

The r-RPA system of equations has practically the same form as the
standard RPA one, but the matrix elements for a Hamiltonian
$H=H_0+ V$ are given by (we suppose that we work in the canonical basis where 
the s.p. density matrices are diagonal)

\bea
&&\sum_{k'_1k'_2}\tilde{\mathcal S}_{k_1k_2k'_1k'_2}N^{-1}_{k'_1k'_2}\chi_{k'_1k'_2}^{\nu}=\nonumber\\ 
&&\sum_{k'_1k'_2}\epsilon_{k_1k_2}\delta_{k_1k'_1}\delta_{k_2k'_2}
+
N_{k_1k_2}\bar v_{k_1k'_2k_2k'_1}\chi_{k'_1k'_2}^{\nu} = \Omega_{\nu}\chi_{k_1k_2}^{\nu}~,\nonumber\\
\label{Cat-RPA}
\eea
with $\epsilon_{k_1k_2}=\epsilon_{k_1}-\epsilon_{k_2}$.
More explicitly in terms of the matrices defined in (\ref{SCRPA-gen}) we can write
\bea
\label{AB}
      {\cal A}_{mi,m'i'}&=&{\oh}\left(
      N^{1/2}_{mi}N^{-1/2}_{m'i'}+N^{1/2}_{m'i'}N^{-1/2}_{mi}
      \right)
\nn&\times&\left(\e_{m'm}\d_{ii'}-\e_{ii'}\d_{mm'}\right)
      +N^{1/2}_{mi}N^{1/2}_{m'i'}
      \la im' |v| mi'\ra
      \nn
         {\cal B}_{mi,m'i'}&=&N^{1/2}_{mi}N^{1/2}_{m'i'}
         \la ii'|v|mm'\ra~,
         \eea
         where
         the single-particle mean-field (MF) energies are given by
         \bea
         \e_{m'm}&=&\la m'|H_0|m\ra+\sum_k n_k\la m'k|v|mk\ra
         \nn
         \e_{ii'}&=&\la i|H_0|i'\ra+\sum_k n_k\la ki|v|ki'\ra~,
         \eea
         and where $N_{mi}$ is the metric matrix
         written in terms of one-body densities $n_m$
         \bea
         \label{metric0}
         N_{mi}=\la 0|\left[A_{im},A_{mi}\right]|0\ra=n_i-n_m~.
         \eea
         The one-body quasiparticle density can be expressed in terms of RPA
         amplitudes up to a fourth order precision, by using the number
         operator
         method \cite{Row68,Cat96}, i.e.
         \bea
         \label{dens1}
         n_p&=&\la 0|n^{\dag}_pn_p|0 \ra
\nn&\approx&
         \sum_{h\nu\nu'}\left(\delta_{\nu\nu'}-
         \oh\sum_{p'h'}N_{p'h'}X^{\nu'}_{p'h'}X^{\nu*}_{p'h'}\right)
         N_{ph}Y^{\nu}_{ph}Y^{\nu'*}_{ph}
         \nn
         n_h&=&\la 0|n^{\dag}_hn_h|0 \ra \approx 1-
\nn&&
         \sum_{p\nu\nu'}\left(\delta_{\nu\nu'}-
         \oh\sum_{p'h'}N_{p'h'}X^{\nu'}_{p'h'}X^{\nu*}_{p'h'}\right)
         N_{ph}Y^{\nu}_{ph}Y^{\nu'*}_{ph}~.
\nn
\eea

It consists of working only with ph configurations like in the standard RPA and in retaining only in a systematic way the single-particle density matrices. The latter are expressed in a simple way by the $Y$-amplitudes of the r-RPA what constitutes a relatively easy to handle self-consistency problem. It is described in several publications and we will skip the details here referring the reader to  examples, where the r-RPA method has been applied, in Sect. III.D. Let us only mention here that the r-RPA amplitudes can sustain all indices as SCRPA besides diagonal configurations. In this case r-RPA keeps all desirable properties of standard RPA intact.

\subsection{The correlation energy and the boson aspect of the Self-Consistent Random Phase Approximation (SCRPA)}

We now come to an important aspect of the SCRPA approach as given in (\ref{SCRPA-gen}). It namely turns out that, like with standard RPA, also SCRPA is equivalent to a bosonisation. This stems from the fact that (\ref{SCRPA-gen}) has exactly the same mathematical structure as standard RPA \cite{Rin80}. Let us sketch shortly how this boson aspect can be made manifest.
Since, as said, the structure of (\ref{SCRPA-gen}) is exactly the same as the one of standard RPA \cite{Rin80}, the former can also be represented by a boson Hamiltonian

\bea
  H_B&=& E_{\rm HF} - \frac{1}{2}\sum_{m>i}{\mathcal A}_{mimi} 
\nn&+& \frac{1}{2}
  \begin{pmatrix}B^{\dag}&B\end{pmatrix}
  \begin{pmatrix}{\mathcal A}&{\mathcal B}\\{\mathcal B}^*&{\mathcal A}^*\end{pmatrix}
    \begin{pmatrix}B\\B^{\dag}\end{pmatrix}~,
    \label{H_B}
\eea
where $B^{\dag},B$ are ideal boson operators. This boson Hamiltonian can be diagonalized with a Bogoliubov transformation

\begin{equation}
  {\mathcal O}^{\dag}_{\nu} = \sum_{m>i}[X^{\nu}_{mi}B^{\dag}_{mi} -
    Y^{\nu}_{mi}B_{mi}]~,
  \label{Bogo}
\end{equation}
what yields

\begin{equation}
  H_{\rm B} = E_{\rm RPA} + \sum_{\nu}\Omega_{\nu}{\mathcal O}^{\dag}_{\nu}{\mathcal O}_{\nu}~,
  \label{H_Bdiag}
\end{equation}
with

\begin{eqnarray}
  E_{\rm RPA} &=& E_{\rm HF} - \sum_{\nu}\Omega_{\nu}\sum_{k>k'}|Y^{\nu}_{kk'}|^2\nonumber\\
  &=&-\frac{1}{2}{\rm Tr}{\mathcal A}+\frac{\hbar}{2}\sum_{\nu > 0}\Omega_{\nu}= -\frac{1}{2}\sum_{\nu > 0}(E^{\rm TDA}_{\nu} - \Omega_{\nu})~,
\nn
  \label{E-RPA}
\end{eqnarray}
where $E^{\rm TDA}_{\nu}$ is the corresponding excitation energy in Tamm-Dancoff approximation \cite{Rin80}.
It is interesting to transform the $X,Y$ amplitudes into position $Q$ and momentum $P$  amplitudes via, see \cite{Rin80}

\bea
Q^{\nu}_{mi} &=& \sqrt{\frac{\hbar}{2M_{\nu}\Omega_{\nu}}}(X - Y^*)^{\nu}_{mi}
\nn
P^{\nu}_{mi} &=& i\hbar\sqrt{\frac{M_{\nu}\Omega_{\nu}}{2\hbar}}(X+Y^*)^{\nu}_{mi}~,
\eea
where $M_{\nu}$ is the mass parameter defined in \cite{Rin80}. With this the correlation energy is written as

\be
E_{\rm RPA}= \sum_{mi\nu}[\frac{{P^{\nu}_{mi}}^2}{2M_{\nu}} + \frac{M_{\nu}}{2}\Omega_{\nu}^2{Q_{mi}^{\nu}}^2]~.
\ee
  For example, in the case of the spurious translational mode where $\Omega_{\nu}=0$ and the $X,Y$ amplitudes diverge, the correlation energy becomes

  \be
  E_{\rm RPA}= -\sum_{mi}\frac{|\langle m|\hat p|i\rangle|^2}{2Am}~,
  \ee
  where $\hat p$ is the momentum operator and $A$ the total number of nucleons. The correlation energy corresponding to the translational mode is thus just the kinetic energy of the whole system.

The corresponding ground state wave function is

\begin{equation}
  |Z) = \exp\left[\sum_{m>i,n>j}z_{kk'll'}B^{\dag}_{mi}B^{\dag}_{nj}\right]|0)~,
    \label{Z0}
\end{equation}
with $B|0)=0$ defining the simple boson vacuum and $O_{\nu}|Z) = 0$ the RPA boson vacuum with $2z_{minj} = [YX^{-1}]_{minj}$.

Details of the derivation of (\ref{H_Bdiag}) can be found in \cite{Rin80}.
The correlation functions in the double commutators of ${\mathcal A}$ and ${\mathcal B}$ matrices can also be evaluated with the bosonisation. Most importantly, one obtains for the occupation numbers as with standard RPA

\be
n_{p_1} = \sum_{k_2<p_1,\nu}|Y^{\nu}_{p_1k_2}|^2~;~~n_{h_2} = 1 -\sum_{k_1>h_2,\nu}|Y^{\nu}_{k_1h_2}|^2~.
\label{RPA-occs}
\ee
We will give the derivation of this formula later in the Sect. IV of the Green's functions.
Actually it is known since long that from boson expansion theory we  immediately find \cite{Rin80}

\begin{equation}
  \langle |c_hc^{\dag}_{h'}\rangle \rightarrow \sum_p(0|{ B}^{\dag}_{\
    ph}{ B}_{ph'}|0) = \sum_{p,\nu}Y^{\nu}_{ph}Y^{\nu*}_{ph'}~,
  \label{dn-h}
\end{equation}
and

\begin{equation}
  \langle |c^{\dag}_pc_{p'}\rangle \rightarrow \sum_h(0|{ B}^{\dag}_{\
    ph}{ B}_{p'h}|0) = \sum_{h,\nu}Y^{\nu}_{ph}Y^{\nu*}_{p'h}~.
  \label{dn-p}
\end{equation}
\\
Also the other correlation functions figuring in the RPA-matrix can be expressed via the bosonisation by the  RPA-amplitudes.

Let us trace back from where the fact that we end up with a boson theory took its origin. It clearly is rooted in the fact that with our operator (2.8) we cannot find a ground state wave function which fulfills the annihilating condition (2.2). If there existed a fermionic ground state wave function which fulfills the annihilating condition, the Pauli principle would {\it not} be violated. We, therefore, will refer to the approximation that we take the annihilating condition as fulfilled,  where it is not, as the {\it boson approximation}. A crucial consequence of this boson approximation is the form, in which the correlation energy (\ref{E-RPA}) is given, which again is unaltered from the standard RPA expression. In all of our applications with SCRPA we will use this expression. It is also important, as already shortly mentioned, to realize that the generalized RPA operator with {\it all possible indices} is necessary to maintain all the appreciated qualities of standard RPA as there are fulfillment of the sum-rule, appearance of the Goldstone mode in case of spontaneously broken symmetries, Ward identities, etc. We will come back to this later in Sect. VII.\\

\subsection{ SCRPA in the particle-particle channel}

It also shall be clear that the SCRPA approach which we sketched above
in the channel of fluctuations of the density operator can, in a very
analogous way, also be developed in the particle pair fluctuation channel, i.e.
in the particle-particle ($pp$) channel, where the $pp$ ladders are 
summed. This leads e.g. to the Feynman-Galitskii T-matrix \cite{Fet71},
as well as to the Thouless criterion for the onset of superfluidity \cite{Tho61}. In this section we will restrict the range of indices to particle states ($p$) and hole states ($h$), despite the fact that a more general domain of indices, analogous to the ph channel, is certainly possible. However, in the pp-channel this is not studied so far and we will refrain from this generalisation.

The starting point is the definition of the so-called two particle
addition operator
\bea
\label{Addition}
A^{\dag}_{\alpha}=
 {\oh}\sum_{p_1p_2}X^{\alpha}_{p_1p_2}a^{\dag}_{p_1}a^{\dag}_{p_2}
-{\oh}\sum_{h_1h_2}Y^{\alpha}_{h_1h_2}a^{\dag}_{h_1}a^{\dag}_{h_2}~,
\nn
\eea
where $p,h$ again refer to the particle and hole states corresponding
to an optimal single-particle basis yet to be defined.
The $X^{\alpha},~Y^{\alpha}$ amplitudes can, as before, be determined
from the extremal condition of the generalised sum rule
\bea
\label{pairsrule}
\Omega_{\alpha}=\frac{\la 0|[A_{\alpha},[H,A^{\dag}_{\alpha}]]|0\ra}
{\la|[A_{\alpha},A^{\dag}_{\alpha}]\ra}~,
\eea
which leads to
\bea
\label{RPApair}
\left(\begin{matrix}{\cal A} & {\cal B} \cr -{\cal B} & -{\cal C}
\end{matrix}\right)
\left(\begin{matrix} X^{\rho} \cr Y^{\rho}\end{matrix}\right)=
\Omega_{\rho}
\left(\begin{matrix} X^{\rho} \cr Y^{\rho}\end{matrix}\right)~,
\eea
with
\bea
\label{ABC}
{\cal A}_{p_1p_2p'_1p'_2}&=&\la 0|[\delta P_{p_1p_2},[H,
\delta P^{\dag}_{p'_1p'_2}]]|0\ra
\nn
{\cal B}_{p_1p_2h_1h_2}&=&\la 0|[\delta P_{p_1p_2},[H,
\delta P^{\dag}_{h_1h_2}]]|0\ra
\nn
{\cal C}_{h_1h_2h'_1h'_2}&=&\la 0|[\delta P_{h_1h_2},[H,
\delta P^{\dag}_{h'_1h'_2}]]|0\ra~,
\eea
and
\bea
\label{dP}
\delta P^{\dag}_{p_1p_2}&=&\frac{a^{\dag}_{p_1}a^{\dag}_{p_2}}
{\sqrt{1-n_{p_1}-n_{p_2}}}
\nn
\delta P^{\dag}_{h_1h_2}&=&\frac{a^{\dag}_{h_1}a^{\dag}_{h_2}}
{\sqrt{|1-n_{h_1}-n_{h_2}|}}~.
\eea
As one verifies, the eigenvalues correspond to those, where one adds or removes
two particles from the original ground state $|0\ra$ with $N$ particles.
We again have to assume that the ground state is the vacuum to the
addition operators, i.e. $A_{\rho}|0\rangle =0$ (however, an exact annihilating condition can again be found with an extended RPA operator as in Sect. II.C). Also the $X^{\rho},~$Y$^{\rho}$
amplitudes have the orthonormality and completeness relations of standard
pp-RPA, as described in textbooks \cite{Rin80}, so we do not repeat them here.
Quite analogously we can define the removal operators
\bea
\label{Removal}
R^{\dag}_{\rho}=
 {\oh}\sum_{h_1h_2}X^{\rho}_{h_1h_2}a_{h_2}a_{h_1}
-{\oh}\sum_{p_1p_2}Y^{\rho}_{p_1p_2}a_{p_2}a_{p_1}~.
\nn
\eea
Again amplitudes can be determined from the stationarity of the corresponding sum rule. 
The resulting RPA equations have a similar structure as in Eqs. (\ref{RPApair})
and (\ref{ABC}).
Actually, the content of RPA equations for removal is the same as the
one for addition. Only the amplitudes $X^{\alpha},~Y^{\alpha}$ and
$X^{\rho},~Y^{\rho}$ have subtle relations involving interchange of
$p\leftrightarrow h$ indices and relative phases. There exist quite extended applications to the pairing Hamiltonian of this self-consistent particle-particle RPA (SCppRPA), where things are explained in detail and which we shortly will review in the Application Sect. III.D.\\

In analogy to the particle-hole case, there also exists an exact annihilator of the CCD wave function $|Z\rangle$ in the particle-particle case. We write the $Z$-operator (\ref{Z-op}) in a different but equivalent form

\begin{equation}
\hat Z = \frac{1}{4}\sum_{p_1p_2h_1h_2}z_{p_1p_2h_1h_2}P^{\dag}_{p_1p_2}P_{h_1h_2}
\label{ppZ}
\end{equation}
with the pair operators $P^{\dag}_{k_1k_2} = a^{\dag}_{k_1}a^{\dag}_{k_2}$.
The annihilation operator then writes

\begin{eqnarray}
  A_{\alpha} &=& \frac{1}{2}\sum_{p_1p_2}X^{\alpha}_{p_1p_2}P_{p_2p_1}-\frac{1}{2}\sum_{h_1h_2}Y^{\alpha}_{h_1h_2}P_{h_1h_2}\nonumber\\
  &+& \frac{1}{2}\sum_{p_1p_2h_1h_2}\eta^{\alpha}_{p_1p_2h_1h_2}S_{p_1p_2}P_{h_1h_2}
\end{eqnarray}
with $S_{p_1p_2} = a^{\dag}_{p_1}a_{p_2}$.

The relations between the various amplitudes are

\begin{eqnarray}
  &&\frac{1}{2}\sum_{p_1p_2}X^{\alpha}_{p_1p_2}z_{p_1p_2h_1h_2}=Y^{\alpha}_{h_1h_2}\nonumber\\
  &&\sum_{p_1}X^{\alpha}_{p_2p_1}z_{p_1p_3h_1h_2}= \eta^{\alpha}_{p_3p_2h_1h_2}
\end{eqnarray}

Similar to the SCRPA correlation energy, an analogous expression can be derived for the pp-case

\bea
E_{\rm corr}^{\rm RPA}&=&  - \frac{1}{2}\sum_{\alpha}\Omega_{\alpha}\sum_{hh'}|Y^{\alpha}_{hh'}|^2 - \frac{1}{2}\sum_{\rho}\Omega_{\rho}\sum_{pp'}|Y^{\rho}_{pp'}|^2\nonumber\\
&=& \sum_{\alpha}\Omega_{\alpha} - {\rm Tr} {\mathcal A}= \sum_{\rho}\Omega_{\rho} + {\rm Tr}{\mathcal C}
\nonumber\\
&=&\frac{1}{2}[ \sum_{\alpha}\Omega_{\alpha} - {\rm Tr} {\mathcal A} ]-\frac{1}{2}[ \sum_{\rho}\Omega_{\rho} + {\rm Tr}{\mathcal C} ]~.
\label{gs-pp-energ}
\eea
An application of these equations to the pairing model will be given in Sect. III.

In conclusion of the Sects. II.A-II.D, we explained in some detail how 
two body correlations can be calculated from the establishment of generalised 
self-consistent RPA equations, which can be paraphrased as resulting from a 
Bogoliubov approach for fermion pair operators. As the standard RPA equations, 
the self-consistent ones are of the Schr\"odinger type, they, therefore, 
may be numerically tractable. We should, however, point out that, in spite of 
the analogy with Bogoliubov theory for ideal bosons, the present approach for 
fermion pairs is not based on an explicit many-body ground state wave function 
and, therefore, is not a truly Raleigh-Ritz variational principle. Also the 
Pauli principle, though certainly much better treated than in standard RPA, is 
not rigorously satisfied. This also stems from the fact that, in order to make 
the SCRPA equations fully self-contained, some approximations had to be 
introduced, which, for example, for the occupation numbers involve an expansion 
in powers of the RPA amplitudes. We will below present some applications to 
model cases, where we will show the progress, which has been achieved with 
respect to standard RPA. An important aspect of SCRPA also is that 
conservation laws and Goldstone theorem in case of broken symmetries can be
conserved, as this is the case with standard RPA. With Schr\"odinger type of 
extensions of RPA theory, this is not at all evident. Also the Raleigh-Ritz 
variational aspect can still be improved as we will show in Sect. VII, that 
with an extended RPA operator one can solve (\ref{anih1}) and give the 
corresponding ground state wave function explicitly in full generality 
for interacting Fermi systems.\\

\subsection{Self-consistent Quasiparticle RPA (SCQRPA)}

It is relatively evident how to generalise SCRPA to the superfluid case, where we want to call it self-consistent quasiparticle RPA (SCQRPA). We pose the same RPA operator as for standard QRPA

\be
Q^{\dag}_{\nu} = \sum_{k>k'}[X^{\nu}_{kk'}\alpha^{\dag}_k\alpha^{\dag}_{k'}
  - Y^{\nu}_{kk'}\alpha_{k'}\alpha_k]~,
\label{qp-RPA}
\ee
where $\alpha^{\dag}$ and $\alpha$ are the usual quasiparticle (q.p.) creation and destruction operators \cite{Rin80}. Formally the SCQRPA equations also are obtained from a minimisation of the energy weighted sum-rule (\ref{sumrule}) with, however, the Hamiltonian written with quasiparticles. The self-consistency for the $X,Y$ amplitudes can be established as in the non-superfluid case. A point of discussion can be whether one should include to the RPA operator the scattering states. This can be done in adding a $\alpha^{\dag}_k\alpha_{k'}$ term to the operator. The problem of the ground state wave function also can be solved with a further extension. Let us consider the following Coupled Cluster Doubles state

\be
|Z\rangle = \exp\left[\frac{1}{4!}\sum_{k_1k_2k_3k_4}z_{k_1k_2k_3k_4}\beta^{\dag}_{k_1}\beta^{\dag}_{k_2}\beta^{\dag}_{k_3}\beta^{\dag}_{k_4}\right]|\rm BCS\rangle~.
\label{Z-bcs}
\ee
The corresponding exact annihilator can be given as follows

\bea
Q_{\nu}  &=& \sum_{k>k'}[X^{\nu}_{kk'}\beta^{\dag}_k\beta^{\dag}_{k'}
- Y^{\nu}_{kk'}\beta_{k'}\beta_k] 
\nn&+&\sum_{k_1<k_2<k_3} \eta^{\nu}_{k_1k_2k_3k_4}
\beta^{\dag}_{k_1}\beta^{\dag}_{k_2}\beta^{\dag}_{k_3}\beta_{k_4}~,
\label{bcs-killer}
\eea
with $X,Y$ antisymmetric in $k,k'$ and $\eta$ antisymmetric in first three indices. Applying this operator on our CCD state, we find $Q_{\nu}|Z\rangle = 0$,
where the relations between the various amplitudes turn out to be

\bea
Y^{\nu}_{ll'} &=& \sum_{k<k'}X^{\nu}_{kk'}z_{kk'll'} 
\nn
\eta^{\nu}_{l_2l_3l_4k'} &=& \sum_kX^{\nu}_{kk'}z_{kl_2l_3l_4}~.
\label{z....}
\eea
As before, to work with the extended operator is not much studied and remains, in general, a task for the future. However, some indications of how to tackle this problem at least approximately will be given in Sect. VII.

SCQRPA has only been applied to a very simple two-level pairing model \cite{Rab02}:
\be
H= \frac{\epsilon}{2}\sum_j j N_j - g\Omega\sum_{jj'}A^{\dag}_jA_{j'}~,~~j=\pm 1~,
\label{H-pair}
\ee
where $\Omega$ is the degeneracy of upper and lower levels and $\epsilon$ is the level spacing. The operators $ N_j, A^{\dag}_j,A_j$ are the usual ones of the pairing Hamiltonian given below in the Application section III.D.
The results are quite encouraging, but more realistic applications have to wait.
An instructive result may be how the gap equation becomes renormalized

\be
\Delta_i =\sum_j\tilde g_{ij}\frac{\Delta_j}{2\sqrt{\xi_j^2 + \Delta_j^2}}~,
\label{renorm-D}
\ee
where $\tilde g_{ij}$ is the renormalized pairing force containing $X$ and $Y$ amplitudes, what is also the case for the single-particle energies $\xi_i$. For the detailed expressions the reader may look up the original paper \cite{Rab02}.
This gap equation with effective constants is equivalent to the EOM which determines the mean-field: $\langle [H,\alpha_i\alpha_j]\rangle =0$, what is the analogue to the generalized mean-field equation (\ref{gen-mf}).

\subsection{Number conserving ph-RPA (NCphRPA) in superfluid nuclei}

Quasiparticle RPA has, of course, the drawback that it violates particle
number conservation. Particle-number projection at the RPA level has been earlier proposed 
in Refs. \cite{Federschmidt1985, Kyotoku1990}. 
The procedure requires the projection of two-quasiparticle states and a subsequent reorthogonalization, 
mixing particle-hole excitations in the A system with particle-particle in the A-2 system and hole-hole in the A+2 system. For these reasons, it has been seldom used in $\beta$ and double-$\beta$ decay calculations 
\cite{Civitarese1991, Suhonen1993}. It is, thus, very interesting that one can build a ph-RPA on a number projected HFB ground state where the latter is given in the canonical basis by

\begin{equation}
|PHFB\rangle =\Gamma ^{\dag}|\mathrm{vac}\rangle ~,~~
\Gamma^{\dag}=\sum_{i=1}^{L}z_{i}a_{i}^{\dag }a_{\bar{\imath}}^{\dag }~,
\label{PHFB}
\end{equation}%
where the $\bar{\imath}$ are the conjugate orbitals to the s.p. states $i$.
For axially deformed systems, $J_{z}$ is conserved and $\bar{\imath}$ has
the opposite spin to $i$. The pair condensate (\ref{PHFB}) is the vacuum of
a complete set of annihilator operators \cite{Duk19}. The subset of annihilators that conserves spin is

\begin{equation}
C_{ij}=z_{i}a_{i}^{\dag }a_{j}-z_{j}a_{\bar{j}}^{\dag }a_{\bar{\imath}%
}~,i\neq j  ~.
\label{phfb-killer}
\end{equation}%
Since $[C_{ij},\Gamma ^{\dag }]=0$, it follows immediately

\begin{equation}
C_{ij}|PHFB\rangle =0  \label{pbhf-kill}~.
\end{equation}%
We are now exactly in an analogous situation to the HF-RPA approach, where
the $a_{h}^{\dag }a_{p}$ hp-operators annihilate the HF ground state. Therefore,
we now will build a ph-RPA approach, which has $|PHFB\rangle $ as a reference
state

\begin{equation}
Q_{\nu }^{\dag }=\sum_{i>j}X_{ij}^{\nu }C_{ij}^{\dag }-Y_{ij}^{\nu }C_{ij}~.
\label{Q-C}
\end{equation}

Notice that for $z_{i}=\Theta \left( M-i\right) $, the pair condensate
reduces to a HF Slater determinant and (\ref{Q-C}) is the standard ph-RPA
operator. In the general case of a superfluid pair condensate, this
definition of the RPA operators allows us to launch the usual EOM machinery
and establish the RPA equations with PHFB as the reference state. Of course,
it is clear that no particle number violation has occurred. The price to pay
is that we must have a correlated PHFB state as input. However, particle
number projection is relatively easy and is now performed mostly routinely.
Similarly, there are powerful techniques to evaluate the expectation values
of the two-body operators in the ${\mathcal A}$ and ${\mathcal B}$ matrices. The complete NCphRPA formalism developed in \cite{Duk19} is an adaptation to nuclear physics of the generalized RPA theory proposed in quantum chemistry \cite{Sangfelt1987}. The lack of superconducting correlations made the theory inefficient in quantum chemistry, though it could find a fertile area for applications in open shell nuclei.

\begin{figure}[tbp]
\centering
\includegraphics[width=7cm,height=6.5cm]{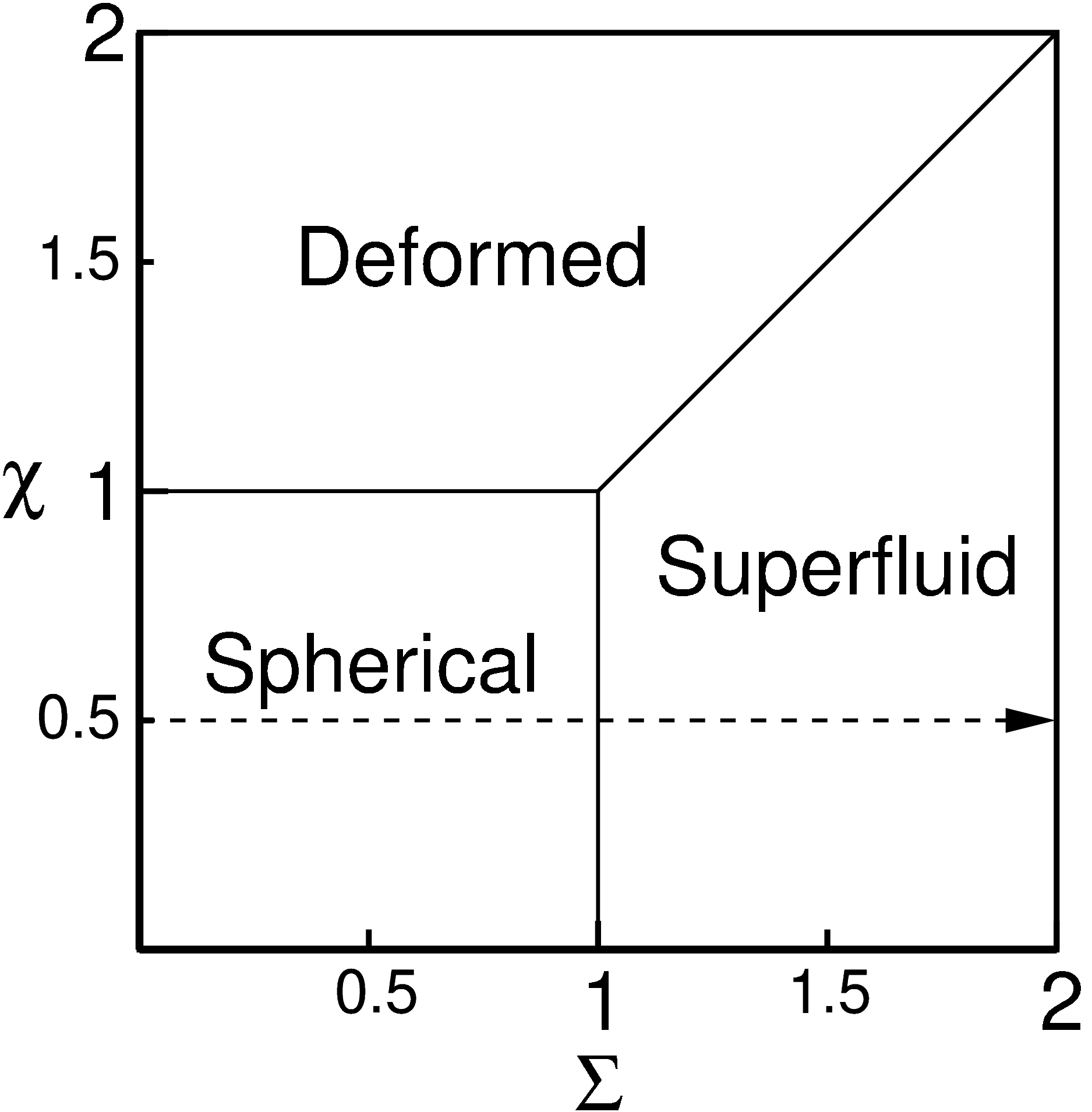}
\caption{Phase diagram of the Agassi Hamiltonian (\ref{Agassi}). The dotted line with $\chi=1/2$ will be used to benchmark the NCphRPA.}
\label{fig:phasediagram}
\end{figure}

The Agassi model \cite{Agassi1968} was chosen for a pilot application of the NCphRPA theory, since it is the simplest model that mixes particle-hole and pairing correlations. The Agassi Hamiltonian combines the Lipkin model with the two-level pairing model

\begin{equation}
H=J_{0}-\frac{\Sigma }{2j-1}\sum_{\sigma \sigma ^{\prime }}A_{\sigma }^{\dag
}A_{\sigma ^{\prime }}-\frac{\chi }{2(2j-1)}[J_{+}^{2}+J_{-}^{2}]~,
\label{Agassi}
\end{equation}%
where $\sigma =\pm 1$ labels each of the two single-particle levels and $%
\Sigma $ and $\chi $ are the coupling constants in the pairing, respectively
ph-channels. The pair creation operators are

\begin{eqnarray}
A_{\sigma }^{\dag } &=&\sum_{m=1}^{j}a_{\sigma ,m}^{\dag }a_{\sigma
,-m}^{\dag }  \nonumber \\
A_{0}^{\dag } &=&\sum_{m=1}^{j}\bigg (a_{-1,m}^{\dag }a_{1,-m}^{\dag
}-a_{-1,-m}^{\dag }a_{1,m}^{\dag }\bigg )  ~,
\label{Ag-pairs}
\end{eqnarray}%
and the ph operators are

\begin{eqnarray}
J_{+} &=&\sum_{m=-j}^{j}a_{1m}^{\dag }a_{-1m}=(J_{-})^{\dag }  \nonumber \\
J_{0} &=&\frac{1}{2}\sum_{m,=-j}^{j}\bigg (a_{1m}^{\dag
}a_{1m}-a_{-1m}^{\dag }a_{-1m}\bigg )  ~.
\label{L-ops}
\end{eqnarray}%

\begin{figure}[tbp]
\centering \includegraphics[width=7cm,height=6cm]{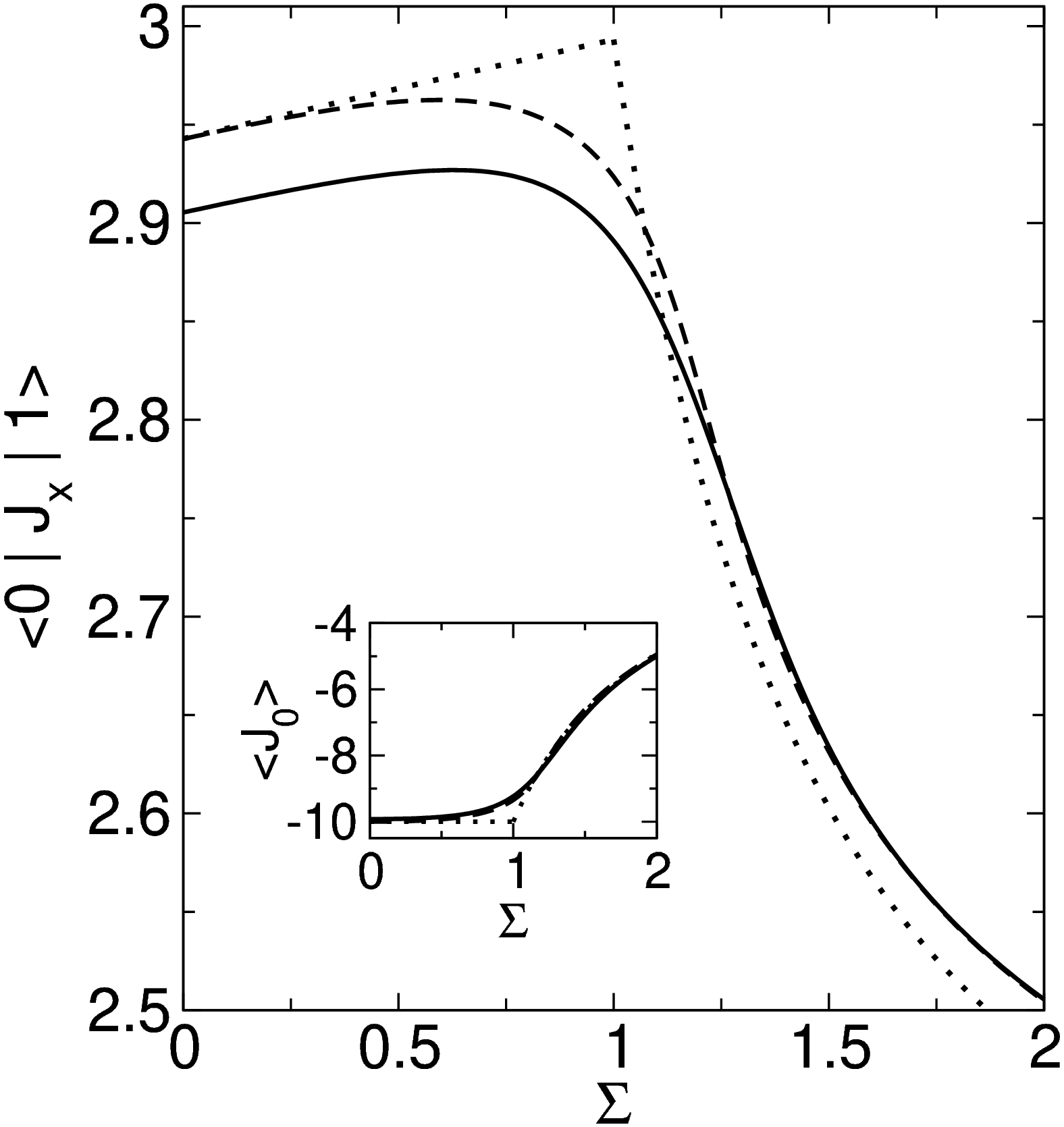}
\caption{Transition matrix element of $J_{x}$ between the
excited state and the GS for level degeneracy $j=10$ and $\chi = 1/2$ as a function of $\Sigma$. The inset shows the ground state expectation value of the operator $J_0$. The solid line depicts the exact rsults, the dashed line NCphRPA, and the dotted line (Q)RPA. }
\label{fig:Transition}
\end{figure}

The Agassi model has a rich phase diagram that has been studied in \cite{Davis1986, Ramos2018} within the HFB approximation. Fig. \ref{fig:phasediagram} shows the phase diagram at half filling. It displays a normal (spherical) phase for $\chi <1$ and $\Sigma <1$,
a ph parity broken (Deformed) phase for $\chi >1$ and $\chi
>\Sigma $,
and a superconducting (Superfluid) phase for $\Sigma >1$ and $\Sigma >\chi $. The horizontal dotted line at $\chi =1/2$ represents an ideal path to test the NCphRPA since it has important ph correlations and a phase transition from normal to superconducting. As expected, in NCphRPA the collective ph-RPA excitation shows a smooth behavior across the transition, as opposed to (Q)RPA with the usual kink at the transition point (see \cite{Duk19}). The differences between both approaches can be more readily seen in the transition probabilities that are more sensitive to the wave functions. Fig. \ref{fig:Transition} shows the transition matrix element of the $J_x$ operator between the first excited state and the ground state for a finite system with $j=10$; the inset shows the expectation value of the $J_0$ operator in the ground state. In both cases the NCphRPA improves over (Q)RPA  overcoming the abrupt change at the phase transition of the (Q)RPA for $\Sigma=1/2$. 
The theory could be extended to describe large amplitude collective motion within a particle-number projected adiabatic time-dependent HFB theory for nuclear fission studies \cite{Bender2020}. However, since the theory is very recent, no other applications, e.g., for realistic systems exist so far.

\subsection{Odd-particle number random phase approximation}

The EOM can also be applied to obtain RPA-type of equations for systems with an odd number of particles \cite{Toh13}. We again consider the CCD state of (\ref{Z}).
We study the following two quasiparticle operators which can be classified, as for the ppRPA, as addition and removal operators

\bea
q^{\dag}_{\alpha}&=& \sum_p u_p^{\alpha}c^{\dag}_p - \frac{1}{2}\sum_{hh'p}V^{\alpha}_{hh'p}c^{\dag}_hc^{\dag}_{h'}c_p\nonumber\\
q^{\dag}_{\rho}&=& \sum_h u_h^{\rho}c_h -\frac{1}{2}\sum_{pp'h}V^{\rho}_{pp'h}c^{\dag}_hc_pc_{p'}~.
\label{Z-qp}
\eea
It can easily be shown that the corresponding destruction operators
$q_{\alpha}$  and $q_{\rho}$ annihilate the $|Z\rangle$ state of (\ref{Z}) under the conditions

\bea
\sum_p u^{\alpha *}_pz_{pp'hh'} &=& V^{\alpha *}_{hh'p'}\nonumber\\
\sum_h u^{\rho *}_h z_{pp'hh'}  &=& V^{\rho *}_{pp'h'}~.
\eea
With our usual EOM technique one obtains the following secular equation for the amplitudes of the, e.g., $q_{\alpha}$ mode

\begin{equation}
  \begin{pmatrix}{\mathcal  H}_{00}&{\mathcal H}_{01}\\{\mathcal H}_{10}\
    &{\mathcal H}_{11}\end{pmatrix}
  \begin{pmatrix}u\\V \end{pmatrix}
  =\lambda \begin{pmatrix}n_{00}&n_{01}\\n_{10}&n_{11}\end{pmatrix}
  \begin{pmatrix}u\\V \end{pmatrix}~,
  \label{odd-eigen}
\end{equation}
with
\bea
{\mathcal H}_{00}&=& \langle \{c_p,[H,c^{\dag}_{p_1}]\}\rangle\nonumber\\
{\mathcal H}_{01}&=&{\mathcal H}_{10} = \langle \{c_p,[H,c^{\dag}_{h_1}c^{\dag}_{h'_1}c_{p_1}]\}\rangle \nonumber\\
{\mathcal H}_{11}&=&\langle \{c^{\dag}c_{h'}c_h,[H,c^{\dag}_{h_1}c^{\dag}_{h'_1}c_{p_1}]\}\rangle~,
\label{Hij}
\eea
and

\bea
n_{00} &=&1\nonumber\\
n_{01}&=&n_{10} = \langle \{c_p,c^{\dag}_{h_1}c^{\dag}_{h'_1}c_{p_1}\}\nonumber\\
n_{11}&=&\langle \{c^{\dag}c_{h'}c_h,c^{\dag}_{h_1}c^{\dag}_{h'_1}c_{p_1}\}\rangle~,
\eea
where $\{..,..\}$ is the anticommutator
and analogous equations hold for the $q_{\rho}$ mode.
How this goes in detail, we can see in Ref. \cite{Toh13} from where the matrix elements in (\ref{Hij}) can be deduced, see also \cite{Jem19},
and in the single-particle Green's function section V, since it is evident that this scheme has a direct relation with the s.p. Dyson equation and a specific form of the self-energy.  Below, in Sect. V.C we will present an application to the Lipkin model. It is worth mentioning that if Eqs.(\ref{odd-eigen}) are solved in the full space, they show the appreciable property to fulfill the Luttinger theorem for the s.p. occupation numbers \cite{Urb14}.

\section{Applications of SCRPA}
\setcounter{equation}{0}
\renewcommand{\theequation}{3.\arabic{equation}}

In this section, we will show on concrete examples how to go beyond standard 
RPA in taking into account the fact that the whole medium is correlated and not 
only the two fermions are explicitly under consideration. 
This extension is the SCRPA introduced in 
Sect. II. We first will present the pairing model which has been treated with 
high dimensional configurations. As a second example, we will consider the three-level Lipkin 
model which has the interesting feature of a spontaneously broken symmetry. 
Therefore, the question of the appearance of a Goldstone mode, important 
for the fulfillment of conservation laws, can be studied. 
As a third model, the 1D Hubbard model with a finite number of sites is 
presented. Various applications of the r-RPA will also be discussed at the end of this section.

\subsection{Picket Fence (Pairing) Model}
\label{sec:pair}

As a first example we treat the picket fence model \cite{Duk99}.
It is defined as the standard pairing Hamiltonian, specialised, 
however, to equidistant levels and each level can accommodate only one pair, 
let us say spin up/down. This model was exactly
solved by Richardson many years back \cite{Ric63}. 

The picket fence (PF) Hamiltonian is given by
\bea
\label{HPF}
H=\sum_{i=1}^{\Omega}(\epsilon_i-\lambda)N_i-G\sum_{i,j=1}^{\Omega}
P^{\dag}_iP_j~,
\eea
with
\bea
N_i=c^{\dag}_ic_i+c^{\dag}_{-i}c_{-i}~,~~~
P^{\dag}_{i}=c^{\dag}_ic^{\dag}_{-i}~,
\eea
where $c^{\dag}_i$ creates a fermion particle in the i-th level
with spin projection $m={\oh}$ and $c^{\dag}_{-i}$ with $m=-{\oh}$.
$\Omega$ is the total number of levels, $G$ is the pairing interaction
strength and the single-particle levels are equally spaced, i.e.
$\epsilon_i=i\epsilon$. The chemical potential $\lambda$ will be defined
such that the system is completely symmetric with respect to particles and 
holes.

{\it Application of EOM to the pairing model.}
First we will show how to treat the system by the Equation of Motion Method.
We will assume that the system is half filled with number of pairs
$N=\Omega/2$. The particle and hole states are defined by
\bea
N_h|HF\ra=2~,~~~N_p|HF\ra=0~,
\eea
where $|HF\ra$ simply stands for the uncorrelated Slater determinant
with $2N$ particles. The particle states $p$ correspond to $\epsilon_p>\lambda$
and the hole states $h$ to $\epsilon_h<\lambda$.

\begin{figure}[ht]
\begin{center}
\includegraphics[width=7cm]{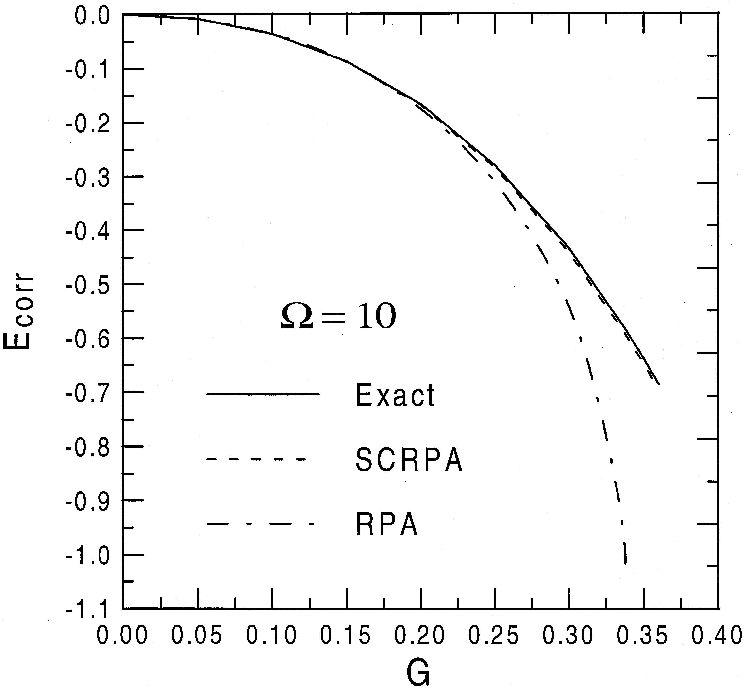}
\caption{Ground state correlation energies of the system with
$\Omega=10$ as a function of the pairing strength $G$.}
\label{Duk01}
\end{center}
\end{figure}
In this case, with no single-particle occupations allowed, 
the following relation is fulfilled
\bea
P^{\dag}_iP_i+P_iP^{\dag}_i=1~,
\eea
which implies
\bea
N_i=2P^{\dag}_iP_i~.
\eea
We can write (\ref{HPF}) in a more convenient $ph$ symmetric way defining
operators as
$P^{\dag}_h=-Q_h,~P^{\dag}_p=Q^{\dag}_p,~
N_h=2-M_h,~N_p=M_p$.
Using for the chemical potential
\bea
\mu=\epsilon(N+{\oh})-\frac{G}{2}~,
\eea
we arrive at a redefinition of the Hamiltonian (\ref{HPF}) in the
following way
\bea
\label{HPF1}
H&=&-\epsilon N^2+\sum_{p=h=1}^N
\left[\epsilon\left(p-{\oh}\right)+\frac{G}{2}\right]\left(M_p+M_h\right)
\nn
&-&G\sum_{pp'}Q^{\dag}_pQ_{p'}-G\sum_{h'}Q^{\dag}_hQ_{h'}
\nn&-&
G\sum_{ph}\left(Q^{\dag}_pQ^{\dag}_h+Q_hQ_p\right)~.
\eea
In this form the complete symmetry between particle and hole states becomes
evident \cite{Duk99}.

\begin{figure}[ht]
\begin{center}
\includegraphics[width=7cm]{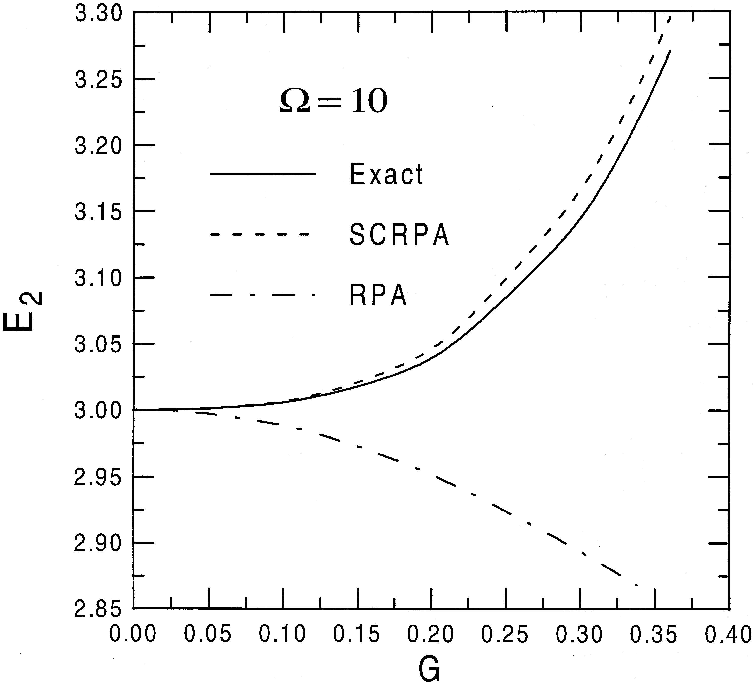}
\vspace{1cm} 
\caption{Second excited state energy of the system with $\Omega=10$
and $N=12$ particles relative to the ground state of the
system with $\Omega=N=10$.}
\label{Duk04}
\end{center}
\end{figure}
Following the definitions given in Eqs. (\ref{Addition}) and 
(\ref{Removal}), let 
us now write out the RPA addition and removal operators corresponding
to this model,
\bea
\label{Addition1}
A^{\dag}_{\rho}=
\sum_p X^{\rho}_p\ov{Q}^{\dag}_p-
\sum_h Y^{\rho}_p\ov{Q}_h~,
\eea
being the addition operator and
\bea
\label{Removal1}
R^{\dag}_{\alpha}=
\sum_h X^{\alpha}_p\ov{Q}^{\dag}_h-
\sum_p Y^{\alpha}_p\ov{Q}_p~,
\eea
the removal operator where
$\ov{Q}_p=Q_p/\sqrt{1-\la M_p\ra}$ and
$\ov{Q}_h=Q_h/\sqrt{1-\la M_h\ra}$.
The matrix elements ${\cal A,B,C}$ of (\ref{ABC}) can fully be expressed by 
the RPA amplitudes with the help of the techniques outlined in Sect. II. 
Since they  are given by Eqs. (20)-(35) in \cite{Hir02}, we will not 
repeat this here.

In order to fully close the set of SCRPA equations, we still must express
the correlation functions $\la M_iM_j\ra$ through the RPA amplitudes, which is 
the usual somewhat difficult point with SCRPA. In this model,
this can also be done exactly, though it is relatively involved.
It is explained in Ref. \cite{Hir02}. Here, we will confine ourselves in a first 
application with the often used approximation
$\la M_iM_j\ra\approx \la M_i\ra\la M_j\ra$, which in this model is
very good. We show the results in Fig. \ref{Duk01} for the ground state correlation
energy and in Fig. \ref{Duk04} for the second excited state \cite{Duk99}. 
In these figures we see the dramatic improvement of SCRPA over standard RPA. 
Indeed, standard RPA shows the usual collapse of the first excited state at the 
critical value of the coupling strength. On the contrary, in this case of ten 
levels the first and second, see \cite{Duk99}, excited states of SCRPA show, in agreement with the exact 
solution, an upward trend signaling that the 
original attractive force has been overscreened and converted into a repulsive one. This is a 
very strong feature of the present solution showing that the screening of the 
force (here actually over-screening) is very well taken into account in SCRPA. The physical 
origin of the repulsion stems from the very strong action of the Pauli 
principle in this model, since each level can only be occupied by zero or two particles.

\begin{center}
\begin{table}
\caption{Excitation energy of the first addition mode as a function of G obtained with
exact calculation, with the RPA, $SCRPA_1$ , and SCRPA methods, for $\Omega=10$, see \cite{Hir02}.}
\vskip5mm
\begin{tabular}{|c|c|c|c|c|}
\hline
~~~G~~~ & ~~~Exact~~~ & ~~~RPA~~~ & ~~~${\rm SCRPA}_1$~~~ & ~~~SCRPA~~~ \cr
\hline
.00 & 1.0000 & 1.0000 & 1.0000 & 1.0000 \cr 
.05 & 1.0003 & 0.9940 & 1.0005 & 1.0003 \cr
.10 & 1.0011 & 0.9732 & 1.0034 & 1.0014 \cr
.20 & 1.0053 & 0.8604 & 1.0279 & 1.0119 \cr
.30 & 1.0143 & 0.5257 & 1.0970 & 1.0539 \cr
.33 & 1.0184 & 0.2574 & 1.1266 & 1.0758 \cr
.34 & 1.0199 &   ***  & 1.1372 & 1.0840 \cr
.35 & 1.0216 &   ***  & 1.1481 & 1.0927 \cr
.36 & 1.0233 &   ***  & 1.1592 & 1.1018 \cr
\hline
\end{tabular}
\end{table}
\end{center}

In Table I we show the quality of the various approximations. SCRPA$_1$ means that the above mentioned factorization approximation of \cite{Duk99} is applied, while SCRPA stands for the full SCRPA solution without approximation of \cite{Hir02}. 
One point to be mentioned here is the following, see  Ref. \cite{Hir02}. 
Since in this model the SCRPA could be pulled through 
without any approximations, it shows the possibility to study the fulfillment 
of the Pauli principle.  In Ref. \cite{Hir02} it was shown in studying certain 
two-body correlation function that the Pauli principle is slightly violated 
(remember that the Pauli principle acts very strongly in this model). This 
feature stems from the fact that the annihilating condition (\ref{anih1}) has no 
solution for the ground state with the present RPA operators (\ref{Addition1}) 
and (\ref{Removal1}) and its use, therefore, implies an approximation in
the SCRPA  scheme. In this model we, however, see that the deviation from 
the solution of (\ref{anih1}) for the ground state is very 
mild. 
Since, as mentioned, the Pauli principle plays a crucial role here, 
we can surmise that the  features of SCRPA found in this model can be 
transposed also to more general cases.

\begin{figure}[ht]
\begin{center}
\includegraphics[width=8cm]{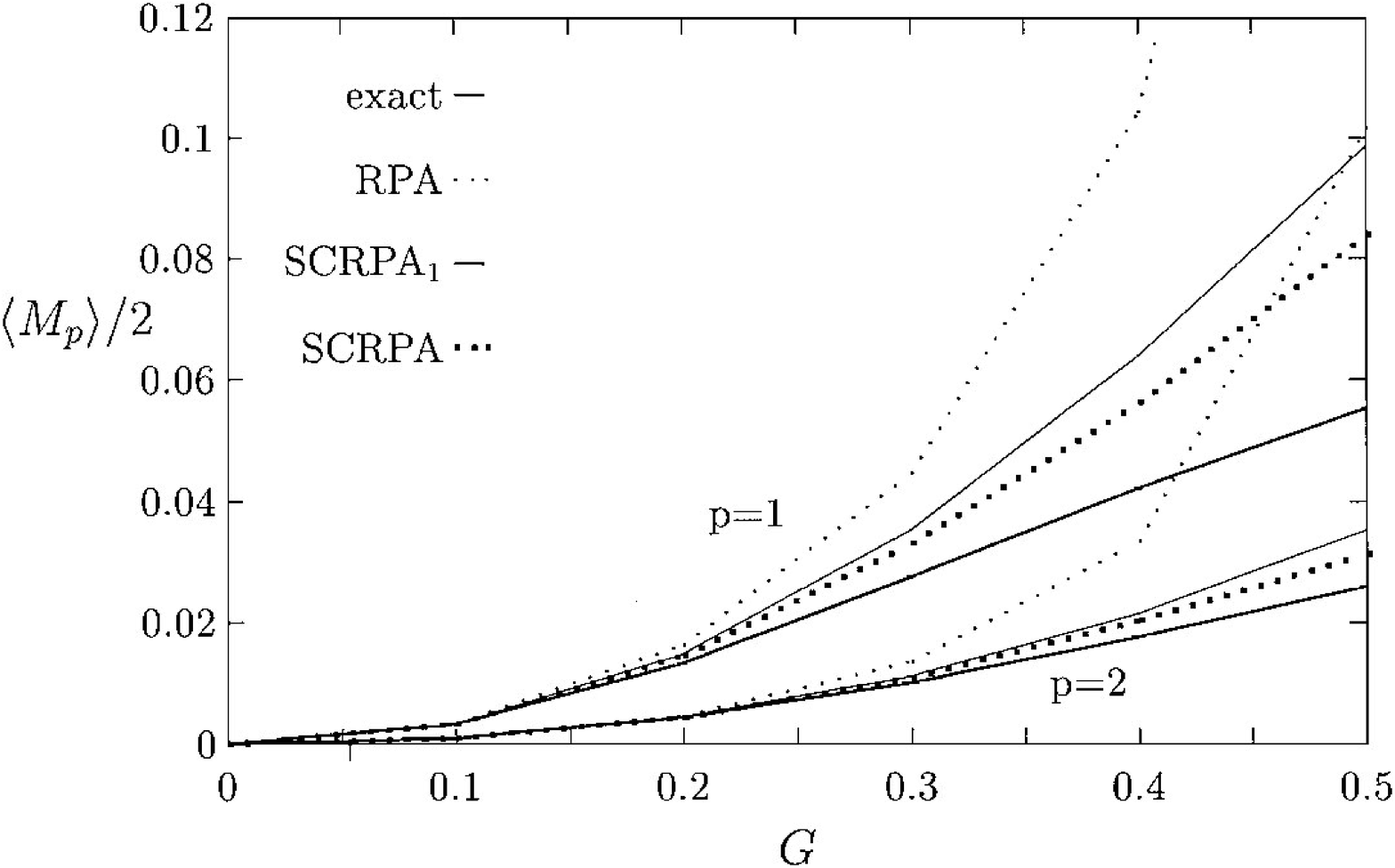}
\caption{Occupation numbers $\la M_p\ra/2$ for the first two particle states $p=1, 2$ as a function of $G$ for $\W=4$. Full thin line: exact result; dotted line: standard ppRPA; thick full line: SCRPA1 (meaning with the factorization approximation); thick dots: full SCRPA. }
\label{Hir01}
\end{center}
\end{figure}

It is also interesting to show some results concerning the
occupation numbers like in Fig. \ref{Hir01} \cite{Hir02}. Again, one notices a very strong improvement over standard RPA results.


\subsection{Three-level Lipkin Model}
\label{sec:numeric}
\setcounter{equation}{0}

We have chosen as a next numerical application the three-level
Lipkin model, corresponding to an SU(3) algebra \cite{Del05}.
This model has been widely used in order to test different
many-body approximations \cite{Li70,Mes71,Sam99,Gra02,Hag00}.
In analyzing this model we have used a particular form of the Hamiltonian,
namely
\bea
\label{Hamilt}
H=\sum_{\a=0}^2\epsilon_{\a}K_{\a\a}
-\frac{V}{2}\sum_{\a=1}^2(K_{\a 0}K_{\a 0}+K_{0\a}K_{0\a})
~,
\eea
with

\bea
\label{Kab}
K_{\a\b}=\sum_{\mu=1}^Nc^{\dag}_{\a\mu}c_{\b\mu}~,
\eea
which are the generators of the SU(3) algebra.
\noindent
By $\epsilon_{\a}$ we denoted the single-particle energies.
According to Ref. \cite{Hag00}, for the three-level Lipkin model
the HF transformation matrix defined by
\bea
\label{ac}
a^{\dag}_{k\mu}=\sum_{\a=0}^2C_{k\a}c^{\dag}_{\a\mu}~,
\eea
can be written as a product of two rotations, in terms of two angles
$(\phi,\psi)$.
The expectation value of the Hamiltonian (\ref{Hamilt}) on the
HF vacuum has a very simple form
\bea
\label{Hmed}
\la HF|H|HF\ra &=& N\epsilon 
[e_0 {\rm cos}^2\phi + e_1 {\rm sin}^2\phi {\rm cos}^2\psi 
\nn&+&
 e_2 {\rm sin}^2\phi {\rm sin}^2\psi
-\chi {\rm sin}^2\phi
 {\rm cos}^2\phi
 ]~,
\eea
where we introduced the following dimensionless notations
\bea
e_k=\frac{\epsilon_k}{\epsilon},~~~\chi=\frac{V(N-1)}{\epsilon}~.
\eea

The Hamiltonian (\ref{Hamilt}) has 
two kinds of HF minima, namely a 'spherical' minimum and a 'deformed' one
\bea
\label{HFstat}
&1)&~~~\phi=0,~~~\psi=0~,~~~\chi<e_1-e_0~,
\nn
&2)&~~~{\rm cos}~2\phi=\frac{e_1-e_0}{\chi},~~~\psi=0,~~~\chi>e_1-e_0~.
\eea
According to Ref. \cite{Del05}, for any mean field (MF) minimum one obtains $\psi=0$,
independent of which kind of vacuum (correlated or not)
we use to estimate the expectation values.
We remark, however, that for $e_1=e_2$ (\ref{Hmed}) becomes
independent of $\psi$ and therefore we can expect a Goldstone mode
in the symmetry broken phase.

For the above mentioned minima one obtains that the standard
RPA matrix elements have very simple expressions \cite{Hag00}
and the ${\cal A}$ and ${\cal B}$ RPA matrices are diagonal.
The RPA frequencies are easy to evaluate
\bea
\label{RPAfre}
\omega_k^2= {\cal A}_{kk}^2-{\cal B}_{kk}^2~,~~~k=1,2~,
\eea
where the indices $k=1, 2, 3$ shall be identified with the following configurations
\be
10 \rightarrow 1,~~~20 \rightarrow 2, ~~~21 \rightarrow 3 .
\ee
For the ph-amplitudes one gets
\bea
\label{RPAamp}
\left(\begin{matrix} X_k^{\nu}\cr Y_k^{\nu} \end{matrix} \right)=
\frac{1}{\sqrt{2}}
\left[\frac{{\cal A}_{kk}}{\omega_k} \pm 1\right]^{1/2} \delta_{k\nu} .
\eea
We fix the origin of the particle spectrum at $e_0=0$.
Then for a spherical vacuum with $\phi=0$ the RPA energies are given by
\bea
\label{RPAfre0}
\omega_{\nu}=\epsilon_{\nu}
\left[ 1-\left(\frac{\chi}{e_{\nu}}\right)^2\right]^{1/2}~,
~~~\nu=1,2~,
\eea
with the corresponding RPA amplitudes
\bea
\label{RPAamp0}
\left(\begin{matrix} X_k^{\nu}\cr Y_k^{\nu} \end{matrix} \right)=
\frac{1}{\sqrt{2}}
\left[\frac{\epsilon_k}{\omega_k} \pm 1\right]^{1/2} \delta_{k\nu}
~.
\eea
As it was shown in Ref. \cite{Hag00}, if the upper single-particle
levels are degenerate, i.e. $\Delta\epsilon\equiv\epsilon_2-\epsilon_1=0$,
for the values of the strength $\chi>e_1$, in the "deformed region",
i.e. with $\phi\neq 0$ given by HF minimum, one obtains a Goldstone mode.
In this case by considering $e_1=1$ one obtains for the excitation energies
\bea
\label{RPAfre1}
\omega_1&=&\epsilon\sqrt{2(\chi^2-1)}~,
\nn
\omega_2&=&0~.
\eea

{\it Application of Self-consistent RPA (SCRPA) to the three-level Lipkin model.}
The SCRPA operator including {\it scattering terms}, see Sect. II, is given by
\bea
Q^{\dag}_{\nu}=\sum_{m>i}\left(X^{\nu}_{mi}
\frac{A_{mi}}{\sqrt{n_{i}-n_{m}}}-
Y^{\nu}_{mi}\frac{A_{mi}}{\sqrt{n_{i}-n_{m}}}\right)
\eea
in terms of the pair operators in the MF basis
\bea
\label{Akk}
A_{mi}=\sum_{\mu=1}^Na^{\dag}_{m\mu}a_{i\mu}
=\sum_{\a\b}C_{m\a}C_{i\b}K_{\a\b}~.
\eea
Let us first discuss the SCRPA results in the spherical region,
i.e. the region where the generalised mean field equation
has only the trivial solution $\phi=0$.
In comparison with standard HF this region is strongly extended.
The content of the spherical region depends on the particle number.
For $N=20$ the spherical region is typically extended
by a factor of 1.5. This comes from the self-consistent coupling
of the quantal fluctuations to the mean field and actually corresponds to a weakening (screening) of the force.

\begin{figure}[ht]
\begin{center}
\includegraphics[width=6cm]{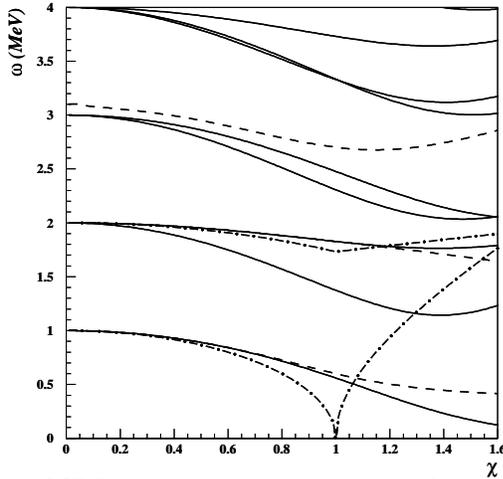}
\vspace{-1cm} 
\caption{
SCRPA excitation energies versus the strength parameter $\chi$,
for $N=20$ and $e_0=0,~e_1=1,~e_2=2$ (dashed lines).
By solid lines are given the lowest exact eigenvalues and
by dot-dashes the standard RPA energies.}
\label{fig01}
\end{center}
\end{figure}

Let us now consider the definite example $e_0=0,~e_1=1,~e_2=2$
for $N=20$. In Fig. \ref{fig01} we show by dashed lines the SCRPA
results for the excitation energies, compared with the exact ones
(solid lines) and to standard RPA (dot-dashes).
We see that SCRPA strongly improves
over standard RPA and in fact first and second excited states
are excellently reproduced up to $\chi$-values of about $\chi\approx 1.2$.
The third state has no analogue in standard RPA and it must therefore be
attributed to the {\it scattering configuration} $A_{21}$.
The SCRPA solution for the the third eigenvalue
approximates rather well the fifth exact eigenvalue in the range $0 < \chi\leq1.0$.
Concerning the SCRPA result, this seems quite surprising, since naively
one would think that for vanishing interaction the SCRPA eigenvalue
corresponding to the $(21)$ component should approach to the value
$\omega_3\rightarrow e_2-e_1=1$. 
In Ref. \cite{Del05} it is shown that this mode indeed corresponds to 
the fifth exact eigenvalue $\nu=5$ as long as $\chi > 0$. At exactly $\chi=0$ the solution jumps to $\omega_3 =1$.




\begin{figure}[ht]
\begin{center}
\includegraphics[width=6cm]{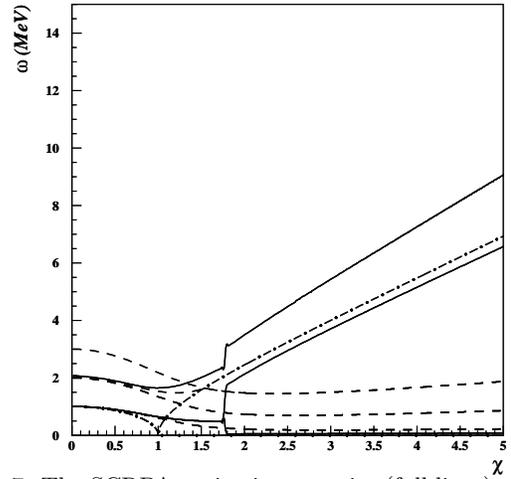}
\vspace{-1cm} 
\caption{
The SCRPA excitation energies  (full lines) compared with
the exact solution (broken lines) for $N=8$. Dot-dashed line
gives standard RPA values.}
\label{fig03}
\end{center}
\end{figure}



In the deformed region a particular situation arises in our model
for $\Delta\epsilon=\epsilon_2-\epsilon_1=0$, since, as already mentioned,
a spontaneously broken symmetry occurs in this case. Here the standard HF-RPA
exhibits its real strength because, as shown in (\ref{RPAfre1}),
a zero mode appears, which signifies that the broken symmetry is partially
restored, i.e. the conservation laws are fulfilled 
\cite{Bla86,Rin80,For75}. 
This property is also fulfilled in SCRPA under the condition that the
scattering terms are included.
In this context we mention that the operator
\bea
\label{L0}
\hat{L}_0&=&i(K_{21}-K_{12})=i[A_{20}-A_{02}){\rm sin}\phi
\nn&+&
(A_{21}-A_{12}){\rm cos}\phi]~,
\eea
where $A_{ij}$ are the pair operators (\ref{Akk}) in the HF basis,
commutes with the Hamiltonian, i.e. $[H,\hat{L}_0]=0$.

It is therefore a symmetry operator which can be identified with the
z-component of the rotation operator.
The existence of this symmetry operator is the reason why equation (\ref{QHQ})
produces in the deformed region a Goldstone mode at zero energy (see also
Eq. (\ref{RPAfre1})). 

Indeed, in this case the deformed RPA equations
possess a particular solution $Q^{\dag}=\hat{L}_0$ and one has
$\la 0|[\delta \hat{L}_0,[H,\hat{L}_0]]|0\ra=0$.
This means that in the deformed region $\hat{L}_0$ can be considered
as an RPA excitation operator with $|0\ra$ not being an eigenstate
of $\hat{L}_0$ and producing a zero excitation energy, i.e.
the Goldstone mode. On the contrary, in the spherical region
the ground state is an eigenstate of $\hat{L}_0$ and therefore
it cannot be used as an excitation operator.


In standard RPA, where the expectation values of the (double) commutators
are evaluated over the deformed HF state, the scattering terms
$A_{12}$ and $A_{21}$ in $\hat{L}_0$ automatically decouple from the
$ph$ and $hp$ space and that is the reason why only $ph$ ($hp$) components
of the symmetry operator suffice to produce the Goldstone mode.
On the other hand, if one works with a deformed correlated ground state
as in SCRPA, the scattering terms do {\it not} decouple from the $ph$ ($hp$)
space and therefore 
the full $ph$, $hp$, $hh$ and $pp$ space must be taken into account to produce
the Goldstone mode. Since in the latter case the symmetry operator
(\ref{L0}) is entirely taken into account, this property
follows again automatically. 

The numerical verification of
this desirable quality of SCRPA must, however, be undertaken with care.
Indeed, a zero mode contains diverging amplitudes which, injected
into the SCRPA matrix, may not lead to self-consistency.
The way to overcome this difficulty is to start the calculation with
a finite small value of $\Delta\epsilon$, i.e. with a slight explicit
symmetry breaking, and then to diminish its value step by step.
We, in this way, could verify with very high accuracy that the zero
eigenvalue occurs in the deformed region for all values of the
interaction strength $\chi$. This is shown in Fig. \ref{fig03} by the solid line which parallels very closely the horizontal axis.
Here we considered the value $\Delta\epsilon=0.001$, but we were
able to reach the value $\Delta\epsilon=10^{-6}$.
We, therefore, see that our theoretical expectation is fully
verified by the numerical solution.
In the same way we checked that the energy weighted sum rule is fulfilled
in SCRPA in the symmetry broken phase with the Goldstone mode present.\\

We also should comment about the other features seen in Fig. \ref{fig03}. 
Up to about $\chi =1.8$, in the spherical region, the first excited state is two-fold degenerate. After that value the degeneracy becomes suddenly lifted and one state goes into the Goldstone mode and the other more or less joins the upgoing RPA state. If one chooses a larger particle number, both states will become closer and join the second band head of the model. This kind of first order phase transition is an artefact of the theory and does not happen in the exact solution. One probably should include second RPA correlations to cure this, see
Sect. VII and also Sect.V.C.
However, the appearance of a Goldstone mode is a quite remarkable feature which we will comment upon in more detail below.

As usual with a continuously broken symmetry, also in the present
model a clear rotational band structure is revealed.
The exact solution found by a diagonalisation procedure
has a definite angular momentum projection $L_0$.
Moreover, the expectation value of the $L_0^2$ operator
has integer values, namely
\bea
\label{band}
\sqrt{\langle L_0^2\rangle}=J=0,1,2,...
\eea
The ground state "rotational band" $J=0,1,2,...$ is built on top
of the RPA excitation with a vanishing energy (Goldstone mode).

\begin{figure}
  \includegraphics[width=6cm]{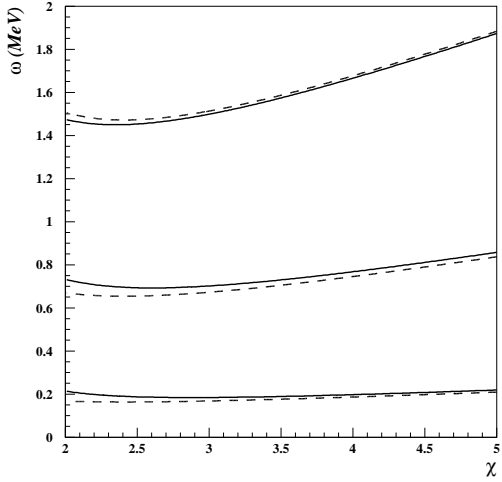}
  \caption{SCRPA rotational spectrum (dashed lines) and exact energies (solid lines) for $N$ = 20. 
The three levels correspond to $J$ = 1, 2, 3. 
These levels correspond to the three levels between $\chi=2$ and $\chi = 5$ in Fig.\ref{fig03}.}

  \label{rot-sp}
\end{figure}

As customary in RPA theory, one also can evaluate the mass
parameter of the rotational band within SCRPA.
By a straightforward generalisation we obtain for the
moment of inertia (see e.g. Ref. \cite{Bla86,Rin80})
\bea
\label{mass}
M=2L_0^*\left({\cal A}-{\cal B}\right)^{-1}L_0~,
\eea
where ${\cal A,B}$ are the SCRPA matrices and
$L_0$ is the angular momentum
operator (\ref{L0}),
which should be written in terms of normalised generators
$\delta Q^{\dag}$, i.e.
\bea
\label{L01}
L_0=i\left(0,~N_{20}^{1/2}sin\phi,~N_{21}^{1/2}cos\phi
\right)~.
\eea
For the standard RPA case, by using the corresponding matrix elements, one obtains an analytical solution, namely

\bea
M=\frac{N(\chi-1)}{\epsilon\chi(\chi+1)}~,
\eea
where $N$ is the particle number. The SCRPA mass is also obtained from (\ref{mass}) but using the SCRPA expressions for the ${\mathcal A}$ and ${\mathcal B}$ matrices. The spectrum of the first three states is shown in Fig. \ref{rot-sp} in the range between $\chi =2$ and $\chi = 5$.  We see that the exact spectrum is very well approximated. These states correspond to the ones seen in
Fig. \ref{fig03} in the same range of the coupling constant. The first rotational state for $\langle L_0\rangle$ matches rather well with the lowest excited state of SCRPA in the spherical region. However, this is not the case for the higher-lying excitations.\\

In conclusion of this section, we can say that SCRPA reproduces very well the 'spherical' region of the three-level Lipkin model. What is new is that the inclusion of the scattering configurations allowed to obtain the Goldstone mode in the 'deformed' region where a clear rotational spectrum appears. The calculation of the SCRPA moment of inertia then allowed to get a very accurate reproduction of the rotational ground state band in this model. We would like to point out that the appearance of the Goldstone mode with a theory, which takes into account strong correlation beyond the ones of the standard RPA theory, is highly non-trivial. To the best of our knowledge, we are not aware of any other fully microscopic extension of the RPA approach which has numerically achieved this taking into account self-consistently screening of the interaction. The Kadanoff and Baym formalism would lead even in this very simple model to numerically almost inextricable complications.

\subsection{Hubbard Model}
\label{sec:Hub}

In this Section we apply the SCRPA scheme to the Hubbard model of strongly 
correlated electrons, which is one of the most wide spread models to 
investigate strong electron correlations and high $T_c$ superconductivity.
Its Hamiltonian is given by
\be
  H \,= -t\, \sum \limits_{<i j> \sigma}\,c^{\dagger}_{i \sigma} c_{j \sigma}
  \,+\, U \sum \limits_{i} \hat{n}_{i \uparrow} \hat{n}_{i \downarrow}
~\label{HubbardCoordSpace}
\ee
where $~c^{\dagger}_{i \sigma}$, $~c_{i \sigma}~$ are the electron creation
and destruction operators at site `$i$' and the
$\hat{n}_{i\sigma}=c^{+}_{i\sigma}\,c_{i\sigma}$ are the number operators for
electrons at site `$i$' with spin projection $\sigma$. As usual $t$, is the
nearest neighbour hopping integral and $U$ the on site Coulomb matrix element.


As an example, we will consider the $1$-dimensional 6 -sites case at half filling.
With the usual transformation to plane waves (we are considering periodic boundary conditions, that is a ring)
$~c_{j,\sigma}=\frac 1{\sqrt{N}} \sum\limits_{\bf{\vec{k}}}
 a_{\bf{\vec{k}},\sigma} e^{-i\bf{\vec{k}\,\vec{x}_{j}}}$.
This leads to the standard expression for a zero range two body interaction
\bea
H &=& \sum_{{\bf{\vec{k}}}, \sigma} \left (\epsilon_{k} - \mu \right )
 \hat{n}_{{\bf{\vec{k}}}, \sigma}
\nn&+&
 \frac U{2\,N}\sum_{\bf{\vec{k},\vec{p},\vec{q}},
 \sigma} a_{{\bf{\vec{k}}}, \sigma}^{\dagger} \, a_{{\bf{\vec{k}+\vec{q}}},
 \sigma} \,a_{{\bf{\vec{p}}}, -\sigma}^{\dagger} \, a_{{\bf{\vec{p}-\vec{q}}},
 -\sigma}
\label{hamiltonian_imp}~,
\eea
where $\hat{n}_{\bf{\vec{k}}, \sigma} = a^{\dagger}_{\bf{\vec{k}},
\sigma} a_{\bf{\vec{k}}, \sigma}$ is the occupation number operator
of the mode ($\bf{\vec{k}}, \sigma$) and the single-particle energies are 
given by $\epsilon _{\bf{\vec{k}}}= -2\,t\sum\limits_{d=1}^{D }
cos\left(k_{d}\right)$ with the lattice spacing set to unity.

\begin{figure}[ht]
  \includegraphics[width=7cm]{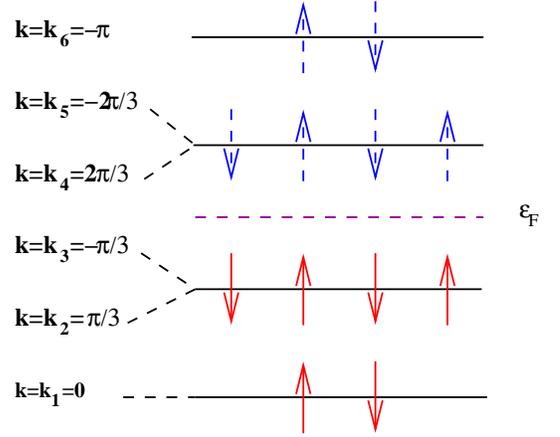}
      \caption{\label{fig08} Excitation spectrum of HF at $U=0$ for the 
chain with 6 -sites at half filling and projection of spin $m_{s} = 0$. 
The occupied states are represented by the full arrows and those not occupied are represented by the dashed arrows.}
\end{figure}

In the first Brillouin zone 
$-\pi\leq k < \pi$ we have for $N=6$ the following wave numbers 
\ba
k_{1}&=&0~,~~~~k_{2}=\frac{\pi}{3}~,~~~~k_{3}=-\frac{\pi}{3}~,
\nonumber \\
k_{4}&=&\frac{2\pi}{3}~,~~~~k_{5}=-\frac{2\pi}{3}~,~~~~k_{6}=-\pi~.
\ea
With the HF transformation 
\be
a_{h, \sigma} = b^{\dagger}_{h,\sigma} \mbox{,}~~~~~~~~ 
a_{p, \sigma} = b_{p, \sigma} ~\mbox{,}
\ee
such that $b_{k, \sigma}|HF\rangle =0$ for all $k$, we can write the
Hamiltonian in the following way (normal order with respect to 
$b^{\dagger}$, $b$)
\be
H= H_{HF} + H_{|q|=0} +  H_{|q|=\frac{\pi}{3}} + H_{|q|=\frac{2\pi}{3}} + 
H_{|q|=\pi}~,
\label{ham_quasip-6sit}
\ee
where the notations are given in \cite{Jem05}.
The level scheme is shown in Fig. \ref{fig08}. 
The hole states are labeled $h=\{1,2,3\}$ and the particle states 
$p=\{4,5,6\}$. The HF groundstate is
\be
|HF\rangle = a^{\dagger}_{1,\uparrow} \, a^{\dagger}_{1,\downarrow} \, 
a^{\dagger}_{2,\uparrow}\, a^{\dagger}_{2,\downarrow} \, 
a^{\dagger}_{3,\uparrow}\, a^{\dagger}_{3,\downarrow}
| - \rangle ~.\label{HF-groundstate}
\ee

There are three different absolute values of momentum transfers, as shown
in Table \ref{table1}.
\begin{table}
\caption{\label{table1} The various momentum transfers in the 6 -sites case.}
\begin{tabular}{|c|c|c|}
\hline
$|q| =\frac{2\pi}{3}$~~~~~~~  & $|q|  =\pi$~~~~~~~ & $|q| = \frac{\pi}{3}$
\\ \hline
$51\rightarrow q_{51} =-\frac{2\pi}{3}$~~~~~~~  & $61\rightarrow q_{61}  =-\pi$~~~~~~~ & $42\rightarrow q_{42} = + \frac{\pi}{3}$
\\ \hline
$41\rightarrow q_{41} =+\frac{2\pi}{3}$~~~~~~~ & $52\rightarrow q_{52}  =-\pi$~~~~~~~ & $53\rightarrow q_{53} =-\frac{\pi}{3}$
\\ \hline
$62\rightarrow q_{62} =+\frac{2\pi}{3}$~~~~~~~ & $43\rightarrow q_{43}  = +\pi$~~~~~~~ &
\\ \hline
$63\rightarrow q_{63} =-\frac{2\pi}{3}$~~~~~~~ & ~~~~~~~ &
\\ \hline
\end{tabular}
\end{table}
Since the momentum transfer $|q|$ is a good quantum number, the RPA equations 
are block diagonal and can be written down for each $|q|$ -value separately. 
For example, for $|q|=\frac{\pi}{3}$ we have the following RPA operator 
for charge and longitudinal spin excitations
\bea
Q^{\dagger}_{|q|=\frac{\pi}{3},\nu} &=&X^{\nu}_{2\uparrow, 4\uparrow}\, 
K^{+}_{4\uparrow,2\uparrow} +X^{\nu}_{2\downarrow, 4\downarrow}\,K^{+}_{4\downarrow, 2\downarrow} 
\nn&+&X^{\nu}_{3\uparrow, 5\uparrow}\, K^{+}_{5\uparrow,3\uparrow} +
X^{\nu}_{3\downarrow, 5\downarrow}\,K^{+}_{5\downarrow, 3\downarrow}
\nonumber \\
&-&Y^{\nu}_{2\uparrow, 4\uparrow}\, K^{-}_{2\uparrow,4\uparrow} - 
Y^{\nu}_{2\downarrow, 4\downarrow} \,K^{-}_{2\downarrow, 4\downarrow} 
\nn&-&Y^{\nu}_{3\uparrow, 5\uparrow}\, K^{-}_{5\uparrow,3\uparrow} - 
Y^{\nu}_{3\downarrow, 5\downarrow}\,K^{-}_{3\downarrow, 5\downarrow}~,
\label{op-dexqp6s4}
\eea
where
\bea
K^{\pm}_{p\sigma, h\sigma} &=& \frac{J^{\pm}_{p\sigma, h\sigma}}
{\sqrt{1 - \langle 0|M_{p\sigma, h\sigma}|0\rangle}}
\nn
M_{p\sigma, h\sigma} &=& \tilde{n}_{p,\sigma} +\tilde{n}_{h,\sigma}~.
\eea
Here $~J^{-}_{\sigma} = b_{1,\sigma}\,b_{2,\sigma}$,
$~ J^{+}_{\sigma} = \left(J^{-}_{\sigma}\right)^{+}$,
$~\tilde{n}_{k_i,\sigma} = b^{\dagger}_{i,\sigma } \,b_{i,\sigma}~$.
The operators $J^{\pm}_{\sigma}$ and $1-M_\sigma$ form a $SU(2)$ algebra of spin
 --$\frac 1 2$ operators and, therefore, using the Casimir relation we obtain
\be
M_\sigma = 2\, J^{+}_{\sigma}\, J^{-}_{\sigma}~.
\ee

We write this RPA operator in short hand notation as 
\be
Q^{\dagger}_{\nu}= \sum\limits_{i=1}^{4}\frac {1}{\sqrt{1 -\langle 0| 
M_{i}|0\rangle}} \left(X_{i}^{\nu}\;J^{+}_{i} - Y_{i}^{\nu}\; J^{-}_{i} \right)~,
 \label{op-ex-6sit}
\ee
with the usual properties.
The matrix elements in the SCRPA equation are then of the form
\sba
\label{eleltsmatix-RPAa}
\ba
\A_{i , i'} &=& \frac {\left\langle 0|\left[ J^{-}_{i'} 
\left[ H, J^{+}_{i}\right] \right]|0\right\rangle} 
{\sqrt{(1-\langle 0|M_{i'}|0\rangle)( 1-\langle 0|M_{i}|0\rangle )}}~\mbox{,}
\label{AA}
 \\
\B_{i, i'} &=& -\frac {\left\langle 0|\left[ J^{-}_{i'} 
\left[ H, J^{-}_{i}\right] \right] |0\right\rangle} 
{\sqrt{(1-\langle 0|M_{i'}|0\rangle)( 1-\langle 0|M_{i}|0\rangle )}}~.
\label{BB}
\ea
\sea
The expectation values are given in \cite{Jem05}.
Let us add that the matrices $\A$ and $\B$ are symmetric.

\begin{figure}[ht]
\includegraphics[width=7cm]{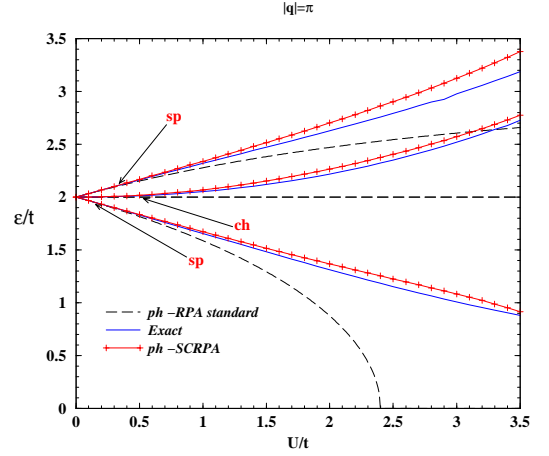}
\caption{\label{fig09} 
Energies of excited states in standard RPA, 
SCRPA, and exact cases as a function of $U$ for 6 -sites with spin projection $m_{s} = 0$ and for $|q|=\pi$.
States of the charge response and those of the longitudinal spin response are denoted by $ch$ and $sp$,
respectively.}
\end{figure}

In Fig. \ref{fig09} we display the excitation energies in the channel
$|q|=\pi$, as a function of $U/t$. The other cases are similar.
The exact values are given by the continuous lines, the SCRPA ones by
crosses and the ones corresponding to standard RPA by the broken lines.
We see that SCRPA results are excellent and strongly
improve over standard RPA. As expected, this is particularly important at
the phase transition points where the lowest root of standard RPA goes to 
zero, indicating the onset of a staggered magnetisation on the mean-field
level. It is particularly interesting that SCRPA allows to go beyond the
mean-field instability point. However,  
 at some values $U>U_{cr}$ the system still ``feels" the phase transition and
SCRPA stops to converge and also deteriorates in quality. Up to these values
of $U$ SCRPA shows very good agreement with the exact solution and, in
particular, it completely smears the sharp phase transition point of standard
RPA, which is an artefact of the linearisation.

\begin{figure}
\includegraphics[width=7cm]{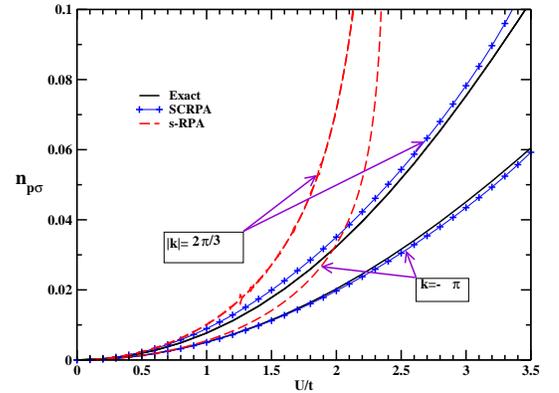}
\caption{ Occupation numbers for the 5th and 6th state with the SCRPA and the Catara approximation
(crosses); full lines represent the exact results and the dots (red) are obtained with the standard 
RPA approach.}
\label{Hub-occs}
\end{figure}

A further quantity which crucially tests the ground state correlations are the
occupation numbers. We used
the so-called Catara approximation, see Sect. II.A, for their evaluation \cite{Cat96}, 
showing an excellent performance of SCRPA, see Fig. \ref{Hub-occs}.

In this section we gave a very short summary of the achievement of SCRPA concerning the Hubbard model. Notably we only considered the symmetry unbroken phase of the model, where SCRPA gives a strong improvement over the standard RPA leading to energies of excited states, which are in close agreement with the exact values even somewhat beyond the critical $U$, where standard RPA has a break down.

\subsection{Various applications and extensions of the renormalized RPA}
\label{sec:rRPA}

\begin{figure}
\includegraphics[width=10cm]{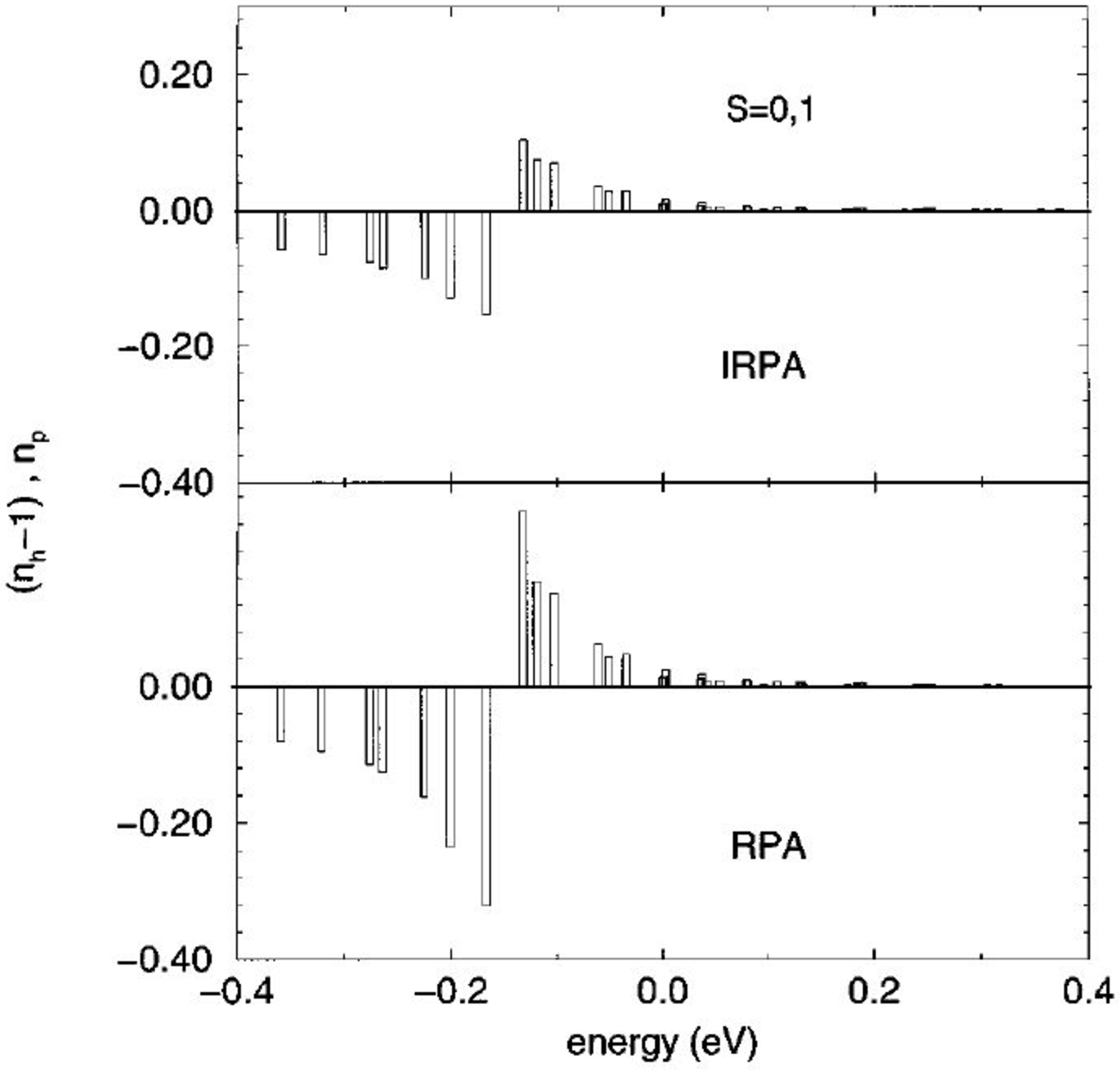}
\caption{Occupation numbers $n_p$ and $1-n_h$ for particle states and the opposite 
of depletion numbers $n_h$ 21 for hole states, respectively. In the lower and in
the upper panel the RPA and IRPA results, respectively, are reported. 
Both spin S=0 and S=1 states are included in the calculations. In the
abscissa the single-particle energies in eV are indicated \cite{Cat96}.}
\label{catara}
\end{figure}

The renormalized RPA (r-RPA) was applied by many authors to various
strongly correlated systems. In Ref. \cite{Cat96}  
a version of the r-RPA called improved RPA (IRPA) is introduced and applied 
to the description of the electronic gas in metallic clusters treated within the jellium model. 
The one- and two-body densities are derived by using
the so-called number operator method proposed long time ago by
D. Rowe \cite{Row68}. The corresponding relations are given by
Eq. (\ref{dens1}).
IRPA, being considerably simpler from the numerical point of view than SCRPA,
represents a significant improvement with respect to standard RPA where uncorrelated occupation numbers are used.
The Authors have found that a better treatment of
correlations leads to important modifications of single-particle occupation 
numbers and strength distributions, see Fig. \ref{catara} \cite{Cat96}.
In Fig. \ref{catara} the upper panel shows the fully self-consistent r-RPA (IRPA). With respect to a non-self-consistent solution, i.e., only first iteration (lower panel), the effect is somewhat damped but still strong, maximally 20 percent! Spin singlet and triplet states are included together with orbital angular momenta up to $L=6$. The $S=1$ channel gives by far the strongest contribution to the renormalisation.
 Renormalised RPA has recently also been applied to superfluid strongly polarised Fermi gases with 
strongly improved results with respect to standard RPA \cite{Dur20}.

On the other hand,
the two-dimensional electron gas provides a valuable testing
ground for microscopical theories of interacting Fermi systems. 
Compared to the three-dimensional case, the two-dimensional
correlations play a more important role, demanding higher precision
to theoretical approaches and exposing their weaknesses.
In Ref. \cite{Sch00} the Authors propose an iterative procedure,
using the momentum distributions for the recalculation of polarization
and vertex functions within the RPA, similar to the r-RPA scheme.
It is shown that the inclusion of the physical momentum distributions 
for the calculation of polarization and vertex functions yields significant contributions to the resulting field correction and correlation energies.

An important application of the r-RPA concerns the double beta decay 
process with emission of two neutrinos ($2\nu\beta\beta$).
The second leg of the $2\nu\beta\beta$ process is very 
sensitive to changing the relative strength of the particle-particle
interaction strength $g_{pp}$.
It is worth mentioning that the two-body interaction of $ph$ type is 
repulsive while that of $pp$ nature is attractive.
Due to this feature there is a critical value for $g_{pp}$ for which 
the first root of the proton-neutron QRPA (pnQRPA) equation vanishes.
Actually, this is the signal that the pnQRPA approach breaks down.
Moreover, the $g_{pp}$ value, which corresponds to a transition amplitude 
agreeing with the corresponding experimental data, is close to the 
mentioned critical value.
That means that the result is dependent on  adding corrections to the standard 
RPA picture.
The first improvement for the pnQRPA was achieved in Ref. \cite{Rad91}
by using a boson expansion (BE) procedure. 
It is interesting to mention that within the BE formalism transitions 
to excited state, forbidden in the pnQRPA approach, become possibe. 
A systematic analysis of the double beta transtions for 18 nuclei has 
been performed.

\begin{figure}
\begin{center}
\includegraphics[height=6cm]{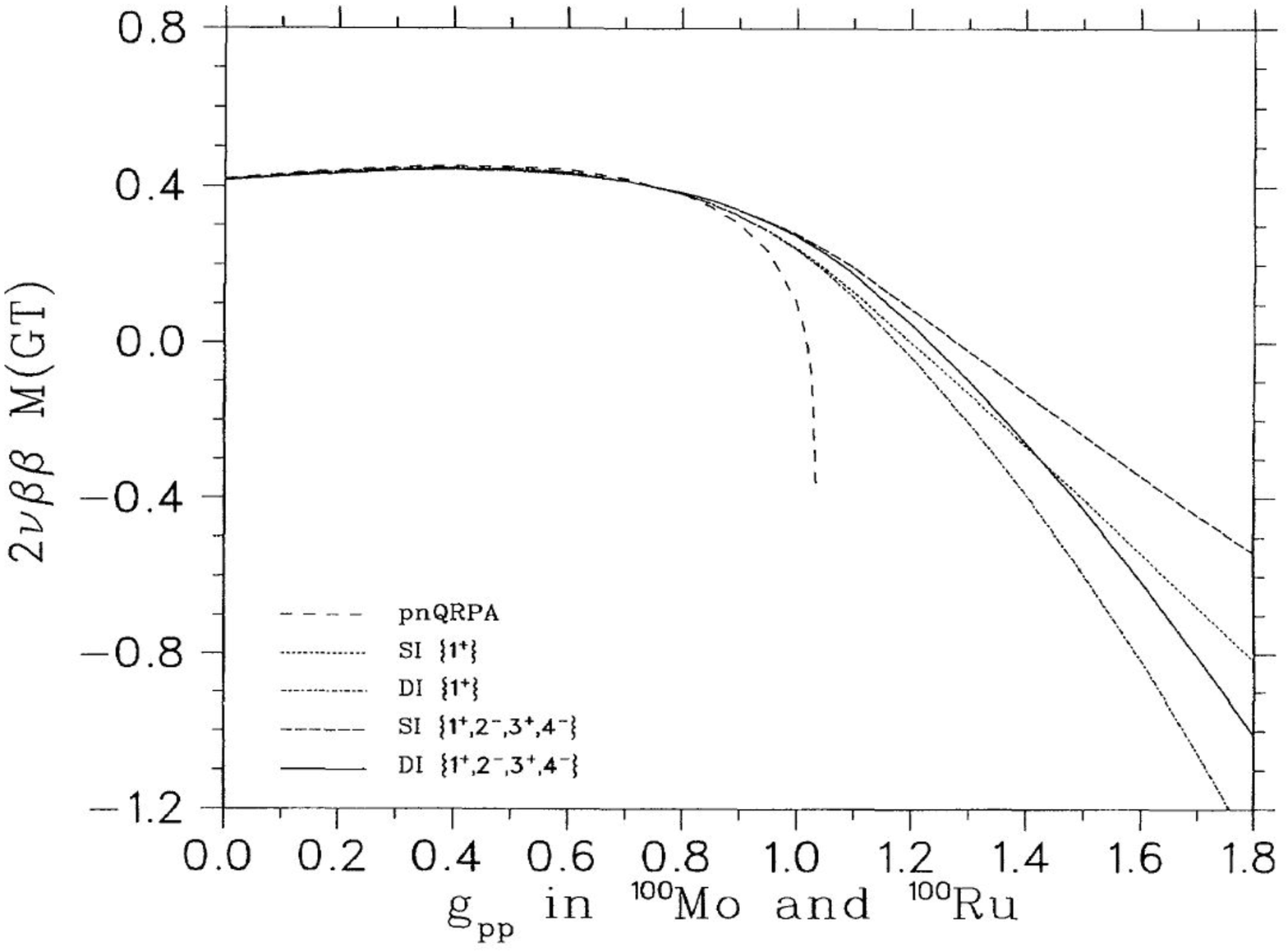}
\end{center}
\caption{
The transition amplitude of the $2\nu\b\b$ process 
$^{100}{\rm Mo (gs)}\rightarrow ^{100}{\rm Ru (gs)}$ versus the particle-particle strength
for various pnQRPA versions. The dashed line corresponds to the standard
pnQRPA, while the other lines to various intermediate $pn$ excitations
considered within r-pnQRPA are described in Ref. \cite{Toi95}.
}
\label{suhonen3}
\end{figure}

Later  another procedure showed up, which renormalized the dipole 
two-quasiparticle operators by replacing the scalar components of their 
commutators by their average values. 
In Ref. \cite{Toi95} this renormalisation
procedure is applied to proton-neutron quasiparticle RPA (r-pnQRPA)
in order to describe $\beta$ and double $\b$ ($\b\b$) decay processes.
The one-body quasiparticle density for protons (p) and neutrons (n)
is expressed in terms of QRPA amplitudes by using
the Catara method \cite{Cat94}, i.e.
\bea
\left[a^{\dag}_p a_p\right]_{00}&=&
\sum_{JMn'}A^{\dag}_{JM}(pn')A_{JM}(p'n)
\nn
\left[a^{\dag}_n a_n\right]_{00}&=&
\sum_{JMp'}A^{\dag}_{JM}(p'n)A_{JM}(p'n)~.
\eea
This ansatz leads to the following equalities for the expectation
value on the QRPA vacuum
\bea
\label{Dpn}
N_{pn}=1
&-&\hat{j}_p^{-1}\sum_{n'}N_{pn'}\sum_{J\n}\hat{J}^2[Y^{J\nu}_{pn'}]^2
\nn&-&
\hat{j}_n^{-1}\sum_{p'}N_{p'n}\sum_{J\n}\hat{J}^2[Y^{J\nu}_{p'n}]^2~,
\eea
where $\hat{j}=\sqrt{2j+1}$ and the $N_{pn}$ metric (norm) matrix elements
are defined as usual by
\bea
\label{metric}
N_{pn}=\la 0|\left[A_{JM}(pn),A^{\dag}_{JM}(pn)\right]|0\ra~.
\eea
In this way the r-pnQRPA system of equations for several
multipolarities $J$ is solved together with (\ref{Dpn}).
In calculations the matrix elements of the realistic Bonn interaction \cite{Mac87} 
are used to describe the two-neutrino (2$\n$) $\b\b$
decay process $^{100}{\rm Mo}(gs)\rightarrow ^{100}{\rm Ru}(gs)$.
It was shown that this kind of calculation reduces
ground state correlations and prevents the collapse
of the standard pnQRPA in the region of physical interest
for the particle-particle channel of the interaction,
which is very important for double-beta decay processes.
This is shown in Fig. \ref{suhonen3} for the transition amplitude of the $2\nu\b\b$ process.
The dashed line describes the collapse of the transition amplitude
versus the particle-particle interaction strength for $g_{pp}\sim 1$ 
within the standard pnQRPA, while the other lines give the same dependence 
within r-pnQRPA. Different lines correspond to various intermediate $pn$ excitations
in the expansion (\ref{Dpn}) and they are described in Ref. \cite{Toi95}.
Notice that the matrix element acquires values close to the experimental ones
for $g_{pp}\sim 1$ for all considered versions of r-pnQRPA.

In Ref. \cite{Toi97} this procedure is applied for several
2$\n\b\b$ emitters like
$^{76}$Ge, $^{78}$Kr, $^{82}$Se, $^{96}$Zr, $^{106}$Cd and $^{130}$Te
to ground states as well as to excited one- and two-phonon
states in daughter nuclei.
Later on, $^{76}$Ge, $^{82}$Se and $^{128-130}$Te emitters were investigated in
\cite{Sch96}. A review on this approach is given in Ref. \cite{Fae98}.

In Ref. \cite{Civ97} r-RPA is applied to the two-level proton-neutron
Lipkin model. The results are compared with the exact diagonalisation
procedure, as well as with  a boson mapping. It turned out that
in spite of the good agreement of r-RPA eigenvalues compared
with the exact solution, the wave function is quite
different from the exact one in the region where standard RPA collapses.

In Refs. \cite{Hir96,Hir97} r-QRPA is applied to
the simplest single j-shell Hamiltonian, describing Fermi 
$\beta$-transitions, namely
\bea
\label{schem}
H&=&e_pN_p+e_nN_n-G_pS^{\dag}_pS_p-G_nS^{\dag}_nS_n
\nn&+&
2\chi\b^-\b^+-2\kappa P^-P^+~,
\eea
where the usual notations were used
\bea
N_i&=&\sum_{m_i}c^{\dag}_{m_i}c_{m_i}~,~~~
S_i = \sum_{m_i}c^{\dag}_{m_i}c^{\dag}_{\ov{m}_i}~,
\nn
\b^-&=&\sum_{m_p=m_n}c^{\dag}_{m_p}c_{m_n}~,~~~
P^- = \sum_{m_p=m_n}c^{\dag}_{m_p}c^{\dag}_{\ov{m}_n}~.
\nn
\eea
In spite of its simplicity, this Hamiltonian
is able to describe the main features of a realistic Hamiltonian,
describing $\b$ and $\b\b$ processes.
In particular, the collapse of standard RPA is obtained by
increasing the particle-particle strength $\kappa$.
It was shown that the collapse of QRPA correlates with the presence
of an exact eigenvalue at zero energy. It was shown that
r-QRPA prevents this collapse.
The role of scattering terms was discussed and they were
shown to be relevant in getting excitation energies closer
to the exact values.
$\b\b$-decay amplitudes were evaluated and compared with
other formalisms.
 
Later on, in Refs. \cite{Hir99,Hir99a} a more complete
analysis of this Hamiltonian was performed.
The Dyson boson representation ($b^{\dag},b$),
together with a coherent representation
\bea
\label{coh}
|\a\ra=N_0\sum_{l=0}^{2\W}\frac{\a^l}{l!}
\left(b^{\dag}\right)^l|0\ra~,
\eea
was used to describe phase transitions by using the complex
order parameter $\a$.
The spontaneous breaking of the proton-neutron-pair
symmetry was induced by the particle-particle strength $\kappa$
and it manifests in the appearance of the zero-energy mode.
It was shown that the coherent state representation
is able to describe the phase transition, while
r-QRPA is unable to describe correctly the energy and, for instance, 
the inclusion of the Pauli principle at the Hamiltonian level
is crucial to describe the phase transition.
 
We mention here that the ground state instability versus the 
particle-particle strength $\kappa$ of the Hamiltonian 
(\ref{schem}) is analyzed in Ref. \cite{Rad02}
within the time-dependent formalism.
 
In Ref. \cite{Bob99} the r-pnQRPA system of equations
is solved consistently with the BCS equations,
derived by considering the same RPA vacuum.
The normal and pairing densities are written in terms
of the quasiparticle density as follows
\bea
\rho_k&=&\la 0|c^{\dag}_k c_k|0\ra=v^2_k+(u^2_k-v^2_k)n_k
\nn
\kappa_k&=&\la 0|c^{\dag}_k c^{\dag}_k|0\ra=u_kv_k(1-2n_k)~.
\eea
It is shown that the system of equations (\ref{Dpn}) can
be written as a linear system of equations for quasiparticle
densities. The case of two neutrino $\b\b$-decay
of $^{76}$Ge is analyzed and it is shown that the inclusion
of self-consistency in solving BCS equations leads to more
reduction of ground states correlations.
In a later Ref. \cite{Bob00} this self-consistent
procedure is applied to several emitters in medium-heavy
mass region in order to show the systematic reduction of
$\beta\beta$-decay matrix elements.

In spite of the fact that the Pauli principle is taken
into account, the sum rules, in particular the Ikeda sum rule
\bea
S_{-}-S_{+}&=&\sum_{\m}(-)^{\m}
\la 0|\left[\hat{\b}^{+}_{\m},\hat{\b}^{-}_{-\m}\right]|0\ra
\nn&=&3(N-Z)
\eea
is not fulfilled within r-pnQRPA, due to the fact that the $\beta$ decay operators 
$\hat{\b}^{\pm}_{\m}$ do not contain the scattering terms.
This drawback is cured in Refs. \cite{Rad98,Rad05} within the 
fully renormalized pnQRPA (FR-pnQRPA), by considering the scattering terms 
in the structure of the QRPA phonon.
Later on, a new method of restoring the gauge symmetry breaking was formulated 
to cure this drawback within the so-called  GPFR-pnQRPA \cite{Rad10}.

Let us mention that in Ref. \cite{Dan00} the scattering terms were also included 
in the phonon operator for a schematic four-level model Hamiltonian. 
The Authors called this model extended r-QRPA (er-QRPA). 
The analysis showed that the contribution of $pp$ and $hh$ terms to the spectrum 
and sum rules becomes important and prevent the collapse of standard RPA,
if their magnitudes are comparable with the $ph$ ones.

Unfortunately, small spurious eigenvalues generated within FR-rQRPA by scattering terms
are mixed with those of physical eigenstates.
An approach to separate these low-lying spurious modes is proposed in Ref. \cite{Rod02}. 
The idea is to consider the phonon operator not in terms of quasiparticles, but of original
particle operators, i.e.
\bea
Q^{\dag}_{JM}(\n)=\sum_{\t\t'}
X_{\t\t'}^{J\n}\left[C^{\dag}_{JM}(\t\t')-
Y_{\t\t'}^{J\n}\tl{C}_{JM}(\t\t')\right]~,
\nn
\eea
where $C^{\dag}_{JM}(\t\t')=\left[c^{\dag}_{\t}\tl{c}_{\t'}\right]_{JM}$
and $\t$ denote single-particle quantum numbers including isospin.
Thus, this phonon operator written in terms
of quasiparticles contains scattering terms.
It is shown that the spectrum does not contain spurious low-lying
states and Ikeda sum rule is analytically fulfilled.
A numerical application of the model to  2$\n\b\b$
emitters $^{76}$Ge, $^{82}$Se, $^{100}$Mo, $^{116}$Cd, $^{128}$Te
and $^{130}$Xe is performed in \cite{Pac03}.
This model was called by its authors also fully renormalised pnQRPA
(fr-pnQRPA).





In all these approaches the expectation values were evaluated
without the explicit form of the ground state wave function.
It turns out that it can approximately be given only for a limited
class of two-level models, like the O(5) model for Fermi transitions,
as a superposition of even number of protons and neutrons
similar to (\ref{coh}), i.e.
\bea
\label{gs0}
|0\ra=\sqrt{N_0}\sum_{l=0}^{\W}a_lz^l\left(A^{\dag}\right)^{2l}|BCS\ra~,
\eea
where $A^{\dag}$ is the proton-neutron monopole creation operator,
$\W=j+1/2$, $z=Y/X$ and the coefficients $a_l$ are given in \cite{Krm98}.
A similar approach is discussed in reference \cite{Sim00}.

An extended RPA phonon of the type
\bea
Q^{\dag}=X_1A^{\dag}-Y_aA+X_3AA^{\dag}A^{\dag}-Y_3AAA^{\dag}
\eea
is discussed in \cite{Sim03} for the two-level proton-neutron
Lipkin model, considering the explicit form of the ground state
of the form (\ref{gs0}). It turned out that the inclusion of
nonlinear terms in the phonon operator leads to a very good
agreement with the exact solution.

In Ref. \cite{Sim01} the ground state is determined as a solution
of a variational equation for the proton-neutron Lipkin model mapped onto 
the boson operators $B^{\dag}$ by using the Marumori  method.
The QRPA phonon operator
\bea
Q^{\dag}=X(B^{\dag}+t^*)-Y(B+t)
\eea
depends upon the variational parameters $t,t^*$, which results in
a non-vanishing expectation values of $Q$ on the "deformed ground state"
\bea
|0\ra=\exp\left[t^*B-tB^{\dag}\right]\exp\left[zB^{\dag}B^{\dag}-z^*BB\right]|-\ra~,
\nn
\eea
where $|-\ra$ is the vacuum state for the boson operator $B$.
The variational parameters are determined in an optimal way
by minimizing the ground state energy.\\

As a general conclusion, let us point out that the inclusion of the Pauli principle 
within the renormalized pn-QRPA, describing beta decay processes, reduces the ground state correlations 
by shifting the collapse of the standard pn-QRPA to larger values of the particle-particle strength, 
i.e. outside the physical region. Thus, at physical values of the particle-particle strength
one obtains a realistic description of the $2\nu\beta\beta$ transition probability.


\section{ The Green's function formalism}
\label{sec:GF}
\setcounter{equation}{0}
\renewcommand{\theequation}{4.\arabic{equation}}

As is well known, each eigenvalue problem has an analogous Green's function formulation. For convenience, let us start with the particle-particle channel, for which we consider the corresponding two-times propagator

\begin{equation}
  G^{t-t'}_{k_1k_2k'_1k'_2} = -i\langle 0|{\rm T}(c_{k_1}c_{k_2})_t(c^+_{k'_2}c^+_{k'_1})_{t'}|0\rangle~.
  \label{G-2}
\end{equation}
Here T is the time ordering operator,  $|0\rangle $ stands for the exact ground state and the time dependence of the fermion pair operators is given by the two body Hamiltonian

\begin{eqnarray}
  &H&= H_0 + V \equiv \nonumber\\
  \sum_{kk'}&e_{kk'}&c^+_kc_{k'} + \frac{1}{4}\sum_{k_1k_2k_3k_4}
  \bar v_{k_1k_2k_3k_4}c^+_{k_1}c^+_{k_2}c_{k_4}c_{k_3}~,
  \label{H}
\end{eqnarray}
where $e_{kk'}$ is the single-particle matrix comprising kinetic energy and external potential, and the antisymmetrised matrix element of the interaction is given after Eq.  (\ref{H-0-V}).

The two-body propagator obeys the following exact integral equation \cite{Duk98,Ole19}


\begin{eqnarray}
  &(& i \partial_t -\tilde e_{k_1} - \tilde e_{k_2})G^{t-t'}_{k_1k_2,k'_1k'_2} = N_{k_1k_2k'_1k'_2}\delta(t-t')\nonumber\\
  &+& \sum_{k_3k_4}\int dt_1 [K^{pp,0}\delta(t-t_1) + K^{pp,{\rm dyn.},t-t_1}]_{k_1k_2k_3k_4}\nonumber\\
  &&N^{pp-1}_{k_3k_4}G^{t_1-t'}_{k_3k_4k'_1k'_2}~,
  \label{ppBSE}
\end{eqnarray}
where with $\delta_{k_1k_2,k'_1k'_2} =\delta_{k_1k'_1}\delta_{k_2k'_2} - \delta_{k_1k'_2}\delta_{k_2k'_1}$
\begin{equation}
  N^{pp}_{k_1k_2k'_1k'_2} = \delta_{k_1k_2,k'_1k'_2}N^{pp}_{k_1k_2};~~~N^{pp}_{k_1k_2} =1-n_{k_1}-n_{k_2}~,
  \label{pp-occ}
\end{equation}
and we supposed that we work in the canonical basis, where the density matrix is diagonal, that is
\bea
\langle 0| c^+_{k_1}c_{k'_1}|0\rangle = \delta_{k_1k'_1}n_{k_1}~.
\eea
Furthermore, the s.p. energies in (\ref{ppBSE}) are given by

\begin{equation}
  \tilde e_k = e_k + V^{\rm MF}_k~,
  \label{HF-e}
\end{equation}
where the mean-field shift

\begin{equation}
  V_k^{{\rm MF}} = \langle 0|\{c_k,[H,c^+_k]\}|0\rangle = \sum_{k'}\bar v_{kk'kk'}n_{k'}
  \label{MF}
\end{equation}
is included
and where $\{..\}$ stands for the anticommutator. We assumed that mean-field energies and density matrix can be diagonalized simultaneously.
The integral kernel is given by

\begin{eqnarray}
  K^{pp}_{k_1k_2k'_1k'_2} &=& \langle [A_{k_1k_2},[V,A^+_{k'_1k'_2}]]\rangle\delta(t-t')\nonumber\\
  &+& (-i)\langle {\rm T}  J_{k_1k_2}(t) J^+_{k'_1k'_2}(t')\rangle_{\rm irr.}\nonumber\\
  &\equiv& K^{pp,0}_{k_1k_2k'_1k'_2}\delta(t-t') +  K^{pp,{\rm dyn.}, t-t'}_{k_1k_2k'_1k'_2}~,
  \label{kernel20}
\end{eqnarray}
where we abreviated

\begin{equation}
  A_{k_1k_2} = c_{k_1}c_{k_2}~,
\end{equation}
and

\begin{equation}
  J^{pp}_{k_1k_2}=[A_{k_1k_2}, V]~~ = j_{k_1}c_{k_2} +c_{k_1}j_{k_2}~,
  \label{pp-current}
\end{equation}
with
\begin{equation}
  j_k =
  [c_k,V] = \frac{1}{2}\sum_{k_2k_3k_4}\bar v_{kk_2k_3k_4}c^+_{k_2}c_{k_4}c_{k_3}~.
  \label{j-k}
\end{equation}

Please note that the $K$-matrix, which after Fourier transform depends only on one frequency can be interpreted as a self-energy for the motion of a fermion pair. As usual, the self-energy is split into a frequency independent, static part and a truly frequency dependent, dynamic part. The latter must be two-line irreducible, hence the index  'irr.'. 
In \cite{Sch19}
(see also \cite{Duk98, Ole19}) we named Eq.(\ref{ppBSE}) with the interaction kernel of Eq. (\ref{kernel20}) Dyson-Bethe-Salpeter Equation (Dyson-BSE). We want again to point out that this Dyson-BSE is an exact equation with a single frequency kernel. Its existence is not often recognized. In fact, it is entirely equivalent to the EOM method. 




\subsection{Static part of the BSE kernel}

Let us now discuss the $K^{pp,0}$ term of the BSE kernel. To establish an explicit form for $K^{pp.0}$, we have to evaluate the double commutator contained in the pair 
mean-field part of $ K^{pp,0}$
see Eq. (\ref{RPApair}) above. 
(It is actually the same as given in pp-SCRPA, section II.D).

\begin{eqnarray}
&& K^{pp,0}_{k_1k_2k'_1k'_2} = N^{pp}_{k_1k_2}\bar v_{k_1k_2k'_1k'_2}N^{pp}_{k'_1k'_2}
  \nonumber\\
  &+&\biggl \{\bigg [ \bigg (\frac{1}{2}\delta_{k_1k'_1}\bar v_{l_1k_2l_3l_4}C_{l_3l_4k'_2l1}
   + \bar v_{l_1k_2l_4k'_2}C_{l_4k_1l_1k'_1}\bigg)
\nn&-&(k_1 \leftrightarrow k_2)\bigg] - (k'_1 \leftrightarrow k'_2]\biggr \}~,
\label{Kpp-0}
\end{eqnarray}
where

\begin{equation}
  C_{k_1k_2k'_1k'_2} = \langle 0|c^+_{k'_1}c^+_{k'_2}c_{k_2}c_{k_1}|0\rangle -n_{k_1}n_{k_2}
  \delta_{k_1k_2,k'_1k'_2}~,
  \label{C-2}
\end{equation}
which is the fully correlated two-body, or cumulant, form of the density matrix.\\

We see that $K^{pp,0}$ involves, besides occupation numbers, static two-body correlation functions. They are of two types: there are single-line corrections with one of the two s.p. motions unaffected by the correlations (those with the Kronecker symbols) and there are exchange terms, where a two-body correlation is exchanged between the two particles. Since our starting point is a two-body propagator, a self-consistent scheme can be established. This is similar to the self-consistency involved with  the s.p. mean field, only here, naturally, two-body correlation functions have to be iterated rather than s.p. densities in the case of the s.p. mean field. We, therefore, call $K^{pp,0}$ the 'fermion pair mean field'. Of course, there appear also s.p. densities in $K^{pp,0}$ and we will later show how they can be consistently obtained from the s.p. Green's function.

A closer investigation of the exchange kernel, however, shows that the exchange is rather of the $ph$ type. At least to lowest order, that is to second order, the exchange is given by a static $ph$ exchange bubble. It is well known that this $ph$ exchange  screens the pairing force by almost a factor of two as has first been evaluated by Gorkov, Melik-Barkhudarov (GMB) \cite{Gor61,Str18}.
Actually for systems like the nuclear ones, where there are more than two species of fermions (that is, four species), the screening can eventually also become anti-screening \cite{Ram18,Pet02}.
Also GMB did not use strict second order, but replaced the vertices by the scattering length, that is the vertices have been dressed to $T$-matrices in the low-energy limit. Though this resummation can be understood easily by graphical analysis, how this can be derived more analytically will be discussed below in sect. IV.B. 

The first term on the r.h.s. of Eq. (\ref{Kpp-0}) is the usual two-body matrix element of the interaction modified with correlated occupation numbers (the standard particle-particle RPA as described in \cite{Rin80} uses HF occupation numbers). One can pre-sum this term, what leads to the so-called renormalized pp-RPA (see Sect.II.A for renormalized RPA)

\begin{eqnarray}
  G^{{\rm r-ppRPA}}_{k_1k_2 k'_1k'_2} &=& 
G^{0,{\rm r-ppRPA}}_{k_1k_2k'_1k'_2} 
\nn&+&
\sum_{k_3k_4} G^{0,{\rm r-ppRPA}}_{k_1k_2k_3k_4}\bar v_{k_3k_4k'_3k'_4}
G^{{\rm r-ppRPA}}_{k'_3k'_4k'_1k'_2}~,
\nn
  \label{r-ppRPA}
\end{eqnarray}
with

\begin{equation}
  G^{0,{\rm r-ppRPA}}_{k_1k_2k'_1k'_2} = \frac{1 -n_{k_1}-n_{k_2}}{\omega - {\tilde e}_{k_1}-{\tilde e}_{k_2}}
  \delta_{k_1k_2,k'_1k'_2}~.
  \label{0-r-ppRPA}
\end{equation}
where we omitted the infinitesimal imaginary part(s) in the denominator for brevity. Either one treats a discrete system where they are not needed or one has to add a $+i\eta$ in the case a retarded Green's function or split the propagator into advanced or retarded parts with alternating signs for the imaginary parts in the case of chronological propagators \cite{Fet71}.
The Dyson-BSE can then be written in the following way:

\begin{eqnarray}
  G_{k_1k_2k'_1k'_2} &=& G^{{\rm r-ppRPA}}_{k_1k_2 k'_1k'_2}
  \nn
  &&+ G^{{\rm r-ppRPA}}_{k_1k_2 k_3k_4}[ N^{-1}(K^{pp,0} - N\bar vN
    \nn
    &&+ K^{pp,{\rm dyn.}})N^{-1}]_{k_3k_4k'_3k'_4}G_{k'_3k'_4k'_1k'_2}~.\nonumber\\
  \label{dressed-BSE}
\end{eqnarray}
For brevity, we have replaced $N^{pp}$ by $N$ in this equation,
Let us stress again that all quantities in (\ref{dressed-BSE}) only depend on a single frequency. Single frequency two-particle propagators have also been considered in \cite{Kra86}.

\subsection{Dynamic part of the BSE kernel}

\begin{figure}\includegraphics[width=7cm]{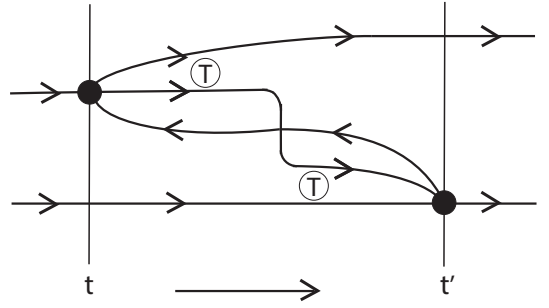}
  \caption{ Second order $T$-matrix approximation to the three-particle propagator contained in the 3p-1h correlation function. Together with the first order contribution contained in the uncorrelated 3p-1h propagator, this can be summed to one full $T$-matrix. Time flows from left (t) to right (t').}
  \label{GMB-T}
\end{figure}

Let us now discuss the time-dependent, {\it  dynamic} part $K^{\rm dyn.}$ of the interaction kernel

\begin{equation}
  K^{pp,\rm dyn.}_{k_1k_2k'_1k'_2} = -i \langle 0|{\rm T}J_{k_1k_2}(t)J^+_{k'_1k'_2}(t')|0\rangle_{\rm irr.}~.
  \label{K-dyn}
\end{equation}
From (\ref{pp-current}) we see that this expression involves four different contributions: two contributions contain the two interaction vertices on the same line and two contributions on opposite lines. The latter, therefore, contain exchange processes while the former are responsible for s.p. self-energy corrections. Approximating the 3p-1h propagator involved in (\ref{K-dyn}) by a product of a hole propagator and the three-body propagator in second order $T$-matrix approximation, we give a schematic graphical representation of the term in
Fig. \ref{GMB-T}. This illustrates how one can replace in the second order the screening term discussed above, the bare vertices by ladder $T$-matrices and then eventually by the scattering lengths as done in 
GMB \cite{Gor61,Str18}.
 We see that the exchange contributions are of the screening (or anti-screening) type 
whereas the other two contributions renormalize the s.p. by particle-vibration couplings. 
Of course, in general, all four lines are correlated.

As a matter of fact the 3p-1h propagator in (\ref{K-dyn}) lends itself to several ``natural'' approximations other than the one we just discussed. For example, instead of considering an uncorrelated ph-propagator exchange, one could take the ph-response function in standard RPA or in SCRPA as discussed in Sect. II. For applications with standard RPA, see \cite{Wen19,Urb20}.

\subsection{The ph-channel}

Since we see that $ph$ and $pp$ channels are coupled, we immediately also give the Dyson-BSE in the ph-channel 
\cite{Ole19}:


\begin{eqnarray}
  &&
  ( \omega -\tilde \epsilon_{k_1} + \tilde \epsilon_{k_2})R_{k_1k_2k'_1k'_2}(\omega) = N^{ph}_{k_1k_2k'_1k'_2}
  \nonumber \\ &&
  + \sum_{k_3k_4} [K^{ph,0}_{k_1k_2k_3k_4}  + K^{ph,\mathrm{dyn}}_{k_1k_2k_3k_4} (\omega)]N^{ph -1}_{k_3k_4} R_{k_3k_4k'_1k'_2}(\omega)
\nn
  \label{BSE}
\end{eqnarray}
with

\begin{equation}
  N^{ph}_{k_1k_2k'_1k'_2} = \delta_{k_1k'_1}\delta_{k_2k'_2}N^{ph}_{k_1k_2}\equiv \delta_{k_1k'_1}\delta_{k_2k'_2}(n_{k_2}-n_{k_1})
\end{equation}
and
the two-time response function defined by (with $k_1 \ne k_2$ and $k_1' \ne k_2'$)
\begin{equation}
  R_{k_1k_2k'_1k'_2}(t-t') =  -i \langle 0| \mathrm{T}\{ c^\dag_{k_2}(t)c_{k_1}(t)
  c^\dag_{k'_1}(t')c_{k_2'}(t') \} |0\rangle~.
  \label{response}
\end{equation}
The static part of the integral kernel is given by (again this is same expression as in the ph-SCRPA equation 
(\ref{cal-S}))
\begin{eqnarray}
  &&
  K^{ph,0}_{k_1k_2k_3k_4} = N^{ph}_{k_1k_2}\bar v_{k_1k_4k_2k_3}N^{ph}_{k_3k_4} +
  \nonumber\\ && \quad
  \Big[- \frac{1}{2}\sum_{ll'l''}(\delta_{k_2k_4} \bar v_{k_1ll'l''}C_{l'l''k_3l} +\delta_{k_1k_3}\bar v_{ll'k_2l''}C_{k_4l''ll'})
    \nonumber\\ && \quad
    + \sum_{ll'}(\bar v_{k_1lk_3l'}C_{k_4l'k_2l} + \bar v_{k_4lk_2l'}C_{k_1l'k_3l})
    \nonumber\\ && \quad
    - \frac{1}{2}\sum_{ll'}(\bar v_{k_1k_4ll'}C_{ll'k_2k_3} + \bar v_{ll'k_2k_3}C_{k_1k_4ll'})\Big]~,
  \label{K-ph0}
\end{eqnarray}
and the dynamic part

\begin{eqnarray}
  K^{ph,\rm dyn}_{k_1k_2k'_1k'_2}(t-t')&=&
  -i\langle 0 | \mathrm{T} \{  J^{ph}_{k_1k_2}(t)  J^{ph\dag}_{k'_1k'_2}(t')\} | 0 \rangle^\mathrm{irr}~.\nonumber\\
  \label{kernel2}
\end{eqnarray}
with

\begin{eqnarray}
  J^{ph}_{k_1k_2} &=&[c^{\dag}_{k_2}c_{k_1},V]
  =  c^\dag_{k_2}j_{k_1} + j^\dagger_{k_2}c_{k_1}~,
  \label{J-kk'}
\end{eqnarray}
As in the pp-channel, we can introduce a renormalized ph-propagator
\bea
 && R^{{\rm r-phRPA}}_{k_1k_2 k'_1k'_2} = R^{0,{\rm r}}_{k_1k_2k'_1k'_2} 
\nn&+&
\sum_{k_3k_4} R^{0,{\rm r}}_{k_1k_2k_3k_4} T_{k_3k'_4k_4k'_3}R^{{\rm r-phRPA}}_{k'_3k'_4 k'_1k'_2}~,
  \label{r-ppRPA1}
\eea
with

\begin{equation}
  R^{0,{\rm r}}_{k_1k_2k'_1k'_2} = \frac{n_{k_2}-n_{k_1}}{\omega - {\tilde e}_{k_1}+{\tilde e}_{k_2}}
  \delta_{k_1k'_1}\delta_{k_2k'_2}~,
  \label{1-r-ppRPA}
\end{equation}
where with repect to the infenitesimal imaginary parts in the denominator the same remarks hold as for the analog expression in the pp-case, see Eq.(\ref{O-r-ppRPA}),
and where we introduced the ladder $T$-matrix

\begin{eqnarray}
&&  T_{k_1k_4k_2k_3} = \bar v_{k_1k_4k_2k_3} + N^{-1}_{k_1k_2}
\nn&\times& \Big[
    - \frac{1}{2}\sum_{ll'}(\bar v_{k_1k_4ll'}C_{ll'k_2k_3}
+ \bar v_{ll'k_2k_3}C_{k_1k_4ll'})\Big]N^{-1}_{k_3k_4}~,
\nn
  \label{K-0}
\end{eqnarray}
which is important when dealing with systems with a hard-core potential.\\

\begin{figure}
  \includegraphics[width=7cm]{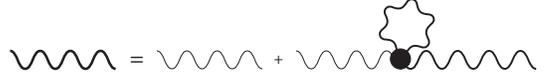}
  \caption{Mean-field ph-propagator with tad-pole self-interaction.}
  \label{Bo-tadpole}
\end{figure}

Keeping from the ph-kernel only the remaining instantaneous part, one arrives at a self-consistent mean-field equation for the ph-propagation

\begin{eqnarray}
&& R^{{\rm MF-RPA}}_{k_1k_2 k'_1k'_2} = R^{{\rm r-phRPA}}_{k_1k_2k'_1k'_2}
\nn&+&
\sum_{k_3k_4k'_3k'_4} R^{{\rm r-phRPA}}_{k_1k_2k_3k_4}[N^{-1}\tilde K^{ph,0}N^{-1} ]_{k_3k_4k'_3k'_4}
R^{{\rm MF-RPA}}_{k'_3k'_4 k'_1k'_2}~,
\nn
\label{mf-phRPA}
\end{eqnarray}
where $\tilde K^{ph,0}$ is the part of (\ref{K-ph0}), where only the second, third, fourth, and fifth terms on the r.h.s. are kept.\\
The eigenvalue form of this equation, given below, is the same as the self-consistent RPA (SCRPA) equation (\ref{SCRPA-gen}). A schematic graphical representation of 
Eq. (\ref{mf-phRPA}) is given in Fig. \ref{Bo-tadpole}.

We want to give the spectral representation of the response function, since it may be important for the following when we discuss the particle-vibration coupling (PVC).

\begin{eqnarray}
  R_{k_1k_2k'_1k'_2}(\omega) &=& \sum_{\nu}\frac{\begin{pmatrix}X^{\nu}\\Y^{\nu}\end{pmatrix}_{k_1k_2}(X^{\nu \dag}~~Y^{\nu ~~\dag})_{k'_1k'_2}}{\omega - \Omega_{\nu} + i\eta}\nonumber\\
  &-&\frac{\bigg [ \begin{pmatrix}Y^{\nu}\\X^{\nu}\end{pmatrix}_{k_1k_2}(Y^{\nu \dag} ~~X^{\nu \dag})_{k'_1k'_2}\bigg ]^*}{\omega + \Omega_{\nu} - i\eta}~.
\nn
  \label{spectral}
\end{eqnarray}
On the pole we get with (\ref{spectral}) again the  eigenvalue equation (SCRPA equation (\ref{SCRPA-gen})).

\subsection{The 
particle-vibration-coupling (PVC) approach}

The Dyson-BSEs in the $pp$- and $ph$-channels are a convenient starting point to study the particle-vibration coupling effects, which play a very important role for nuclear and condensed-matter systems
\cite{Lit18a,Lit19a}.
Let us consider the response function in the $ph$-channel, which is the quantity describing excitations caused by external perturbations without particle transfer. The model-independent spectral image of this function given by Eq. (\ref{spectral}) illustrates its direct connection to the observables: its poles correspond to the exact excitation energies and the residues to the respective transition amplitudes.

\begin{figure}
\includegraphics[width=8cm]{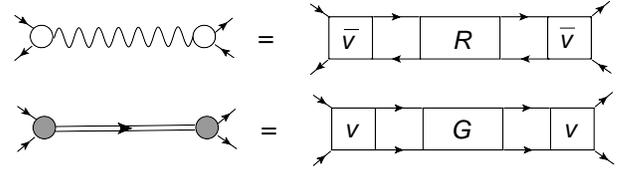}
\caption{The emergence of the phonon-exchange interaction: phonon vertices are denoted by circles, and their propagators by wavy lines and double lines for the normal and pairing phonons, respectively. Straight lines stand for the fermionic propagators. Rectangular blocks are associated with the particle-hole ($R$) and particle-particle ($G$) correlation functions, and the square blocks represent the bare interaction. Top: normal (particle-hole) phonon, bottom: pairing (particle-particle) phonon. See Refs. \cite{Lit19a,LitvinovaSchuck2019a} for more details.}
\label{Mapping}%
\end{figure}

\begin{figure}
\includegraphics[width=8cm]{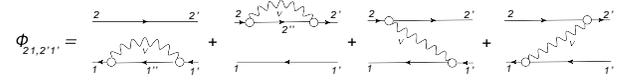}
\caption{$1p1h\otimes phonon$ approach to the dynamic kernel of the Dyson-BSE for the response function.}
\label{Phi_i}
\end{figure}

\begin{figure}
\includegraphics[width=8cm]{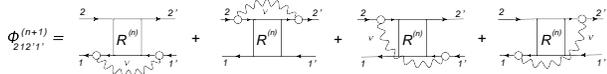}
\caption{Generic contributions to the dynamical kernel of the Dyson-BSE response function including one phonon and a full response function $R^{(n)}$, where the upper index $n$ indicates the number of phonons it contains. See also \cite{Rop80} where similar processes are discussed.}
\label{Phi_ii}
\end{figure}

Already in the static part of the interaction kernel for this response function, see, e.g., (\ref{cal-S}), one discovers an instantaneous exchange of a density-density correlation (vibration). The dynamical part of the kernel defined in Eq. (\ref{kernel2}), after calculating the commutators,  leads to four terms with the formally exact two-time two-particle-two-hole ($2p2h$) correlation function contracted with two matrix elements of the bare interaction in four possible ways.  The direct way of calculating this $2p2h$ correlation function would require the solution of the equation of motion for it, however, the latter equation contains even a higher-rank correlation function in its dynamical kernel. Obviously, such a procedure  generates a hierarchy of equations for increasing-rank correlation functions, making  the many-body problem practically intractable. A variety of approximations to the $2p2h$ correlation function, however, are known, and the choice of the specific approximation is dictated by the physical constraints.  The simplest approximation is given by the replacement of the correlated $2p2h$-propagator by the uncorrelated one. This comprises the second random phase approximation, which is quite appropriate for weakly-coupled systems. Other possibilities  are given by the cluster expansions, where one can limit the $2p2h$ propagator by the products of two $1p1h$ or $2p$ and $2h$-propagators. Here the options are (i) one  pair of fermions  propagates without interaction, while the other retains correlations and (ii) both fermionic pair configurations form correlated quasiboson  states. These states  can be associated with the phonons. After identifying the two-fermion correlation functions contracted with the interaction matrix elements with the phonon propagators and vertices, as shown in Fig. \ref{Mapping}, the interaction kernel of type (i) can be illustrated by Fig. \ref{Phi_i}, while the kernel of type (ii) with two correlation functions corresponds to the one displayed in Fig. \ref{Phi_ii}. Note that in these figures we show explicitly only the normal particle-hole phonons, however, the terms with the pairing phonons look analogously with the only difference that they carry couplings between $ph$ and $hp$ correlation functions in the Dyson-BSE. 
Note that in Figs. \ref{Phi_i} and \ref{Phi_ii} we retain the notation $\Phi$ for the dynamical kernels, which was used in the original works. It corresponds to various approximations to the kernel of Eq. (\ref{kernel2}).
The approaches to the dynamical kernels in the phonon-coupling form as those shown in Figs. \ref{Phi_i}, \ref{Phi_ii} are commonly known in the literature as (quasi)particle-phonon or (quasi)particle-vibration coupling (PVC) \cite{BohrMottelson1969,BohrMottelson1975,Broglia1976,BortignonBrogliaBesEtAl1977,BertschBortignonBroglia1983,
Soloviev1992,Bortignon1978,Bortignon1981a,MahauxBortignonBrogliaEtAl1985,Bortignon1986,Bortignon1997,ColoBortignon2001,
Tselyaev1989,KamerdzhievTertychnyiTselyaev1997,Ponomarev2001,Ponomarev1999b,LoIudice2012,Litvinova2016,
RobinLitvinova2016,RobinLitvinova2018,Robin2019,Bor19}.
It is worth mentioning that very similar expressions as those shown in Fig. \ref{Phi_i} have also been used in chemical physics for the purpose of screening the bare electron-electron force \cite{Rom09}. In addition explicit expressions corresponding to Fig. \ref{Phi_i} can also be found in \cite{Ole19}, Eq. (58).

Similar studies were performed for the $pp$-channel \cite{LitvinovaSchuck2019a}, where analogous treatments of the static and dynamical kernels were suggested.
Considering together the EOM's for both $ph$ and $pp$ correlation functions and the factorizations discussed above allows for a truncation of the EOM series on the two-body level.  Thus, the many-body problem is reduced to the closed system of coupled equations for one-fermion and two-fermion propagators in the $pp$ and $ph$-channels. The phonon vertices and propagators, which enter  the dynamical kernels of these equations, are, in general, model-independent and, ideally, should contain the same correlation functions, which constitute the main variables of these EOM's. As this system of equations is essentially non-linear, it should be solved by iterations with a good initial guess for the sought-for correlation functions. Those could be, in principle, the correlation functions calculated with the only static kernel. However, in strongly-coupled systems, such as atomic nuclei, calculations of the correlation functions with bare interactions and only static kernels can be quite far from the realistic ones \cite{PapakonstantinouRoth2009}. This raises the question about convergence of this approach. Another possibility could be employing the correlation functions calculated in  RPA-like approaches with one of the well-known effective interactions at the entrance point. That could potentially provide better convergence, however, this strategy has not been explored yet. 

\begin{figure}
\includegraphics[width=8cm]{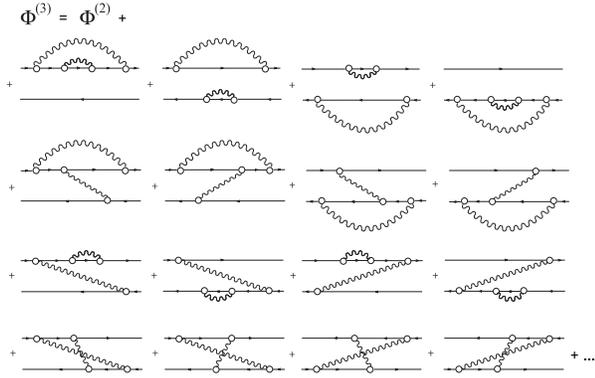}
\caption{The simplest time-ordered $1p1h\otimes 2phonon$, or $2q\otimes 2phonon$ (two quasiparticles coupled to two phonons), diagrams taken into account in 
the dynamical kernel. The ellipsis stands for multiple PVC exchange and self-energy contributions as well as for the correlated particle-hole configurations in the internal particle-hole propagators.}
\label{3loop}
\end{figure}

PVC studies of the $ph$-response in nuclear systems are dominated so far by those based on effective interactions. While in the older works the phonon characteristics were extracted from experimental data, in the last couple of decades the trend has been shifted to fully self-consistent calculations \cite{LitvinovaRingTselyaev2007,LitvinovaRingTselyaev2008,LitvinovaRingTselyaev2010,LitvinovaRingTselyaev2013,Tselyaev2016,Tselyaev2018,NiuNiuColoEtAl2015,Niu2018,Shen2020}. The density functional theory (DFT) allows for calculations of the phonon vertices and energies within the self-consistent RPA, that does not imply any non-linearity as it is confined by the static kernel defined by effective interactions. The advantage of this approach is that, although it is quite simple, the phonon characteristics come out reasonably close to the realistic ones, at least for the most important low-energy phonons. The phonons computed in this way are then used to calculate the dynamical kernels with the subsequent solution of the Dyson-BSE for the full response function. Strictly speaking, this response function with all its components from various channels should be recycled in the dynamical kernel until convergence is achieved.  This procedure generates higher-order configurations as it is illustrated in Fig. \ref{Phi_ii} for the case of $n=3$, where $n$ stands for the total number of both correlated and non-correlated $ph$-pairs of fermions.
The latter approach to the dynamical kernel was first proposed in \cite{Litvinova2015} and implemented numerically in Ref. \cite{Lit19a}. The higher-order correlations generated in this way were found to be necessary for nuclear structure calculations, if a certain level of accuracy is required. For instance, astrophysical applications to r-process nucleosynthesis and to core-collapse supernova require extraction of the radiative neutron capture, beta decay and electron capture rates from the calculated spectra of excitations. These rates are extracted mostly from the low-energy fractions of the dipole, Gamow-Teller and spin-dipole strength distributions, which are very sensitive to the higher-order correlations. It turned out that taking into account only (correlated) 2p-2h configurations, one can not always reproduce the observed spectra of excitations with the required accuracy. Although the results are greatly improved as compared to the RPA calculations and while the gross features of the giant resonances and soft modes can be approximately described by the approaches like $1p1h\otimes phonon$ or $phonon\otimes phonon$, these approaches call for further upgrades in terms of higher-order correlations.

The last major upgrades made recently in Ref. \cite{Lit19a} and Ref. \cite{Robin2019} 
included explicitly the $1p1h\otimes 2phonon$ configurations.
More precisely, Ref. \cite{Lit19a} 
included the $2q\otimes 2phonon$ (two quasiparticles coupled to two phonons) configurations, 
where 'q' stands for 'quasiparticle' in the Bogoliubov's sense, while Ref. \cite{Robin2019} analyzed
 the complex ground state correlations caused by the time-reversed PVC loops (GSC-PVC). The former allowed for significant improvements of the description of the nuclear dipole response in both high and low-energy sectors, while the latter demonstrated how the GSC-PVC's bring the Gamow-Teller strength distribution to a considerably better agreement with data for both proton-neutron and neutron-proton branches.

\begin{figure}
\includegraphics[width=8cm]{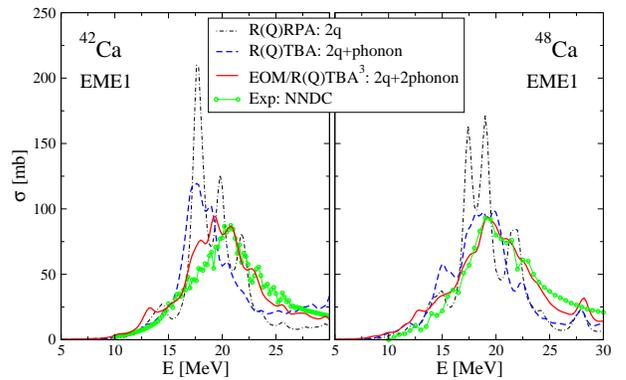}
\caption{Giant dipole resonance in $^{42,48}$Ca nuclei calculated within R(Q)RPA, R(Q)TBA and EOM/R(Q)TBA$^3$, in comparison to experimental data of Ref. \cite{Erokhova2003}. See text and Ref. \cite{Lit19a} for details.}
\label{GDR}
\end{figure}

As an illustration, we display in Fig. \ref{GDR} the results of the calculations performed in Ref. \cite{Lit19a} for the nuclear dipole response with the dynamical kernels of increasing complexity. The dipole response is given in the form of the full photoabsorption cross section for the three approaches: the relativistic (quasiparticle) random phase approximation (R(Q)RPA), the relativistic (quasiparticle) time blocking approximation (R(Q)TBA) with $2q\otimes phonon$ configurations and the upgraded R(Q)TBA, where the $2q\otimes 2phonon$ configurations were included (EOM/R(Q)TBA$^3$). It turned out that even the use of a fairly large model space of the $2q\otimes phonon$ configurations underestimates the experimental value of the width of the giant dipole resonance (GDR).  Another problem can be identified on the high-energy shoulder of the GDR above its centroid, where the cross section is systematically underestimated. A similar situation was reported in Ref. \cite{Tselyaev2016}, which revealed a systematic downshift of the non-relativistic QTBA strength distributions, with respect to the RPA ones, in the calculations with various Skyrme forces.  In the EOM/RQTBA$^3$ calculations with more complex $2q\otimes 2phonon$ configurations we see that adding these configurations can resolve those problems. Indeed, Fig. \ref{GDR} shows that these higher-complexity configurations present in the EOM/RQTBA$^3$ approach induce a stronger fragmentation of the GDR and reinforce spreading toward both higher and lower energies. A visible shift of the main peak toward higher energies can be associated with the appearance of the new higher-energy complex configurations, which are responsible for higher-energy poles in the resulting response functions. In this way, the energy balance of the overall strength distribution is rearranged, however, with the conservation of the energy-weighted sum rule. 
The low-energy tail of the GDR was also studied in detail in Ref. \cite{Lit19a}, which showed that the new correlations bring some important improvements to the description of experimental data also in the low-energy sector.

As already mentioned, in the PVC studies discussed above the static part of the interaction kernel is expressed by phenomenological effective  interactions. They are adjusted to nuclear masses and radii on the Hartree or Hartree-Fock level, which means that 
the effects from dynamical kernels are implicitly included in the parameters of these interactions. The latter indicates that adding the dynamical kernel to the EOM's should be supplemented by a proper subtraction of the double-counting effects. This procedure was developed and discussed in detail in Ref. \cite{Tselyaev2013}, and since then it is systematically applied in the PVC approaches based on the effective interactions. The fully microscopic approach based on one of the bare nucleon-nucleon interactions remains a task for the future.

More on beyond mean-field and RPA approaches will be discussed below in Sects. VII and VIII.
Presently several other groups are also working on multi-configurational extensions of single-particle and RPA states with interesting developments \cite{Shen2020,Pap10,Rai19,Gra20}. A promising approach on improving on the Pauli principle could be the concept of the generalized optical theorem elaborated in \cite{Ada89,Dan94}.
The whole is a very active field of nuclear physics.

\subsection{Application of the Green's Function Approach to the pairing  model at finite temperature}

In order to apply the Green's Function technique for the
particle-particle problem at finite temperature to the pairing Hamiltonian (\ref{HPF}), we define the following set of two-body Matsubara GF's,
see Sect. II.D and Sect. III.A for definitions
\bea
G^{\t}_{ji}=-\la {\rm T}_{\t}\ov{P}_j(\t)\ov{P}^{\dag}_i(0)\ra~,
\eea
where ${\rm T}_{\tau}$ is the time ordering operator for imaginary times \cite{Fet71} and
\bea
\ov{P}_j=\frac{P_j}{\sqrt{\la|1-N_j|\ra}}~.
\eea
By using the static kernel of the Dyson-BSE defined
in (\ref{Kpp-0}) one obtains
\bea
i\w_nG^{SCRPA}_{ji}=\d_{ji}+\sum_k{\cal H}^{(0)}_{jk}G^{SCRPA}_{ki}~,
\eea
with
\bea
    {\cal H}^{(0)}_{jk}&=&2\d_{jk}
    \left(e_j+\frac{G}{\la 1-N_j\ra}\sum_{j'}\la P^{\dag}_jP_{j'}\ra
    \right)
\nn &-&G\frac{\la (1-N_j)(1-N_k)\ra}{\sqrt{|\la 1-N_j\ra\la 1-N_k\ra|}}~.
    \eea
    As remarked in Sect. \ref{sec:pair}, a good approximation of the
    two-body correlation function in the above relation
    is given by the factorisation
    $\langle M_iM_j\rangle \simeq \langle M_i\rangle \langle M_j\rangle$,
    One then obtains as solution for the particle-particle Green's function

    \bea
    \label{GSCRPA}
&&  G^{SCRPA}_{ji}=\d_{ji}\frac{1}{z-C_j}
\nn&-&
    \frac{G\sqrt{|D_jD_i|}}{(z-C_j)(z-C_i)}
    \left[1+G\sum_k\frac{D_k}{z-C_k}\right]^{-1}~,
    \eea
    where
    \bea
    z&=&i\w_n
    \nn
    D_i&=&\la 1-N_i\ra
    \nn
    C_j&=&2\left(e_j-Gn_j+\frac{G}{D_j}\sum_{j'\neq j}\la P^{\dag}_jP_{j'}\ra
    \right)~.
    \eea
    By equating to zero the denominator of (\ref{GSCRPA}) one obtains
    the excitation spectrum. Knowing the poles $E_{\m}$ of the GF one obtains
    its spectral representation with the corresponding residua.
    They allow us to obtain also the two-body correlation functions \cite{Fet71}
    in terms of RPA amplitudes for particle (p) and hole (h) states,
    defined as follows
    \bea
    X^{\m}_{p}&=&\frac{\sqrt{ GD_p}}{|C_p|-E_{\m}}F_{\m}~,~~~
    Y^{\m}_{h} = \frac{\sqrt{-GD_h}}{|C_h|+E_{\m}}F_{\m}~,
    \nn
    X^{\m}_{h}&=&\frac{\sqrt{-GD_h}}{|C_h|-E_{\m}}F_{\m}~,~~~
    Y^{\m}_{p} = \frac{\sqrt{ GD_p}}{|C_p|+E_{\m}}F_{\m}~,
    \eea
    with
    \bea
    F^{-2}_{\m}=\frac{\pa}{\pa z}\left[1+G\sum_k\frac{D_k}{z-C_k}
      \right]_{z=E_{\m}}~.
    \eea

    In order to close SCRPA equations it is necessary to determine the
    occupation numbers $n_k = \la c^{\dag}_kc_k\ra$. To achieve this
    one has to find a single-particle GF, $G^{\t}_k$, consistent with
    the SCRPA scheme. The single-particle self-energy $\Sigma_k$
    has the exact
    representation in terms of the two body T-matrix (see e.g. \cite{Duk98})
    and then an appropriate approximation for the $G^{\t}_k$ can be obtained.
    It consists in using the self-energy $\wtl{\Sigma}_k$ calculated through
    the T-matrix found by SCRPA. As the relation between T-matrix and
    the sum of all irreducible Feynman graphs in the $pp$ channel is also
    known, one obtains
    \bea
    \label{M}
    \wtl{\Sigma}_k=G\sum_{k_1k_2}G^{0(\t'_1-\t_1)}_{\ov{k}}
    G^{(\t'_1-\t_1)}_{k_1k_2}\wtl{\cal H}_{k_2k}^{(0)}~,
    \eea
    where
    \bea
    \wtl{\cal H}_{kk'}^{(0)}&=&{\cal H}_{kk'}^{(0)}-2\d_{kk'}\e_k
    \nn
    \e_k&=&e_k-Gf_k\frac{f_k}{n_k}\frac{D_k}{D^0_k}
    \eea
    with $f_k=\frac{1}{1+e^{\e_k\b}}$ being the Fermi-Dirac distribution and
    $D^0_k=1-f_k-f_{\ov{k}}$.
    The expectation value of the Hamiltonian can be computed
    by using the two-body GF's and, alternatively, the following
    relation involving the single-particle Green's function
    \bea
    \la H\ra=-{\oh}\lim_{\t'-\t\rightarrow 0^+}\sum_k
    \left(\frac{\pa}{\pa\t}-e_k\right)
    \left(G^{(\t-\t')}_k+G^{(\t-\t')}_{\ov{k}}\right)~.
\nn
    \eea
    In order to get the same expressions with both procedures, in Ref. \cite{Urb14}
    it was shown that one has to expand the single-particle GF to first order
    in the self-energy (\ref{M}), i.e. in the above relation one has to use (see also discussion below in Sect.V.B)
    \bea
    G_k&=&G^0_k+G^0_k\Sigma^{SCRPA}_kG^0_k
    \nn
    \Sigma^{SCRPA}_k&=&\frac{\sqrt{D_k}}{D^0_k}\wtl{\Sigma}_k~.
    \eea
    In principle, this is the same procedure as used by Nozi\`eres and Schmitt-Rink
    \cite{Noz85}, only extended to SCppRPA, see also Sect. V.A.

    Finally, one gets a closed expression for the single-particle self-energy
    $\Sigma_p^{SCRPA}$ and, therefore, for the occupation numbers by using
    this GF
    \bea
    n_k=\la c^{\dag}_kc_k\ra=\lim_{\t\rightarrow 0^-}G^{\t}_k~.
    \eea
    \begin{figure}[ht]
      \begin{center}
        \includegraphics[width=8cm]{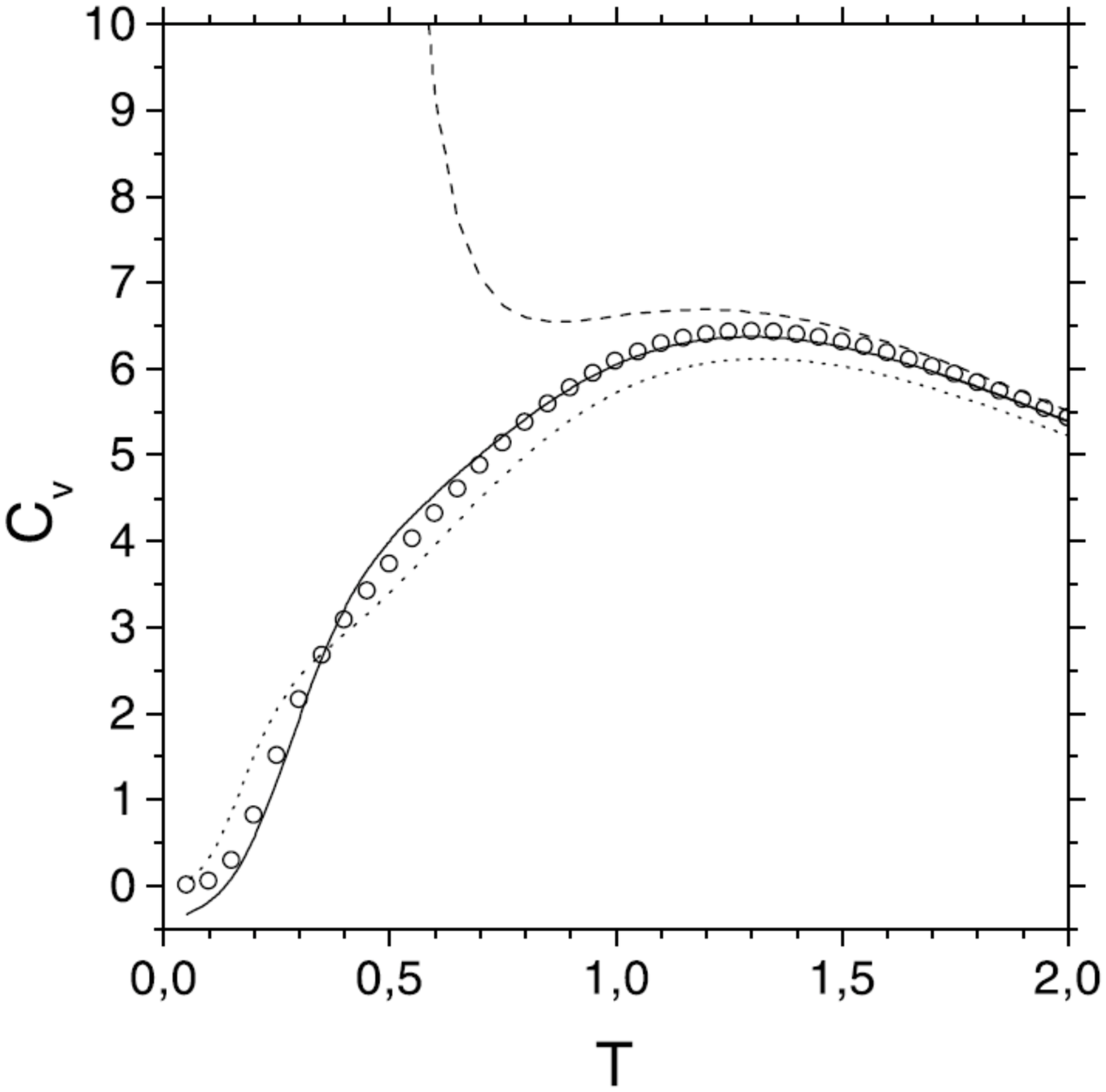}
        \caption{The heat capacity $C_v$ as a function of temperature
          for $\W=N=10$ and $G=0.4$. The exact results - open circles,
          the BCS results - dotted line,  the TRPA results - dashed line
          and the TSCRPA results - solid line.}
        \label{sto09}
      \end{center}
    \end{figure}
    In order to check the accuracy of this theory, in Ref. \cite{Sto03} 
    $\la H\ra$ is plotted as a function of particle number $N$ and
    temperature $T$, and the exact results are compared with the standard
    thermal RPA (TRPA), thermal mean field (TMF) and thermal SCRPA (TSCRPA).
    One concludes that, when the number of levels $\W$ increases
    (together with the number of particles $N$), the agreement with
    the exact solution improves, so that for $N=\W=10$ the
    TSCRPA results practically coincide with the exact ones.
    Concerning the excitation energies, one also obtains a very
    good agreement between exact and SCRPA 
    versus the coupling constant at zero temperature,
    at variance with the lowest RPA mode, which collapses when crossing
    the critical strength from normal to the superfluid phase at
    $G_{cr}\approx 0.33$.
    A similar behaviour is present for finite temperatures: the TSCRPA
    reproducing the exact results, while the TRPA values giving
    qualitatively wrong results.

    In order to analyse the behaviour of the system near the critical
    point, in Fig. \ref{sto09} \cite{Sto03} we plot the heat capacity 
    \bea
    C_v=\frac{\pa \la H\ra}{\pa T}
    \eea
   as a function
    of temperature for $G=0.4$. One sees that TSCRPA values (solid curve)
    follow the exact results (open circles), while TRPA values
    are in a total disagreement around the critical temperature.
    It is remarkable that the TSCRPA results are accurate down to practically
    zero temperature, in spite of the fact that within the standard BCS theory
    one enters the superfluid regime. A quasiparticle formulation
    of SCRPA will only be necessary for stronger $G$ values driving
    the system more deeply into the symmetry broken phase.

    \begin{figure}[ht]
      \begin{center}
        \includegraphics[width=8cm]{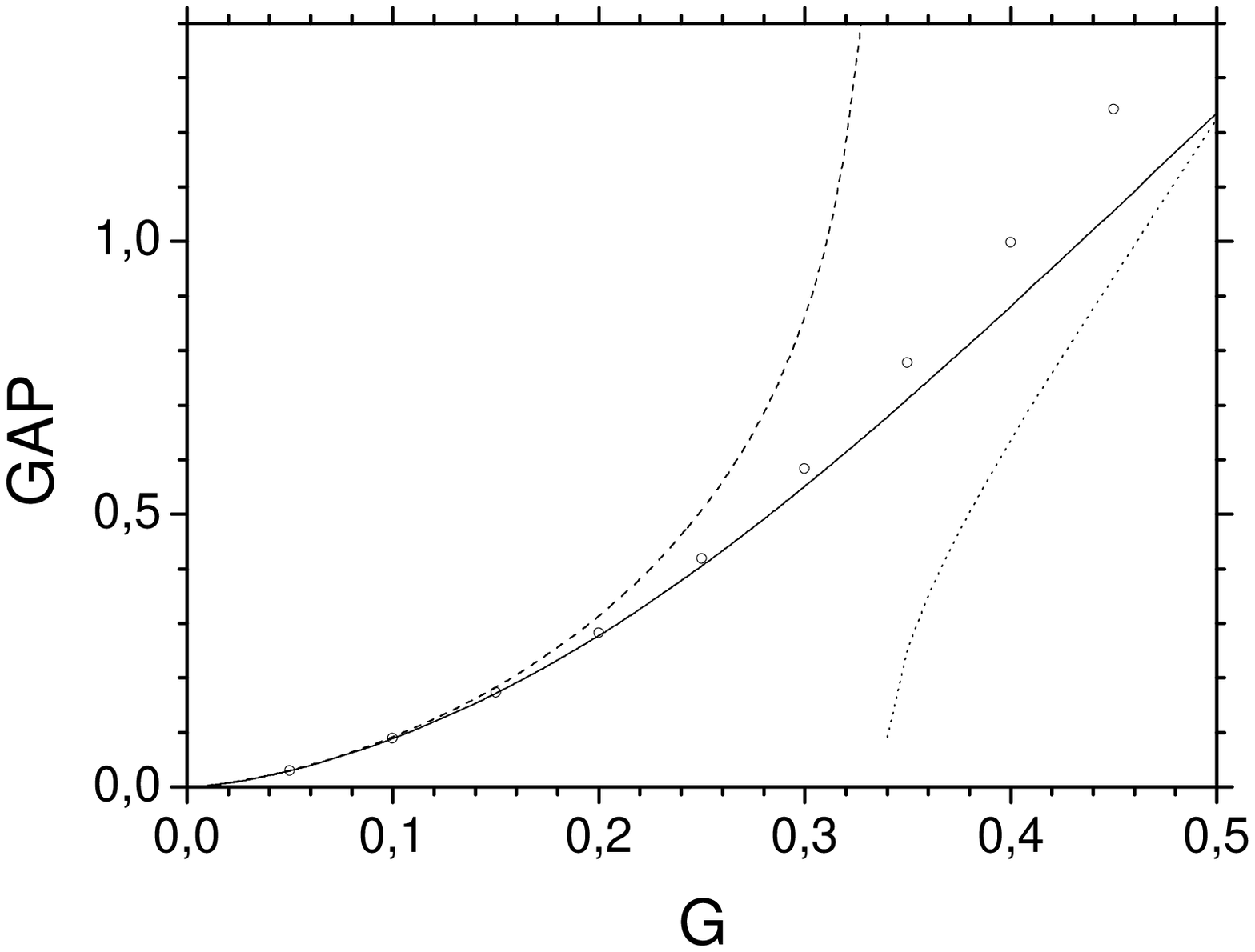}
        \caption{The effective gap $\D$ as a function of the interaction constant
          $G$ for $\W=N=10$ and $T=0$. Same line and symbol assignments as in Fig. \ref{sto09} are applied.}
        \label{sto11}
      \end{center}
    \end{figure}
    It is also instructive to consider the spectral function given in
    \cite{Sto03}.
    The dependence of this function upon $\w$ for different temperatures
    shows
that the distance between the two quasiparticle peaks around Fermi
    energy ($\w=0$) increases by decreasing the temperature. This reminds the
    {\it pseudo gap} feature which will be  discussed in Sect. V.D. Apparently,
    it is quite a generic property that pair correlations diminish the density
    of levels around the Fermi level whereas $ph$ correlations
    give rise to an increase.
    \begin{figure}[ht]
      \begin{center}
        \includegraphics[width=8cm]{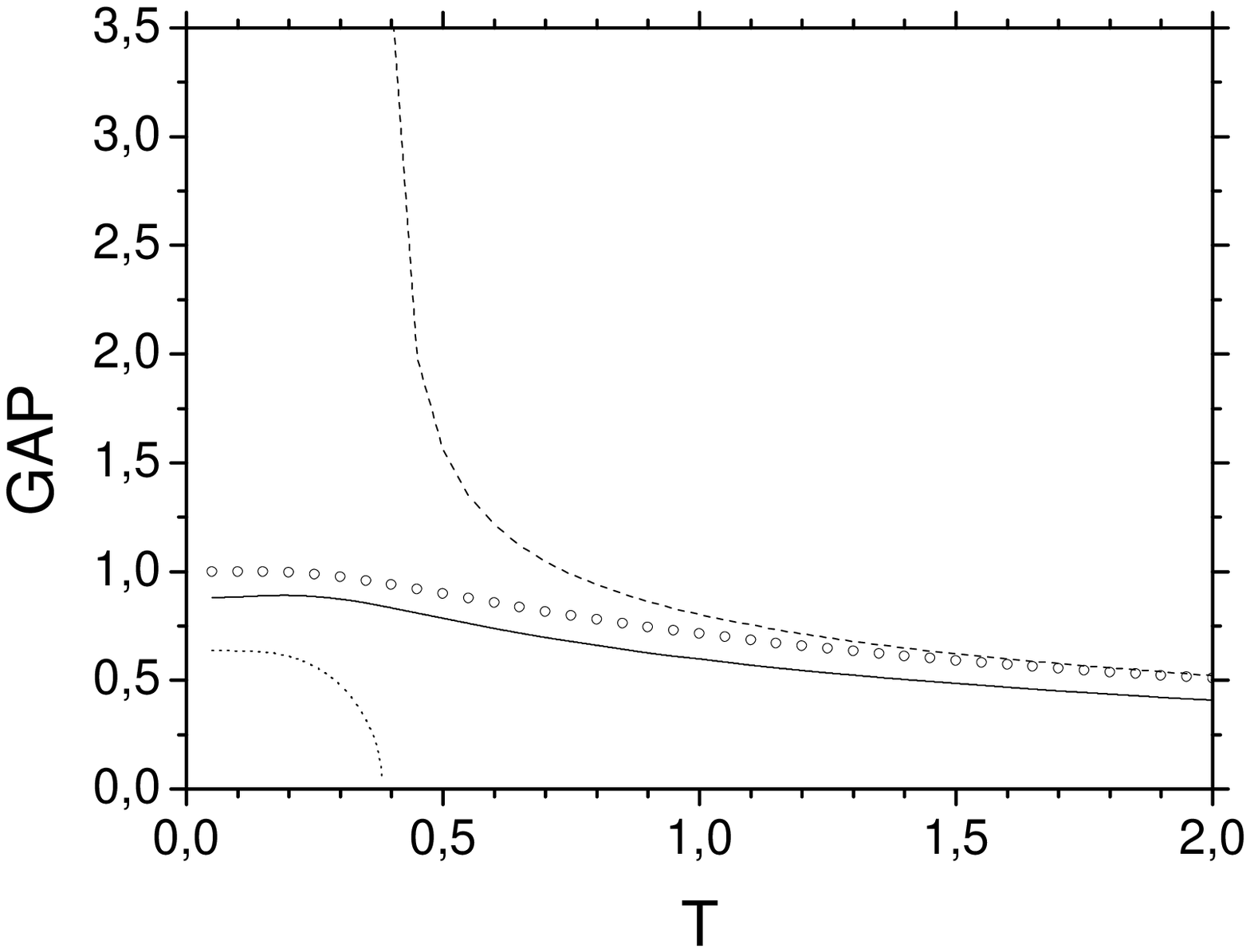}
        \caption{The effective gap $\D$ as a function of temperature
          calculated for $\W=N=10$ and $G=0.4$. Same line and symbol assignments as in Fig. \ref{sto09} are applied.}
        \label{sto12}
      \end{center}
    \end{figure}
    In order to make the temperature dependence of the gap more transparent,
    it is useful to introduce an effective gap as follows
    \bea
    \D=G\sqrt{\sum_{ik}\left(\la P^{\dag}_iP_k\ra-\la c^{\dag}_ic_k\ra
      \la c^{\dag}_{\ov{i}}c_{\ov{k}}\ra\right)}~.
    \eea
    In the BCS approximation the effective gap coincides with the grand
    canonical BCS gap, i.e.
    \bea
    \D_0=G\sum_k\la P_k\ra_{BCS}~.
    \eea
    The dependencies of the effective gap on the interaction strength $G$
    at zero temperature and on temperature $T$ for $G=0.4$ are shown
    in Figs. \ref{sto11} and \ref{sto12}, respectively \cite{Sto03}.
    The SCRPA results give a very good description in both cases
    compared with the exact values. It is clearly seen that the SCRPA
    and exact calculations do not display the phase transition
    at the point where BCS gap vanishes. Anyway, we notice in Fig. \ref{sto11}
    that SCRPA results deteriorate for values of $G$ well beyond
    the critical value. We stress again that in this regime a quasiparticle
    generalisation of the SCRPA is necessary because one enters deeply into the
    superfluid region.

   In conclusion of this section, we see that SCRPA also works very well at finite temperature. To perform similar studies for other models like the Lipkin and Hubbard models remains a task for the future.

\section{Single-particle Green's function, Dyson equation, and applications 
to thermodynamic properties of nuclear matter}
\label{sec:Dys}
\setcounter{equation}{0}
\renewcommand{\theequation}{5.\arabic{equation}}

    So far we have considered fermion pair problems with SCRPA employing secular equations or Dyson-Bethe-Salpeter equations. Both formulations are equivalent at zero temperature, but the Dyson-BSE has an advantage at finite temperature. However, together with the two-body problems one must also consider the single-particle Green's function and its Dyson equation. At zero temperature we use the chronological Green's function

    \begin{equation}
      G^{t-t'}_{kk'} = -i\langle T c_k(t)c^{\dag}_{k'}(t')\rangle~,
      \label{spGF}
    \end{equation}
    where we used the same notation as for the two-particle GF in Sect. IV. 
    The s.p. Green's function (\ref{spGF}) obeys the Dyson equation

    \be
       [i\partial_t - e_k]G_{kk'} =\delta_{kk'}\delta(t-t') + 
\sum_{k_1} \int dt_1 \Sigma^{t-t_1}_{kk_1}G^{t_1-t'}_{k_1k'}~,
       \label{D-eq}
       \ee
       where $\Sigma $ is the single-particle self-energy and $e_k$ is a  s.p. energy supposedly in an external field. The self-energy can be split into two terms

       \begin{equation}
         \Sigma_{kk'}^{t-t'} = V_{kk'}^{{\rm MF}}\delta(t-t') 
-i\langle 0|{\rm T}j_k(t)j_{k'}^+(t' )|0\rangle_{{\rm irr.}}~,
         \label{self4}
       \end{equation}
       with
       \begin{equation}
         j_k =
         [c_k,V] = \frac{1}{2}\sum_{k_2k_3k_4}\bar v_{kk_2k_3k_4}c^+_{k_2}c_{k_4}c_{k_3}~,
         \label{j-k1}
       \end{equation}
       and  where the index 'irr.'  indicates that the corresponding correlation function should be one-line irreducible.

       The time-dependent part of the self-energy contains the following 3-body propagator:

       \bea
&&  G_{k_2k_3k_4;k'_2k'_3k'_4}^{t-t'} =
\nn&& -i \langle 0|{\rm T}(:c^{\dag}_{k_2}c_{k_4}c_{k_3}:)_t
(:c^{\dag}_{k'_3}c^{\dag}_{k'_4}c_{k'_2}:)_{t'}|0\rangle_{\rm irr.}~,
       \eea
       where $:c^{\dag}_{k_2}c_{k_4}c_{k_3}: = c^{\dag}_{k_2}c_{k_4}c_{k_3} - [c_{k_3}\langle 0|c^{\dag}_{k_2}c_{k_4}|0\rangle -(k_3 \leftrightarrow k_4)]$. This is to avoid self-contractions at the same time-level which is to be eliminated.
       We want to establish an integral equation for this propagator. As usual, we employ the equation of motion and approximate the integral kernel in first approximation by the static part. We obtain (since the following equation will have an irreducible kernel, we will drop the index 'irr.')

       \begin{eqnarray}
         (i\partial_t &-&e_{k_3}-e_{k_4} + e_{k_2})G_{k_2k_3k_4;k'_2k'_3k'_4}^{t-t'} = N_{k_2k_3k_4;k'_2k'_3k'_4}\nonumber\\
         &+& \sum\limits_{l_2l_3l_4,l'_2l'_3l'_4} K^0_{k_2k_3k_4;l_2l_3l_4}N^{-1}_{l_2l_3l_4;l'_2l'_3l'_4}G_{l'_2l'_3l'_4;k'_2k'_3k'_4}^{t-t'}~,\nonumber\\
         \label{2p1h-prop}
       \end{eqnarray}
       where

       \begin{equation}
         N_{k_2k_3k_4;k'_2k'_3k'_4} = \langle 0|\{\{:c^{\dag}_{k_2}c_{k_4}c_{k_3}:,:c^{\dag}_{k'_3}c^{\dag}_{k'_4}c_{k'_2}:\}|0\rangle~,
         \label{3-body-norm}
       \end{equation}
       and

       \begin{equation}
         K^0_{k_2k_3k_4;k'_2k'_3k'_4} = \langle 0|\{:c^{\dag}_{k_2}c_{k_4}c_{k_3}:,[V,:c^{\dag}_{k'_3}c^{\dag}_{k'_4}c_{k'_2}:]\}|0\rangle~.
         \label{3b-K}
       \end{equation}
       The approach outlined above is the Green's function equivalent to Sect.II.G, see also \cite{Toh13}.
       A slightly subtle point is that, if we had started the EOM in differentiating with respect to $t'$ instead of $t$, we would not necessarily get the the same integral kernel. Of course, this should not be so. This asymmetry is naturally eliminated by demanding that the time derivative of the two-particle density matrix must vanish at equilibrium. It turns out that this equation is just the difference of the present kernel with its hermitian conjugate form. See for this also the Sect. VII on second RPA equations (\ref{stationarity2+3}) . 
       If we want to be consistent with our ground state wave function (\ref{Z}), we have to restrict the indices of $K^0$ to particles or holes like in Sect. II.D. Evaluating the norm matrix and the interaction kernel in the HF ground state leads to the following eigenvalue equation for the three-fermion transition amplitudes $\chi^{\nu}$ \cite{Sch73a}


       \bea
&&(\omega_{\nu}-\tilde e_n-\tilde e_m+\tilde e_l) \chi^{\nu}_{l,nm}
= (n^0_l-n^0_n)\bar v_{nrlk}\chi^{\nu}_{r,mk} 
\nn&-& (n^0_l - n^0_m){\bar v}_{mrlk}\chi^{\nu}_{r,nk} 
+ \frac{1}{2}(\bar n^0_n\bar n^0_m - n^0_nn^0_m)\bar v_{nmpq}\chi^{\nu}_{l,pq}~,
\nn
       \label{2p-h-HF}
       \eea
       where $\bar n = 1-n$ and repeated indices shall be summed over. The $n^0_i$ are the mean-field occupancies. They can be replaced by the correlated occupations, similar to the renormalized RPA. However, the use of the HF ground state in this equation is inconsistent, because it does not correspond to a linearized time-dependent equation with the time-dependent HF wave function. This fact leads to some difficulties which have been pointed out in \cite{Rij96}.
It is then better to split the equation into two independent TDA equations, one for 2p-1h amplitudes, and one for 2h-1p ones. This is also the case if we want to be consistent with the $|Z\rangle$ ground state wave function (\ref{Z}). 
       In this case we can write the Dyson equation in the following form, see \cite{Toh13}

       \begin{eqnarray}
         G^{\omega}_{kk'} = G^0_k\delta_{kk'} + G^0_k\sum_{k_1}\Sigma^{\omega}_{kk_1}G^{\omega}_{k_1k'}~,
         \label{Dyson-z}
       \end{eqnarray}
       where
       \begin{eqnarray}
         G^0_k = \frac{1 - n_k}{\omega - \tilde e_k + i \eta}+
         \frac{n_k}{\omega - \tilde e_k - i \eta}
         \label{g0}
       \end{eqnarray}
       is the  Green's function which sums up the static mean-field contribution with the occupation numbers $n_k$ containing in principle ground state correlations.
       The self-energy is given by
\begin{widetext}
       \begin{eqnarray}
         \label{Sigma-Z}
         \Sigma_{kk'} &=& \sum_{\rho hh'p_1p_2p'_1p'_2} \langle kh|v|p_1p_2\rangle
         \frac{\tilde V^{\rho}_{p_1p_2:h}\tilde V^{\rho *}_{p_1'p_2':h'}}{\omega - (E_{\rho}^{N+1}-E_0^N)  + i\eta}
       \langle p_1'p_2'|v|p'k'\rangle
\nn&+&
         \sum_{\alpha pp'h_1h_2h'_1h'_2}\langle kp|v|h_1h_2\rangle
         \frac{\tilde V^{\alpha}_{h_1h_2:p}\tilde V^{\alpha *}_{h_1'h_2':p'}}{\omega - (E_0^N-E_{\alpha}^{N-1}) - i\eta}
        \langle h_1'h_2'|v|p'k'\rangle~,
         \end{eqnarray}
\end{widetext}
       where $\tilde V^{\alpha,\rho}$  are the correctly normalised 2p-1h and 2h-1p amplitudes and  $\lambda^+= E^{N+1}_{\rho}-E^N_0,~\lambda_- = E^N_0-E^{N-1}_{\alpha}$
       are the eigenvalues obtained from the corresponding equations (\ref{odd-eigen}). In the case, where we strictly work with the odd-RPA corresponding
       to the ground state Eq. (\ref{Z}), the coupled system of particle and hole propagators in above Dyson equation
       decouples into two separate Dyson equations, one for the particles
       and one for the holes.
       It may certainly be rewarding to work with an approach which is based on a correlated ground state wave function as in (\ref{Z}). Applications of the general formalism to $^{16}$O are given in \cite{Toh13}.\\

\begin{figure}
         \includegraphics[width=7cm]{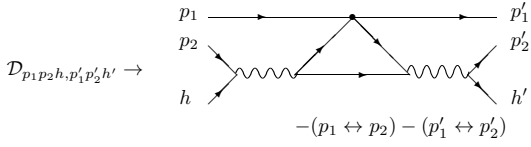}
         \caption{Schematic view of the PVC interaction vertex which contains itself a selfconsistent PVC process. The full dot stands for the antisymmetrized matrix element and the wiggly line for the vibration. Please note that contrary to the graphical impression, the vertex is instantaneous.}
         \label{PVC-vertex}
       \end{figure}

  In the double commutator of (\ref{3b-K}), each of the two triples of 2p-1h fermion operators contracts a particle state to the interaction. Naturally, from each triple then only a $ph$ pair remains. We can express those $ph$ pairs via the $Q^{\dag}, Q$ operators of (\ref{Qdag2}) using the inverse relation (\ref{aa}). Commuting then the destructors to the right, we exploit the annihilating property (\ref{killing}) and then $K^0$ is expressed by (self-consistent) RPA amplitudes $X, Y$ and occupation numbers. A graphical representation of this PVC vertex is shown 
  in Fig. \ref{PVC-vertex}.

       On the other hand, the occupation numbers are directly related to the s.p. Green's function (\ref{g0}) and then via the dynamical part of the s.p. self-energy which is related to the solution of (\ref{2p1h-prop}), we have a closed system of equations for the SCRPA amplitudes, via the SCRPA equations, and the s.p. occupation probabilities. We want to call this system of equations the even-odd-SCRPA (eo-SCRPA). It has been solved for the Lipkin and 1D Hubbard model with very good success \cite{Jem19}.

\begin{figure}
         \includegraphics[width=7cm]{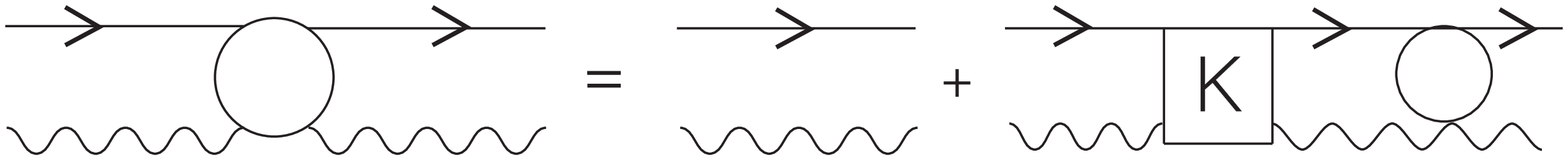}
         \caption{ Representation of the fermion-vibration scattering equation. The kernel is the one of Fig. \ref{PVC-vertex} with all Pauli exchanges included.}
         \label{pvc}
       \end{figure}

       Of course, the solution of the $2p-1h (2h-1p)$ equations will in general be quite demanding because of the eventually large configuration space. However, presently in nuclear physics quite routinely in the so-called second RPA huge configuration $2p-2h$ spaces are considered, so that a $2p-1h$ space should be a less difficult problem. We should also point out that the $2p-1h$ integral equation (\ref{2p1h-prop}) can be interpreted as a particle-vibration scattering equation with full respect of the Pauli principle. This is schematically shown in Fig. \ref{pvc}. The vibrations (wiggly lines) here are the solutions of the SCRPA equation.


\subsection{ Relation of the s.p. Green's function to the ground state energy. The tadpole,  perturbative particle-vibration coupling, and the spurious mode}

      The s.p. Green's function has the important property that it is related to the ground state energy $E_0$ of the system in the following way \cite{Fet71,Rin80}

      \be
     -\frac{i}{2} \lim_{t' \rightarrow t}{\rm Tr}[i\partial_t + \tilde e_k]G_{kk}(t-t') = E_0~.
     \label{E0}
     \ee
     where we supposed that the interaction term of the Hamiltonian does not contain any mean field contribution, since it is already included in $\tilde e_k$.
     A useful variant of the relation (\ref{E0}) is a formula deduced from the Hellman-Feynman theorem, see \cite{Fet71}

     \begin{equation}
       E_0 - E^{\rm HF}_0 = -\frac{i}{2}\int_0^1\frac{d\lambda}{\lambda}\lim_{t'\rightarrow t} {\rm Tr}[i\partial - \tilde e_k]G^{\lambda}_{kk}(t-t')
       \label{lambda-int}
     \end{equation}
     where the $\lambda$ dependence of $G^{\lambda}$ stems from multiplying the matrix element $\bar v_{k_1k_2k_3k_4}$ in the Hamiltonian with the factor $\lambda$.
     We have seen in Sect. II.A that SCRPA also gives an expression for $E_0$. Then, there exists an important consistency property: the ground state energy which one obtains from SCRPA and from the s.p. Green's function should be the same. This entails, for instance, that one should find an expression of the self-energy which is conform with this property. However, since, as we have seen, SCRPA corresponds to a linear boson theory, we must expect that the self-energy also must be linear in this boson and additionally that the self-energy in the solution of the Dyson equation also should only be kept to linear order. We will elaborate on this now.


As just mentioned, keeping with a theory which is linear in the ph-correlations, we only should consider the self-energy to linear order. We will then see that this leads to the RPA correlation energy in (\ref{E0}) and, thus, the above demanded consistency in the two ways to calculate the ground state energy will be fulfilled.

Therefore, we postulate the following form of the s.p. Green's
function, where for simplicity we suppose that the Green's function is diagonal

     \be
     G_{k} = G^{(0)}_{k} + G^{(0)}_{k}\Sigma^{\rm PVC}_{k}G^{(0)}_{k}~,
     \label{self-PVC}
     \ee
     with

     \be
     \Sigma^{\rm PVC}_{k}=-\sum_{l}G^{(0)}_{l}(t'_1-t_1)(0|T\hat F_{kl}(t'_1)\hat F^{\dag}_{lk}(t_1)|0)~,
           \label{Sigma}
\end{equation}
and 
\begin{equation}
\hat F_{kl}(t)=\sum_{nn'}[{\mathcal B}_{klnn'} B_{nn'}(t) + {\mathcal A}'_{klnn'}
  B^{\dag}_{nn'}(t) ]
\label{PVC}
\end{equation}
with ${\mathcal A}'$ being the matrix element in ({\ref{RPAeq}), where the MF part is missing. The fermion propagators correspond actually to ideal fermions, that is

\be
G^{(0)}_{k}(t-t')=-i[\Theta(t-t')\bar n^0_k-\Theta(t'-t)n^0_k]e^{-i\tilde e_k(t-t')}~,
\label{ideal-ferm}
\ee
where the occupation numbers $n^0_k~ ({\bar n}^0_k = 1-n^0_k)$ are step functions serving 
only to distinguish states from below and above the Fermi level.
The fact that those fermion propagators are ideal ones can already be seen in considering the SCRPA equations (\ref{RPAeq}). Let us take the equation for $Y$. Then, schematically,

\be
Y^{\nu}_{mi}=-\frac{1}{\Omega_{\nu} + \tilde e_{mi}}F^{\nu}_{mi}~,
\label{Y}
\ee
with $\tilde e_{mi} = \tilde e_m-\tilde e_i$. We see that the ph-propagator is the Fourier transform of the product

\bea
 G^{(0)}_{m}(t-t')G^{(0)}_{i}(t'-t)~.
\eea
The vertex function is obtained as

\bea
F_{mi}^{\nu} &=& [BX^{\nu} + A'Y^{\nu}]_{mi} ~ = ~ \langle 0|\hat F_{mi}(0)|\nu\rangle 
\nn&=&
\langle 0|[\hat F_{mi}(0)~,{\mathcal O}^{\dag}_{\nu}]|0\rangle~,
\label{ph-vertex}
\eea
where we used the inversion (\ref{Bogo}), that is, $ B_{mi}(t) = \sum_{\nu}[X^{\nu}_{mi}{\mathcal O}_{\nu}(t) + Y^{\nu}_{mi}{\mathcal O}^{\dag}_{\nu}(t)] $
with
${\mathcal O}(t) = {\mathcal O}(0)e^{-i\Omega_{\nu}t}$ and the annihilating condition ${\mathcal O}_{\nu}|0) = 0$.

The upper index PVC on the self-energy in Eq. (\ref{Sigma}) shall indicate that this approximation corresponds to (a lowest order) particle-vibration coupling scheme.
We then obtain for the correlated parts of the occupation numbers

\begin{eqnarray}
  -i\lim_{t'\rightarrow t}G_p(t-t') &=& \delta n_p =\sum_{k'<p}|Y^{\nu}_{pk'}|^2~~\nonumber\\
   i\lim_{t\rightarrow t'}G_h(t-t') &=& \delta n_h =\sum _{k>h}|Y^{\nu}_{kh}|^2~,
  \label{lim-G}
\end{eqnarray}
as well as the following relation for the correlated part of the ground state energy

\begin{eqnarray}
E_{corr}&=& -\sum_h\tilde e_h  
-\frac{i}{2}  \lim_{t'\rightarrow t}\sum_{k}[i\partial_t + \tilde e_k] G_{k}(t-t')
\nn&=&-\frac{i}{2}\lim_{t'\rightarrow t}\sum_{k} [(i\partial_t -
    \tilde e_{k}) G_{k}(t-t')+2\tilde e_{k}G_{k}(t-t')]\nonumber\\
  &=&- \sum_{\nu,k>k'}\Omega_{\nu}|Y^{\nu}_{kk'}|^2~,
  \label{E-corr-G}
\end{eqnarray}
which is the same as obtained from SCRPA. The inverse of the free s.p. GF is  
$i\partial_t - \tilde e_{k} = {G_k^{0}}^{-1}$. 
Using Eq. (\ref{self-PVC}) for $G_{k}$  and since

\be
-\frac{i}{2}\lim_{t' \rightarrow t}G_k^{0^{-1}}G_k = 
\sum_{l\nu}\frac{|F^{\nu}_{kl}|^2  }{\Omega_{\nu}+ \tilde e_k - \tilde e_l}~,
\ee
we obtain with Eq.(\ref{Y}) the result (\ref{E-corr-G}) for the correlation energy.

So, since we have from the s.p. Green's function the same ground state energy as obtained from SCRPA in Sect. II.A, we conclude that the form (\ref{self-PVC}) with (\ref{Sigma}) of $G_{kk'}$ is also consistent with the occupation numbers (\ref{lim-G}).
We also realise that the Luttinger theorem, that is $\sum_hn_h + \sum_p n_p =N$, is fulfilled. \\

It is worth mentioning that the boson-fermion scheme just outlined can also be derived from the following boson-fermion coupling Hamiltonian

\be
H_{\rm B-F}= \sum_k\tilde e_k a^{\dag}_ka_k
+ H_B + \sum_{kk'}(\hat F_{kk'}a^{\dag}_ka_{k'} + h.c.)~,
\label{H-BF}
\ee
where $H_B$ is given in (\ref{H_B}).\\
We now can also read off a more explicit form of the self-energy given in frequency space

\bea
&&\Sigma^{\rm PVC}_k(\omega) = \sum_{k'\nu}|F^{\nu}_{kk'}|^2
\nn&\times&
\bigg [\frac{\bar n^0_{k'}}{\omega - \tilde e_{k'} -\Omega_{\nu}+i\eta}+\frac{n^0_{k'}}{\omega -\tilde e_{k'}+\Omega_{\nu} -i\eta} \bigg ]~.
\label{self-omega}
\eea

Though it will not be consistent, it may be tempting to use the above
expression (\ref{Sigma}) with (\ref{self-omega}) of the self-energy in the full solution of the Dyson equation. We then not only have to consider the self-energy, but also the correction to the HF s.p. energies coming from the correlated part of the occupation numbers

\begin{eqnarray}
  \delta  \tilde e_{q}  &=&
\sum_p v^{eff}_{qpqp}\delta n_p - \sum_h v^{eff}_{qhqh}\delta n_h\nonumber\\
 &=&\sum_{kk'} v^{eff}_{qkqk}(n^0_{k'}-n^0_k)\sum_{\nu}|Y^{\nu}_{kk'}|^2~,
\end{eqnarray}
where the effective force is, e.g., obtained from a second variation with respect to the density of a density functional of the Skyrme or Gogny type. In addition, we supposed that the correction to the occupation numbers also stays diagonal.
The above correction to the mean-field potential is known as the so-called Tad-pole 
\cite{Kho82,Gne14,Bor19}.
It is an important contribution to the PVC scheme because it weakens the contribution of the 
energy-dependent part $\Sigma^{\rm PVC}$. It is for instance crucial for the '{\it spurious}' translational mode at
zero energy, which has diverging amplitudes $X,Y$. As a matter of fact the tad-pole and PVC 
contributions exactly cancel.
Let us dwell on this important aspect a little more. We shall work in the canonical basis.

{\it The spurious mode.}
For the translational mode, we have $Y_{kk'} = -C\langle k'\
|\hat p|k\rangle $ where $C$ is the diverging normalization constant \cite{Rin80}. Then

\begin{equation}
 \delta \tilde e_{q} =\sum_{kk'} v^{eff}_{qkqk}(n^0_{k'}-n^0_k)\langle k|\hat p|k' \
  \rangle \langle k'|\hat p|k\rangle C^2~,
\end{equation}
with $\langle k|\hat \rho |k'\rangle = \delta_{kk'}n^0_k$, further
\begin{eqnarray}
 \delta \tilde e_{q} &=& \sum_{kk'} v^{eff}_{qkqk}\langle k|[\hat p, \hat \rho]|k'\
  \rangle \langle k'|\hat p|k\rangle C^2\nonumber\\
 \delta \tilde e_{q} &=&\sum_{k} v^{eff}_{qkqk}\langle k|[\hat p, \hat \rho]\hat 
  p|k\rangle C^2~.
\end{eqnarray}
For simplicity but without loss of generality, we discard the exchange term and
suppose a translational-invariant finite-range force. Then
\begin{widetext}
\begin{equation}
 \delta \tilde e_{q} =  N^2 \int d^3r \phi^*_q(r)\phi_{q'}(r)\int d^3r_1
  v^{eff}({\bf r}-{\bf r}_1)\langle {\bf r}_1|[\hat p,\hat \rho]\hat p|{\bf r}_1\rangle~,
\end{equation}

\begin{equation}
 \delta \tilde e_{q} =  N^2 \int d^3r \phi^*_q(r)\phi_{q'}(r)\int d^3r_1
  v^{eff}({\bf r}-{\bf r}_1)\frac{\partial\rho({\bf r}_1)}{\partial {\bf r}_1}\frac{\partial\
  }{\partial {\bf r}_1}~.
\end{equation}
\end{widetext}
With a simple mean field $U({\bf r})$, we have $v({\bf r}-{\bf r}_1)= \frac{ \delta U({\bf r})}{\delta \rho({\bf r}_1)}$ and, thus, with a partial integration, we get

\begin{equation}
 \delta \tilde e_{q} = -C^2\langle q|\Delta U|q\rangle~.
  \label{diverging-qp}
\end{equation}
From Baldo et al. \cite{Bor19}, we know that part of the dynamic self energy has the same
expression with opposite sign. So, the diverging term cancels as can be seen from the following.

From (\ref{self-omega}) we have (as long as there is no resonance)

\be
\Sigma_k^{\rm PVC}(\omega) = \sum_{k'\nu}|F^{\nu}_{kk'}|^2\frac{\omega -\tilde e_{k'} + (1-2n^0_{k'})\Omega_{\nu}}{(\omega -\tilde e_{k'})^2-\Omega_{\nu}^2}~.
\ee
Let us now consider the spurious translational mode with $\Omega_0=0$. Then

\bea
&&\Sigma_q^{\rm PVC}(\omega = \tilde e_q)=\sum_{k'\nu=0}\frac{|F^{0}_{qk'}|^2}
{\tilde e_q - \tilde e_{k'}} 
\nn&+& 
\lim_{\Omega_0 \rightarrow 0}\sum_{k'\nu}|F^{0}_{kk'}|^2\frac{(1-2n^0_{k'})\Omega_0}{(\tilde e_q - \tilde e_{k'})^2}
\nn&=&
\sum_{k'\nu=0}\frac{|F^{0}_{qk'}|^2}
{\tilde e_q - \tilde e_{k'}}+\sum_{k'}\frac{|\langle k'|\hat p|q\rangle|^2}{2Am}(\bar n^0_{k'}-n^0_{k'})~,
\nn
\eea
          where $A$ is the total nucleon number and we again used (\ref{Y}) to simplify the second term. It is  the recoil term which is  finite, since $C^2 = \frac{1}{2Am\Omega_0}$. It is positive for particles and negative for holes.
          The first part cancels (\ref{diverging-qp}) and this has to be explained a little. The vertex (\ref{ph-vertex}) can be written for the spurious mode where $X_{kq} = C\langle k|\hat p|q\rangle$ and $Y_{kq} = - C\langle q|\hat p|k\rangle$ as the commutator of the effective force with the momentum operator, as long as we consider the ${\mathcal A}'$ and ${\mathcal B}$ matrices based on this effective force, which we already have introduced above (the inverse way seems more difficult).
          Introducing the force as originating from the variational derivative with respect to the density of the mean field potential $U(r)$ leads then to the expression (see also \cite{Kho82,Gne14,Bor19}):

\bea
F_{kq} &=& - \langle k|\nabla U|q\rangle=-\langle k|[U,\nabla]|q\rangle
\nn&=&
 -\langle k|[H_0,\nabla]|q\rangle = (\tilde e_k-\tilde e_q)\langle k|\nabla|q\rangle~.
\eea
With this and the second vertex function we obtain that

\bea
\sum_{k'\nu=0}\frac{|F^{0}_{qk'}|^2}
{\tilde e_q - \tilde e_{k'}}= C^2\langle q|\Delta U|q\rangle~,
\eea
that is the opposite of expression (\ref{diverging-qp}).

          This shows that the PVC self-energy, which is by itself diverging, stays together with the tadpole finite. It was not clear since long what to do with the spurious mode in the optical potential for elastic nucleon-nucleus scattering, strongly related to the self-energy, because of it divergency. Most of the time it just was not considered despite it is a member of the complete set of states. We now see that this prescription is almost correct. Only the recoil term is missed. It is positive for particle states and negative for hole states.
Nevertheless, we should remember that our self-energy used in the full Dyson equation is not completely consistent and, e.g., the Luttinger theorem is violated.
Only in SCRPA (r-RPA) scheme this deficiency is overcome. It would be interesting to repeat this derivation in the case of collective rotation.

\subsection{An application to the Lipkin model of the coupling constant integration with SCRPA}

The Lipkin model will be introduced in Sect.V.D just below. The SCRPA is trivially applied, with good success \cite{Duk90}. It is interesting  to apply the coupling constant integration of
(\ref{lambda-int})
to this case. The general formula can be deduced from (\ref{E-corr-G})

\begin{equation}
  E_{\rm corr} = -\sum_{ph,\nu}\int_0^1 \frac{d\lambda}{\lambda}|Y^{\nu}_{ph}|^2[\Omega_{\nu}(\lambda) + \tilde e_p - \tilde e_h]
  \label{lambda-int-scrpa}
\end{equation}
It is immediately verified that if in this formula the standard RPA expressions are used, this leads back to the usual, see \cite{Rin80}. On the contrary, if we use the SCRPA approach, this can lead to substantial improvements. This is shown for the Lipkin model in
Fig. \ref{corr-lambda-Lipkin}.
One, indeed sees a quite strong improvement over the SCRPA result via the $\lambda$ integration.

\begin{figure}
  \includegraphics[width=7cm]{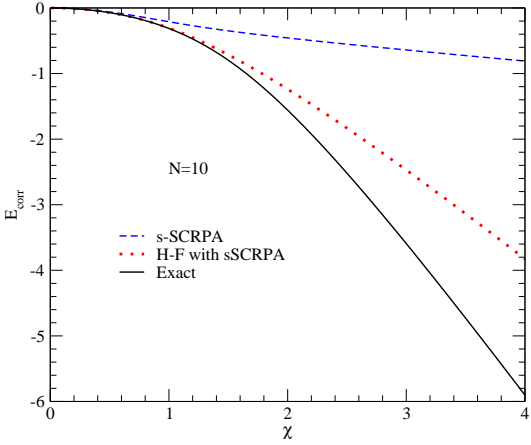}
  \caption{Demonstration of the positive effect of the coupling constant integration for the correlation energy $E_{\rm corr}$. s-SCRPA, broken line: standard SCRPA; H-F with sSCRPA, dotted line: coupling constant integration; full line: exact result.}
  \label{corr-lambda-Lipkin}
  \end{figure}

This method id also applied in theoretical chemistry, see, e.g., \cite{Per18,Nam19}, 
          

\subsection{Inclusion of particle-particle RPA correlations into the self-energy. The $T$-matrix approximation}

We just have treated the self-energy including ph-RPA correlations. Of course, there exists also the analogous situation for pp-RPA correlations. This will lead to the so-called $T$-matrix approximation of the self-energy which has been elaborated in a work by Nozi\`eres-Schmitt-Rink 
\cite{Noz85}. We write for the s.p. Green's function

\be
G_{kk'} = G^{(0)}_{kk'} + \sum_{k_1}G^{(0)}_{kk_1}\Sigma^{pp}_{k_1k'_1}G^{(0)}_{k'_1k'}~,
\label{self-pp}
\ee
with for the self-energy

\be
\Sigma^{pp}_{kk'}(\omega) = 
-i\sum_{k_1}\int \frac{d\omega'}{2\pi}G_{k_1}^{(0)}(\omega')T_{kk_1k'k_1}(\omega + \omega')~,
\label{T-matrix-self}
\ee
where we assumed that $G^{(0)}_{kk'}$ is diagonal and the $pp$ T-matrix is related 
to the $pp$ Green's function (\ref{G-2}) by

\be
T_{k_1k_2k'_1k'_2}(\omega) = \frac{1}{4}\sum_{k_3k_4k'_3k'_4}\bar v_{k_1k_2k_3k_4}
G_{k_3k_4k'_3k'_4}(\omega)\bar v_{k'3k'4k'_1k'_2}~.
\label{pp-T}
\ee
where the two-partilcle Green's function is given in (\ref{G-2}) and we assumed that the uncorrelated Green's function contains the HF potential.
All considerations we have outlined above for the $ph$-case shall also apply 
analogously to the $pp$-case. Only $ph$ correlation functions have been exchanged by $pp$ correlations. This scheme has for instance been applied by Nozi\`eres and Schmitt-Rink \cite{Noz85} to treat the fameous BCS-BEC (Bose-Einstein condensation) transition. Below we will give some applications of the $pp$-case to nuclear matter.

     \subsection{Single-particle Green's function from the CCD wave function. Application to the Lipkin model}

     In this section we want to elaborate on the odd-particle number SCRPA, see Sect. II.G. Specifically, we  want to study how far one can go with the s.p. self-energy (\ref{Sigma-Z}) entirely based on the CCD wave function. 
     In order not to develop the whole general formalism, which would be quite lengthy, we immediately switch to the Lipkin model, however. It is sufficiently general so that the full  many-body formalism with, e.g., a realistic two body force can be deduced. The full formalism will be published in a separate work \cite{Jem20}


     \subsubsection{Calculating with the CCD wave function as a quantum vacuum. Application to the Lipkin model}

The standard  two-level Lipkin model, see, e.g., \cite{Rin80}, 
can be seen as a one site quasi-spin model with the following Hamiltonian 
(analogous to the 3-level model of Sect. III.B)


\noindent

\bea
H = eJ_0 -\frac{V}{2}[J_+J_+ + J_-J_-]~,
\label{2-Lipkin}
\eea
where
\bea
J_+ &=& \sum_mc^{\dag}_{1m}c_{0m}~;~ J_-= [J_+]^{\dag}~;~2J_0 = n_1- n_0
\nn 
n_i &=& \sum_mc^{\dag}_{im}c_{im}~,
\eea
with the following SU(2) commutation relations among the quasi-spin operators
\bea
[J_-,J_+] &=& -2J_0~;~~[J_0,J_{\pm}] = \pm J_{\pm}~.
\eea


     According to the spin algebra, the model is easily diagonalisable exactly for any spin value where $\langle {\rm HF}|J_0|{\rm HF}\rangle = -N/2$ and $N$ the particle number or degeneracy of each of the two levels which have a distance of $e$. The model is non-trivial and cited, besides in nuclear physics where it was invented \cite{Lip65},
in several other fields \cite{Vid04,Rib07,Col18,Cam15}.
For instance, the Lipkin model shows in the strong coupling limit a spontaneously (discrete) broken symmetry, what makes it particularly useful for testing new many-body approaches.\\

     In the Lipkin model the CDD wave function writes

     \be
     |Z\rangle = e^{\frac{z}{2}J_+J_+}|{\rm HF}\rangle~,
     \ee
     with $z$ an in principle complex number. This CCD wave function 
      is the vacuum of the following addition ($\alpha$) and retrieval ($\rho $) annihilators, see (\ref{Z-qp})

      \bea
      q_{1,\alpha} &=& \frac{1}{N}u^{\alpha}_1\sum_m c_{1m} + V^{\alpha}_1\frac{1}{N}
     \sum_mJ_+c_{0m}/\sqrt{n_{11}}\nonumber\\ 
     q_{0,\rho}&=& \frac{1}{N}u^{\rho}_0\sum_m c^{\dag}_{0m} + V^{\rho}_0\frac{1}{N}
     \sum_mc^{\dag}_{1m}J_+/\sqrt{n_{11}}~,
\nn
     \label{killer-odd}
     \eea
     with $n_{11}=-\langle \{J_-c_{1m},c^{\dag}_{1m}J_+\}\rangle = \langle \{c^{\dag}_{0m}J_-,J_+c_{0m}\}\rangle $ a norm factor. This yields the following norm relations $|u^{\rho}_0|^2 + |V^{\rho}_0|^2~ = ~|u^{\alpha}_1|^2 + |V^{\alpha}_1|^2 = 1$. The various amplitudes are related by $z =\frac{V^{\rho}_0}{u^{\rho}_0\sqrt{n_{11}}} = -
       \frac{V^{\alpha}_0}{u^{\alpha}_0\sqrt{n_{11}}}$.\\
       Applying the EOM leads to the eigenvalue equation outlined in (\ref{odd-eigen}) equivalent to the Dyson equation of the s.p. Green's function in (\ref{Dyson-z}, \ref{Sigma-Z}). The matrix elements of this equation contain three-body correlation functions. 


     It does not seem to be very easy to work with these operators (\ref{killer-odd}), since they do not represent a canonical transformation and, thus, cannot be inverted and the killing property cannot directly be exploited. However, there exists a way around which allows, due to the annihilating condition,~~
     $q|Z\rangle = 0$,~~
     to reduce the order of the correlations \cite{PS20}.
For example, it is clear that the secular problem (\ref{odd-eigen}) involves, as mentioned, maximally three-body correlation functions. They can be reduced to two-body correlation functions in the following way. For this, we use the annihilating conditions which directly yield

     \bea
     J_0|Z\rangle &=&[-\frac{N}{2} + 2zJ_+^2]|Z\rangle \nonumber\\
     J_-|Z\rangle &=& [2z(N-1)J_+ - 4z^2J_+^3]|Z\rangle~.
     \eea
     These are the basic two relations relating higher operators to lower ones. We multiply the first equation from left with $J^2_+$
     and the second one with $J_+$. This gives two equations containing $J_+^4$. Isolating and equalling leads to an equation for $\langle J_+^2J_0\rangle$ in terms of $\langle J_+^2 \rangle $ and $\langle J_+J_-\rangle$. We multiply first equation with $J_0$ what helps to find the Casimir relation $J_+J_-= \frac{1}{4}N(N+2) - J_0^2 + J_0$. 
     From where we get

     \bea
     2z\langle J_+^2\rangle &=& \frac{N}{2} + \langle J_0\rangle \nonumber\\
     z\langle J_+^2J_0\rangle &=& \frac{1}{2}(N-2)z\langle J_+^2\rangle - \langle J_+J_-\rangle \nonumber\\
     2z\langle J_+J_-J_0\rangle &=& zN(N+2)-\langle J_+^2\rangle + z(N-6)\langle J_+J_-\rangle 
\nn&-&2z(N-4)\langle J_0\rangle~.
     \label{corr-red}
     \eea
     In the last formula also the Casimir relation has been used. Then all the matrix elements in the equation corresponding to (\ref{odd-eigen}) for the Lipkin model can be given as a function of $\langle J_0\rangle $ and $\langle J_0^2\rangle $.

\begin{figure}[h]
\includegraphics[width=8cm,height=8cm]{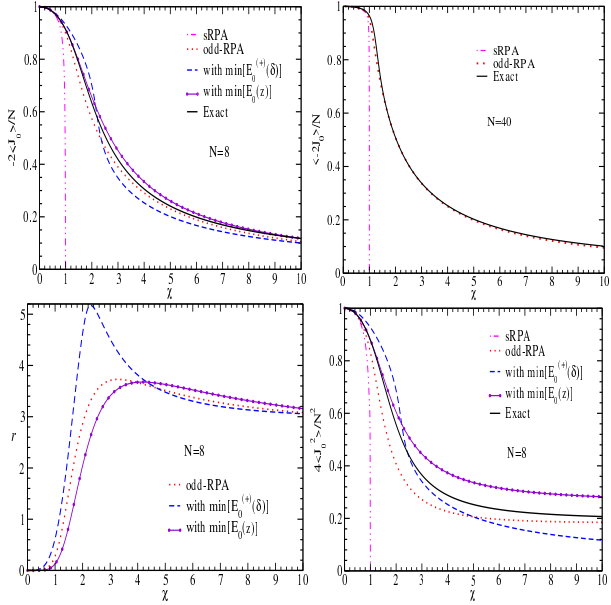}
\caption{Upper left panel: the occupation number difference between upper and lower levels, $\langle -2J_0\rangle $, for $N=8$  with standard RPA (sRPA) (double dot broken line), present odd-RPA (dotted line), projected HF min[$E_0^{(+)}(\delta)$] (broken line), CCD variational wave function min[$E_0(z)$] (continuous line with dots), and exact solution (full line) as function of the intensity of interaction $\chi=\frac{V}{e}(N-1)$. Upper right panel:$\langle -2J_0\rangle $, for $N=40$ with sRPA, odd-RPA, and exact solution. Lower left panel: percentage error of the correlation energy as $r=100\times \frac{(E_0^{odd-RPA} - E_0^{Exact})}{E_0^{Exact}} $ (dotted line), $r=100\times\frac{(min[E_0(z)]-E_0^{Exact})}{E_0^{Exact} }$ (continuous line with dots) and $r=100\times\frac{(min[E^{(+)}_0(\delta)]-E_0^{Exact})}{E_0^{Exact}} $ (broken line) as function of the intensity of interaction $\chi=\frac{V}{e}(N-1)$. Lower right panel: occupation fluctuation $\langle 4J^2_0\rangle $ for $N=8$ with same ingredients as upper left panel.}
\label{Carre}
\end{figure} 

\begin{figure}
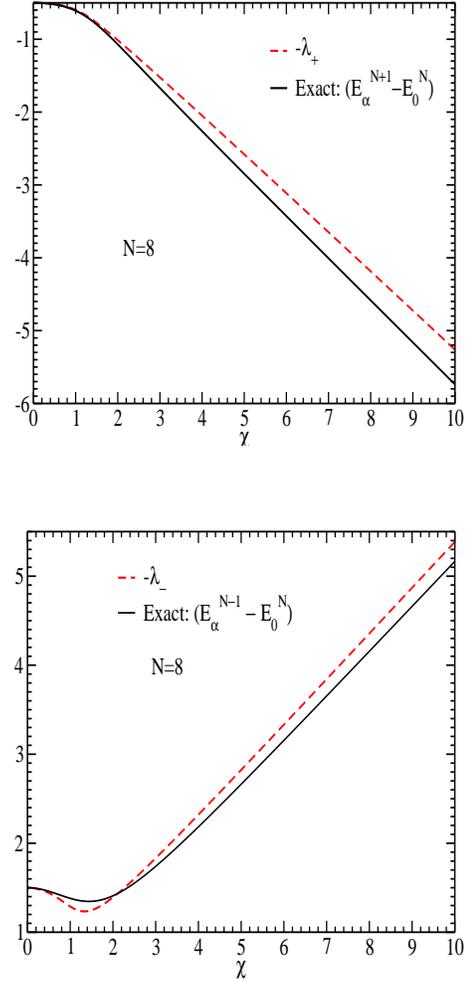

\includegraphics[width=6cm,height=6cm]{LpN8.eps}
\vskip1cm
\includegraphics[width=6cm,height=6cm]{LmN8.eps}
\caption{ The eigenvalues $\lambda_+$ and $\lambda_-$ of the odd-RPA matrix compared 
to the exact values as a function of $\chi$.}
\label{LpLmN10}
\end{figure} 


     Usually, it is a very good approximation to take $\langle J_0^2 \rangle \simeq \langle J_0\rangle^2$ so that the whole becomes simply a self-consistency relation for $\langle J_0\rangle $. It seems to be also very valid here: for $N=4$ the solution becomes exact and for $N=6$ the solution is quite good  throughout the whole domain of couplings. However, we will take this only for the initialisation of the iterative cycle because there exists a still better procedure. One can demand that the time derivative of the two-body density matrix be zero, that is stationary at equilibrium

\bea
\langle[H,J_-J_-]\rangle &=& 2e\langle J_-^2\rangle -V\langle (2J_0 + 4J_0^2 -4J_+J_-J_0)\rangle  
\nn&=& 0~.
\label{stationary}
\eea
     This gives an extra relation linking $\langle J_0\rangle$ with $\langle J_0^2\rangle$ so that, together with (\ref{corr-red}) the system of equations obtained from the EOM is closed together with the relation for the occupation numbers  obtainable from (\ref{odd-eigen}) adapted to the Lipkin model
\bea
\langle J_0\rangle = \frac{N}{2}\frac{4z^2n_{11}-1}{4z^2n_{11}+1}~.
\eea
The results for different quantities are excellent throughout the whole domain of coupling constants as can be seen in Fig. \ref{Carre}. One can say that this approach solves the Lipkin model for the correlation energy for all values of $N$ and $\chi$ with an error not larger than four percent in the transition region and better elsewhere. This  is a very satisfying result. It is quite remarkable that one can work with a nonlinear transformation like in (\ref{Z-qp}) very efficiently. For the collective states in the even systems one may try to perform a Tamm-Dancoff or RPA approach with the new quasiparticle operators (\ref{killer-odd}), but this remains a task for the future.

In Fig. \ref{LpLmN10} we show the eigenvalues $\lambda_{\pm}$ of the odd-RPA matrix. It is seen that they are also reproduced with good quality over the whole range of $\chi$-values.

 One may wonder from where this high performance in the whole parameter space comes. For this we remark that if one replaces in ({\ref{killer-odd}) the operator $J_+$ by its expectation value $\langle J_+\rangle$, the equation for $q$ just represents a HF transformation to a new symmetry broken s.p. basis. To keep the operator $J_+$ means that good parity is kept during this transformation. We want to coin it quantum-mean-field (qu-mf) transformation. In spirit it is very similar to symmetry projected mean-field calculation. The results for the Lipkin model with projected HF are also quite close to the present results, see Fig. \ref{Carre}. Also using CCD as a variational wave function yields results rather close to the present odd-RPA method. The method can be seen as a variant to the usual mean field projection techniques \cite{Rin80}.  It would be very interesting to apply this technique to the case of rotations.

\subsection{Cluster expansion of the single-particle self-energy and applications to infinite matter problems}

In the preceding section we have outlined how to construct a s.p. GF to reproduce the SCRPA correlation energy in the $ph$ and $pp$ channels. However, the self-energy contains also higher than two-body correlations. We will, therefore, use the EOM to include three-body and four-body correlations to the self-energy and construct self-energies tailored to treat cluster phenomena in nuclear physics. We are specially interested in four-body correlations, since in nuclear physics, at low densities, besides the triton (t, $^3$H) and the helion (h, $^3$He), the $\alpha$ particles play a particularly strong role. This implies that we will have to treat not only more body correlation, but also corresponding cluster bound states.

Within the present section, starting from the s.p. Green's function, we are in addition interested in the equation of state of 
symmetric
nuclear matter expressing the nucleon density $\rho (T, \mu)$ as a
function of temperature $T= 1/ k_B\beta$ and chemical potential $\mu$. Note that for asymmetric nuclear matter different chemical potentials for neutrons ($\mu_n$) and protons ($\mu_p$) have to be introduced.
In the Green's functions approach, the density of the system is 
given by
\begin{equation}
\label{eos}
\rho(\beta, \mu) = \sum_{k_1} \int \frac{d \omega}{2 \pi} \frac{1}{e^{\beta
     (\hbar \omega - \mu_1)}+1}  S(k_1,\omega)~,
\end{equation}
where the {\it spectral function} $S(k,\omega)$, depending upon
single-particle coordinate $k_1=\{p_1,\sigma_1,\tau_1\}$ describing momentum, 
spin, and isospin, is defined by \cite{Mah81,Neg88,Fet71}
\bea
\label{spectf}
S(k_1,\w)=\frac{1}{i}\left[
G^{\w_n}_{k_1}\vert_{i\w_n\rightarrow\w-i\eta}-
G^{\w_n}_{k_1}\vert_{i\w_n\rightarrow\w+i\eta}
\right]~.
\nn
\eea
Starting from the equation of state $\rho (\beta, \mu)$, all thermodynamic
properties can be calculated after determining the thermodynamic
potentials $f(T,\rho) = \int_0^\rho \mu(\beta, \rho') d\rho'$ or
$p(T,\mu) = \int_{-\infty}^\mu \rho(\beta, \mu') d\mu'$. 

The evaluation of the spectral function or the single-particle Matsubara 
Green's function, respectively, can be 
performed with  the single-particle 
self-energy $\Sigma(k,\omega)$) according to
\begin{widetext}
\begin{equation}
S(k,\omega) = \frac{2 {\rm Im} \Sigma(k,\omega -i0) }{(\omega - (\hbar k)^2/2
  m - {\rm Re} \Sigma(k,\omega))^2 + ( {\rm Im} \Sigma(k,\omega
  -i0))^2}~.
\end{equation}
\end{widetext}
For the self-energy, under certain assumptions, see below, a cluster 
decomposition \cite{Kra86} is possible,

\begin{figure}[ht]
\centerline{
\includegraphics[width=7cm]{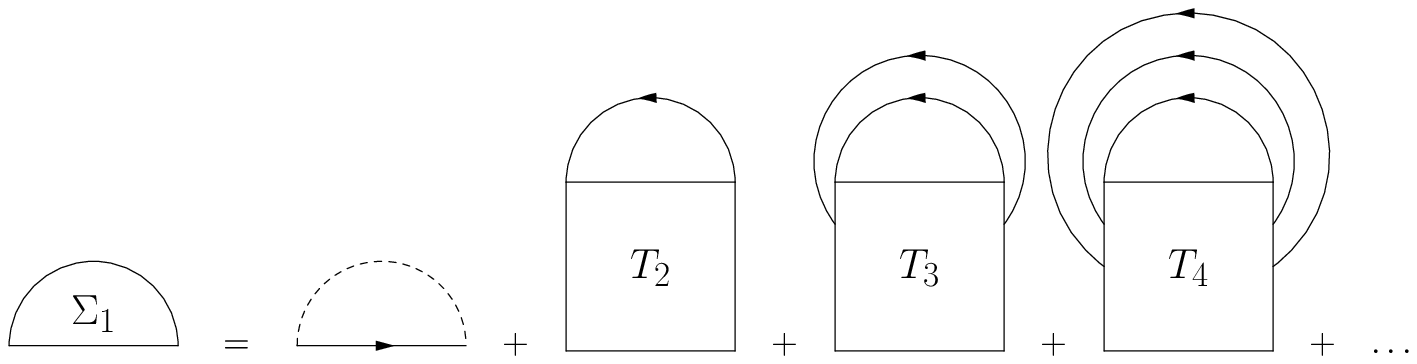}
}
\end{figure}

The ${\rm T}_A$ matrices are related to the $A$-particle Green functions 
obtainable from the EOM. They, therefore only depend on one energy
\begin{widetext}
\begin{eqnarray}
{\rm T}_A(k_1\dots k_A, k'_{1} \dots k'_{A}, z) &=&  V_A(k_1\dots k_A, k'_{1} \dots k'_{A})
+V_A(k_1\dots k_A, k''_{1} \dots k''_{A})
G_A(k''_{1}\dots k''_{A}, k'''_{1} \dots k'''_{A}, z) 
\nn&\times&
V_A(k'''_{1}\dots k'''_{A}, k'_{1} \dots k'_{A})~,
\end{eqnarray}
\end{widetext}
with the potential $ V_A(k_1\dots k_A, k'_{1} \dots k'_{A}) = \sum_{i<j}
v_{ij,i'j'} \prod_{k \neq i,j} \delta_{k,k'}$, and subtraction of double
counting diagrams when inserting the T matrices into the
self-energy. The solution of the $A$-particle propagator in the
low-density limit is given by
\bea
&&G_A(k_1\dots k_A, k'_{1} \dots k'_{A}, z)= 
\nn&&\sum_{n,P} 
\frac{\psi_{A,n,P}(k_1\dots k_A)\psi^*_{A,n,P}(k'_{1}\dots k'_{A})}{z- E_{A,n,P}}~,
\eea
using the eigenvalues $E_{A,n,P}$ and wave functions
$\psi_{A,n,P}(k_1\dots k_A)$ of the $A$-particle Schr\"odinger equation,
$P$ denotes the total momentum, and the internal quantum number $n$
covers bound as well as scattering states. 

The evaluation of the equation of state in the low-density limit is
straightforward.  Considering only the bound-state contributions, we
have the result 
\begin{equation}
\label{rho}
\rho(\beta, \mu) = \sum_{A,n,P} \frac{1}{ e^{\beta(E_{A,n,P}-A\mu)} -
    (-1)^A}~,
\end{equation}
which is an {\it ideal mixture of components} obeying Fermi or Bose statistics. 
This equation of state which describes nuclear matter in the low-density limit
is denoted as nuclear statistical equilibrium (NSE) and is a standard
approach to heavy ion collisions and the astrophysics of compact objects,
where the composition of dense nuclear matter and the yields of clusters
is of relevance. 

In the nondegenerate (classical) limit, the integrals over $P$ can be 
carried out for the respective bound states, and
one obtains the mass-action law that determines the matter composition
at given temperature and total particle density.  At low densities,
quantum effects become relevant.  The most dramatic is Bose-Einstein
condensation (BEC), which occurs for bound states in the channels with 
even $A$ when
$E_{A,n,P}-A\mu = 0$. Usually this happens first for $P=0$ and $n=0$ 
(ground state). At low, but fixed temperature, with  increasing density 
BEC occurs first for those clusters with the 
largest binding energy per particle. If
we consider the problem of clustering in nuclear matter, the
two-particle (deuteron) binding energy per nucleon is $1.11$ MeV,
while the four-particle (alpha) binding energy is $7.1$ MeV.  One,
therefore, anticipates that a quantum condensate of $\alpha$ particles
is formed first. At very low temperatures heavier nuclei can be formed
with even a higher binding energy per nucleon, so that in the low 
density regime considered here nuclei of the iron region would form 
the condensate at low temperatures. However, the formation of iron may be retarded with respect to $\alpha$ particles because of the much more complex structure of the former. In any case, we will restrict ourselves only to 
two-nucleon and four-nucleon correlations in the present review article.

An important problem is the modification of the single-particle and
bound-state properties at higher nucleon densities, when {\it medium
effects} have to be taken into account \cite{Rop80,Rop20}. 
The change of the cluster energies due to self-energy and Pauli-blocking contributions will be
investigated in the following subsections. 
In addition, the inclusion of scattering states \cite{Sch90} would
slightly change the composition as well as the equation of state,
see also \cite{Hor06}.  

Concluding, we emphasize that the calculation of the single-particle 
self-energy,
e.g. using the technique of cluster expansion and the EOM, allows to evaluate 
any thermodynamic property.  It should be mentioned that 
cluster expansions can also be developed for other quantities, 
such as the polarization function \cite{Kra86}, that are of interest 
in calculating transport properties. Also the hole lines in the cluster expansion of the self-energy can themselves correlate to clusters. For example, the three-hole lines in the four-particle $T$-matrix contribution can form a triton or a helion 
\cite{Sch08}.
We will expose $\alpha$ clustering in nuclear matter in Sect.VI.

We have shown how the formation of correlations, in particular bound states, can be implemented in the equation of state via the cluster decomposition of the single-particle self energy. A well-known approach to nuclear matter is the Fermi liquid theory of Landau, Migdal and Pomeranchuk  considering the linear response function, which is related to the dynamical structure factor. This also allows to derive an equation of state and, in particular, the search for instabilities. 
The inclusion of correlations was an open problem for a long time but has been considered recently \cite{Rop18},
so that the equivalence of both approaches has been shown. The approach to nuclear matter properties via the dynamical structure factor is based on an extension of the RPA including correlations, see \cite{Kra86} and \cite{Rop79},
which has been worked out for partially ionized plasmas to investigate the dielectric function. 
Further considerations of how to incorporate small nuclear clusters into the equation of state of nuclear matter 
will be discussed in Sect.VI.

\subsection{Applications of the in-medium two-nucleon problem and the $T$-matrix approximation for the s.p. self-energy}

\subsubsection{Equation of state and critical temperature}

With increasing density of nuclear matter, medium modifications of
single-particle states as well as of few-nucleon states become of
importance. The self-energy of an $A$-particle cluster can, in principle, 
be deduced from contributions describing the single-particle
self-energies as well as medium modifications of the interaction and
the vertices. A guiding principle in incorporating medium effects is
the construction of {\it consistent} (``conserving'') approximations, which
treat medium corrections in the self-energy and in the interaction
vertex at the same level of accuracy.  This can be achieved in a
systematic way using the two-times cluster Green's functions 
formalism presented 
in this article.  At the mean-field
level, we have only the Hartree-Fock self-energy $\Gamma^{\rm HF} (k_1)=
\sum_{k_2} \ov{v}_{k_1k_2,k_1k_2} f(k_2)$ together with the Pauli blocking
factors, which modify the interaction from $\ov{v}_{k_1k_2,k'_{1}k'_{2}}$ to
$\ov{v}_{k_1k_2,k'_{1}k'_{2}}[1 - f(k_1) - f(k_2)]$.  In the case of the in-medium two-nucleon system
( $A=2$), the resulting effective wave equation which includes the
mean-field corrections reads
\begin{widetext}
\begin{equation}
\label{two_part_bind}
\left[\e_{k_1}+\e_{k_2}-E_{2,n,P}\right] \psi_{2,n,P}(k_1k_2) +
\sum_{k'_{1}k'_{2}}[1-f(k_1)-f(k_2)]\,\,\ov{v}_{k_1k_2,k'_{1}k'_{2}} 
 \psi_{2,n,P}(k'_{1}k'_{2})=0~.
\end{equation}
\end{widetext}
Here $\e_{k_1}$ denotes the single-nucleon quasiparticle energy, which in 
Hartree-Fock approximation reads $\e_{k_1}= p_1^2/2m +\Gamma^{\rm HF} (k_1)$. 
Both the self-energy and the Pauli blocking have a similar structure, 
a product of the interaction 
with the distribution function, and have to be considered together to find a 
consistent approximation. Note that correlations in the surrounding nuclear
matter are neglected. How those correlations can be considered has been shown 
in Sect. III.A within SCppRPA applied to the pairing model. 
Contributions of two-nucleon bound states (deuterons) to the effective 
Hamiltonian
are taken into account within the Cluster Mean-Field Approximation \cite{Rop80}.

This {\it effective wave equation} (\ref{two_part_bind}) is of great interest. 
It describes not
only bound states, but as well
scattering states in nuclear matter at arbitrary density, in particular, the 
merging of a bound state with the continuum of scattering states due to Pauli 
blocking
(Mott effect). The Gor'kov equation describing the appearance of a quantum 
condensate is
reproduced when the deuteron ($d$) binding energy $E_{d,P=0}$ coincides 
with $2 \mu$. 
The investigation of Eq. (\ref{two_part_bind}) yields also the crossover from 
BEC of deuterons to BCS pairing if the nucleon density increases.

\begin{figure}[t]
\centerline{
\includegraphics[width=7cm]{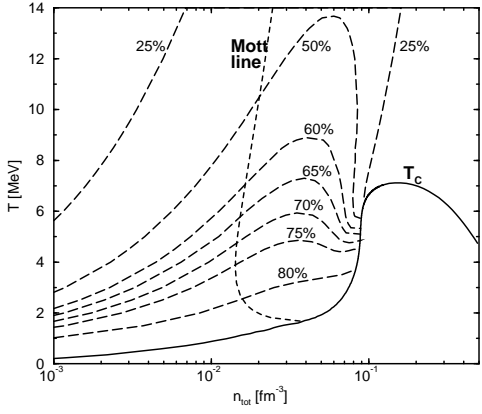}
}
\caption{Phase diagram of symmetric nuclear matter showing lines of
  equal concentration $\rho_{\rm corr}/\rho$ of correlated nucleons, the
  Mott line, and the critical temperature $T_c$ of the onset of superfluidity.}
\label{phase_diagram}
\end{figure}

Solutions of the effective two-nucleon wave equation (\ref{two_part_bind})
have been investigated for separable nucleon-nucleon interactions \cite{Sch90,Men10}
 in nuclear physics. With the corresponding T matrix, 
the self-energy  ${\rm Im} \Sigma (k_1, \omega )$,
the spectral function and the equation of state $\rho (\beta, \mu)$ (\ref{eos})
can be calculated. We are interested in the virial expansion that follows
in the limit where the imaginary part of the self-energy is small.

For small ${\rm Im} \Sigma (k_1, \omega )$, there will be a contribution to 
the density from the quasiparticle peak at 
$\omega - p_1^2/2m - {\rm Re} \Sigma (k_1,\omega )=0$. 
In addition, contributions to the total nucleon density arise from the 
bound cluster states.  

This approach was followed in Refs.~\cite{Sch90,Men10} under the 
restriction to two-particle contributions (through the 
${\rm T}_2$-matrix), but otherwise implementing a full 
treatment including scattering states.  In this way a 
generalized Beth-Uhlenbeck formula was derived, namely
\index{Beth-Uhlenbeck formula}
\bea
\label{Beth-Uhlenbeck}
\rho(\beta, \mu) &=& \rho_{\rm qp}(\beta, \mu) +\rho_{\rm corr}(\beta, \mu)
\nn&=&
\rho_{\rm qp}(\beta, \mu) + \rho_{\rm bound}(\beta, \mu) + \rho_{\rm
  scatt}(\beta, \mu)~,
\nn
\eea
with (see   Refs.~\cite{Sch90,Men10})
\begin{eqnarray}
\label{rho1}
\rho_{\rm qp}(\beta, \mu) &=& \sum_{k_1} f_1(E_{\rm qp}(k_1))\nonumber\\
\rho_{\rm bound}(\beta,\mu) 
&=& \sum_{P>P^{\rm Mott}} f_2(E_{d,P})\nonumber\\
\rho_{\rm scatt}(\beta, \mu) 
&=& \sum_{P} \int \frac{dE}{\pi} f_2\Bigl(\frac{P^2
  }{ 4 m} +E\Bigr) 
\nn&\times&\frac{d }{ dE}\Bigl[\delta_{2,P}(E) -\sin \delta_{2,P}(E)
\cos\delta_{2,P}(E) \Bigr]\,. \nonumber\\
\end{eqnarray}
The distribution functions appearing here are 
\be
f_1(E) = [\exp\,\beta (E - \mu) +1]^{-1}~~~(Fermi)~,
\ee 
and
\be
f_2(E) = [\exp\,\beta (E - 2 \mu) -1]^{-1}~~~(Bose)~.
\ee
The quasiparticle energy is determined implicitly from 
\be
E_{\rm qp}(k_1)=  p_1^2/2m + {\rm Re} \Sigma (k_1,E_{\rm qp}(k_1))~. 
\ee
As already mentioned,
evaluation of the Beth-Uhlenbeck formula (\ref{Beth-Uhlenbeck}) 
including two-particle
correlations has been carried out in Ref.~\cite{Sch90,Hor06,Men10} 
based on a
separable nucleon-nucleon potential. The result \cite{Alm95} for the 
composition of nuclear matter as function of density and temperature 
is shown in Fig.~\ref{phase_diagram}.  Two aspects of this study 
of two-particle condensation deserve special attention.

(i) The contribution of the correlated density, which is determined both
from deuterons as bound states in the isospin-singlet channel and
from scattering states, is found to increase with decreasing 
temperature, in accordance with the law of mass action.  This
law also predicts the increase of correlated density with 
increasing nucleon density (as also seen in Fig.~\ref{phase_diagram} 
for the low-density limit). 

However, with increasing density, the binding energy of the bound
state (deuteron) decreases due to Pauli blocking ({\it Mott effect}).  
At the Mott density $\rho^{\rm Mott}_{A,n,P(T)}$, the bound states with vanishing 
center-of-mass (c.o.m.) momentum are dissolved into the continuum of
scattering states, see discussion below in Sect. V.D and \cite{Rop15}.
Bound states with higher c.o.m. momentum merge 
with the continuum at higher densities.  According to Levinson's 
theorem, if a bound state merges with the continuum, the scattering 
phase shift in the corresponding channel exhibits a discontinuity by 
a phase jump of 180 degrees, such that no discontinuity appears in the 
equation of state.  Accordingly, the contribution of the
correlated density will remain finite at the Mott density, 
but will be strongly reduced at somewhat higher densities. 

Thus, one salient result is the disappearance of bound states 
and correlated density already well below the saturation density of
nuclear matter. The underlying cause of the Mott effect is 
Pauli blocking, which prohibits the formation of bound states 
if the phase space is already occupied by the medium (Fermi sphere), 
and hence no longer available for the formation of the wave 
function of the bound state (momentum space).  This effect
holds also for higher-$A$ bound states, such as the triton, 
helion, and $\alpha$ particle, which disappear at corresponding 
densities (see Fig.~\ref{shifts} below).

(ii) The Bose pole in the correlated density signals the onset 
of a quantum condensate.  As is well known, for the bound-state 
(deuteron channel) contribution the ${\rm T}$-matrix approach 
breaks down, when the pole corresponding to the bound-state energy 
coincides with twice the chemical potential.  This is a consequence of the 
Thouless condition \cite{Tho61} embodied in the in-medium two-body T-matrix equation 
(equivalent to (\ref{two_part_bind}) for $E_{2,n,P}= 2\mu$)
\begin{widetext}
\begin{equation}
{\rm T}(k_1,k_2,k''_{1},k''_{2};2 \mu)= \frac{1-f(k_1)-f(k_2)}{ 2\mu- \e(k_1) -\e(k_2)}
\sum_{k'_{1}k'_{2}} v_{k_1k_2k'_{1}k'_{2}} {\rm T}(k'_{1},k'_{2},k''_{1},k''_{2};2 \mu)~.
\end{equation}
\end{widetext}
This Thouless condition is not restricted to the presence of bound states, but
also holds for the contribution of scattering
states, describing BCS pairing of interacting nucleons in single-particle 
states.  
Consequently, the transition temperature for the onset of a
quantum condensate appears as a smooth function of density, as shown in
Fig.~\ref{fig:TC}.

At low densities, where the two-body bound states 
(deuterons) are well-defined composite particles, the mass action 
law implies that the deuterons will dominate the composition in 
the low-temperature region.  In this region, the critical 
temperature for the transition to the quantum condensate 
coincides with the Bose-Einstein condensation of deuterons as 
known for ideal Bose systems.  At high densities, where bound 
states are absent, the transition temperature coincides with the
solution of the Gor'kov equation describing the formation of Cooper
pairs.  Thus, BEC and BCS scenarios characterize the low- and 
high-density regimes, respectively.  We observe a smooth {\it crossover
transition} from BEC to BCS behavior -- as predicted generally
for fermion systems by Nozi\`eres and Schmitt-Rink 
in Ref.~\cite{Noz85}. It should be mentioned that the work in Ref.~\cite{Noz85} 
exactly follows what was discussed in the preceding section, namely that the 
single-particle mass operator in $T$-matrix approximation only should be 
evaluated in first order perturbation theory in order to respect 
conservation laws, when the $T$-matrix itself is evaluated with the ppRPA approach.

\begin{figure}[bth]
\centerline{
\includegraphics[width=8cm]{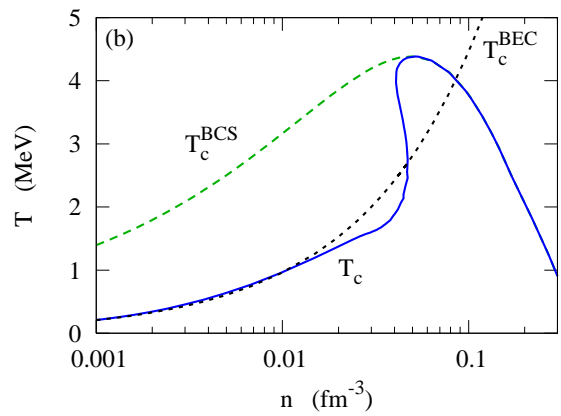}
}
\caption{Superfluid critical temperature as a function of the (total) density. 
The solid line is the full calculation, while the long dashes correspond to the BCS result. 
The short dashes show the critical temperature of Bose-Einstein condensation of a deuteron gas. 
Demonstration of the crossover BEC to BCS. Figure adapted from Fig. 7 in \cite{Men10}, see also \cite{Ste95}.}
\label{fig:TC}
\end{figure}

Below the transition temperature, the ${\rm T}$-matrix approach is no
longer applicable.  However, a mean-field approach becomes possible
in this regime after performing a Bogoliubov transformation.  Even
so, the proper inclusion of correlations below the critical 
temperature remains a challenging problem.  To date, only the first 
steps have been undertaken \cite{Rop94,Boz99,Hau09} 
toward solving this problem for general quantum many-particle systems.

Already at the mean-field level, the calculation of the transition
temperature should include the quasiparticle shifts in the Hartree-Fock
approximation.  In general, the {\it effective-mass approximation} in nuclear
matter will reduce the transition temperature \cite{Lom01,Men10}.
A noteworthy case is the reduction of the transition temperature in the
isospin-singlet channel for asymmetric matter \cite{Sog10}
We shall not discuss  further effects that can be described in
mean-field approximation, such as the shift and/or deformation of the Fermi 
surfaces \cite{Sed05}, e.g. the LOFF phases.

\subsubsection{Pseudo gap in nuclear matter}

Going beyond the mean-field approximation, the first remarkable
feature \cite{Sch99a} emerging at the two-particle level is 
the formation of a {\it pseudogap} in the density of states (DOS) 
above the critical temperature $T_c$.  Compared with the 
orthodox BCS solution, for which a gap opens in the DOS 
below $T_c$, a quite different situation is present in strongly 
correlated Fermi systems.  The full treatment of the (two-body) 
${\rm T}$-matrix in the single-particle self-energy with full solution of the Dyson equation and
self-consistently calculated single-particle 
spectral functions leads to a reduction of the single-particle DOS near the 
Fermi energy already {\it above} $T_c$, within an energy 
interval of the same order as the BCS gap at zero temperature. 
This behavior may be traced to pair fluctuations above $T_c$
that presage the transition to the superfluid state.  Similar
precursor behavior is known to occur in other systems of
strongly correlated fermions.  In the Hubbard model, for 
example, the formation of local magnetic moments already 
begins above the critical temperature, at which long-range order 
of the moments becomes manifest.  The pseudogap phenomenon 
is, of course, a widely discussed aspect of compounds exhibiting 
high $T_c$ superconductivity \cite{Ran97}, see also Sect. IV.D, where we mention that a pseudogap also forms in the pairing model at finite temperature.
\begin{figure}[bth]
\centerline{
\includegraphics[width=7cm]{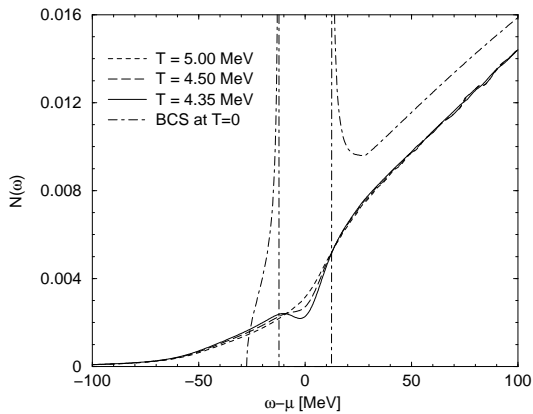}
}
\caption{Nucleon level density as a function of the energy $\w-\m$,
relative to the chemical potential $\m$, given for the density
$\r=\r_0/3$ and three values of the temperature. For comparison,
the BCS result at zero temperature is shown.}
\label{level}
\end{figure}
In the context of nuclear matter, the occurrence of a pseudogap 
phase was first considered by Schnell et al.~\cite{Sch99a} in
the quasiparticle approximation, as noted above.  They 
showed that this effect is partially washed out if a self-consistent 
approximation for the spectral function is implemented, but a full
description should take vertex corrections into account.  A similar
assessment applies to a recent self-consistent solution of the Gor'kov
equation in terms of the spectral function \cite{Dic04}, 
which shows a reduction of the transition temperature 
for quantum condensation.  
However, vertex corrections should also be included 
in this case and may partially compensate the self-energy effects.

In general, it is necessary to take account {\it all bosonic clusters} to
gain a complete picture of the onset of superfluidity.  The picture
developed in the preceding section only includes the effects of
two-particle correlations leading to deuteron clusters.
However, as is well known, the deuteron (binding energy 2.225 MeV) 
is weakly bound compared to
other nuclei.  Higher-$A$ clusters can arise that are more stable.  In
the following subsection, we will consider the formation of $\alpha$ particles,
which are of special importance because of their large binding energy
per nucleon (7 MeV).  We will not include tritons or helions, which
are fermions and not so tightly bound.  Moreover, we will not consider
nuclei in the iron region, which have even larger binding energy per
nucleon than the $\alpha$ and thus comprise the dominant component at
low temperatures and densities. The latter are complex structures of
many particles and are strongly affected by the medium as the density
increases. We assume, that  in the temperature and density region considered here, 
these more complex condensates are not of relevance.

\section{Quartetting and $\alpha$ particle condensation}
\label{sec:moreRPA}
\setcounter{equation}{0}
\renewcommand{\theequation}{6.\arabic{equation}}

The thermodynamic Green's function method described above can be
extended from the two-particle problem to arbitrary numbers of particles in a 
cluster, i.e. to incorporate few-body correlations with $A > 2$. Here we will
proceed with the method of equation of motion and focus on the inclusion of 
four-particle correlations.

The self consistent RPA scheme may be reliable numerically for the case of
two body correlation functions. i.e. for $ph$ or $pp$ ($hh$) - RPA's.
For higher correlation functions the self consistent scheme is at
present not feasible in its full generality and one must be satisfied if higher and more
particle in-medium equations can be solved on the standard RPA level or with relatively strong other  approximations.
Actually, the solution of in-medium equations for more than two particles
is far from trivial. Even on the purely theoretical side the structure
of the equations often has not been worked out completely in the past.
We will discuss this in the context of second RPA, see Sect.VII, 
in the symmetry broken
phase, where the appearance of a double spurious mode at zero energy
is one of the issues of importance.
We here want to consider the in-medium four-particle problem.

In-medium four-particle correlations, for example, appear, if one adds
an $\a$-particle on top of a doubly-magic nucleus, such as $^{208}$Pb and $^{100}$Sn \cite{Cha16,Cha17},
or in semiconductors, where in the gas of excitons, i.e. $ph$ bound
states, may appear bi-excitations, i.e. bound states of two $ph$ pairs.
The effective wave equation contains in mean-field approximation the 
Hartree-Fock self-energy shift of the single-particle energies as well as 
the Pauli blocking of the interaction.

Let us try to set up an
analogous BCS procedure for quartets. Obviously we should write for the wave
function

\begin{equation}
  \label{quartet-gs}
  |Z\rangle = \exp\left[\frac{1}{4!}\sum_{k_1k_2k_3k_4}Z_{k_1k_2k_3k_4}
    c^{\dag}_{k_1}c^{\dag}_{k_2}c^{\dag}_{k_3}c^{\dag}_{k_4}\right]|\mbox{vac}\rangle~,
\end{equation}

\noindent
where the quartet amplitudes $Z$ are fully antisymmetric (symmetric)
with respect to
an odd (even) permutation of the indices.
The task will now be to find a annihilating operator for this quartet condensate
state. Whereas in the pairing case the partitioning of the pair
operator into a linear combination of a fermion creator and a fermion
destructor is unambiguous, in the quartet case there exist two ways to
partition the quartet operator, that is into a single plus a triple or into
two doubles. Let us start with the superposition of a single and a triple. As
a matter of fact it is easy to show that (in the following, we always will assume that all amplitudes are real)

\bea
\label{q-killer}
q_{\nu} &=& \sum_{k_1}u^{\nu}_{k_1}c_{k_1} -\frac{1}{3!}\sum v^{\nu}_{k_2k_3k_4}c^{\dag}_{k_1}c^{\dag}_{k_2}c^{\dag}_{k_3} 
\nn&=&
\sum_{k_1}u^{\nu}_{k_1}[c_{k_1}-\frac{1}{6}\sum_{k_2k_3k_4}Z_{k_1k_2k_3k_4}]~,
\eea
annihilates the quartet state under the condition

\begin{equation}
  \label{Z=v/u}
  Z_{k_1k_2k_3k_4} = \sum_{\nu}(u^{-1})^{\nu}_{k_1}v^{\nu}_{k_2k_3k_4}~.
\end{equation}

\noindent
However, so far, we barely have gained anything, since the above quartet destructor
contains a non-linear fermion transformation which, a priory, cannot be
handled (see, however, Sect. V.C). Therefore, let us first try with a superposition of two fermion pair
operators which is, in a way, the natural extension of the Bogoliubov
transformation in the pairing case, i.e. with $Q = \sum [XP - YP^+]$
where $P^+ =
c^{\dag}c^{\dag}$ is a fermion pair creator. We will, however,  find out that such an
operator cannot annihilate the quartet state of Eq.~(\ref{quartet-gs}).
In analogy to the  Self-Consistent RPA (SCRPA) approach in Sect. II.C
\cite{Jem13}, we will
introduce a slightly more general operator, that is

\begin{eqnarray}
  \label{pair-bogo}
  Q_{\nu} &=& \sum_{k<k'}[X^{\nu}_{kk'}c_kc_{k'} - Y^{\nu}_{kk'}c^{\dag}_{k'}c^{\dag}_k]\nonumber \\
  &-&\sum_{k_1<k_2<k_3k_4}\eta^{\nu}_{k_1k_2k_3k_4}c^{\dag}_{k_1}c^{\dag}_{k_2}c^{\dag}_{k_3}c_{k_4}~,
\end{eqnarray}
with $X,Y$ antisymmetric in $k,k'$.

Applying this operator on our quartet state, we find $Q_{\nu}|Z\rangle = 0$
where the relations between the various amplitudes turn out to be

\bea
\label{XYZ-relations}
Y^{\nu}_{ll'} &=& \sum_{k<k'}X^{\nu}_{kk'}Z_{kk'll'}
\nn
\eta^{\nu}_{l_2l_3l_4;k'} &=& \sum_kX^{\nu}_{kk'}Z_{kl_2l_3l_4}~.
\eea
These relations are quite analogous to the ones which hold in the case of the
SCRPA approach discussed in Sect. II.C \cite{Jem13}. One also notices that the
relation between $X, Y, Z$ amplitudes is similar in structure to the
one of BCS theory for
pairing. As with SCRPA, in order to proceed, we have to approximate the
additional
$\eta$-term. The quite suggestive recipe is to replace in the $\eta$-term of Eq.~(\ref{pair-bogo}) the density operator $c^+_{k'}c_k$ by
its mean value $\langle Z|c^{\dag}_{k'}c_k|Z\rangle/\langle Z|Z\rangle \equiv
\langle c^{\dag}_{k'}c_k\rangle  =
\delta_{kk'}n_k$, i.e.  $c^{\dag}_{k_1}c^{\dag}_{k_2}c^{\dag}_{k_3}c_{k_4} \rightarrow c^{\dag}_{k_1}c^{\dag}_{k_2}n_{k_3}\delta_{k_3k_4}$,  where we supposed that we work in the basis where the single
particle density matrix is diagonal, that is, it is given by the occupation
probabilities $n_k$. This approximation, of course, violates the Pauli
principle but,
as it was found in applications of SCppRPA \cite{Kru94} and SCRPA \cite{Jem13}, we suppose that
also here
this violation will be quite mild (of the order of a couple of percent). With
this approximation, we see that the $\eta$-term only renormalises the $Y$
amplitudes and, thus, the annihilating operator boils down to a linear super
position of a fermion pair destructor with a pair creator. This can then be
seen as a Hartree-Fock-Bogoliubov (HFB) transformation of fermion pair
operators, i.e., pairing of 'pairs'. Replacing the pair operators by
ideal bosons as done in RPA, would
lead to a standard bosonic HFB approach (see \cite{Rin80}, Ch. 9 and Appendix).
Here, however, we will stay with the
fermionic description and elaborate an HFB theory for fermion pairs. For this,
we will suppose that we can use the annihilating property $Q_{\nu}|Z\rangle = 0$
even with the approximate $Q$-operator. As already mentioned,
we assume that this violation of consistency is weak (in the future one may try to work with (\ref{pair-bogo}) in a similar way as done in Sect. V.C).

Let us continue with elaborating our just defined frame. We will then use for the
pair-annihilating operator

\begin{equation}
  \label{approxi-Q}
  Q_{\nu} = \sum_{k<k'}[X^{\nu}_{kk'}c_kc_{k'}
    - Y^{\nu}_{kk'}c^{\dag}_{k'}c^{\dag}_k]/N^{1/2}_{kk'}~,
\end{equation}

\noindent
with (the approximate) property $Q|Z\rangle=0$ and the first
relation in (\ref{XYZ-relations}). The normalisation factor $N_{kk'}= \bar n_k\bar n_{k'} - n_kn_{k'} =
1-n_k-n_{k'}$  has been introduced ($\bar n_k = 1 -n_k$), so that
$\langle[Q,Q^+]\rangle=
\frac{1}{2}\sum (X^2-Y^2) =1$,
i.e., the quasi-pair state $Q^+|Z\rangle$ and the $X, Y$ amplitudes
being normalised to one. Of course, the indices $k'$ have to be chosen so that $\sqrt{N_{kk'}}$ stays real. We now
will minimise the following energy weighted sum rule:

\begin{equation}
  \label{pair-sum-rule}
  \Omega_{\nu} = \frac{\langle Z| [Q_{\nu},[H - 2\mu \hat N,Q^+_{\nu}]]|Z\rangle}
        {\langle Z| [Q_{\nu},Q^+_{\nu}]|Z\rangle}~.
\end{equation}
The minimisation with respect to $X, Y$ amplitudes leads to

\begin{equation}
  \label{f-pair-gap}
  \left (
  \begin{array}{cc}
    {\mathcal {\bf H}}       &  {\bf \Delta}^{(22)}    \\
    -{{\bf \Delta}^{(22)}}^+ & -{\mathcal {\bf H}}^*
  \end{array}
  \right)
  \left(
  \begin{array}{c}
    X^{\nu}  \\
    Y^{\nu}
  \end{array}
  \right)
  =\Omega_{\nu}
  \left(
  \begin{array}{c}
    X^{\nu}  \\
    Y^{\nu}
  \end{array}
  \right)~,
\end{equation}
with (we eventually will consider a symmetrized double commutator in ${\bf H}$)

\begin{eqnarray}
  \label{A-matrix}
  &&{\mathcal {\bf H}}_{k_1k_2,k'_1k'_2} \nonumber \\
  &&= \langle [c_{k_2}c_{k_1},[H-2\mu \hat N,c^+_{k'_1}c^+_{k'_2}]]\rangle/(N^{1/2}_{k_1k_2}N^{1/2}_{k'_1k'_2})\nonumber\\
  && = (\xi_{k_1} + \xi_{k_2})\delta_{k_1k_2,k'_1k'_2}\nonumber \\
  &&  +N^{-1/2}_{k_1k_2}N^{-1/2}_{k'_1k'_2}  \{N_{k_1k_2}\bar v_{k_1k_2k'_1k'_2}N_{k'_1k'_2}\nonumber\\
  && +  [ (\frac{1}{2}\delta_{k_1k'_1}\bar v_{l_1k_2l_3l_4}C_{l_3l_4k'_2l_1}+\bar v_{l_1k_2l_4k'_2}C_{l_4k_1l_1k'_1} )\nonumber \\
    && -(k_1 \leftrightarrow k_2) ] - [k'_1 \leftrightarrow k'_2] \}~,
\end{eqnarray}
where

\bea
  \label{corr-fct}
  C_{k_1k_2k'_1k'_2}&=&\langle c^{\dag}_{k'_1}c^{\dag}_{k'_2}c_{k_2}c_{k_1}\rangle 
\nn&-&
n_{k_1}n_{k_2}[\delta_{k_1k'_1}\delta_{k_2k'_2}-\delta_{k_1k'_2}\delta_{k_2k'_1}]~,
\eea
is the two body correlation function and

\begin{eqnarray}
  \label{B-matrix}
  &&{\bf \Delta}^{(22)}_{k_1k_2,k'_1k'_2} \nonumber \\
  && = - \langle [c_{k_2}c_{k_1},[H-2\mu \hat N ,c_{k'_1}c_{k'_2}]]\rangle/(N^{1/2}_{k_1k_2}N^{1/2}_{k'_1k'_2})\nonumber\\
  &&= N^{-1/2}_{k_1k_2}[( \Delta_{k_1k'_2;k'_1k_2} - k_1 \leftrightarrow k_2 ) - (k'_1 \leftrightarrow k'_2) ]N^{-1/2}_{k'_1k'_2}~, \nonumber \\
\end{eqnarray}
with the quartet order parameter field
\begin{equation}
  \label{a-gap}
  \Delta_{k_1k'_2;k'_1k_2}=
  \sum_{l<l'}\bar v_{k_1k'_2ll'}\langle c_{k'_1}c_{k_2}c_{l'}c_l\rangle~.
\end{equation}
In (\ref{f-pair-gap}) the matrix multiplication is to be understood
as $\sum_{k'_1<k'_2}$
for restricted summation (or as
$\frac{1}{2}\sum_{k'_1k'_2}$ for unrestricted summation ) . We see
from (\ref{B-matrix}) and (\ref{a-gap}) that the bosonic gap
${\bf \Delta}^{(22)}$ involves the quartet order parameter $\langle c_{k'_1}c_{k_2}c_{l'}c_l\rangle$  quite in
analogy to the
usual gap field in the BCS case. The ${\bf H}$ operator in (\ref{f-pair-gap})
has already been discussed in Sect. II.D in connection with SCRPA in
the particle-particle channel. Equation (\ref{f-pair-gap}) has the typical
structure of a bosonic HFB equation but, here, for fermion pairs, instead of
bosons. It remains the task to close those HFB equations in expressing all
expectation values involved in the ${\bf H}$ and ${\bf \Delta}^{(22)}$ fields
by the
$X, Y$ amplitudes. This goes in the following way. Because of the HFB
structure of (\ref{f-pair-gap}), the $X, Y$ amplitudes obey the usual
orthonormality relations, see \cite{Rin80}. Therefore, one can invert
relation (\ref{approxi-Q})
to obtain (see the corresponding inversion relation (\ref{aa}))

\begin{equation}
  \label{inversion}
  c^{\dag}_{k'}c^{\dag}_k = N^{1/2}_{kk'}\sum_{\nu}[X^{\nu}_{kk'}Q^{\dag}_{\nu} + Y^{\nu}_{kk'}Q_{\nu}]~~~(k<k')~,
\end{equation}

\noindent
and by conjugation the expression for $cc$.
With this relation, we can calculate all two-body correlation functions
in (\ref{B-matrix}) and (\ref{A-matrix}) in terms of $X, Y$ amplitudes.
This is achieved in commuting the
destruction operators $Q$ to the right hand side and use the annihilating property.
For example, the quartet order parameter in the gap-field (\ref{a-gap}) is
obtained as $\langle c_{k'_1}c_{k_2}c_{l'}c_l\rangle =
N^{1/2}_{k'_1k_2}\sum_{\nu}X^{\nu}_{k_2k'_1}Y^{\nu}_{ll'}N^{1/2}_{ll'}$.
The task remains how to link the occupation numbers
$n_k = \langle c^+_kc_k\rangle$ to the $X, Y$ amplitudes. Of course, that is
where our partitioning of the quartet operator into singles and triples comes
into play. Therefore, let us try to work with the operator (\ref{q-killer}).
First, as a side-remark, let us notice that if in (\ref{q-killer}) we replace
$c^+_{k_1}c^+_{k_2}$ by its expectation value which is the pairing tensor, we are
back to the standard Bogoliubov transformation for pairing. Here we want to
consider quartetting and, thus, we have to keep the triple operator fully.
Minimising, as in (\ref{pair-sum-rule}). an average single-particle energy, we
arrive at the following equation for the amplitudes $u, v$ in (\ref{q-killer})

\begin{equation}
  \label{sp-eq}
  \left(
  \begin{array}{cc}
    \xi & {\bf \Delta}^{(13)}  \\
        {{\bf \Delta}^{(13)}}^+ & -{\mathcal N}{\mathcal H}^*
  \end{array}
  \right)
  \left(
  \begin{array}{c}
    u \\
    v
  \end{array}
  \right)
  =E
  \left(
  \begin{array}{cc}
    1 &  0             \\
    0 & {\mathcal N}
  \end{array}
  \right)
  \left(
  \begin{array}{c}
    u \\
    v
  \end{array}
  \right)~,
\end{equation}
with (we disregard pairing, i.e., $\langle cc\rangle$ amplitudes)

\begin{equation}
  \label{13-gap}
        {\bf \Delta}^{(13)}_{k;k_1k_2k_3}= \Delta_{kk_3;k_2k_1} -
        [(k_2 \leftrightarrow k_3) - (k_1 \leftrightarrow k_2)]~,
\end{equation}
and
\bea
&&\hspace{-0.5cm} ({\mathcal N}{\mathcal H}^*)_{k_1k_2k_3;k'_1k'_2k'_3}=
\langle \{ c^{\dag}_{k_3}c^{\dag}_{k_2}c^{\dag}_{k_1},[H-3\mu\hat N,
c_{k'_1}c_{k'_2}c_{k'_3}]\}\rangle \label{3body-H} 
\nonumber \\
&&\hspace{-0.5cm} {\mathcal N}_{k_1k_2k_3;k'_1k'_2k'_3}=
\langle \{ c^{\dag}_{k_3}c^{\dag}_{k_2}c^{\dag}_{k_1},
c_{k'_1}c_{k'_2}c_{k'_3}\}\rangle \label{3body-N} ~,
\nonumber \\
\eea
with $\{..,..\}$ the anticommutator.
We will not give ${\mathcal H}$ in full because it is a very complicated
expression involving self-consistent determination of three-body densities \cite{Toh13}.
To lowest order in the interaction it is given by
\begin{eqnarray}
  \label{H33}
  &&{\mathcal H}_{k_1k_2k_3;k'_1k'_2k'_3}= (\xi_{k_1} +\xi_{k_2}+\xi_{k_3})\delta_{k_1k_2k_3,k'_1k'_2k'_3} \nonumber \\
  && + [(1 - n_{k_1}-n_{k_2})\bar v_{k_1k_2k'_1k'_2}\delta_{k_3k'_3} + ~~\mbox{permutations}]~, \nonumber \\
  \label{4body-H}
\end{eqnarray}
where $\delta_{k_1k_2k_3,k'_1k'_2k'_3}$ is the fully antisymmetrised three-fermion Kronecker symbol.
Even this operator is still rather complicated for numerical applications and
mostly one will replace the correlated occupation numbers by their free 
Fermi-Dirac function steps, i.e., $n_k \rightarrow n^{0}_k$. To this order the three-body
norm in (\ref{3body-N}) is given by
\begin{eqnarray}
&&\hspace{-0.3cm}{\mathcal N}_{k_1k_2k_3;k'_1k'_2k'_3} \simeq [\bar n^0_{k_1}\bar n^0_{k_2}\bar n^0_{k_3}
+n^0_{k_1} n^0_{k_2} n^0_{k_3}]\delta_{k_1k_2k_3,k'_1k'_2k'_3}~, \nonumber \\
\end{eqnarray}
with $\bar n^0 = 1-n^0$.
In principle, this effective
three-body Hamiltonian leads to three-body bound and scattering states. In our
application to nuclear matter given below, we will make an even more drastic
approximation and completely neglect the interaction term in the three-body
Hamiltonian. Eliminating under this condition the $v$-amplitudes from
(\ref{sp-eq}), one can write down the following effective single-particle
equation
\begin{eqnarray}
  \label{eff-sp-field}
  && \xi_ku_k^{(\nu)} + \nonumber \\
  &&\hspace{-0.6cm} \sum_{k_1<k_2<k_3k'}\hspace{-0.2cm}\frac{{\bf \Delta}^{(13)}_{kk_1k_2k_3}(\bar n^{0}_{k_1}\bar n^{0}_{k_2}\bar n^{0}_{k_3} + n^{0}_{k_1}n^{0}_{k_2}n^{0}_{k_3}){\bf \Delta}^{{(13)}^*}_{k_3k_2k_1k'}}{E_{\nu} + \xi_{k_1}+\xi_{k_2}+\xi_{k_3}}u_{k'}^{(\nu)} \nonumber \\
  &&= E_{\nu}u_k^{(\nu)}~.
\end{eqnarray}
The occupation numbers are given by
\begin{equation}
  \label{occ-numbers}
  n_k =1-\sum_{\nu}|u^{(\nu)}_k|^2~.
\end{equation}
Of course, one could also replace the uncorrelated occupation numbers in the three-body term by the correlated ones, but we want to refrain from this complication here.

\noindent
The effective single-particle field in (\ref{eff-sp-field}) is graphically interpreted in Fig.~\ref{self-quartet} below in Sect.VI.B.
The gap-fields in (\ref{eff-sp-field}) are then to be calculated as
in (\ref{13-gap}) and (\ref{a-gap})
with (\ref{inversion}) and the system of equations is fully closed.
This is quite in parallel to the pairing case. It is clear that the solution of
this complicated self-consistent quartet problem is at present hardly envisageable.
In cases,
where the quartet consists out of four different fermions, which, in addition, is
rather strongly bound, as this will be the case for the $\alpha$-particle in
nuclear physics, one still can make a very good, but drastic simplification:
one writes the quartic order parameter as a translationally invariant product
of four times the same single-particle wave function in momentum space. We
will see right below, how this goes  when we apply our theory to $\alpha$ particle
condensation in nuclear matter for the simpler case of the determination of the critical temperature. Comparing the effective single-particle field
in (\ref{eff-sp-field}) with the one of standard pairing ~\cite{Fet71}, we
find strong analogies, but also several
structural differences. The most striking is that in the quartet case Pauli
factors figure in the numerator of (\ref{eff-sp-field}), whereas this is not the case for pairing.
In principle, in the pairing case, they are also there, but since
$\bar n_k +n_k = 1$, they drop out. This difference between the pairing and the quartetting cases has quite dramatic
consequences. Namely when the
chemical potential $\mu$ changes from negative (binding) to positive, the
implicit three-hole level density
passes through zero at $\omega-3\mu=0$, because phase space constraints and
energy conservation cannot be fulfilled simultaneously at that point. So, we will see that as soon as with increasing density the chemical potential turns positive, quartet condensation is suppressed. More details can be found in Sect. VI.B below.\\
It is also rather evident that the three hole lines in Fig. \ref{self-quartet} can further be correlated, eventually also to bound states of helions ($^{3}$He) or tritons ($^{3}$H). This is illustrated in Fig. \ref{d,t-clusters} where also the possiblity of deuteron pairing is indicated. It remains a task for the future to include all these processes in a coherent approach of $\alpha$ particle condensation.

\begin{figure}
  \includegraphics[width=8cm]{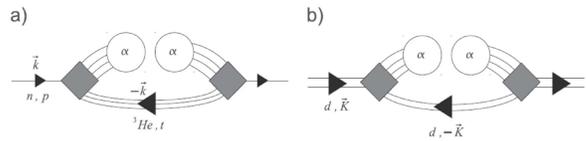}
  \caption{Inclusion of two and three body clusters in the formation of $\alpha$ particle condensation.}
  \label{d,t-clusters}
\end{figure}

\subsection{Critical temperature for $\alpha$ condensation}

We are interested in an example of nuclear physics, where the $\a$-particle
constitutes a particularly strongly bound cluster of four nucleons.
One can ask the question how, for a fixed temperature, the binding energy
of the $\a$-particle varies with increasing temperature.

The effective wave equation has been solved using separable potentials
for $A=2$ by integration. For $A=3,4$ we can use a 
{\it Faddeev approach} \cite{Sog09}.
The shifts of binding energy can 
also be calculated approximately via perturbation theory.  
In Fig.~\ref{shifts} we show the shift of the binding energy of the light 
clusters ($d, t/h$ and $\alpha$) in symmetric nuclear matter as a function 
of density for temperature $T$ = 10 MeV~\cite{Bey00}.
\begin{figure}[t]
\begin{center}
\includegraphics[width=7cm]{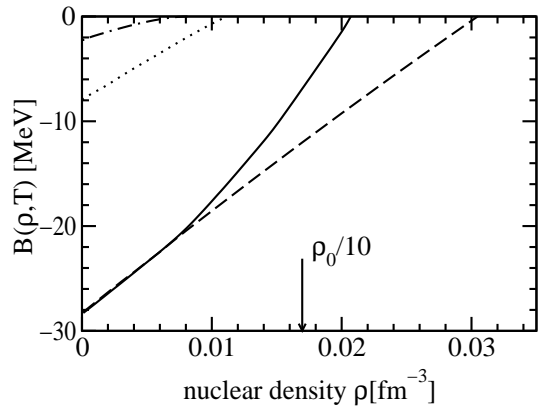}
\caption{Shift of binding energy of the light clusters 
($d$ - dash dotted, $t/h$ - dotted, and $\alpha$ - dashed: perturbation theory,
full line:~non-perturbative Faddeev-Yakubovski equation) in symmetric nuclear 
matter as a function of density for given temperature 
$T = 10$ MeV~\cite{Bey00}.}
\label{shifts}
\end{center}
\end{figure}
It is found that the cluster binding energy decreases with increasing density.
Finally, at the {\it Mott density} $\rho_{A,n,P}^{\rm Mott}(T)$ the bound 
state is dissolved. The clusters are not present at higher densities,  
merging into the nucleonic medium.  
It is found that the $\alpha$ particle at $T=10$ MeV already dissolves at a density  
$\rho_\alpha^{\rm Mott} \approx \rho_0/10$,  
see Fig.~\ref{shifts}.   For a given cluster type characterized 
by $A,n$, we can also introduce the Mott momentum 
$P^{\rm Mott}_{A,n}(\rho,T)$ in terms of the ambient temperature $T$ and 
nucleon density $\rho$, such that the bound states exist only for 
$P \ge P^{\rm Mott}_{A,n}(\rho,T)$.  We do not present an example here, 
but it is intuitively clear that a cluster with high c.o.m. momentum with 
respect to the medium is less
affected by the Pauli principle than a cluster at rest.

  
Since Bose condensation only is of relevance for $d$ and $\alpha$, and the 
fraction of $d$, $t$ and $h$ becomes low compared with that of $\alpha$ with 
increasing density, we can neglect the contribution of them to the equation 
of state in a first approximation. Consequently, if we further neglect the contribution of the 
four-particle scattering phase shifts in the different channels, we can now 
construct an equation of state 
\be
\rho(T, \mu) =\rho^{\rm free}(T, \mu) + \rho^{{\rm bound}, d}
(T, \mu) +\rho^{{\rm bound}, \alpha}(T, \mu)~,
\ee
such that $\alpha$-particles determine the behavior of symmetric nuclear matter at 
densities below $\rho_\alpha^{\rm Mott}$ and temperatures below the 
binding energy per nucleon of the $\alpha$-particle. The formation of 
deuteron clusters alone gives an incorrect description because the 
deuteron binding energy is small, and, thus, the abundance of $d$-clusters 
is small compared with that of $\alpha$-clusters. In the low density region 
of the phase diagram, $\alpha$-matter emerges as an adequate model for 
describing the nuclear matter equation of state.
 
With increasing density, the medium modifications, especially Pauli 
blocking, will lead to a deviation of the critical temperature 
$T_c(\rho)$ from that of an ideal Bose gas of $\alpha$-particles 
(the analogous situation holds for deuteron clusters, i.e., in the 
isospin-singlet channel), see Sect. IV.B.
 
Symmetric nuclear matter is characterized by the equality of the proton 
and neutron chemical potentials, i.e., $\mu_p=\mu_n=\mu$. Then an extended 
Thouless condition based on the relation for the four-body T-matrix 
(in principle, equivalent to (\ref{4-order-eq}) below) 
\begin{widetext}
\begin{eqnarray}
  {\rm T}_4(k_1k_2k_3k_4,k''_{1}k''_{2}k''_{3}k''_{4}, 4 \mu)& =& \sum_{k'_{1}k'_{2}k'_{3}k'_{4}} \Biggl\{
  \frac{v_{k_1k_2,k'_{1}k'_{2}}[1-f(k_1)-f(k_2)] }{ 4
    \mu - \epsilon_{k_1}-\epsilon_{k_2}-\epsilon_{k_3}-\epsilon_{k_4} }
\delta(k_3,k'_{3})\delta(k_4,k'_{4})\nonumber\\
&& \qquad \qquad + {\rm permutations} \Biggr\}
{\rm T}_4(k'_{1}k'_{2}k'_{3}k'_{4},k''_{1}k''_{2}k''_{3}k''_{4}, 4 \mu)
\label{eq:4body_T_matrix}
\end{eqnarray}
\end{widetext}
serves to determine the onset of Bose condensation of $\alpha$-like clusters, 
noting that the existence of a solution of this relation signals a divergence 
of the four-particle correlation function. An approximate solution has been 
obtained by a variational approach, in which the wave function is taken as 
Gaussian incorporating the correct solution for the two-particle 
problem~\cite{Rop98}.
 
\begin{figure}[t]
\begin{center}
\includegraphics[width=7cm]{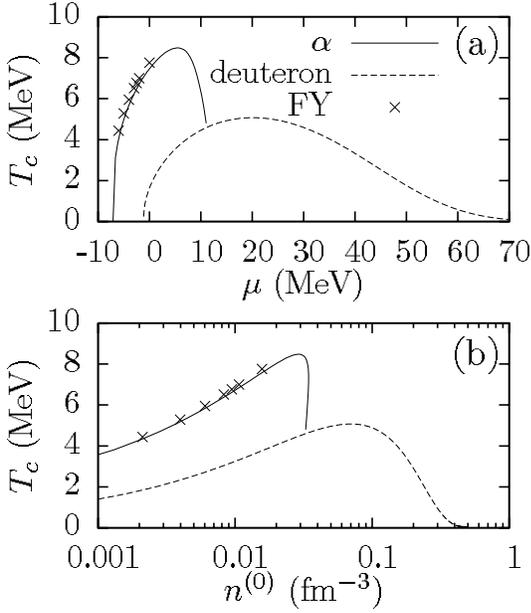}
\caption{\label{fig2}
{Critical temperature of alpha and deuteron condensations 
as functions of (a)~chemical potential and (b)~density of free nucleon
{~\cite{Sog09}}. Crosses ($\times$) correspond to calculation 
of Eq.~(\ref{eq:4body_T_matrix}) with the Malfliet-Tjon interaction (MT I-III) using the 
Faddeev-Yakubovski method.}}
\end{center}
\end{figure}
 
 Eq. (\ref{eq:4body_T_matrix}) at eigenvalue $4\mu$ has been solved numerically 
exactly by the Faddeev-Yakubovsky method employing the Malfliet-Tjon 
force {~\cite{Mal69}}.  The results for the critical temperature 
of $\alpha$-condensation is presented in Fig.~\ref{fig2} as a function of the 
chemical potential {$\mu$}. The exact solution is a numerical challenge and could only be 
obtained for negative {$\mu$}, i.e.~when there exists a bound 
cluster. It is, therefore, important to try another, approximate, solution 
of the in-medium four-body equation. Since the $\alpha$-particle is strongly 
bound, we make a momentum projected mean-field ansatz \cite{Rin80} for the quartet wave 
function \cite{Kam05}
\begin{equation}
\Psi_{1234}= (2\pi)^3 \delta^{(3)}({\bf k}_1 +{\bf k}_2 + {\bf k}_3 + 
{\bf k}_4) \prod_{i=1}^4\varphi({\bf k}_i)\chi^{ST}~,
\label{eq3}
\end{equation}
where $\chi^{ST}$ is the spin-isospin function which we suppose to be the 
one of a scalar ($S=T=0$). We will not further mention it from now on. 
We work in momentum space and $\varphi({\bf k})$ is the as-yet unknown 
single-particle $0S$ wave function. In position space, this leads to the 
usual formula \cite{Rin80} 
$\Psi_{1234} \rightarrow \int d^3R \prod_{i=1}^4 \tilde\varphi({\bf r}_i - 
{\bf R})$ where $\tilde\varphi({\bf r}_i)$ is the Fourier transform of 
$\varphi({\bf k}_i)$. If we take for $\varphi({\bf k}_i)$ a Gaussian shape, 
this gives: 
$\Psi_{1234} \rightarrow \exp[-c\sum_{1\leq i<k \leq 4} ({\bf r}_i - 
{\bf r}_k)^2]$ which is the translationally invariant ansatz often used 
to describe $\alpha$-clusters in nuclei. For instance, it is also employed 
in the $\alpha$-particle condensate wave function of Tohsaki, Horiuchi, Schuck,
R\"opke (THSR) in {Ref.~\cite{Toh01}}.
 
Inserting the ansatz (\ref{eq3}) into the 4-body wave equation equivalent to (\ref{eq:4body_T_matrix}) and integrating over 
superfluous variables, or minimizing the energy, we arrive at a Hartree-Fock 
type of equation for the single-particle $0S$ wave function 
$\varphi(k)=\varphi(|{\bf k}|)$ which can be solved. However, for a general 
two body force 
${v_{{\bf k}_1 {\bf k}_2, {\bf k}'_1 {\bf k}'_2}}$, 
the equation to be solved is still rather complicated. We, therefore, 
proceed to the last simplification and replace the two-body force by a 
unique separable one, that is
\begin{equation}
{v_{{\bf k}_1 {\bf k}_2, {\bf k}'_1 {\bf k}'_2}} = 
\lambda e^{-k^2/k_0^2}e^{-k'^2/k_0^2} (2\pi)^3\delta^{(3)}({\bf K}-{\bf K}')~,
\label{eq05}
\end{equation}
where 
${\bf k}=({\bf k}_1-{\bf k}_2)/2$, ${\bf k}'=({\bf k}'_1-{\bf k}'_2)/2$, 
${\bf K}={\bf k}_1+{\bf k}_2$, and ${\bf K}'={\bf k}'_1+{\bf k}'_2$. 
This means that we take a spin-isospin averaged two-body interaction and 
disregard that, in principle, the force may be somewhat different in the 
$S,T = 0, 1$ or $1, 0$ channels. It is important to remark that for a mean 
field solution the interaction only can be an effective one, very different 
from a bare nucleon-nucleon force. This is contrary to the usual gap equation 
for pairs, to be considered below, where, at least in the nuclear context, 
a bare force can be used as a reasonable first approximation.
 
We are now ready to study the solution of (\ref{eq:4body_T_matrix}) with (\ref{eq3}) 
for the critical temperature $T_c^{\alpha}$, defined by the point where 
the eigenvalue equals $4\mu$. For later comparison, the deuteron (pair) 
wave function at the critical temperature is also deduced from 
Eqs.~(\ref{two_part_bind})
and (\ref{eq05}) to be
\begin{equation}
\phi(k)= -\frac{1-2f(\varepsilon)}{k^2/m-2\mu}\lambda e^{-k^2/k_0^2}
\int \frac{d^3k'}{(2\pi)^3} e^{-k'^2/k_0^2} \phi(k')~,
\label{eq08}
\end{equation}
where $\phi(k)$ is the relative wave function of two particles given by 
$\Psi_{12} \to {\phi(|\frac{{\bf k}_1-{\bf k}_2}{2}|)}$ 
${\delta^{(3)}({\bf k}_1+{\bf k}_2)}$, and $\varepsilon=k^2/(2m)$. 
We also neglected the momentum dependence of the Hartree-Fock mean field 
shift in {Eq.~(\ref{eq08})}. It, therefore, can be 
incorporated into the chemical potential $\mu$. With Eq.~(\ref{eq08}), 
the critical temperature of pair condensation is obtained from the following 
equation:
\begin{equation}
1=-\lambda \int \frac{d^3k}{(2\pi)^3}
\frac{1-2f(\varepsilon)}{k^2/m-2\mu} e^{-2k^2/k_0^2}~.
\label{eq010}
\end{equation}
 
In order to determine the critical temperature for $\alpha$-particle 
condensation, we have to adjust the temperature so that the eigenvalue 
of (\ref{eq:4body_T_matrix}) equals $4\mu$. The result 
is shown in Fig. \ref{fig2}(a). In order to get an idea how this converts 
into a density dependence, we use for the moment the free gas relation 
between the density $n^{(0)}$ of uncorrelated nucleons and the chemical 
potential
\begin{equation}
n^{(0)}=4\int \frac{d^3k}{(2\pi)^3} f(\varepsilon)~.
\label{eq-density}
\end{equation}
We are well aware of the fact that this is a relatively gross simplification, 
for instance, at the lowest densities and we intend to generalize our theory 
in the future so that correlations are included into the density. This may be 
done along the work of Nozi\`eres and Schmitt-Rink \cite{Noz85}. 
The two open constants $\lambda$ and $k_0$ in 
Eq. (\ref{eq05}) are determined so that binding energy ($28.3$ MeV) and 
radius ($1.71$ fm) of the free ($f_i=0$) $\alpha$-particle come out right. 
The adjusted parameter values are: $\lambda=-992$ MeV fm$^{3}$, and 
{$k_0=1.43$} fm$^{-1}$. The results of the calculation are 
shown in Fig.~\ref{fig2}.
 
 
In Fig.~\ref{fig2}, the maximum of critical temperature 
$T^{\alpha}_{c, {\rm max}}$ is at $\mu=5.5$ MeV, and the 
$\alpha$-condensation can exist up to  $\mu_{\rm max}=11$ MeV.  
It is very remarkable that the results obtained with (\ref{eq3}) 
for $T_c^{\alpha}$ very well agree with the exact solution of (\ref{eq:4body_T_matrix}) using the Malfliet-Tjon interaction 
(MT I-III)~\cite{Mal69}
with the Faddeev-Yakubovski method also shown by 
crosses in Fig.~\ref{fig2} (the numerical solution only could be obtained 
for negative values of $\mu$). This indicates that $T_c^{\alpha}$ is 
essentially determined by the Pauli blocking factors.
 
In Fig.~\ref{fig2} we also show the critical temperature for deuteron 
condensation derived from Eq.~(\ref{eq010}). In this case, the bare force 
is adjusted with $\lambda= -1305$ MeV fm$^3$ and $k_0 = 1.46$ fm$^{-1}$ 
to get experimental energy ($-2.2$ MeV) and radius ($1.95$ fm) of the 
deuteron. It is seen that at higher densities deuteron condensation wins 
over the one of $\alpha$-particle. The latter breaks down rather abruptly 
at a critical positive value of the chemical potential. Roughly speaking, 
this corresponds to the point where the $\alpha$-particles start to overlap. 
This behavior stems from the fact that Fermi-Dirac distributions in the 
four-body case, see (\ref{eq:4body_T_matrix}), can never become 
step-like, as in the two-body case, even not at zero temperature, since 
the pairs in an $\alpha$-particle are always in motion. Therefore, 
no threshold effect occurs as with pairing for Cooper pairs at rest. 
As a consequence, $\alpha$-condensation generally only exists as a BEC 
phase and the weak coupling regime is absent, see also discussion in 
{Sec.~\ref{sebsec:gap_equation}}.
 
 An important consequence of this study is that at the lowest temperatures, 
Bose-Einstein condensation occurs for $\alpha$-particles rather than for 
deuterons.  As the density increases within the low-temperature regime, 
the chemical potential $\mu$ first reaches $-7$ MeV, where the $\alpha$-particles 
start to Bose-condense.  In contrast, Bose condensation of deuterons would not occur 
until $\mu$ rises to $-1.1$ MeV.
 
The {\it ``quartetting''} transition temperature sharply drops as the rising 
density approaches the critical Mott value, at which the four-body bound 
states disappear.  At that point, pair formation in the isospin-singlet 
deuteron-like channel comes into play, and a deuteron condensate will exist 
below the critical temperature for BCS pairing up to densities above the 
nuclear-matter saturation density $\rho_0$, as described in the previous 
Section. Of course, also isovector {\it n-n} and 
{\it p-p} pairing develops. The critical (Mott) density, at which 
the $\alpha$ condensate disappears is estimated to be $\rho_0/3$. 
Therefore, $\alpha$-particle condensation primarily only exists in the 
Bose-Einstein-Condensed (BEC) phase and there exist no 
phase, where the quartets acquire a large extension as Cooper pairs do in 
the weak coupling regime.  However, the variational approaches of 
Ref.~\cite{Rop98} and of Eq.~(\ref{eq3}), on which this conclusion is 
based, represent only first attempts at the description of the transition 
from quartetting to pairing.  The detailed nature of this fascinating 
transition remains to be clarified. Many different questions arise in 
relation to the possible physical occurrence and experimental manifestations 
of quartetting: Can we observe the hypothetical ``$\alpha$ condensate'' 
in nature?  What about thermodynamic stability?  What happens with 
quartetting in asymmetric nuclear matter?  Are more complex quantum 
condensates possible?  What is their relevance for finite nuclei?  
As discussed, the special type of microscopic quantum correlations 
associated with quartetting may be important in nuclei, its role in these 
finite inhomogeneous systems being similar to that of pairing 
\cite{Toh01}.
 
On the other hand, if at all, $\alpha$-condensation in compact stars occurs 
at strongly asymmetric matter. It is, therefore, important to generalize the 
above study for symmetric nuclear matter to the asymmetric case. This can be 
done straightforwardly again using our momentum projected mean-field ansatz 
(\ref{eq3}) generalized to the asymmetric case. This implies to introduce 
two chemical potentials, one for neutrons and one for protons. We also have to 
distinguish two single-particle wave functions in our product ansatz. 
For the results, we refer the reader to \cite{Sog10} and simply state that 
as a function of asymmetry the $\alpha$ wins over the deuteron because the latter's 
binding is much weaker than the one of the $\alpha$-particle. 

 {
In conclusion the $\alpha$-particle (quartet) condensation was investigated 
in homogeneous symmetric nuclear matter as well as in asymmetric nuclear 
matter.  We found that the critical density at which the $\alpha$-particle 
condensate appears is estimated to be around ${\rho_0}/3$ in the symmetric 
nuclear matter, and the $\alpha$-particle condensation can occur only at low 
density. This result is consistent with the fact that the Hoyle state ($0^+_2$)
of $^{12}$C, which is considered as a three-$\alpha$ condensed state also has a very low density $\rho \sim \rho_0/3$. 
On the other hand, in the asymmetric nuclear matter, the critical temperature 
$T_c$ for the $\alpha$-particle condensation was found to decrease with 
increasing asymmetry. However, $T_c$ stays relatively high for very strong 
asymmetries, a fact of importance in the astrophysical context. The asymmetry 
affects deuteron pairing more strongly than $\alpha$-particle condensation. 
Therefore, at high asymmetries, if at all, $\alpha$-particle condensate seems 
to dominate over deuteron-like pairing at all possible densities.
}

\subsection{'Gap' equation for quartet order parameter}
\label{sebsec:gap_equation}
In the preceding section, we considered $\alpha$-particle condensation at finite temperature, specifically we determined the critical temperature for the onset of condensation. We now want to consider $\alpha$ condensation at zero temperature, where the full non-linear order parameter equation has to be solved, similar to the solution of the gap equation in the case of pairing. 
For macroscopic $\alpha$ condensation it is, of course, not conceivable to 
work with a number projected $\alpha$-particle condensate wave function as 
we did, when in finite nuclei only a couple of $\alpha$ particles were present
\cite{Toh01}. 
We rather have to develop an analogous procedure to BCS theory, but 
{generalized} for quartets. In principle, a wave function of the already mentioned type 
{$|\alpha\rangle = \exp[\sum_{1234}z_{1234}c_1^+c_2^+c_3^+c_4^+]|
{\rm vac}\rangle$} would be the ideal {generalization} of the BCS wave function
for the case of quartets. However, unfortunately, it is unknown so far 
(see, however, Ref.~\cite{Jem13} and Sect. V.C) how to treat such a 
complicated many-body wave function mathematically in a reasonable way. So, we rather attack 
the problem from the other end, that is with a Gor'kov type of approach, 
well known from pairing, but here extended to the quartet case. 
Since, naturally, the formalism is complicated, we only will outline the 
main ideas and refer for details to the literature. 
 
Actually, one part of the problem is written down easily. Let us guide from a 
particular form of the gap equation in the case of pairing. We have at zero 
temperature (see also eq.(\ref{two_part_bind}))
\bea
&&(\varepsilon_{k_1}+\varepsilon_{k_2})\kappa_{k_1k_2}+ (1-n_{k_1} - n_{k_2})
\frac{1}{2}\sum_{k_1'k_2'} {\bar{v}_{k_1k_2k_1'k_2'}} 
\kappa_{k_1'k_2'} 
\nn&=& 
2\mu\kappa_{k_1k_2}~,
\label{eq:gap_eq_two_particles}
\eea
where $\kappa_{k_1k_2} = \langle c_{k_1}c_{k_2}\rangle$ is the pairing tensor, 
$n_i = \langle c_i^+c_i \rangle$ are the BCS occupation numbers, and 
{$\bar{v}_{k_1k_2k_1'k_2'}$ denotes the antisymmetrized matrix element of the 
two-body interaction.} The $\varepsilon_i$ are the usual mean-field energies. 
Equation {(\ref{eq:gap_eq_two_particles})} is equivalent to the usual gap 
equation in the case of zero total momentum and opposite spin, i.e. in short 
hand: $k_2=\bar k_1$, where the bar stands for 'time reversed conjugate'. With the EOM,
the extension of {(\ref{eq:gap_eq_two_particles})} to the quartet case is 
formally written down without problem. It is a direct consequence of our self-consistent quartet equations derived above. As in the pairing case, we then linearize to large extent the problem keeping only the occupation numbers coupled to the quartet condensate in the self-consistent cycle. We write for zero temperature
\begin{widetext}
\begin{eqnarray}
(\varepsilon_{1234} - 4\mu)\kappa_{1234} &=& (1-n_1-n_2)
\frac{1}{2}\sum_{k_1'k_2'} {\bar{v}_{k_1k_2k_1'k_2'}} 
\kappa_{k_1'k_2'k_3k_4} 
\nonumber\\
&+& (1-n_{k_1}-n_{k_3})\frac{1}{2}\sum_{k_1'k_3'} {\bar{v}_{k_1k_3k_1'k_3'}} 
\kappa_{k'_1k_2k'_3k_4} + {\rm all~permutations}~,
\label{4-order-eq}
\end{eqnarray}
\end{widetext}
with $\kappa_{k_1k_2k_3k_4} = \langle c_{k_1}c_{k_2}c_{k_3}c_{k_4} \rangle $ the quartet order 
parameter.
This is formally the same equation as Eq.~(\ref{eq:4body_T_matrix}) with, 
however, 
the Fermi-Dirac occupation numbers replaced by the zero temperature quartet 
correlated single-particle occupation numbers, similar to the BCS case. 
For the quartet case, the crux lies in the determination of those occupation 
numbers. However, as we have seen, they can very naturally be determined from the generalized s.p. self-energy in (\ref{eff-sp-field}).

 

Let us discuss some properties of the s.p. self-energy in the case of $\alpha$-particles with respect to the one of pairing, see Fig.
\ref{self-quartet}.
Put aside the
difficulty 
to derive a manageable 
expression for this 'quartet' s.p. self-energy, what immediately strikes 
is that instead of only one 'backward going line' with $(-{\bf p},-\sigma)$ 
as in the pairing case, we now have three backwards going lines. 
As a consequence, the three momenta ${\bf k}_1$, ${\bf k}_2$, ${\bf k}_3$ 
in these lines are only constrained so that their sum is equal to 
${\bf k}_1 + {\bf k}_2 + {\bf k}_3 = - {\bf p}$ in order that the total 
momentum of the order parameter is zero and, thus, the remaining 
freedom has to be summed over. This is in strong contrast to the pairing 
case, where the single backward-going line is constrained by momentum 
conservation to have momentum $-{\bf p}$, so that together with the incoming particle at ${\bf p}$ the total momentum of the gap function is zero, see also Fig. \ref{self-quartet}. So, no internal summation 
occurs in the self-energy belonging to pairing. The consequence of this additional momentum
summation in the mass operator for quartetting leads, with respect to pairing, 
to a completely different analytic structure of the self-energy in case of 
quartetting. This is best studied with the so-called three-hole level density 
$g_{3h}(\omega)$, which is related to the imaginary part of the three-hole 
Green's function 
${G^{3h}(k_1, k_2,k_3; \omega)} = 
({\bar f_1} {\bar f_2} {\bar f_3} + f_1f_2f_3)/(\omega + 
\varepsilon_{123})$ with $\varepsilon_{123}=\varepsilon_{1}+\varepsilon_{2}+
\varepsilon_{3}$ figuring in the self-energy, see Fig. \ref{self-quartet}
\begin{widetext}
\begin{eqnarray}
g_{3h}(\omega) &=&-\int \frac{d^3k_1}{(2\pi)^3}\frac{d^3k_2}{(2\pi)^3}
\frac{d^3k_3}{(2\pi)^3} {\rm Im} G^{(3h)}(k_1,k_2,k_3;\omega+i\eta)
\nonumber \\
&=&\int \frac{d^3k_1}{(2\pi)^3}\frac{d^3k_2}{(2\pi)^3}\frac{d^3k_3}{(2\pi)^3}
(\bar f_1 \bar f_2 \bar f_3+ f_1 f_2 f_3)
\pi\delta(\omega+\varepsilon_1
                +\varepsilon_2
                +\varepsilon_3)~.
\label{eq-ld}
\end{eqnarray}
\end{widetext}
 
\begin{figure}[ht]
  \begin{center}
 \includegraphics[width=7cm]{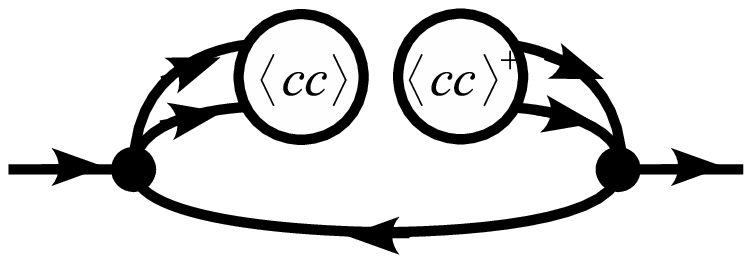}   
\includegraphics[width=7cm]{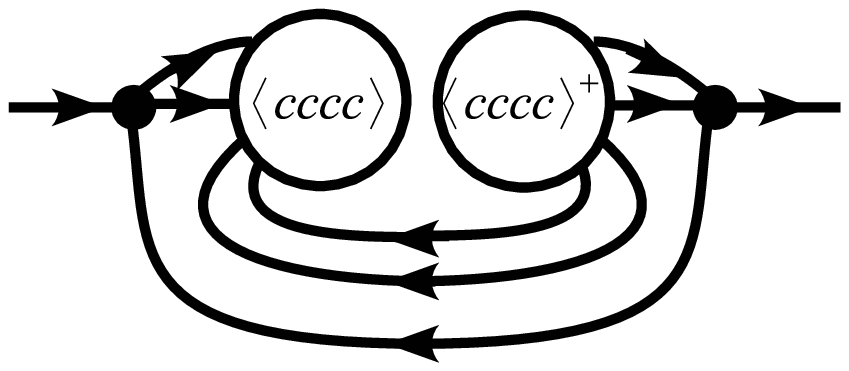}
\caption{
\label{self-quartet}
Graphical representation of the BCS self-energy (top) and the approximate $\alpha$-BEC self-energy
$M^{\rm quartet}$ (bottom) of Eq.~(\ref{eff-sp-field}). }
\end{center}
\end{figure}

In Fig.~\ref{fig-ld} we show the level density at zero temperature 
($f(\omega)=\theta(-\omega)$), where it is calculated with the proton mass 
$m=938.27$~MeV 
{~\cite{Sog10}}. Two cases have to be considered, chemical potential 
$\mu$ positive or negative. In the latter case we have binding of the quartet. 
Let us first discuss the case $\mu>0$. We remark that, in this case, the 
{$3h$} level density goes through zero at $\omega=0$, i.e., since we are 
measuring energies with respect to the chemical potential $\mu$, just in the 
region where the quartet correlations should appear. This is at strong 
variance with the pairing case where the {$1h$} level density, 
$g_{1h}(\omega)=\int \frac{d^3k}{(2\pi \hbar)^3} (\bar f_k + f_k)
\delta(\omega+\varepsilon_k) = \int \frac{d^3k}{(2\pi \hbar)^3}
\delta(\omega+\varepsilon_k)$, does not feel any influence from the medium 
and, therefore, the corresponding level density varies (neglecting the mean 
field for the sake of the argument) like in free space with the square root  
of energy. In particular, this means that the level density is {\it finite} 
at the Fermi level. This is a dramatic difference with the quartet case and 
explains why Cooper pairs can strongly overlap whereas for quartets this is 
impossible as we will see below. We also would like to point out that the 
{$3h$} level density is just the mirror to the {$3p$} level density, which has 
been discussed in Ref.~\cite{Bli86}.
 
For the case $\mu<0$, where anyway the $f_i$'s are zero at $T$=0,  
there is nothing very special, besides the fact 
that the level density is non-vanishing only for negative values of $\omega$ and that the 
upper boundary is given by $\omega = 3\mu$. Therefore, the level density of 
Eq.~(\ref{eq-ld}) is zero for $\omega>3\mu$. Therefore, in the BEC 
regime ($\mu < 0$), there is no marked difference between the pairing and 
quartetting cases.

\begin{figure}[ht]
\begin{center}
\includegraphics[width=6cm]{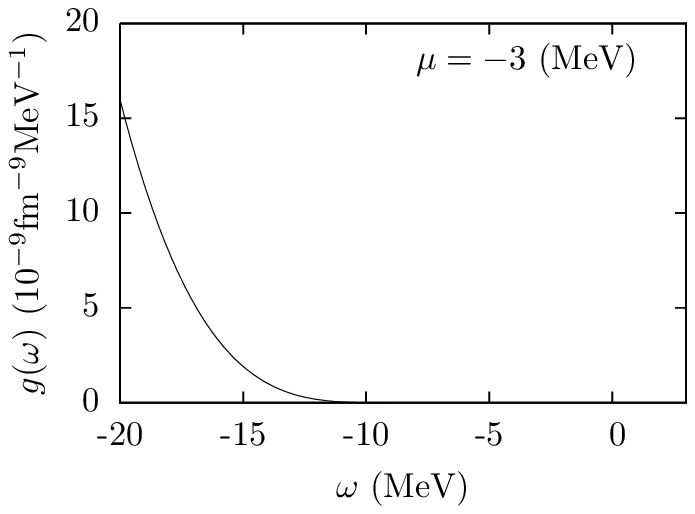}
\includegraphics[width=6cm]{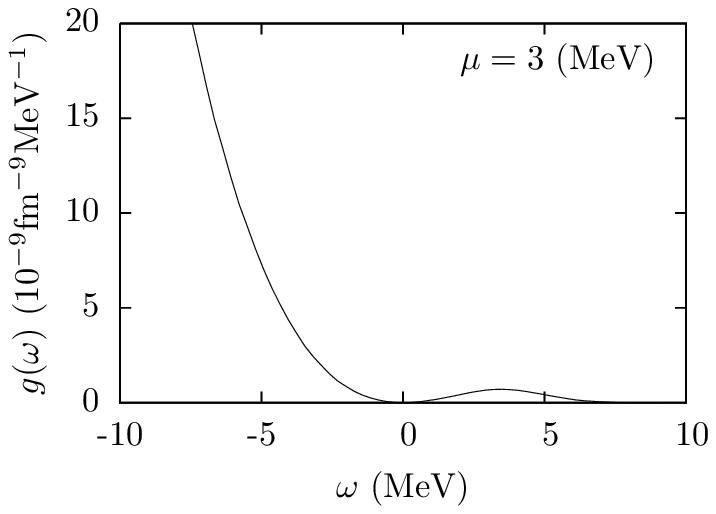}
\includegraphics[width=6cm]{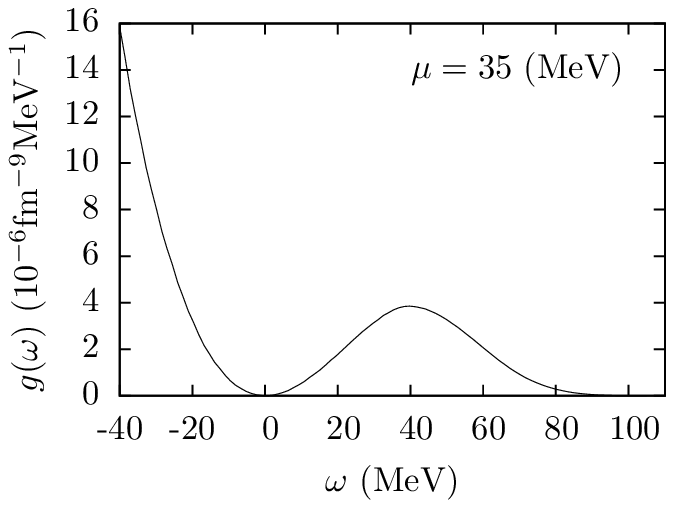}
\caption{\label{fig-ld}
{$3h$} level densities defined in Eq.~(\ref{eq-ld}) for various values of 
the chemical potential $\mu$ at a zero temperature
{~\cite{Sog10}}. }
\end{center}
\end{figure}
 
\begin{figure}[ht]
\begin{center}
\includegraphics[width=7cm]{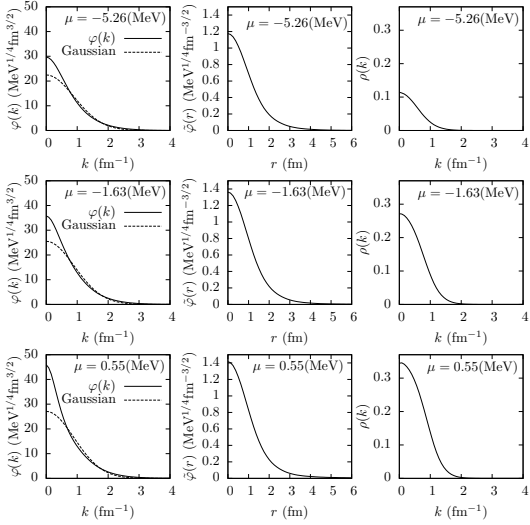}
\caption{\label{fig-spwf_1}
Single-particle wave function $\varphi(k)$ in $k$-space (left) and
$r$-space $\tilde \varphi(r)$ (middle), and occupation numbers (right) at 
$\mu=-5.26$ (top), $-1.63$ (middle) and $0.55$ (bottom). 
The $r$-space wave function $\tilde \varphi(r)$ is derived from the 
Fourier transform of $\varphi(k)$ by 
$\tilde \varphi(r)=\int d^3k e^{i\vec k \cdot \vec r}\varphi(k)/(2\pi)^3$. 
The dashed line in the left panels correspond to the Gaussian with same norm 
and {r.m.s.} momentum as $\varphi(k)$
{~\cite{Sog10}}.}
\end{center}
\end{figure}


The complexity of the calculation in {Eq.~(\ref{4-order-eq})} 
is  much reduced using for the order parameter {${\langle cccc \rangle}$} 
our mean-field ansatz projected onto zero total momentum, as it was already 
very successfully employed with Eq.~(\ref{eq3}),
{
\begin{eqnarray}
{\langle c_1c_2c_3c_4 \rangle} &\rightarrow& 
\phi_{\vec k_1 \vec k_2,\vec k_3 \vec k_4}\chi_0, \nonumber \\
\phi_{\vec k_1 \vec k_2,\vec k_3 \vec k_4}
&=&
\varphi(|\vec k_1|)\varphi(|\vec k_2|)\varphi(|\vec k_3|)\varphi(|\vec k_4|)
\nonumber \\
&&\times
(2\pi)^3\delta(\vec k_1+\vec k_2+\vec k_3+\vec k_4)~,
\label{eq-PHF4bwf}
\end{eqnarray}
where $\chi_0$ is the spin-isospin singlet wave function.} 
It should be pointed out that this product ansatz with four identical $0S$ 
single-particle wave functions is typical for a ground state configuration of 
the $\alpha$ particle. Excited configurations with wave functions of 
higher nodal structures may eventually be envisaged for other physical 
situations. We also would like to mention that the momentum conserving 
$\delta$-function induces strong correlations among the four particles 
and (\ref{eq-PHF4bwf}) is, therefore, a rather non-trivial variational 
wave function.

For the two-body interaction  
${v}_{k_1k_2,k_3k_4}$ in Eq.~(\ref{4-order-eq}), 
we employ the same separable form  as done already for the quartet 
critical temperature.
 
 
As already mentioned, in this pilot application of our self-consistent 
quartet theory, we only will consider the zero temperature case. 
As a definite physical example, we will treat the case of nuclear physics 
with the particularly strongly bound quartet, the $\alpha$ particle. 
It should be pointed out, however, that, if scaled appropriately, all energies 
and lengths can be transformed to other physical systems. For the nuclear 
case it is convenient to measure energies in Fermi energies 
$\varepsilon_F = 35$~MeV and lengths in inverse Fermi momentum 
$k_F^{-1} = 1.35^{-1}$~fm which are the values at nuclear saturation. 
 
We are now in a position to solve, as in the BCS case, the  coupled 
equations (\ref{occ-numbers}, \ref{4-order-eq}) for the quartet order parameter and the single-particle 
occupation numbers from the single-particle Dyson equation with the self-energy (\ref{eff-sp-field}) self-consistently.
The single-particle wave functions and occupation numbers obtained from 
the above cycle are shown in Fig.~\ref{fig-spwf_1}. We also insert the 
Gaussian wave function with the same r.m.s. momentum as the single-particle 
wave function in the left figures in Fig.~\ref{fig-spwf_1}. 
As shown in Fig.~\ref{fig-spwf_1}, the single-particle wave function is 
sharper than a Gaussian.

We could not obtain a convergent solution for $\mu>0.55$~MeV. 
This difficulty has precisely its origin in the fact that the three-hole 
level density goes through zero at 3$\mu>0$, just where the four-body 
correlations should build up, as this was discussed above. 
In the r.h.s. panels of Fig.~\ref{fig-spwf_1} we also show the corresponding 
occupation numbers. We see that they are very small. However, they increase 
for increasing values of the chemical potential. For $\mu = 0.55$~MeV 
the maximum of the occupation still only attains 0.35 what is far away from 
the saturation value of one. What really happens for larger values of the 
chemical potential, is unclear. Surely, as discussed above,
the situation for the quartet 
case is completely different from the standard pairing case. This is due to 
the just mentioned particular behavior of the 3h level density. Due to this fact, the inhibition 
to go into the positive $\mu$ regime is here even stronger than in the case 
of the critical temperature~\cite{Sog09}. This transition is akin to a Quantum Phase Transition (QPT), where the density plays the role of a control parameter \cite{Ebr20}.

In conclusion of this quartet condensation section, let us say that we could build with the help of the EOM method a self-consistent scheme for the quartet order parameter, quite in analogy to the pairing case. Strong differences with the latter have been revealed. In first place comes the fact that quartet condensation occurs only as long as the quartet is bound, that is as long as the corresponding chemical potential $\mu$ is negative. Contrary to the pairing case, there exists no quartet condensation effect for states in the continuum. This stems from a radically different behavior of the level densities involved in both cases, as explained above. It is very rewarding that in finite nuclei, there seems to exist the $\alpha$-particle condensation phenomenon. As we have investigated, the famous Hoyle state of $^{12}$C seems to be well described by a three-$\alpha$ condensate type of wave function (THSR). It remains a challenge to find other quartet condensates in nature \cite{Toh17}.

\section{Second RPA and extensions}
\label{sec:ERPA}
\setcounter{equation}{0}
\renewcommand{\theequation}{7.\arabic{equation}}

RPA and SCRPA are designed to describe the excitation of one-phonon states.
For the description of properties of two-phonon states and the coupling of 
one-phonon states to two-phonon states,
we need to use extensions of RPA, which enable us to calculate
two-body transition amplitudes in addition to one-body transition amplitudes. 
One of such extended RPA theories is the second RPA (SRPA)
\cite{Saw62} which has been used for the study of decay properties
of giant resonances \cite{Dro90}. However, the original formulation of SRPA is 
not complete as will be demonstrated in this section. Also the effects of 
ground-state correlations are
not included in SRPA. Recently, we have developed a more general Extended RPA 
(ERPA) 
which contains the effects of ground-state correlations and meets the 
requirement of hermiticity \cite{Toh04}.
In this section we present the formulation of our ERPA and show recent 
applications. 

We consider the Hamiltonian
\begin{eqnarray}
H=\sum_{kk'} t_{kk'} a^{\dag}_k a_{k'}
+\frac{1}{2}\sum_{k_1k_2k_1'k_2'} v_{k_1k_2k_1'k_2'} a^{\dag}_{k_1}a^{\dag}_{k_2}
a_{k_2'}a_{k_1'}~.
\nn
\end{eqnarray}

\subsection{ Ground state}

The ground state $|0\rangle$, which is used to evaluate various
matrices in ERPA, is given by the stationary conditions of the occupation matrix $n_{\alpha\alpha'}$, the two-body correlation matrix
$C_{klk'l'}$ and the three-body correlation matrix
$C_{klmk'l'm'}$
defined as
\begin{eqnarray}
n_{kk'}&=&\langle0|a^{\dag}_{k'}a_{k}|0\rangle
\nn
C_{klk'l'}&=&\langle0|a^{\dag}_{k'}a^{\dag}_{l'}a_{l}a_{k}|0\rangle
-{\cal A}(n_{kk'}n_{ll'})
\nn
C_{klmk'l'm'}&=&
\langle0|a^{\dag}_{k'}a^{\dag}_{l'}a^{\dag}_{m'}a_{m}a_{l}a_{k}|0\rangle
\nn&-&
          {\cal A}(n_{kk'}n_{ll'}n_{mm'}+{\cal S}(n_{kk'}C_{lml'm'}))
          \label{stationarity2+3}~,
\nn
\end{eqnarray}
where ${\cal A}$ and ${\cal S}$ mean that the products in the parentheses
are properly antisymmetrized and symmetrized under 
the exchange of single-particle indices \cite{Wan85}.
The stationary conditions whose significance will become clear immediately 
below are written as
\begin{eqnarray}
F_1(kk')=\langle0|[a^{\dag}_{k'}a_{k},\hat{H}]|0\rangle&=&0 
\nn
F_2(klk'l')=\langle0|[a^{\dag}_{k'}a^{\dag}_{l'}a_{l}a_{k},\hat{H}]|0\rangle&=&0 
\nn
F_3(klmk'l'm')=\langle0|[a^{\dag}_{k'}a^{\dag}_{l'}a^{\dag}_{m'}a_{m}a_{l}a_{k},
\hat{H}]|0\rangle&=&0~.
\nn
\label{gs3}
\end{eqnarray}
Since a four-body correlation matrix is neglected, the expectation values of four-body operators 
in Eq.(\ref{gs3}) are approximated by the products of 
$n_{kk'}$, 
$C_{klk'l'}$ and 
$C_{klmk'l'm'}$.
Evaluation of the above conditions is straightforward and 
is given in Ref. \cite{Toh04} using the single-particle states which satisfy the 
HF-like mean-field hamiltonian
\begin{eqnarray}
h(\rho)\phi_k(1)=\epsilon_k\phi_k(1)~.
\label{hfeq}
\end{eqnarray}
Here, $h(\rho)$ is the mean-field Hamiltonian and $\rho$ is the one-body density matrix given by 
$\rho(1,1')=\sum_{kk'}n_{kk'}\phi_k(1)\phi^*_k(1')$ and numbers 
indicate spatial, spin and isospin coordinates.
The three-body correlation matrix 
$C_{klmk'l'm'}$ is necessary to make ERPA hermitian, as will be
discussed below.
To obtain the ground state implies that all quantities $n_{kk'}$, 
$C_{klk'l'}$, $C_{klmk'l'm'}$
and $\phi_k$ are determined under the conditions
(\ref{gs3}-\ref{hfeq}). It was found \cite{Toh04a} that this task can be 
achieved  
using the gradient method: Starting from a simple ground state, such as the 
HF ground state, where $n_{kk'}$, 
$C_{klk'l'}$ and $C_{klmk'l'm'}$ can be easily evaluated, we iterate
\begin{eqnarray}
&&\left(
\begin{array}{c}
n(i+1)\\
C_2(i+1)\\
C_3(i+1)
\end{array}
\right)
=
\left(
\begin{array}{c}
n(i)\\
C_2(i)\\
C_3(i)
\end{array}
\right)
\nn&-&\alpha\left(
\begin{array}{ccc}
\frac{\delta F_1}{\delta n} &\frac{\delta F_1}{\delta C_2} & 0 \\
\frac{\delta F_2}{\delta n} & \frac{\delta F_2}{\delta C_2} & \frac{\delta F_2}{\delta C_3} \\
\frac{\delta F_3}{\delta n} & \frac{\delta F_3}{\delta C_2} & \frac{\delta F_3}{\delta C_3}
\end{array}
\right)^{-1}
\left(
\begin{array}{c}
F_{1}(i)\\
F_{2}(i)\\
F_{3}(i)
\end{array}
\right)~,
\label{gradient}
\end{eqnarray}
until convergence is achieved. Here, $n$, $C_2$ and $C_3$ imply
$n_{kk'}$, $C_{klk'l'}$, 
$C_{klmk'l'm'}$, respectively.
Equation (\ref{gradient}) is coupled to
$h\phi_k=\epsilon_k\phi_k$, because $h$ depends on $n_{kk'}$. 
The convergence process is controlled using a small parameter ($\alpha$).
The matrix consisting of the functional derivatives of 
$F_1, ~F_2$ and $F_3$ has a close relation with the hamiltonian matrix of ERPA 
\cite{Toh04a}. Therefore, the ground state is not independent of the excited 
states in ERPA. Of course, the equations (\ref{gs3}) can also be written out explicitly and one can try to solve for the implied correlation functions. How this can be achieved is demonstrated in Sect.V.C with the very simple Lipkin model.


\subsection{Extended RPA (ERPA) equation}

We consider an excitation operator consisting of one-body and two-body operators
\begin{eqnarray}
\hat{Q}^{\dag}_{\mu}&=&\sum_{kk'}{x^\mu_{kk'}:a^{\dag}_k a_{k'}:}
\nn&+&
\sum_{k_1k_2k_1'k_2'}{X^\mu_{k_1k_2k_1'k_2'}:a^{\dag}_{k_1}a^{\dag}_{k_2}a_{k_2'}a_{k_1'}:}~,
\label{exoper}
\end{eqnarray}
where :~: implies that uncorrelated parts consisting of lower-level operators are to be subtracted; for example, 
\begin{eqnarray}
:a^{\dag}_{k'}a_{k}:&=&a^{\dag}_{k'}a_{k}-n_{kk'}
\nn
:a^{\dag}_{k'}a^{\dag}_{l'}a_{l}a_{k}:&=&a^{\dag}_{k'}a^{\dag}_{l'}a_{l}a_{k}
-{\cal AS}(n_{kk'}:a^{\dag}_{l'}a_{l}:)
\nn&-&
[{\cal A}(n_{kk'}n_{ll'})+C_{klk'l'}]~.
\end{eqnarray}
where ${\cal AS}$ stands for properly antisymmetrized and symmetrized.
The ERPA equations are derived using the EOM method minimizing, as usual, the energy weighted sum-rule (\ref{gensumrule1}). 
In the evaluation of the matrix elements we approximate
the ERPA ground state with $|0\rangle$ which satisfy Eqs. (\ref{gs3})-(\ref{hfeq}). The ERPA equations 
are written as 
\begin{eqnarray}
\left(
\begin{array}{cc}
A & C \\
B & D 
\end{array}
\right)\left(
\begin{array}{c}
x^\mu \\
X^\mu
\end{array}
\right)=\Omega_\mu\left(
\begin{array}{cc}
S_1 & T_1 \\
T_2 & S_2
\end{array}
\right)\left(
\begin{array}{c}
x^\mu \\
X^\mu
\end{array}
\right)~,
\nn
\label{erpa}
\end{eqnarray}
where
the matrix elements are given by
\begin{widetext}
\bea
A(kk':mm')&=&\langle 0|[[:a^{\dag}_{k'}a_k:,\hat{H}],:a^{\dag}_m a_{m'}:]|0\rangle
\nn
B(klk'l':mm')
&=&\langle 0|[[:a^{\dag}_{k'}a^{\dag}_{l'}a_l a_k:,\hat{H}],:a^{\dag}_m a_{m'}:]|0\rangle
\nn
C(kk':m_1m_2m_1'm_2')
&=&\langle 0|[[:a^{\dag}_{k'}a_k:,\hat{H}],
:a^{\dag}_{m_1}a^{\dag}_{m_2}a_{m_2'}a_{m_1'}:]|0\rangle
\nn
D(klk'l':m_1m_2m_1'm_2')
&=&\langle 0|[[:a^{\dag}_{k'}a^{\dag}_{l'}a_l a_k:,\hat{H}],
:a^{\dag}_{m_1}a^{\dag}_{m_2}a_{m_2'}a_{m_1'}:]|0\rangle~,
\eea
\bea
S_1(kk':mm')&=&\langle0|[:a^{\dag}_{k'}a_k:,:a^{\dag}_m a_{m'}:]|0\rangle
\nn
T_1(kk':m_1m_2m_1'm_2')
&=&\langle 0|[:a^{\dag}_{k'}a_k:,
:a^{\dag}_{m_1}a^{\dag}_{m_2}a_{m_2'}a_{m_1'}:]|0\rangle
\nn
T_2(klk'l':mm')
&=&\langle 0|[:a^{\dag}_{k'}a^{\dag}_{l'}a_l a_k:,:a^{\dag}_m a_{m'}:]|0\rangle
\nn
S_2(klk'l':m_1m_2m_1'm_2')
&=&\langle 0|[:a^{\dag}_{k'}a^{\dag}_{l'}a_l a_k:,
:a^{\dag}_{m_1}a^{\dag}_{m_2}a_{m_2'}a_{m_1'}:]|0\rangle~.
\eea
\end{widetext}
When the ground-state correlations are neglected, Eq. (\ref{erpa}) and the one-body section of
Eq. (\ref{erpa}), $Ax^\mu=\Omega_\mu S_1x^\mu$, are equivalent to SRPA and RPA, respectively.
 
Above equations give the ERPA scheme in its most general form. Below we will discuss some of the important properties of these equations, notably that they conserve all desirable and nice properties of standard RPA as conservation laws, sum-rules, Goldstone modes, etc.

\subsection{Hermiticity of ERPA matrix \label{hermiticity}}

The hamiltonian matrix on the left hand side of Eq. (\ref{erpa}) 
is hermitian. This is because the following operator identity for $\hat{A}$ and $\hat{B}$, which are either $:a^{\dag}_{k'}a_k:$ 
or $:a^{\dag}_{k'}a^{\dag}_{l'}a_l a_k:$,
\begin{eqnarray}
&&\langle0|[[\hat{B},\hat{H}],\hat{A}]|0\rangle-\langle0|[[\hat{A},\hat{H}],\hat{B}]|0\rangle
\nn&=&\langle0|[\hat{H},[\hat{A},\hat{B}]]|0\rangle=0~,
\label{id}
\end{eqnarray}
is satisfied due to the ground-state conditions of Eqs. (\ref{gs3}). We show this explicitly for the
matrix $D$ where $\hat{A}$ and $\hat{B}$ are both two-body operators 
$:a^{\dag}_{k'}a^{\dag}_{l'}a_l a_k:$ and 
$:a^{\dag}_{m_1}a^{\dag}_{m_2}a_{m_2'}a_{m_1'}:$, respectively.
Since $[\hat{A},\hat{B}]$ consists of at most three-body operators, 
Eq. (\ref{id}) holds
because of Eqs. (\ref{gs3}). This means
\begin{eqnarray}
D(klk'l':m_1m_2m_1'm_2')=
D(m_1'm_2'm_1m_2:k'l'kl)~.
\nn
\label{D1}
\end{eqnarray}
From its definition the hermitian conjugate of $D$ is 
\begin{eqnarray}
D(m_1m_2m_1'm_2':klk'l')^*=
D(m_1'm_2'm_1m_2:k'l'kl)~.
\nn
\label{D2}
\end{eqnarray}
Eqs. (\ref{D1}) and (\ref{D2}) imply that
$D$ is hermitian, namely,
\begin{eqnarray}
D(klk'l':m_1m_2m_1'm_2')=
D(m_1m_2m_1'm_2':klk'l')^*~.
\nn
\label{D3}
\end{eqnarray}
The following symmetries of other matrices $A$, $B$ and $C$ are shown in a similar way:
\begin{eqnarray}
A(kk':mm')&=&A(m'm:k'k)=A(mm':kk')^*
\nn
C(kk':m_1m_2m_1'm_2')
&=&B(m_1'm_2'm_1m_2:k'k)
\nn&=&
B(m_1m_2m_1'm_2':kk')^*~.
\label{B1}
\end{eqnarray}
Therefore, the hamiltonian matrix in Eq. (\ref{erpa}) 
is hermitian. The three-body correlation matrix is necessary for Eq.  (\ref{D3}), whereas Eqs. (\ref{B1}) 
hold without it.
The matrices $S_1$, $T_1$, $T_2$ and $S_2$ have the following properties
\begin{widetext}
\begin{eqnarray}
S_1(kk':mm')^*&=&S_1(mm':kk')=-S_1(k'k:m'm)\nn
T_1(kk':m_1m_2m_1'm_2')^*
&=&T_2(m_1m_2m_1'm_2':kk')=-T_1(k'k:m_1'm_2'm_1m_2)\nn
T_2(klk'l':mm')^*
&=&T_1(mm':klk'l')=-T_2(k'l'kl:m'm)\nn
S_2(klk'l':m_1m_2m_1'm_2')^*
&=&S_2(m_1m_2m_1'm_2':klk'l')=-S_2(k'l'kl:m_1'm_2'm_1m_2)~.
\end{eqnarray}
\end{widetext}
Therefore, also the norm matrix in Eq. (\ref{erpa}) 
is hermitian.
Taking hermitian conjugate of Eq. (\ref{erpa}) and using the above symmetries, we can show
that when
\begin{eqnarray}
\left(
\begin{array}{c}
x^\mu_{kk'} \\
X^\mu_{klk'l'}
\end{array}
\right)
\nonumber
\end{eqnarray}
is a positive energy solution with $\Omega_\mu(>0)$,
\begin{eqnarray}
\left(
\begin{array}{c}
{x^\mu_{k'k}}^* \\
{X^\mu_{k'l'kl}}^*
\end{array}
\right)
\nonumber
\end{eqnarray}
is a negative energy solution with $-\Omega_\mu$ as in RPA and other 
extended RPA theories \cite{Dro90}.

\subsection{Orthonormal condition}

For a hermitian hamiltonian matrix
the orthonormal condition is given as \cite{Tak88}
\begin{eqnarray}
\left(
{x^\mu}^*~~{X^\mu}^*
\right)\left(
\begin{array}{cc}
S_1 & T_1 \\
T_2 & S_2
\end{array}
\right)\left(
\begin{array}{c}
x^{\mu'} \\
X^{\mu'}
\end{array}
\right)
=\pm\delta_{\mu\mu'}~,
\label{orthogonal}
\end{eqnarray}
where the negative sign is for a negative-energy solution.
Accordingly, the closure relation is written as
\begin{eqnarray}
&&\sum_{\Omega_\mu>0}
\left(
\begin{array}{c}
x^\mu \\
X^\mu
\end{array}
\right)\left({x^\mu}^*~{X^\mu}^*\right)
\left(
\begin{array}{cc}
S_1 & T_1 \\
T_2 & S_2
\end{array}
\right)
\nn&-&
\sum_{\Omega_\mu>0}
\left(
\begin{array}{c}
{x^\mu_{k'k}}^* \\
{X^\mu_{k'l'kl}}^*
\end{array}
\right)\left(x^\mu_{k'k}~X^\mu_{k'l'kl}\right)
\left(
\begin{array}{cc}
S_1 & T_1 \\
T_2 & S_2
\end{array}
\right)
=I~,
\nn
\label{closure}
\end{eqnarray}
where $I$ is the unit matrix.

\subsection{Energy-weighted sum rule}

We discuss the energy-weighted sum rule and show that the Thouless theorem 
\cite{Tho61-b}
is satisfied. We consider
a hermitian operator
\begin{eqnarray}
&&\hat{F}=F_0+\sum_{mm'}{f_{mm'}:a^{\dag}_m a_{m'}:}
\nn&+&
\sum_{m_1m_2m_1'm_2'}{F_{m_1m_2m_1'm_2'}:a^{\dag}_{m_1}a^{\dag}_{m_2}a_{m_2'}a_{m_1'}:}~,
\end{eqnarray}
where $F_0$ is $\langle 0|\hat{F}|0\rangle$.
Since the one-body and two-body transition amplitudes $z^\mu_{kk'}=\langle0|:a^{\dag}_{k'}a_k:|\mu\rangle$
and $Z^\mu_{klk'l'}=\langle0|:a^{\dag}_{k'}a^{\dag}_{l'}a_l a_k:|\mu\rangle$
are given by
\begin{eqnarray}
\left(
\begin{array}{c}
z^\mu \\
Z^\mu
\end{array}
\right)=\left(
\begin{array}{cc}
S_1 & T_1 \\
T_2 & S_2
\end{array}
\right)\left(
\begin{array}{c}
x^\mu \\
X^\mu
\end{array}
\right)~,
\end{eqnarray}
the energy-weighted strength $m_1$ is written as 
\begin{eqnarray}
m_1&=&\sum_{\Omega_\mu>0}\Omega_\mu|\langle0|\hat{F}|\mu\rangle|^2
\nonumber \\
&=&\sum_{\Omega_\mu>0}\Omega_\mu\left(f~~F\right)
\left(
\begin{array}{c}
z^\mu \\
Z^\mu
\end{array}
\right)
\left({z^\mu}^*~~{Z^\mu}^*\right)
\left(
\begin{array}{c}
f \\
F
\end{array}
\right)~.
\nn
\end{eqnarray}
Taking into account the contribution of negative-energy solutions and
using Eqs. (\ref{erpa}) and (\ref{closure}),
we can show \cite{Toh04b} that the Thouless
theorem holds, that is,
\begin{eqnarray}
m_1=\frac{1}{2}\langle0|[[\hat{F},\hat{H}],\hat{F}]|0\rangle~.
\end{eqnarray}
The important point is that $\hat{F}$ can be both one-body and two-body operators.

\subsection{Spurious modes in ERPA and SCRPA}

We show
that ERPA gives zero excitation energy to spurious modes associated with operators $\hat{F}$
which commute with the hamiltonian, that is, $[\hat{F},\hat{H}]=0$.
This actually holds for situations for which the ground state determined by Eqs. (\ref{gs3})-(\ref{hfeq})
has a spontaneously broken symmetry. The operator $\hat{F}$ can be both one-body and two-body
operators 
\begin{eqnarray}
\hat{F}=F_0+\sum_{kk'}f_{k'k}:a^{\dag}_{k'}a_{k}:
+\sum_{klk'l'}F_{k'l'kl}
:a^{\dag}_{k'}a^{\dag}_{l'}a_{l}a_{k}:~.
\nn
\label{SP}
\end{eqnarray}
This means that our discussion holds for spurious modes associated with
two-body operators, such as double excitation of translational motion.
 We consider 
$\Omega_\mu\langle0|\hat{F}|\mu\rangle$, where $\langle0|\hat{F}|\mu\rangle\ne0$
because the ground state has a broken symmetry. Using Eqs. (\ref{erpa}),
we obtain \cite{Toh04b}
\begin{eqnarray}
\Omega_\mu\langle0|\hat{F}|\mu\rangle&=&
\Omega_\mu\left(f~~F\right)\left(
\begin{array}{cc}
S_1&T_1\\
T_2&S_2
\end{array}
\right)\left(
\begin{array}{c}
x^\mu \\
X^\mu
\end{array}
\right)
\nn&=&
\left(f~~F\right)\left(
\begin{array}{cc}
A&C\\
B&D
\end{array}
\right)\left(
\begin{array}{c}
x^\mu\\
X^\mu
\end{array}
\right)
\nonumber \\
&=&\langle 0|[[\hat{F},\hat{H}],\hat{Q}^{\dag}_\mu]|0\rangle=0~.
\label{zeroenergy}
\end{eqnarray}
This implies $\Omega_\mu=0$ for spurious modes. 
Equation (\ref{zeroenergy}) also implies that the transition amplitudes
of the symmetry operator $\hat{F}$ vanish for physical excited states, 
because $\Omega_\mu\neq0$ in such cases.

\subsection{Approximate forms of ERPA}
The calculation of the three-body correlation matrix for realistic hamiltonians is very difficult.
We derive two approximate forms of ERPA without the three-body correlation matrix, which can be used for realistic applications.

\subsubsection{Hamiltonian matrix of ERPA}
We rewrite the hamiltonian matrix in ERPA in a different way using the commutation relations between
the hamiltonian and one-body and two-body operators, which are written as
\begin{widetext}
\begin{eqnarray}
[:a^{\dag}_{k'}a_{k}:,\hat{H}]&=&\sum_{mm'}a(kk':mm'):a^{\dag}_{m'}a_{m}:
+\sum_{m_1m_2m_1'm_2'}c(kk':m_1m_2m_1'm_2')
:a^{\dag}_{m_1'}a^{\dag}_{m_2'}a_{m_2}a_{m_1}:~,
\label{1H}
\end{eqnarray}
\begin{eqnarray}
[:a^{\dag}_{k'}a^{\dag}_{l'}a_l a_k:,\hat{H}]&=&\sum_{mm'}b(klk'l':mm'):a^{\dag}_{m'}a_{m}:
+\sum_{m_1m_2m_1'm_2'}d(klk'l':m_1m_2m_1'm_2')
:a^{\dag}_{m_1'}a^{\dag}_{m_2'}a_{m_2}a_{m_1}:
\nonumber \\
&+&\sum_{m_1m_2m_3m_1'm_2'm_3'}
e(klk'l':m_1m_2m_3m_1'm_2'm_3')
:a^{\dag}_{m_1'}a^{\dag}_{m_2'}a^{\dag}_{m_3'}a_{m_3}a_{m_2}a_{m_1}:~.
\label{2H}
\end{eqnarray}
\end{widetext}
The matrices $a$, $b$, $c$, $d$ and $e$ are explicitly given in Ref. 
\cite{Toh04}. The ground-state conditions of Eqs. (\ref{gs3}) 
and (\ref{hfeq}) are employed in the
derivation of these matrices.
The matrices $a$, $b$, $c$, $d$ and $e$ are equivalent to 
$\delta F_1/ \delta n$, $\delta F_2/ \delta n$, $\delta F_1/ \delta C_2$,
$\delta F_2/ \delta C_2$, $\delta F_2/ \delta C_3$ in Eq. (\ref{gradient}), respectively 
\cite{Toh04}.
Using  $a$, $b$, $c$, $d$ and $e$, the matrices $A$, $B$, $C$ and $D$ in ERPA are written as
\begin{eqnarray}
A&=&\langle 0|[[:a^{\dag}_{k'}a_k:,\hat{H}],:a^{\dag}_m a_{m'}:]|0\rangle 
\nn&=&aS_1+cT_2
\nn
B&=&\langle 0|[[:a^{\dag}_{k'}a^{\dag}_{l'}a_l a_k:,\hat{H}],:a^{\dag}_m a_{m'}:]|0\rangle
\nn&=&bS_1+dT_2+eT_{31}
\nn
C&=&\langle 0|[[:a^{\dag}_{k'}a_k:,\hat{H}],
:a^{\dag}_{m_1}a^{\dag}_{m_2}a_{m_2'}a_{m_1'}:]|0\rangle 
\nn&=&aT_1+cS_2 
\nn
D&=&\langle 0|[[:a^{\dag}_{k'}a^{\dag}_{l'}a_l a_k:,\hat{H}],
:a^{\dag}_{m_1}a^{\dag}_{m_2}a_{m_2'}a_{m_1'}:]|0\rangle
\nn&=&bT_1+dS_2+eT_{32}
\label{DT1}~,
\end{eqnarray}
where
\begin{eqnarray}
T_{31}&=&\langle 0|[:a^{\dag}_{k'}a^{\dag}_{l'}a^{\dag}_{m'}a_m a_l a_k:,
:a^{\dag}_{m}a_{m'}:]|0\rangle
\nn
T_{32}&=&\langle 0|[:a^{\dag}_{k'}a^{\dag}_{l'}a^{\dag}_{m'}a_m a_l a_k:,
:a^{\dag}_{m_1}a^{\dag}_{m_2}a_{m_2'}a_{m_1'}:]|0\rangle~.
\nn
\end{eqnarray}
Since a four-body correlation matrix is not considered,
expectation values of four-body operators in $T_{32}$ are approximated
by the products of $n_{\alpha\alpha'}$, $C_{klk'l'}$
and $C_{klmk'l'm'}$.

\subsubsection{STDDM}

Here we discuss the relation of ERPA and the Small
amplitude limit of the Time-Dependent Density-Matrix (STDDM) theory  
\cite{Toh89}
which has been used for realistic cases \cite{Toh05,Toh07}.
When the three-body correlation matrix $C_{klmk'l'm'}$ and the matrix $e$ in Eq. (\ref{2H})
are neglected, the ERPA equations in this approximation are written as 
\begin{eqnarray}
\left(
\begin{array}{cc}
A&C'\\
B'&D'
\end{array}
\right)\left(
\begin{array}{c}
x^\mu\\
X^\mu
\end{array}
\right)
=\Omega_\mu
\left(
\begin{array}{cc}
S_1&T_1\\
T_2&S_2'
\end{array}
\right)\left(
\begin{array}{c}
x^\mu\\
X^\mu
\end{array}
\right)~,
\nn
\label{stddm0}
\end{eqnarray}
where $S_2'$ does not have the three-body correlation matrix, and $B'$,
$C'$ and $D'$ are given by
$B'=bS_1+dT_2$, $C'=aT_1+cS_2'$ and 
$D'=bT_1+dS_2'$, respectively. Other matrices in Eq.(\ref{stddm0}) are the same as those in ERPA.

Equation (\ref{stddm0})
can be written in a different form as 
\begin{widetext}
\begin{eqnarray}
\left(
\begin{array}{cc}
A&C'\\
B'&D'
\end{array}
\right)\left(
\begin{array}{c}
x^\mu\\
X^\mu
\end{array}
\right)
&=&\left(
\begin{array}{cc}
aS_1+cT_2&aT_1+cS_2'\\
bS_1+dT_2&bT_1+dS_2'
\end{array}
\right)\left(
\begin{array}{c}
x^\mu\\
X^\mu
\end{array}
\right)
\nonumber \\
&=&\left(
\begin{array}{cc}
a&c\\
b&d
\end{array}
\right)
\left(
\begin{array}{cc}
S_1&T_1\\
T_2&S_2'
\end{array}
\right)\left(
\begin{array}{c}
x^\mu\\
X^\mu
\end{array}
\right)
=\Omega_\mu
\left(
\begin{array}{cc}
S_1&T_1\\
T_2&S_2'
\end{array}
\right)\left(
\begin{array}{c}
x^\mu\\
X^\mu
\end{array}
\right)~.
\end{eqnarray}
\end{widetext}
Using the transition amplitude
\begin{eqnarray}
\left(
\begin{array}{c}
z^\mu \\
Z^\mu
\end{array}
\right)=\left(
\begin{array}{cc}
S_1 & T_1 \\
T_2 & S_2'
\end{array}
\right)\left(
\begin{array}{c}
x^\mu \\
X^\mu
\end{array}
\right)~,
\end{eqnarray}
Eq.(\ref{stddm0}) is written as 
\begin{eqnarray}
\left(
\begin{array}{cc}
a&c\\
b&d
\end{array}
\right)\left(
\begin{array}{c}
z^\mu\\
Z^\mu
\end{array}
\right)
=\Omega_\mu
\left(
\begin{array}{c}
z^\mu\\
Z^\mu
\end{array}
\right)~.
\label{stddm}
\end{eqnarray}
This is the form derived as the small amplitude limit of TDDM. The effects of ground-state correlations are included in STDDM, because the matrices
$a$, $b$ and $d$ in Eq. (\ref{stddm}) contain $n_{kk'}$ and $C_{klmk'l'm'}$.

\subsubsection{Modified STDDM}

When we keep the matrix $e$ and neglect
the three-body correlation matrix $C_{klmk'l'm'}$ in Eq. (\ref{erpa}), we obtain 
\begin{eqnarray}
\left(
\begin{array}{cc}
A&C'\\
B'&D''
\end{array}
\right)\left(
\begin{array}{c}
x^\mu\\
X^\mu
\end{array}
\right)
=\Omega_\mu
\left(
\begin{array}{cc}
S_1&T_1\\
T_2&S_2'
\end{array}
\right)\left(
\begin{array}{c}
x^\mu\\
X^\mu
\end{array}
\right)~,
\nn
\label{mstddm}
\end{eqnarray}
where  
$D''=bT_1+dS_2'+eT_{32}'$. Here $T_{32}'$ does not contain the three-body
correlation matrix. The last term in $D''$ gives significant improvement of
the description of two-phonon states \cite{Toh08}.
We refer to  this approximation of ERPA as mSTDDM (modified STDDM).
In STDDM and mSTDDM, the hermiticity of $D'$ and $D''$ is lost. However, 
non-hermiticity does not cause any serious problems in the following applications.

\subsection{particle-particle ERPA equation}
In this subsection we present very briefly  an extended particle-particle RPA which can be formulated in a way similar to the particle-hole RPA using EOM method.
We consider an excitation operator consisting of one-body and two-body parts
\begin{eqnarray}
\hat{Q}^{\dag}_{\mu}=\sum_{kk'}{{x}^\mu_{kk'}a^{\dag}_{k} a^{\dag}_{k'}}
+\sum_{k_1k_2k_2'k_1'}{{\mathcal X}^\mu_{k_1k_2k_2'k_1'}:a^{\dag}_{k_1}a^{\dag}_{k_2}a^{\dag}_{k_2'}a_{k_1'}:}~.
\nn
\label{ppexoper}
\end{eqnarray}
In the evaluation of the matrix elements we approximate
the ppERPA ground state with $|0\rangle$ which satisfy Eqs. (\ref{gs3}-\ref{hfeq}). The ERPA equations 
are written as 
\begin{eqnarray}
\left(
\begin{array}{cc}
{\mathcal A} & {\mathcal C} \\
{\mathcal B} & {\mathcal D} 
\end{array}
\right)\left(
\begin{array}{c}
{x}^\mu \\
{\mathcal X}^\mu
\end{array}
\right)=\Omega_\mu\left(
\begin{array}{cc}
{\mathcal S}_1 & {\mathcal T}_1 \\
{\mathcal T}_2 & {\mathcal S}_2
\end{array}
\right)\left(
\begin{array}{c}
{ x}^\mu \\
{\mathcal X}^\mu
\end{array}
\right)~,
\nn
\label{pperpa}
\end{eqnarray}
where
the matrix elements are given by
\begin{widetext}
\bea
{\mathcal A}(kk':mm')&=&\langle 0|[[a_{k'}a_k,\hat{H}],a^{\dag}_m a^{\dag}_{m'}]|0\rangle
\nn
{\mathcal B}(klk'l':mm')
&=&\langle 0|[[:a^{\dag}_{k'}a_{l'}a_l a_k:,\hat{H}],a^{\dag}_m a^{\dag}_{m'}]|0\rangle
\nn
{\mathcal C}(kk':m_1m_2m_1'm_2')
&=&\langle 0|[[a_{k'}a_k,\hat{H}],
:a^{\dag}_{m_1}a^{\dag}_{m_2}a^{\dag}_{m_2'}a_{m_1'}:]|0\rangle
\nn
{\mathcal D}(klk'l':m_1m_2m_1'm_2')
&=&\langle 0|[[:a^{\dag}_{k'}a_{l'}a_l a_k:,\hat{H}],
:a^{\dag}_{m_1}a^{\dag}_{m_2}a^{\dag}_{m_2'}a_{m_1'}:]|0\rangle~,
\label{ppD}
\eea
\bea
{\mathcal S}_1(kk':mm')&=&\langle0|[a_{k'}a_k,a^{\dag}_m a^{\dag}_{m'}]|0\rangle
\nn
{\mathcal T}_1(kk':m_1m_2m_1'm_2')
&=&\langle 0|[a_{k'}a_k,
:a^{\dag}_{m_1}a^{\dag}_{m_2}a^{\dag}_{m_2'}a_{m_1'}:]|0\rangle
\nn
{\mathcal T}_2(klk'l':mm')
&=&\langle 0|[:a^{\dag}_{k'}a_{l'}a_l a_k:,a^{\dag}_m a^{\dag}_{m'}]|0\rangle
\nn
{\mathcal S}_2(klk'l':m_1m_2m_1'm_2')
&=&\langle 0|[:a^{\dag}_{k'}a_{l'}a_l a_k:,
:a^{\dag}_{m_1}a^{\dag}_{m_2}a^{\dag}_{m_2'}a_{m_1'}:]|0\rangle~.
\eea
\end{widetext}
Similar considerations and approximations as in the $ph$ case can also performed
in this $pp$ case, but nothing has been worked out so far in the literature and, thus, we will also not go further here. Nonetheless, a small application will be given in the next section.

\section{Applications \label{sect.3}}
\label{sec:Appl}
\setcounter{equation}{0}
\renewcommand{\theequation}{8.\arabic{equation}}

\subsection{Self-consistent second RPA in the exactly solvable single shell pairing model}

As a first small, but instructive exercise we want to solve the single shell pairing (seniority) model \cite{Rin80,sen-model}. This model is usually easily solved with angular momentum algebra. However, here we want to solve it with the EOM technique, which is a new rather interesting many-body way of solution. The Hamiltonian of the single shell seniority model is

\be
H= -GS_+S_-~,
\label{H-sen}
\ee
where
\be
S_+ = \sum_{m>0}c^{\dag}_mc_{-m}~,
\ee
with $S_- = (S_+)^{\dag}$ and
$m$ being the magnetic quantum numbers of the $j$-shell.
The sum over the magnetic quantum numbers runs over $\Omega = j + 1/2$ states.
We make the following ppERPA ansatz for the pair addition operator which is a simplified form of (\ref{ppexoper})

\bea
A^{\dag} &=& X_1S_+ + X_2S_+S_0\nonumber\\
A&=& X_1S_- + X_2S_0S_-~,
\label{add-op}
\eea
and for the removal operator

\bea
R^{\dag} &=& Z_1S_- + Z_2S_0S_-\nonumber\\
R&=& Z_1S_+ + Z_2S_+S_0~.
\label{remove-op}
\eea
It is not difficult to see that with certain relations among the variational amplitudes, one can fulfill the exact annihilating conditions

\be
A|0\rangle = R|0\rangle = 0~.
\label{killing-c}
\ee

These operators obey the following orthonormalization conditions

\bea
\langle [A,A^{\dag}]\rangle &=& X_1^2N_{11} + X_1X_2(N_{12} + N_{21}) +
X_2^2N_{22}=1\nonumber\\
\langle [R,R^{\dag}] &=& Z_1^2N_{11} + Z_1Z_2(N_{12}+N_{21} + Z_2^2N_{22} = -1~,
\nn
\label{norm-sen}
\eea
where
\bea
N_{11} &=& \langle [S_-,S_+]\rangle = -2\langle S_0\rangle
\nn
N_{12} &=& N_{21} = \langle S_-,S_+S_0]\rangle = -2\langle S_0^2\rangle + \langle S_+S_-\rangle
\nn
N_{22} &=& \langle S_0S_-,S_+S_0\rangle = -2\langle S_0^3\rangle+
\langle S_+S_-(2S_0-1)\rangle~.
\nn
\eea
  By simple manipulations with backward and forward amplitudes we can invert
  (\ref{add-op}) and (\ref{remove-op})and obtain the following expressions for $S_+$ and $S_+S_0$

  \be
  S_+=\frac{Z_2A^{\dag}-X_2R}{Z_2X_1-X_2Z_1}~;~~~S_+S_0=\frac{Z_1A^{\dag}-X_1R}{Z_1X_2-X_1Z_2}~,
  \label{inv-sen}
  \ee
  and $S_-$ and $S_0S_-$ are obtained by hermitian conjugation. From above relations we obtain

\be
\langle S_+S_-\rangle = \frac{X_2^2}{(Z_2X_1 - X_2Z_1)^2}~.
\ee
  Applying the EOM method to the addition and removal mode in the usual way we obtain

  \be
  \begin{pmatrix} A&B\\B&C \end{pmatrix}
  \begin{pmatrix}X_1\\X_2 \end{pmatrix}= \omega_a
  \begin{pmatrix} N_{11}&N_{12}\\N_{21}&N_{22}\end{pmatrix}
  \begin{pmatrix}X_1\\X_2 \end{pmatrix}~,
  \label{EOM1-sen}
  \ee
  and same for removal mode

  \be
  \begin{pmatrix} A&B\\B&C \end{pmatrix}
  \begin{pmatrix}Z_1\\Z_2 \end{pmatrix}= -\omega_r
  \begin{pmatrix} N_{11}&N_{12}\\N_{21}&N_{22}\end{pmatrix}
  \begin{pmatrix}Z_1\\Z_2 \end{pmatrix}~,
  \label{EOM2-sen}
  \ee
  where
\bea
A &=& \langle [S_-,[H,S_+]]\rangle = 2GN_{12}
\nn
B &=& \langle [S_0S_-,[H,S_+]]\rangle = 2GN_{22}
\nn
C &=& \langle [S_0S_-,[H,S_+S_0]]\rangle 
\nn
&=& 2G\langle -2S_0^4 + S_+S_-(3S_0^2 - 3S_0 +1)\rangle~.
\eea
  The system of equations contains the operator $S_0$ which in the seniority model is expressed as \cite{Rin80} $2S_0 = \hat N -\Omega$. Therefore, in all expressions $\hat N$ can be replaced by the appropriate particle number. 
  At this moment the system of equations (\ref{EOM1-sen}, \ref{EOM2-sen}) is closed and can be solved with the usual methods. It turns out that the exact solution is obtained. It may be worthwhile to notice that the seniority model also can be solved with the corresponding Dyson-BSE. It just corresponds to the transcription of the eigenvalue problem into the Green's function approach.
  
  Actually, the seniority model can also be solved with the s.p. Dyson equation in evaluating the 2p-1h Green's function figuring in the self-energy with the EOM method as described in Sect. V and in \cite{sen-model}. So, in this simple case of the seniority model even and odd SCRPA's are decoupled. However, already on the level of the Lipkin model both equations become coupled and then constitute together a powerful system of equations for the solution of the many-body problem.\\


Next, we present the results for the Lipkin model \cite{Lip65} and the Hubbard model using
the original form of ERPA with the three-body correlation matrix. 
We also review the main results for the quadrupole states in
$^{16}$O and $^{40}$Ca calculated using STDDM.  

\subsection{Lipkin model}

We consider the standard two-level Lipkin model of Eq. (\ref{2-Lipkin}).
The excitation energies of the first and second excited states are displayed in Fig. \ref{lip4} as functions of $\chi=(N-1)|V|/\epsilon$ for $N=4$ \cite{ST16}. 
We see that ERPA (squares) performs extremely well for the first excited state, even far beyond the RPA instability point of $\chi = 1$. SCRPA (circles)  also is very good, but deteriorate after the instability point. 
In the case of the second excited state which can be obtained with ERPA, deviation from the exact solution 
becomes larger with increasing $\chi$. This can be explained either by the neglect of the coupling to higher amplitudes or by the fact that in ERPA non-collective amplitudes are not included.
The occupation probabilities of the upper state are also reasonably well reproduced in ERPA and SCRPA as shown in Fig. \ref{lip4n}.  
\\

\begin{figure}[h]
\begin{center}
\includegraphics[width=8cm,height=5cm]{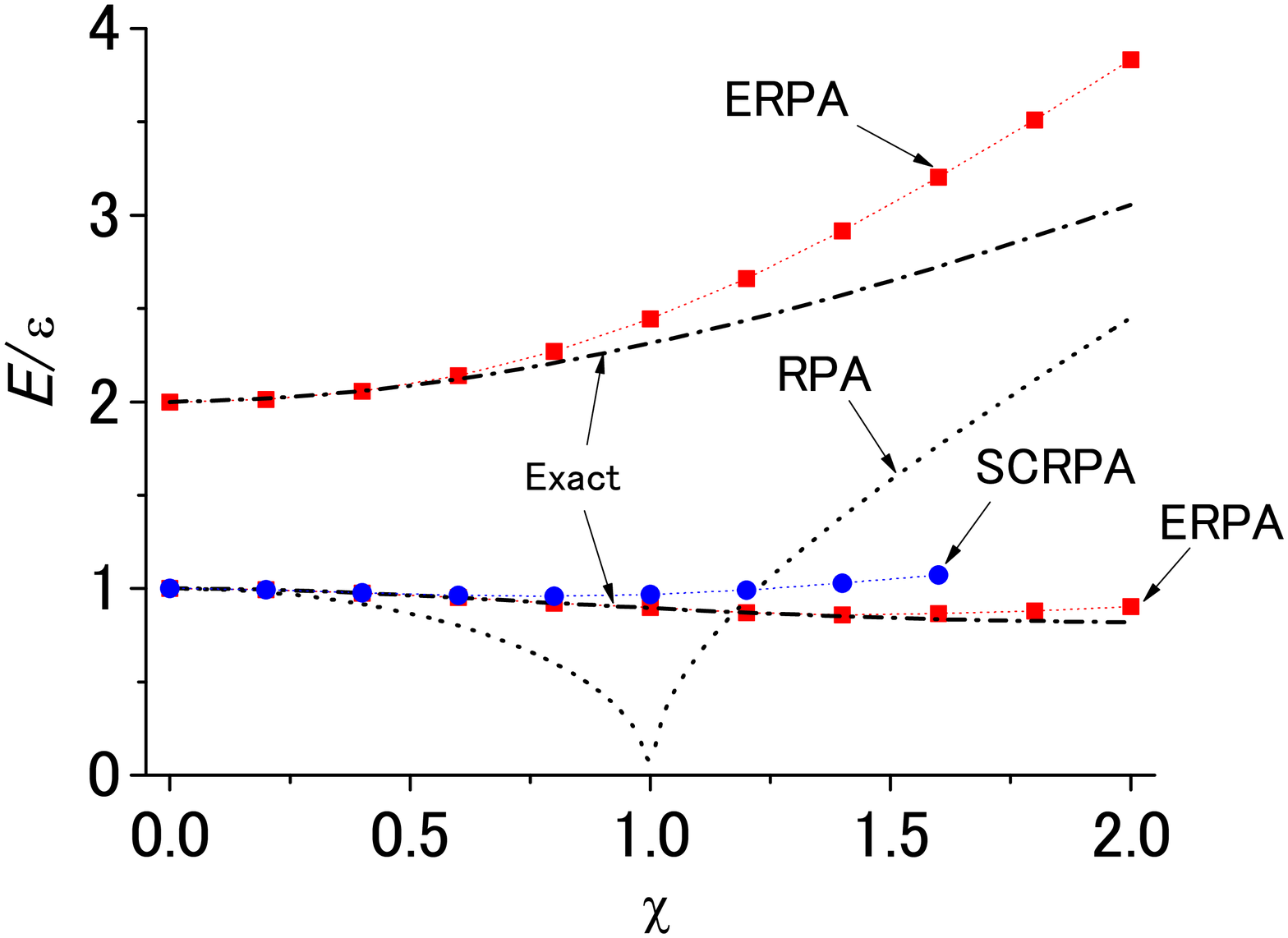}
\end{center}
\caption{Excitation energies of the first and second excited states in the Lipkin model calculated in ERPA (squares), SCRPA (circles) and RPA (dotted line)
as functions of $\chi=(N-1)|V|/\epsilon$ for $N=4$. The exact solutions are shown with the dot-dashed lines. Readapted from Ref. \cite{ST16}.}
\label{lip4}
\end{figure}

\begin{figure}[h]
\begin{center}
\includegraphics[width=8cm,height=5cm]{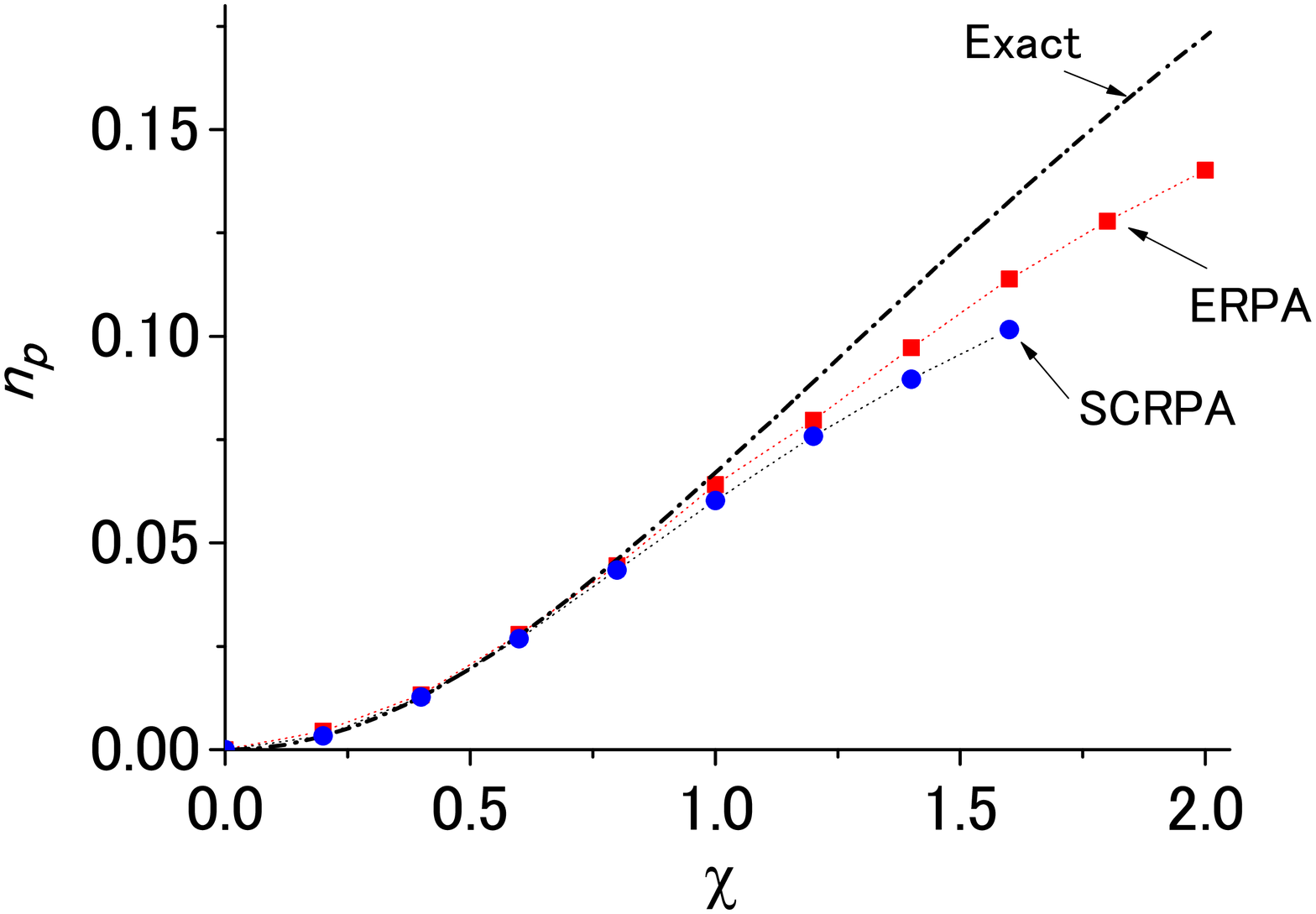}
\end{center}
\caption{Occupation probabilities of the upper state in the Lipkin model calculated in ERPA (squares) and SCRPA (circles)
as functions of $\chi$ for $N=4$. The exact solutions are shown with the dot-dashed lines. Readapted from Ref. \cite{ST16}.}
\label{lip4n}
\end{figure}

\begin{figure}[h]
\begin{center}
\includegraphics[width=8cm,height=5cm]{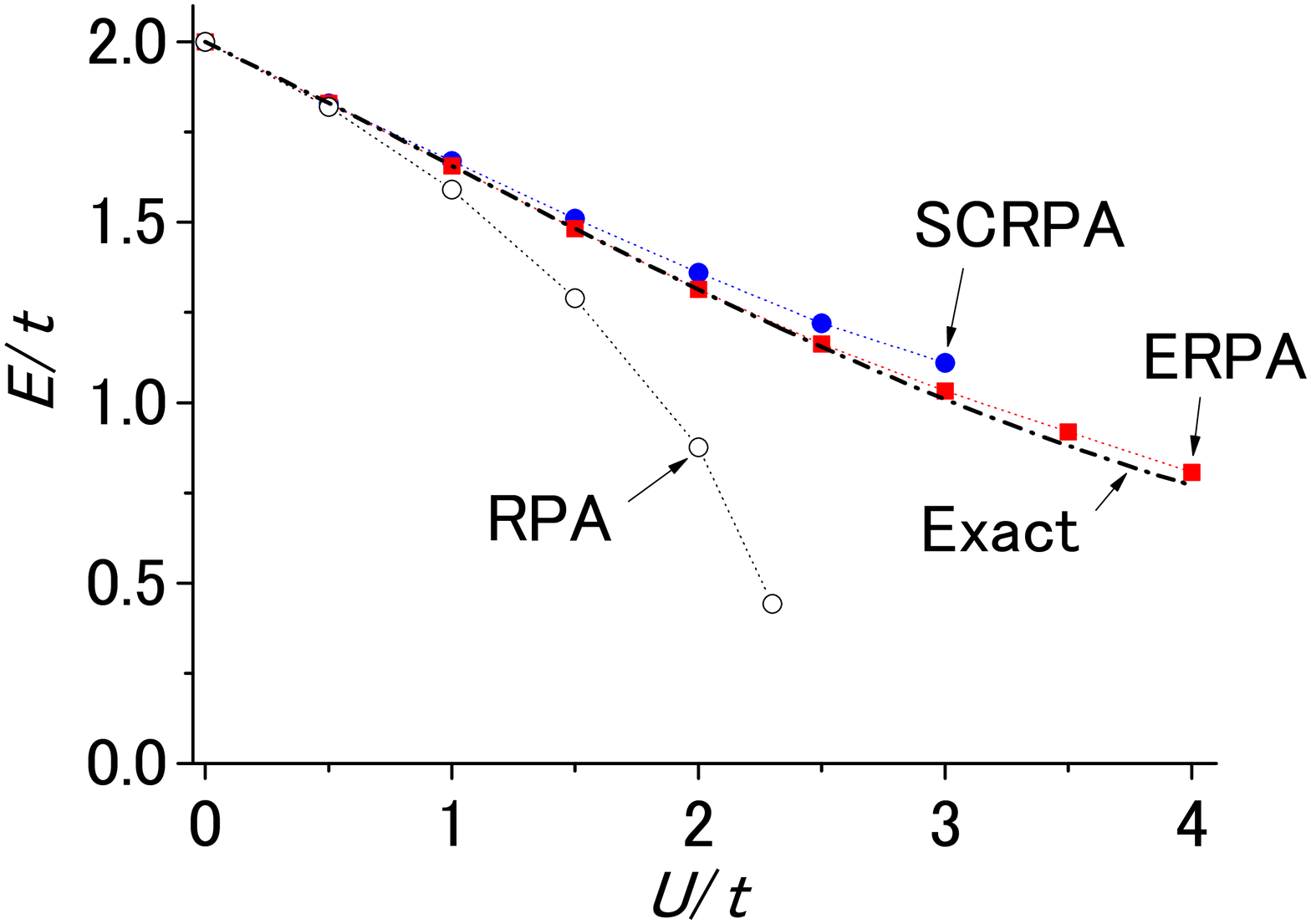}
\end{center}
\caption{Excitation energy of the first excited state (the spin mode with momentum transfer $|q|=\pi$) calculated in ERPA (squares) and SCRPA (circles) as a function of $U/t$ for the 
six-site Hubbard model with half-filling. The exact values are shown with the dot-dashed line. The open circles depict the results in RPA. Readapted from Ref. \cite{ST16}.}
\label{hub6}
\end{figure}

\subsection{Hubbard model}

As the next model we consider the one-dimensional six-site Hubbard model with half-filling (see Eq. (\ref{hamiltonian_imp})).
In the ERPA calculations we take only the 2p-2h and 2h-2p
components of ${X}^\mu_{\alpha\beta\alpha'\beta'}$ to facilitate the numerics.
The excitation energy of the first excited state, which is the spin mode with momentum transfer $|q|=\pi$, is shown in Fig. \ref{hub6} as a function of $U/t$ \cite{ST16}. The results in ERPA (squares) show good agreement with
the exact values (dot-dashed line). The SCRPA results (filled circles) are reasonable and avoid the instability of RPA (open circles). 
The SCRPA results in Fig. \ref{hub6} are, however, less good than the ones in \cite{Jem13}, that is, no SCRPA solutions exist beyond $U/t=3$. 
This is due to the fact that the implicit cross excitation mode couplings involved in $n_{\alpha\alpha}$
and $C_{\alpha\beta\alpha'\beta'}$ are neglected in \cite{Jem13}.

Let us mention that ERPA can only be tackled at the moment for simple models. In realistic cases this approach becomes numerically too complicated and has to be simplified as, e.g., in  STDDM or mSTDDM or even more drastic approximations.

\subsection{Damping of giant resonances}

Applications to the Isoscalar Giant Quadrupole Resonances (ISGQR) in $^{16}$O \cite{Toh07} and $^{40}$Ca \cite{MT17}
have been done using the simplified version of ERPA,
that is, STDDM. 
A simplified Skyrme force has been adapted as the effective interaction for
the mean-field potential and also as the residual interaction. 
Assuming that single-particle states around the Fermi energy are
most important for ground-state correlations, we use
the $1p_{3/2}$, $1p_{1/2}$, $1d_{5/2}$ and $2s_{1/2}$ states to calculate $n_{\alpha\alpha'}$, $C_{\alpha\beta\alpha'\beta'}$ and ${X}^\mu_{\alpha\beta\alpha'\beta'}$ for $^{16}$O 
and the $1d_{5/2}$, $2s_{1/2}$, $1d_{3/2}$ and $1f_{1/2}$ states for $^{40}$Ca, and only the 2p-2h and 2h-2p components of $C_{\alpha\beta\alpha'\beta'}$ and ${X}^\mu_{\alpha\beta\alpha'\beta'}$
are considered to facilitate the numerics.
To satisfy the energy weighted sum rule, we use a
large number of single-particle states to calculate the one-body
transition amplitude: The continuum states are discretized by confining
the single-particle wave functions in a sphere of radius 20 fm and all single-particle states with $\epsilon\le40$ MeV and 
the orbital angular momentum $\ell\le 5\hbar$ are taken.
\begin{figure}[h]
  \begin{center}
    \includegraphics[width=10cm,height=8cm]{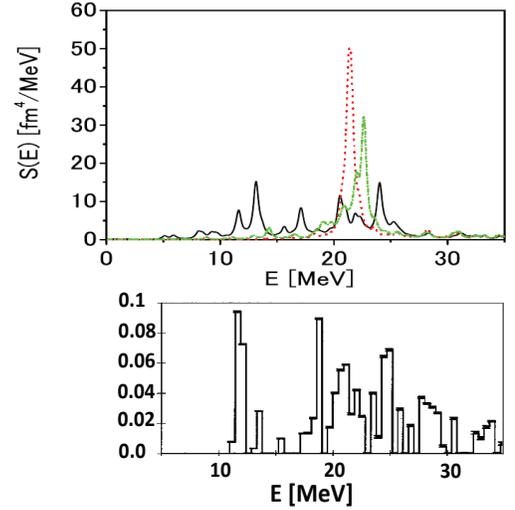}
  \end{center}
  \caption{Upper panel: strength functions for the isoscalar quadrupole states 
in $^{16}$O calculated in STDDM (solid line), SRPA (dot-dashed line) and RPA (dotted line). 
The strength functions are smoothed with $\Gamma$ =0.5 MeV.  Adopted from Ref. \cite{Toh07}; 
lower panel: experimental strength function for the ISGQR from \cite{Lui01}.}
\label{o16}
\end{figure}
\begin{figure}[h] 
\begin{center} 
\includegraphics[width=10cm,height=6cm]{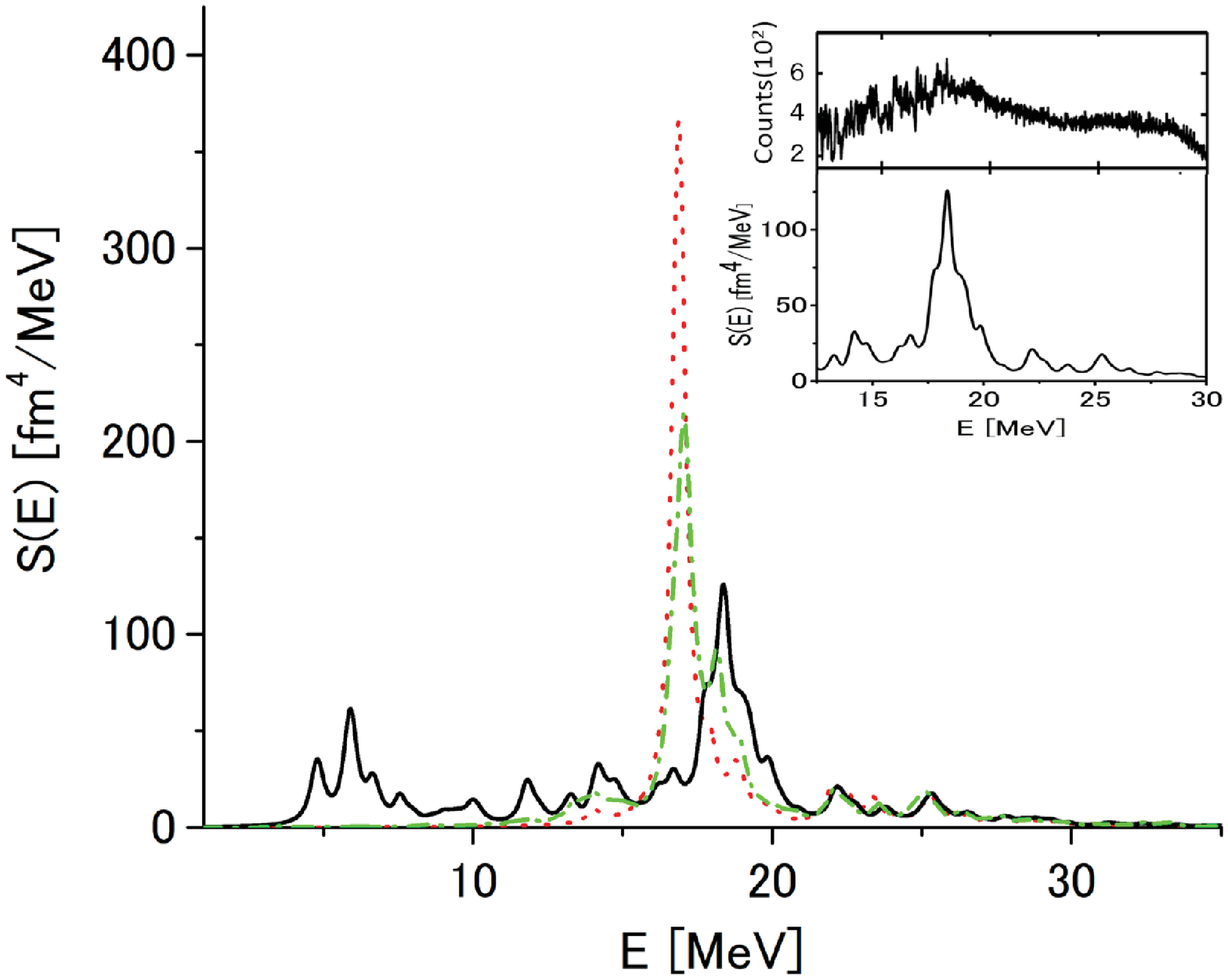}
\end{center}
\caption{Strength functions calculated in STDDM (solid line), SRPA (dot-dashed line) and RPA (dotted line) for the isoscalar quadrupole excitation in $^{40}$Ca. 
The distributions are smoothed with an artificial width $\Gamma=0.5$ MeV. 
In the inset the STDDM strength distribution (lower part) is compared with the experimental data 
from $(p,p')$ experiments at $E_p=200$ MeV 
and $\theta_{\rm Lab}=11^\circ$ \cite{usman} (upper part). Adopted from Ref. \cite{MT17}.} 
\label{ca40} 
\end{figure}
\begin{figure}[h]
\begin{center}
\includegraphics[width=8cm]{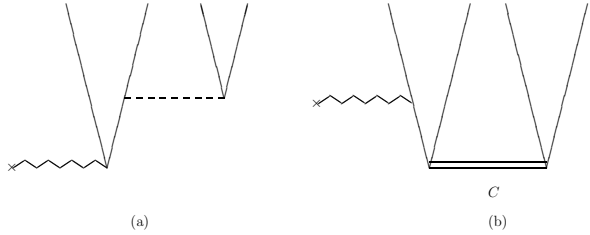}
\end{center}
\caption{(a) Damping process both in STDDM and SRPA. 
(b) Process only in STDDM.
The wavy line means an external field (the quadrupole field in this case), the
dashed line is  the interaction and the vertical lines indicate either particle states or hole states.
The double line with $C$ means $C_{\alpha\beta\alpha'\beta'}$.}
\label{gqrdamp}
\end{figure}
\begin{figure}[h]
\begin{center} 
\includegraphics[width=8cm]{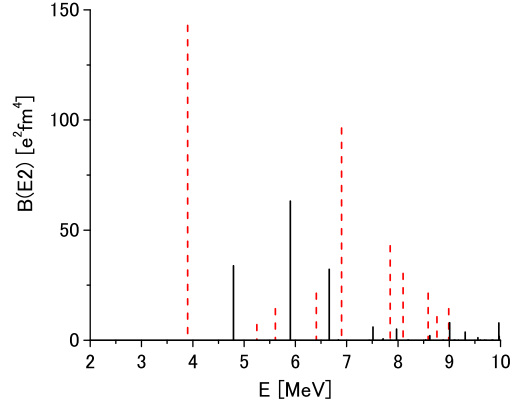}
\end{center}
\caption{Distribution of $B(E2)$ strength below 10 MeV calculated in STDDM for $^{40}$Ca. 
Experimental data (dashed line) are taken from Ref. \cite{hartmann}.} 
\label{ca40low} 
\end{figure} 
In Figs. \ref{o16} and \ref{ca40} the strength functions for ISGQR in $^{16}$O and $^{40}$Ca
calculated in STDDM are compared with the results of RPA and SRPA. The sharp peak in RPA corresponds to the ISGQR.
STDDM gives much larger fragmentation of the quadrupole strength than SRPA.
In the lower panel of Fig.~\ref{o16} the experimental distribution \cite{Lui01}
is reproduced. Good agreement with theory can be seen.
This large fragmentation, especially the concentration of the strength in 
the region below ISGQR, is also consistent with experiment for $^{40}$Ca \cite{Hot74,hartmann}. Main components of the peaks located below ISGQR are 2p-2h configurations. In the case of $^{16}$O, for example,
they are either $[1d_{5/2}(p)1d_{5/2}(n)
(1p_{3/2}(p))^{-1}(1p_{1/2}(n))^{-1}]$ or 
$[1d_{5/2}(p)1d_{5/2}(n)(1p_{1/2}(p))^{-1}(1p_{3/2}(n))^{-1}]$, where $(p)$ and $(n)$ denote proton and
neutron states, respectively. This means that proton-neutron 
correlations play an important role
in the splitting of ISGQR in $^{16}$O and also in $^{40}$Ca. The difference between the STDDM and SRPA
results also indicates the importance of ground-state correlations.
In fact, the process shown diagrammatically in Fig. \ref{gqrdamp}(b), which involves the ground-state
correlations, is responsible for the enhancement of the strength in the low-energy region \cite{Toh07,MT17}.
In the inset of Fig. \ref{ca40} the  STDDM  strength distribution in the ISGQR region (lower part) is compared with the experimental data from $(p,p')$ experiments \cite{usman} (upper part).
Although the peak position in STDDM corresponds to the experimental data, STDDM cannot describe the large fragmentation of ISGQR. 
The result of the large-scale SRPA calculation \cite{Vas18} suggests the importance of higher configurations.
There are 19 sates below 10 MeV in STDDM, which are compared with experiment \cite{hartmann} in Figure \ref{ca40low}.
The first $2^+$ state in $^{40}$Ca cannot be described in RPA and STDDM because it mainly consists of 4p--4h states.
The summed strength below 10 MeV is $166$ $e^2$fm$^4$ in STDDM, which explains about two thirds of the experimental value $263\pm46$ $e^2$fm$^4$, where the
first $2^+$ state is excluded.

In summary of this section, we can say that a simplified version of ERPA, that is STDDM, yields quite encouraging results. However, the numerical challenges are quite enormous and much effort must still be invested in the future to make this promising approach a full success.

\section{Discussion and Conclusions}
\label{sec:concl}
\setcounter{equation}{0}
\renewcommand{\theequation}{9.\arabic{equation}}

In this review we gave a survey of recent developments of the equation of motion (EOM) method as applied to the many-body problem of fermions. This method leads essentially to extended RPA equations, which are mostly applied in nuclear physics, but not only. There is some activity on this also in theoretical chemistry, in condensed matter, and in plasma physics which we mention in the main text. In order to point to the distinctive feature with respect to other many-body approaches, we transcribed the EOM approach into formally exact 'Dyson-Bethe-Salpeter' equations (Dyson-BSEs) for two-fermion propagators. These Dyson-BSEs contain an integral kernel which depends only on one frequency which, e.g., in the case of the response function, corresponds to the one of the external photon. This integral kernel contains a static, i.e., frequency independent and a genuinely frequency dependent, i.e., dynamic part. This is analogous to the single-particle Dyson equation, where the self-energy also splits into the static mean-field and the frequency dependent parts. The static part of the integral kernel in the Dyson-BSE can be interpreted as the mean field for the two-fermion propagation. For example, in the case of $pp$ propagation, it enters in a natural way the optical potential of elastic deuteron scattering on nuclei.
In the many-body community outside nuclear physics it is not well recognized that such formally exact Dyson-BSEs with well-defined single frequency integral kernels in terms of higher correlation functions exist. Bethe-Salpeter equations (BSEs) are mainly constructed on the basis of Green's functions with four time dependencies, that is, as many as there are fermion operators in, e.g., the response function. Taking into account the time homogeneity, the kernels still depend on three frequencies, which makes a numerical solution other than for the most basic approximations practically impossible. The popularity of this multi-frequency approach, probably, stems from the fact that in the past it has been shown that corresponding BSEs can be constructed, which fulfill all basic properties as there are conservation laws and Ward identities. In condensed matter and theoretical chemistry, for example, the GW approximation 
is based on the so-called Hedin equations \cite{Hed65} or else conserving approximations can also be derived from the 
Kadanoff-Baym so-called  $\phi$-derivable functionals \cite{Kad62,Bon96}.
All these techniques lead, besides lowest order approximations, like HF-RPA, to integral kernels which depend on more than one frequency, which in most cases cannot be treated numerically without further quite drastic approximations in order to reduce the integral kernel to a one frequency dependence. However, we have shown that most of the time those one frequency forms can directly be obtained from our explicit expression of the one-frequency kernel, see, e.g., \cite{Ole19}. In addition, as we show in this review, also EOM equations and corresponding Dyson-BSEs can be formulated so as to {\it fulfill conservation laws and Ward identities}. Since the corresponding equations are of the Schroedinger type, the access to numerical solutions is much improved and physically motivated non-trivial approximations are straightforwardly elaborated. We give an example with the three-level Lipkin model, where the numerical solution clearly shows the appearance of a  Goldstone mode in the spontaneously broken symmetry phase. Even in this simple model, to solve the Kadanoff-Baym approach including screening, that is, vertex corrections, would be extremely difficult from the numerical point of view.  In Section VII and in  \cite{Toh04,Del16}
it is shown in complete generality that conservation laws and Goldstone theorem are fulfilled with appropriate EOM equations.

After these general remarks let us be more specific on the objectives and content of this review. Our basic aim is to improve on the standard HF-RPA approach (we always include exchange in our equations, so HF-RPA means linearised time dependent HF equations). It is well known that, besides nice properties, HF-RPA also suffers from sometimes severe shortcomings. One of those is that HF-RPA corresponds to a bosonisation, see Sect.II.B and \cite{Rin80}, of the $ph$ fermion pairs, which stems from the fact that the RPA matrix is constructed with the HF vacuum. RPA treats two-body correlations, so using an uncorrelated ground state is clearly an inconsistency. We remedy to this problem in showing that the Coupled Cluster Doubles (CCD) wave function is actually the vacuum to an extended RPA operator containing a specific two-body term, see Sect.II.C. Because of this extra term the approach is a little difficult to handle, but first results are very promising, see Sect.V.C. It is clear that with the CCD wave function as vacuum the Pauli principle is entirely respected. However, in the recent past, because of this difficulty, the two-body term was neglected, which lead to a (relatively mild) violation of the Pauli principle, see \cite{Hir02}, within Richardson's pairing model which contains one of the most severe Pauli principle constraints possible since each s.p. level is only two fold degenerate. However, also in other models as, e.g., Lipkin and Hubbard models, it turned out that the Pauli principle violations are surprisingly weak \cite{Jem13}.
In this way, with a correlated ground state, the RPA approach turns into a Hartree-Fock Bogoliubov approach for fermion pairs, where the RPA matrix depends in a quite non-linear way on the usual RPA amplitudes $X$ and $Y$. It was named Self-Consistent RPA (SCRPA). Results on some non-trivial models (Richardson model, three-level Lipkin model, Hubbard model) show very strong improvements, most importantly close to a phase transition point. However, around  the phase transition points appeared some slight discontinuities reminiscent of a first order phase transition, which are absent in the exact solutions of those models. In Sect. V.C, in an application to the Lipkin model,  we achieved for the first time to keep the vacuum property of the CCD wave function fully. The results are excellent with, e.g., the correlation energy going smoothly through the phase transition region, showing there a 4 percent error while becoming even better in the weak and strong coupling limits. So, this success opens up quite wide perspectives. Of course, the approach is necessarily of some complexity and for problems, which do not demand such sophistication, more easily accessible variants of improved RPA approaches are available.
The easiest one and, may be, also the most obvious one, is the so-called renormalised RPA (r-RPA). In standard HF-RPA equations one uses HF occupation numbers which are zero or one. Since RPA calculates two-body correlation functions, it is natural to replace the HF occupation numbers by correlated, rounded ones. There exists indeed, in nuclear physics, since long a correction to the HF occupation numbers, which involves in a direct way the RPA two-body correlation density matrix, see Sect.II.B. With this, in a minimal way, the RPA matrix depends on the RPA solution and a self-consistency problem is created. Because of its numerical simplicity, this form of RPA extension has known quite a number of applications, as outlined in Sect. III.D.  
In particular, this approach was extensively applied in the investigation of double beta decay processes.
By the way, this r-RPA also can be formulated in such a way as to conserve all nice properties of the standard HF-RPA.  This simple extension of RPA has in addition the property that the Luttinger theorem is conserved \cite{Urb14}.    The full Hartree-Fock-Bogoliubov approach for fermion pairs was coined, as mentioned, Self-Consistent RPA (SCRPA). It contains much more correlations in the RPA matrix. One of the most important ones is that the bare force becomes screened (or antiscreened). For example, in the Richardson pairing Hamiltonian, for small particle numbers, the Pauli principle is so strong and, thus, screening so important that the originally attractive force is turned into repulsion, see Fig. \ref{Duk04} in Sect.III.A.

Besides the usual two-fermion ($ph$ and $pp$) problems, RPA-like equations can also be applied to the single-particle propagator, where in the self-energy of the Dyson equation the $2p-1h (2h-1p)$ propagator appears. With the EOM method we find a form of this propagator, which uses again the CCD wave function as the vacuum. This approach was coined odd-self-consistent RPA or, in short, odd-RPA. It is by the way in this way that the aforementioned excellent results for the Lipkin model were obtained in Sect.V.C allowing to pass through the phase transition region in a continuous and accurate way. One-fermion and two-fermion RPA approaches are, of course, complementary to each other. The Dyson equation allows one to calculate the correlated occupation numbers which are also needed in the SCRPA equations.

Those s.p. and two-particle Dyson equations lend themselves in an efficient way to tackle the important aspect of nuclear physics, which is the particle-vibration coupling (PVC) approach. It turns out that for the description of the width of giant resonances it is absolutely necessary to consider self-energy of the Dyson equation and dynamic kernel of the response function to rather high order in the particle-vibration (phonon) couplings. An overview of this is given in Sect. IV.D.

An extremely important aspect of nuclear physics is alpha-particle clustering. Again the EOM method has allowed to tackle this difficult subject in a very efficient way. It was shown by an in-medium four-body equation that $\alpha$-particle condensation occurs only below a certain critical density which corresponds to the point where the chemical potential turns from negative (bound state) to positive (scattering states). We explain that this has to do with the four-body level density  which for positive chemical potentials $\mu$ goes through zero at the Fermi-energy. This entails that there is no $\alpha$-particle condensation possible at those higher densities. The density, therefore, is the control parameter of this nuclear quantum phase transition (QPT). We explain why  this is opposite to deuteron condensation or, more generally, to pairing where the BCS solution exists at all densities as long as there is an attractive force. This hinges on the unique situation that for pairs at rest the two-particle level density does not go through zero at the Fermi energy. This quartet condensation approach entailed that the enigmatic structure of the Hoyle state in $^{12}$C at 7.65 MeV, so important for the carbon production in the universe and, thus, for life on earth, could be explained as a state where three $\alpha$-particles move almost freely in an extended volume within their 0S center of mass wave function, being a forerunner of $\alpha$-particle condensation in nuclear matter, see \cite{Toh17}.

In  the last two Sections VII and VIII, we illustrate another extension of RPA theory in including to the density-density (response) function the next higher configurations, which contain the squares of the density ($2p-2h$). The corresponding equations are also consistently constructed on a corresponding correlated ground state. It is demonstrated that the approach, including the SCRPA one, is {\it conserving} and keeps all good properties of standard HF-RPA intact. The numerical applications to simple models show very good accuracy. A pilot calculation in a very restricted space for giant resonances in 
$^{16}$O and $^{40}$C shows interesting spreading effects. In a much more rudimentary, that is, phenomenological form this theory in nuclear physics is known as the 'second RPA' (SRPA). It is still numerically very demanding because of the extremely high dimensions of the matrices to be diagonalized. However, nice successes using phenomenological effective forces have in the mean-time appeared in nuclear physics \cite{Bor19,Robin2019,Vas18}. 
The theory outlined in this review has as final objectives {\it ab initio calculations}. However, because of the numerical complexity of the equations, 
applications are still very rare and remain for the future.

Let us also mention that the EOM can naturally be applied in almost the same way to Bose systems 
\cite{Duk91,Aou91,Aou95},
or to mixtures of bosons and fermions \cite{Wat08,Bar08}.

In summary, we outlined progress in the {\it equation of motion method} and demonstrated its great utility. We, for instance, pointed out that a Bethe-Salpeter equation for, e.g., the response function can be derived with a single frequency kernel which yields, in principle, the exact solution for the response function. The same holds for the two fermion propagator in the pairing channel. It is argued that this Dyson-BSE approach is much more amenable to numerical solutions than the multi-frequency BSE, since the former leads to equations of the Schr\"odinger type. Also the explicit form of the integral kernels in terms of higher correlation functions lends itself to physically well motivated approximations which are difficult to obtain from the multi-frequency formulation. Additionally the advocated equations can be formulated in such a way that all conservation laws, sum-rules, Goldstone theorem and Ward identities are fullfilled, see Sects.VII.D and VII.E. Therefore, we do not see any advantage to start with the much more complicated multi-frequency formulation of the two fermion Green's function which in the past was almost exclusively at the basis of many-body approaches in various fields of physics.

\section*{Acknowledgements}
This review could not have been accomplished without the help of many recent and less recent collaborators. We want in particular thank: V. Olevano, J. Toulouse, G. Strinati, U. Lombardo, M. Baldo, M.Urban, A. Pastore, A. Rabhi, T. Sogo, V.V. Baran, R. Lazauskas, A. Storozhenko, J. G. Hirsch, P. Nozi\`eres, J. Wambach, M. Beyer, D. Janssen, S. Adachi, P. Danielewicz, S. Ethofer. Special thanks go to Markus Holzmann for valuable comments on the manuscript.
D.S.D. was supported by the grant PN-18090101/2019-2021 of the Romanian  Ministry of Education and Research, 
J.D. was supported by the Spanish Ministerio de Ciencia, Innovaci on y Universidades and 
the European regional development fund (FEDER), Project No. PGC2018-094180-B-I00,
and E.L. was supported by the US-NSF Career Grant PHY-1654379.
%



\begin{thebibliography}{99}
\bibitem{Bla86}J.P.Blaizot and G. Ripka, {\it Quantum Theory of Finite Systems} (MIT Press, Cambridge, 1986).
\bibitem{Mah81} G.D. Mahan, {\it Many Particle Physics} (Plenum Press, New York, 1981).
\bibitem{Neg88} J.W. Negele, H. Orland, {\it Quantum Many Particle Systems, Frontiers in Physics} (Addison Weseley, New York, 1988).
\bibitem{Fet71} A.L. Fetter and J.D. Walecka, {\it Quantum Theory of Many particle Systems} (McGraw-Hill, New York, 1971).
\bibitem{Rin80} P. Ring and P. Schuck, {\it The Nuclear Many-body Problem}  (Springer-Verlag, Berlin, 1980).
\bibitem{Bar07} R. J. Bartlett and M. Musial, Rev. Mod. Phys. 79, 291 (2007);
G. Hagen, T. Papenbrock, M. Hjorth-Jensen, and D. J. Dean, 
{\it Coupled-cluster computations of atomic nuclei}, Rep. Prog. Phys. {\bf 77}, 096302 (2014).
\bibitem{Ful} P. Fulde, {\it Electron  Correlations in Molecules and Solids},
 (Springer Series in Solid State Sciences, Gebhardt book, 1995).

\bibitem{Wag16} L. K. Wagner and D. M. Ceperley, 
{\it Discovering correlated fermions using quantum Monte Carlo},
Rep. Prog. Phys. {\bf 79} 094501 (2016).
\bibitem{Car15} J. Carlson, S. Gandolfi, F. Pederiva, Steven C. Pieper, R. Schiavilla, K. E. Schmidt, and R. B. Wiringa,
{\it Quantum Monte Carlo methods for nuclear physics},
Rev. Mod. Phys. {\bf 87}, 1067 (2015).


\bibitem{Cla66} J. W. Clark, P. Westhaus, Phys. Rev. {\bf 141} (1966) 833.
\bibitem{Fab02} A. Fabrocini, S. Fantoni, E. Krotscheck (Editors), 
{\it Introduction to Modern Methods of Quantum Many Body Theory}, Vol.7, World Scientific, Singapore (2002).
\bibitem{Bau19} R. Bauerschmidt, D. C. Bydges, and G. Slade, {\it Lecture Notes in Mathematics}, (Springer 2019).
\bibitem{Bis91} R. F. Bishop, Theor. Chem. Acta {\bf 80} (1991) 95.
\bibitem{Keh} S. Kehrein, {\it The Flow Equation Approach to Many-Particle Systems}, (Springer Tracts in Modern Physics, 2006).
\bibitem{Sch05} U. Schollw\"ock, Rev. Mod. Phys. {\bf 77}, 259 (2005).
\bibitem{Row68} D.J. Rowe, Rev. Mod. Phys. {\bf 40}, 153 (1968); Phys. Rev. {\bf 175}, 1283 (1968).
\bibitem{Rop80} G. R\"opke, T. Seifert, H. Stolz, and R. Zimmermann, Phys. Stat. Sol. (b) {\bf 100}, 215 (1980);\\
G. R\"opke, M. Schmidt,  L. M\"unchow, and H. Schulz, Nucl. Phys. A {\bf 399}, 587 (1983). 
\bibitem{Rop95} G. R\"opke, Z. Phys. {\bf B99}, 83 (1995) and the references therein.
\bibitem{Rop09} G. R\"opke, Phys. Rev. C {\bf 79}, 014002 (2009).
\bibitem{Her16} H. Hergert, S. K. Bogner, T. D. Morris, A. Schwenk, and K. Tsukiyama, Phys. Rep. {\bf 621}, 165 (2016).
\bibitem{Duk90} J. Dukelsky and P. Schuck, Nucl. Phys. A {\bf 512}, 466 (1990).
\bibitem{Duk98} J. Dukelsky, G. R\"opke, and P.Schuck, Nucl. Phys. A {\bf 628}, 17 (1998).
\bibitem{Duk99} J. Dukelsky and P. Schuck, Phys. Lett. B {\bf 464}, 164 (1999).
\bibitem{Bar70} M. Baranger, Nucl. Phys. A {\bf 149} (1970) 225.
\bibitem{Duk91} J. Dukelsky and P. Schuck, Mod. Phys. Lett. A {\bf 26}, 2429 (1991).
\bibitem{Sog13} T. Sogo, P. Schuck, and M. Urban, Phys. Rev. A {\bf 88}, 023613 (2013).
\bibitem{Wat08} T. Watanabe, T. Suzuki, and P. Schuck, Phys. Rev. A {\bf 78}, 033601 (2008).
\bibitem{Sto05} A. Storozhenko, P. Schuck, T. Suzuki, H. Yabu, and J. Dukelsky, Phys. Rev. A {\bf 71}, 063617 (2005).
\bibitem{Kru94} P. Kr\"uger and P. Schuck, Europhy. Lett. {\bf 27} (1994) 395;\\
J. G. Hirsch, A. Mariano, J. Dukelsky, and P. Schuck, Ann. Phys. (NY) {\bf 296}, 187 (2002).
\bibitem{Sch73} P. Schuck, S. Ethofer, Nucl. Phys. A {\bf 212}, 269 (1973);  \\
J. Dukelsky, G. R\"opke, P. Schuck, Nucl. Phys. A {\bf 628}, 17 (1998).
\bibitem{Ada89} S. Adachi and P. Schuck, Nucl. Phys. A {\bf 496}, 485 (1989). 
\bibitem{Sch00} H.-J. Schulze, P. Schuck, and N. Van Giai, Phys. Rev. B {\bf 61}, 8026 (2000);
P. Schuck, H.-J. Schulze, and N. Van Giai, and M. Zverev, Phys. Rev. B {\bf 67}, 233404 (2003).
\bibitem{Cha12} K. Chatterjee and K. Pernal, J. Chem. Phys. {\bf 137}, 204109 (2012). 
\bibitem{Esh12} H. Eshuis, J. Bates, and F. Furche, Theor. Chem. Acc. {\bf 131}, 1084 (2012). 
\bibitem{Per14} K. Pernal, K. Chatterjee, and P. H. Kowalski, J. Chem. Phys. {\bf 140}, 014101 (2014). 
\bibitem{Per18} K. Pernal, Phys. Rev. Lett. {\bf 120}, 013001 (2018).
\bibitem{Shi70} T. Shibuya, V. McKoy, Phys. Rev. A {\bf 2}, 2208 (1970). 
\bibitem{Shi73} T. Shibuya, J. Rose, and V. McKoy, J. Chem. Phys. {\bf 58}, 500 (1973).
\bibitem{Las77} A. C. Lasaga and M. Karplus, Phys. Rev. A {\bf 16}, 807 (1977).
\bibitem{Ter17} J. Terasaki, A. Smetana, F. Simkovic, and M. I. Krivoruchenko, Int. J. Mod. Phys. E {\bf 26}, 1750062 (2017).
\bibitem{Jem11} M. Jemai and P. Schuck, Phys. At. Nucl. {\bf 74}, No 8, 1139 (2011). 
\bibitem{Jem13} M. Jemai, D.S. Delion, and P. Schuck, Phys. Rev. C {\bf 88}, 044004 (2013).
\bibitem{Hir02} J.G. Hirsch, A. Mariano, J. Dukelsky, and P. Schuck, Ann. Phys. (NY) {\bf 296}, 187 (2002).
\bibitem{Sch16} P. Schuck and M. Tohyama, Phys. Rev. B {\bf 93}, 165117 (2016).
\bibitem{Del05} D.S. Delion, P. Schuck, and J. Dukelsky, Phys. Rev. C {\bf 72}, 064305 (2005).
\bibitem{Har64} K. Hara, Progr. Theor. Phys. {\bf 32}, 88 (1964).
\bibitem{Cat96} F. Catara, G. Piccitto, M. Sambataro, and N. Van Giai, Phys. Rev. B {\bf 54}, 17536 (1996);\\
F. Catara, M. Grasso, G. Piccitto, and M. Sambataro, Phys. Rev. B {\bf 58}, 16070 (1998).
\bibitem{Tho61} D. J. Thowless {\it The Quantum Mechanics of Many-Body Systems}, Academic Press Inc., New York and London (1961).
\bibitem{Rab02} A. Rabhi, R. Bennaceur, G. Chanfray, and P. Schuck, Phys. Rev. C {\bf 66}, 064315 (2002).
\bibitem{Federschmidt1985} C. Federschmidt, P. Ring, Nucl. Phys. A {\bf 435}, 110 (1985).
\bibitem{Kyotoku1990} M. Kyotoku, K.W. Schmid, F. Gr\"ummer, and A. Faessler, Phys. Rev. C {\bf 41}, 284 (1990).
\bibitem{Civitarese1991} O. Civitarese, A. Faessler, J. Suhonen, and X.R. Wu, Nucl. Phys. A {\bf 524}, 404 (1991).
\bibitem{Suhonen1993} J. Suhonen, J. Phys. G {\bf 19}, 139 (1993).
\bibitem{Duk19} J. Dukelsky, J. E. Garcia-Ramos, J. M. Arias, P. P\'erez-Fern\'andez, and P. Schuck, 
Phys. Lett. B {\bf 795}, 537 (2019).
\bibitem{Sangfelt1987} E. Sangfelt, R. Roy Chowduri, B. Weiner, Y. Ahrn,  J. Chem. Phys. {\bf 86}, 4523 (1987).
\bibitem{Agassi1968} D. Agassi, Nucl. Phys. A {\bf 116}, 49 (1968).
\bibitem{Davis1986} E. D. Davis, W. D. Heiss, J. Phys. G {\bf 12}, 805 (1986).
\bibitem{Ramos2018} J. E. Garcia-Ramos, J. Dukelsky, P. PÃ©rez-FernÃ¡ndez, and J. M. Arias, Phys. Rev. C {\bf 97}, 054303 (2018).
\bibitem{Bender2020} M. Bender {\it et al} arXiv:2005.10216. 
\bibitem{Toh13} M. Tohyama, and P. Schuck, Phys. Rev. C {\bf 87}, 044316 (2013).
\bibitem{Jem19} M. Jemai and P. Schuck, Phys. Rev. C {\bf 100}, 034311 (2019). 
\bibitem{Ric63} R.W. Richardson, Phys. Lett. {\bf 3}, 277 (1963).
\bibitem{Li70} S.Y. Li, A. Klein and R.M. Dreizler, J. Math. Phys. {\bf 11}, 975 (1970).
\bibitem{Mes71} N. Meshkov, Phys. Rev. C {\bf 3}, 2214 (1971).
\bibitem{Sam99} M. Sambataro, Phys. Rev. C {\bf 60}, 064320 (1999).
\bibitem{Gra02} M. Grasso, F. Catara, and M. Sambataro, Phys. Rev. C {\bf 66}, 064303 (2002).
\bibitem{Hag00} K. Hagino and G.F. Bertsch, Phys. Rev. C {\bf 61}, 024307 (2000).
\bibitem{For75} D. Forster, {\it Hydrodinamic fluctuations-broken symmetry and correlation functions},
Frontiers in Physics: A Lecture Note and Reprint Series, No {\bf 47} (Benjamin, New York, 1975).
\bibitem{Jem05} M. Jemai, P. Schuck, J. Dukelsky, and R. Bennaceur, Phys. Rev. B {\bf 71}, 085115 (2005).
\bibitem{Dur20} D. Durel and M. Urban, Phys. Rev. A {\bf 101}, 013608 (2020).
\bibitem{Rad91} A.A.Raduta, A.Faessler, and S.Stoica, Nucl.Phys.A {\bf 534}, 149 (1991).
\bibitem{Toi95} J. Toivanen and J. Suhonen, Phys. Rev. Lett. {\bf 75}, 410 (1995).
\bibitem{Cat94} F. Catara, N. Dinh Dang, and M. Sambataro, Nucl. Phys. A {\bf 579}, 1 (1994).
\bibitem{Mac87} R. Machleidt, K. Holinde, and C. Elster, Phys. Rep. {\bf 149}, 1 (1987).
\bibitem{Toi97} J. Toivanen and J. Suhonen, Phys. Rev. C {\bf 55}, 2314 (1997).
\bibitem{Sch96} J. Schwieger, F. Simkovic, and Amand Faessler, Nucl. Phys. A {\bf 600}, 179 (1996).
\bibitem{Fae98} Amand Faessler and F. Simcovic, J. Phys. G {\bf 24}, 2139 (1998).
\bibitem{Civ97} O. Civitarese and M. Reboiro, Phys. Rev. C {\bf 57}, 3092 (1997).
\bibitem{Hir96} J.G. Hirsch, P.O. Hess, and O. Civitarese, Phys. Rev. C {\bf 54}, 1976 (1996).
\bibitem{Hir97} J.G. Hirsch, P.O. Hess, and O. Civitarese, Phys. Rev. C {\bf 56}, 199 (1997).
\bibitem{Hir99} J.G. Hirsch, P.O. Hess, and O. Civitarese, Phys. Rev. C {\bf 60}, 064303 (1999).
\bibitem{Hir99a} J.G. Hirsch, O. Civitarese, and M. Reboiro, Phys. Rev. C {\bf 60}, 024309 (1999).
\bibitem{Rad02} A.A. Raduta, Prog. Part. Nucl. Phys. {\bf 48}, 233 (2002).
\bibitem{Bob99} A. Bobyk, W.A. Kaminski, and P. Zareba, Eur. Phys. J. A {\bf 5}, 385 (1999).
\bibitem{Bob00} A. Bobyk, W.A. Kaminski, and P. Zareba, Nucl. Phys. A {\bf 669}, 221 (2000).
\bibitem{Rad98} A.A. Raduta {\it et al.}, Nucl. Phys. A {\bf 634}, 497 (1998).
\bibitem{Rad05} C.M. Raduta and A.A. Raduta, Nucl. Phys. A {\bf 756}, 153 (2005).
\bibitem{Rad10} C. M. Raduta and A. A. Raduta, Phys. Rev. C {\bf 82}, 068501 (2010). 
\bibitem{Dan00} N. Dinh Dang and A. Arima, Phys. Rev. C {\bf 62}, 024303 (2000).
\bibitem{Rod02} V. Rodin and A. Fessler, Phys. Rev. C {\bf 66}, 051303(R) (2002).
\bibitem{Pac03} L. Pacearescu, V. Rodin, F. Simkovic, and A. Faessler, Phys. Rev. C {\bf 68}, 064310 (2003).
\bibitem{Krm98} F. Krmpotic, E.J.V. Passos, D.S. Delion, J. Dukelsky, and P. Schuck, Nucl. Phys. A {\bf 637}, 295 (1998).
\bibitem{Sim00} F. Simkovic, A.A. Raduta, M. Veselsky, and A.Faessler, Phys. Rev. C {\bf 61}, 044319 (2000).
\bibitem{Sim03} F. Simkovic, M. Smotlak, and G. Pantis, Phys. Rev. C {\bf 68}, 014309 (2003).
\bibitem{Sim01} F. Simkovic, M. Smotlak, and A.A. Raduta, J. Phys. G {\bf 27}, 1757 (2001).
\bibitem{Ole19} V. Olevano, J. Toulouse, and P. Schuck,  J. Chem. Phys. {\bf 150}, 084112 (2019).
\bibitem{Sch19} P. Schuck, Eur. Phys. J. A {\bf 55}, 250 (2019).
\bibitem{Gor61} L. P. Gorkov and T. K. Melik-Barkhudarov, Sov. Phys. JETP {\bf 13}, 1018 (1961).
\bibitem{Str18} G. Calvanese Strinati, P. Pieri, G. R\"opke, P. Schuck, and M. Urban, Phys. Rep. {\bf 738}, 1 (2018).
\bibitem{Ram18} S. Ramanan and M. Urban,  Phys. Rev. C {\bf 98}, 024314 (2018).
\bibitem{Pet02} C.J. Pethick and H. Smith,
{\it Bose-Einstein Condensation in Dilute Gases} (Cambridge University Press, 2002).
\bibitem{Kra86} W. D. Kraeft, D. Kremp, W. Ebeling, and G. R\"opke, 
{\it Quantum Statistics of Charged Particle Systems} (Berlin, Akademie-Verlag 1986).
\bibitem{Wen19} Wenmei Guo, U. Lombardo, and P. Schuck, Phys. Rev. C {\bf 99}, 014310 (2019). 
\bibitem{Urb20} M. Urban and S. Ramanan, Phys. Rev. C {\bf 101}, 035803 (2020).
\bibitem{Lit18a} E. Litvinova, C. Robin, and P. Schuck, EPJ Web Conf. {\bf 182}, no. 02075 (2018). 
\bibitem{Lit19a} E. Litvinova and P. Schuck, Phys. Rev. C {\bf 100}, 064320 (2019).

\expandafter\ifx\csname natexlab\endcsname\relax\def\natexlab#1{#1}\fi
\expandafter\ifx\csname bibnamefont\endcsname\relax
  \def\bibnamefont#1{#1}\fi
\expandafter\ifx\csname bibfnamefont\endcsname\relax
  \def\bibfnamefont#1{#1}\fi
\expandafter\ifx\csname citenamefont\endcsname\relax
  \def\citenamefont#1{#1}\fi
\expandafter\ifx\csname url\endcsname\relax
  \def\url#1{\texttt{#1}}\fi
\expandafter\ifx\csname urlprefix\endcsname\relax\def\urlprefix{URL }\fi
\providecommand{\bibinfo}[2]{#2}
\providecommand{\eprint}[2][]{\url{#2}}

\bibitem[{\citenamefont{Bohr and Mottelson}(1969)}]{BohrMottelson1969}
\bibinfo{author}{\bibfnamefont{A.}~\bibnamefont{Bohr}} \bibnamefont{and}
  \bibinfo{author}{\bibfnamefont{B.~R.} \bibnamefont{Mottelson}},
  \emph{\bibinfo{title}{Nuclear structure}}, vol.~\bibinfo{volume}{1}
  (\bibinfo{publisher}{World Scientific}, \bibinfo{year}{1969}).

\bibitem[{\citenamefont{Bohr and Mottelson}(1975)}]{BohrMottelson1975}
\bibinfo{author}{\bibfnamefont{A.}~\bibnamefont{Bohr}} \bibnamefont{and}
  \bibinfo{author}{\bibfnamefont{B.~R.} \bibnamefont{Mottelson}},
  \emph{\bibinfo{title}{Nuclear structure}}, vol.~\bibinfo{volume}{2}
  (\bibinfo{publisher}{Benjamin, New York}, \bibinfo{year}{1975}).

\bibitem[{\citenamefont{Broglia and Bortignon}(1976)}]{Broglia1976}
\bibinfo{author}{\bibfnamefont{R.~A.} \bibnamefont{Broglia}} \bibnamefont{and}
  \bibinfo{author}{\bibfnamefont{P.~F.} \bibnamefont{Bortignon}},
  \bibinfo{journal}{Phys. Lett. B} \textbf{\bibinfo{volume}{65}},
  \bibinfo{pages}{221} (\bibinfo{year}{1976}).

\bibitem[{\citenamefont{Bortignon et~al.}(1977)\citenamefont{Bortignon,
  Broglia, Bes, and Liotta}}]{BortignonBrogliaBesEtAl1977}
\bibinfo{author}{\bibfnamefont{P.~F.} \bibnamefont{Bortignon}},
  \bibinfo{author}{\bibfnamefont{R.}~\bibnamefont{Broglia}},
  \bibinfo{author}{\bibfnamefont{D.}~\bibnamefont{Bes}}, \bibnamefont{and}
  \bibinfo{author}{\bibfnamefont{R.}~\bibnamefont{Liotta}},
  \bibinfo{journal}{Phys. Rep.} \textbf{\bibinfo{volume}{30}},
  \bibinfo{pages}{305} (\bibinfo{year}{1977}).

\bibitem[{\citenamefont{Bertsch et~al.}(1983)\citenamefont{Bertsch, Bortignon,
  and Broglia}}]{BertschBortignonBroglia1983}
\bibinfo{author}{\bibfnamefont{G.}~\bibnamefont{Bertsch}},
  \bibinfo{author}{\bibfnamefont{P.}~\bibnamefont{Bortignon}},
  \bibnamefont{and} \bibinfo{author}{\bibfnamefont{R.}~\bibnamefont{Broglia}},
  \bibinfo{journal}{Rev. Mod. Phys.} \textbf{\bibinfo{volume}{55}},
  \bibinfo{pages}{287} (\bibinfo{year}{1983}).

\bibitem[{\citenamefont{Soloviev}(1992)}]{Soloviev1992}
\bibinfo{author}{\bibfnamefont{V.}~\bibnamefont{Soloviev}},
  \emph{\bibinfo{title}{Theory of Atomic Nuclei: Quasiparticles and Phonons}}
  (\bibinfo{publisher}{Institute of Physics Publishing}, \bibinfo{year}{1992}).

\bibitem[{\citenamefont{Bortignon et~al.}(1978)\citenamefont{Bortignon,
  Broglia, and Bes}}]{Bortignon1978}
\bibinfo{author}{\bibfnamefont{P.~F.} \bibnamefont{Bortignon}},
  \bibinfo{author}{\bibfnamefont{R.~A.} \bibnamefont{Broglia}},
  \bibnamefont{and} \bibinfo{author}{\bibfnamefont{D.~R.} \bibnamefont{Bes}},
  \bibinfo{journal}{Phys. Lett. B} \textbf{\bibinfo{volume}{76}},
  \bibinfo{pages}{153} (\bibinfo{year}{1978}).

\bibitem[{\citenamefont{Bortignon and Broglia}(1981)}]{Bortignon1981a}
\bibinfo{author}{\bibfnamefont{P.~F.} \bibnamefont{Bortignon}}
  \bibnamefont{and} \bibinfo{author}{\bibfnamefont{R.~A.}
  \bibnamefont{Broglia}}, \bibinfo{journal}{Nucl. Phys. A}
  \textbf{\bibinfo{volume}{371}}, \bibinfo{pages}{405} (\bibinfo{year}{1981}).

\bibitem[{\citenamefont{Mahaux et~al.}(1985)\citenamefont{Mahaux, Bortignon,
  Broglia, and Dasso}}]{MahauxBortignonBrogliaEtAl1985}
\bibinfo{author}{\bibfnamefont{C.}~\bibnamefont{Mahaux}},
  \bibinfo{author}{\bibfnamefont{P.}~\bibnamefont{Bortignon}},
  \bibinfo{author}{\bibfnamefont{R.}~\bibnamefont{Broglia}}, \bibnamefont{and}
  \bibinfo{author}{\bibfnamefont{C.}~\bibnamefont{Dasso}},
  \bibinfo{journal}{Phys. Rep.} \textbf{\bibinfo{volume}{120}},
  \bibinfo{pages}{1} (\bibinfo{year}{1985}).

\bibitem[{\citenamefont{Bortignon et~al.}(1986)\citenamefont{Bortignon,
  Broglia, Bertsch, and Pacheco}}]{Bortignon1986}
\bibinfo{author}{\bibfnamefont{P.}~\bibnamefont{Bortignon}},
  \bibinfo{author}{\bibfnamefont{R.}~\bibnamefont{Broglia}},
  \bibinfo{author}{\bibfnamefont{G.}~\bibnamefont{Bertsch}}, \bibnamefont{and}
  \bibinfo{author}{\bibfnamefont{J.}~\bibnamefont{Pacheco}},
  \bibinfo{journal}{Nucl. Phys. A} \textbf{\bibinfo{volume}{460}},
  \bibinfo{pages}{149} (\bibinfo{year}{1986}).

\bibitem[{\citenamefont{Bortignon and Dasso}(1997)}]{Bortignon1997}
\bibinfo{author}{\bibfnamefont{P.~F.} \bibnamefont{Bortignon}}
  \bibnamefont{and} \bibinfo{author}{\bibfnamefont{C.~H.} \bibnamefont{Dasso}},
  \bibinfo{journal}{Phys. Rev. C} \textbf{\bibinfo{volume}{56}},
  \bibinfo{pages}{574} (\bibinfo{year}{1997}).

\bibitem[{\citenamefont{Col{\`o} and Bortignon}(2001)}]{ColoBortignon2001}
\bibinfo{author}{\bibfnamefont{G.}~\bibnamefont{Col{\`o}}} \bibnamefont{and}
  \bibinfo{author}{\bibfnamefont{P.-F.} \bibnamefont{Bortignon}},
  \bibinfo{journal}{Nucl. Phys. A} \textbf{\bibinfo{volume}{696}},
  \bibinfo{pages}{427} (\bibinfo{year}{2001}).

\bibitem[{\citenamefont{Tselyaev}(1989)}]{Tselyaev1989}
\bibinfo{author}{\bibfnamefont{V.}~\bibnamefont{Tselyaev}},
  \bibinfo{journal}{Sov. J. Nucl. Phys.}
  \textbf{\bibinfo{volume}{50}}, \bibinfo{pages}{780} (\bibinfo{year}{1989}).

\bibitem[{\citenamefont{Kamerdzhiev et~al.}(1997)\citenamefont{Kamerdzhiev,
  Tertychny, and Tselyaev}}]{KamerdzhievTertychnyiTselyaev1997}
\bibinfo{author}{\bibfnamefont{S.~P.} \bibnamefont{Kamerdzhiev}},
  \bibinfo{author}{\bibfnamefont{G.~Y.} \bibnamefont{Tertychny}},
  \bibnamefont{and} \bibinfo{author}{\bibfnamefont{V.~I.}
  \bibnamefont{Tselyaev}}, \bibinfo{journal}{Physics of Particles and Nuclei}
  \textbf{\bibinfo{volume}{28}}, \bibinfo{pages}{134} (\bibinfo{year}{1997}).

\bibitem[{\citenamefont{Ponomarev et~al.}(2001)\citenamefont{Ponomarev,
  Bortignon, Broglia, and Voronov}}]{Ponomarev2001}
\bibinfo{author}{\bibfnamefont{V.~{\relax Yu}.} \bibnamefont{Ponomarev}},
  \bibinfo{author}{\bibfnamefont{P.~F.} \bibnamefont{Bortignon}},
  \bibinfo{author}{\bibfnamefont{R.~A.} \bibnamefont{Broglia}},
  \bibnamefont{and} \bibinfo{author}{\bibfnamefont{V.~V.}
  \bibnamefont{Voronov}}, \bibinfo{journal}{Nucl. Phys. A}
  \textbf{\bibinfo{volume}{687}}, \bibinfo{pages}{170} (\bibinfo{year}{2001}).

\bibitem[{\citenamefont{Ponomarev et~al.}(1999)}]{Ponomarev1999b}
\bibinfo{author}{\bibfnamefont{V.~{\relax Yu}.} \bibnamefont{Ponomarev}}
  \bibnamefont{et~al.}, \bibinfo{journal}{Phys. Rev. Lett.}
  \textbf{\bibinfo{volume}{83}}, \bibinfo{pages}{4029} (\bibinfo{year}{1999}).

\bibitem[{\citenamefont{Lo~Iudice et~al.}(2012)\citenamefont{Lo~Iudice,
  Ponomarev, Stoyanov, Sushkov, and Voronov}}]{LoIudice2012}
\bibinfo{author}{\bibfnamefont{N.}~\bibnamefont{Lo~Iudice}},
  \bibinfo{author}{\bibfnamefont{V.~Y.} \bibnamefont{Ponomarev}},
  \bibinfo{author}{\bibfnamefont{C.}~\bibnamefont{Stoyanov}},
  \bibinfo{author}{\bibfnamefont{A.~V.} \bibnamefont{Sushkov}},
  \bibnamefont{and} \bibinfo{author}{\bibfnamefont{V.~V.}
  \bibnamefont{Voronov}}, \bibinfo{journal}{J. Phys. G}
  \textbf{\bibinfo{volume}{39}}, \bibinfo{pages}{043101}
  (\bibinfo{year}{2012}).

\bibitem[{\citenamefont{Litvinova}(2016)}]{Litvinova2016}
\bibinfo{author}{\bibfnamefont{E.}~\bibnamefont{Litvinova}},
  \bibinfo{journal}{Physics Letters B} \textbf{\bibinfo{volume}{755}},
  \bibinfo{pages}{138 } (\bibinfo{year}{2016}).

\bibitem[{\citenamefont{Robin and Litvinova}(2016)}]{RobinLitvinova2016}
\bibinfo{author}{\bibfnamefont{C.}~\bibnamefont{Robin}} \bibnamefont{and}
  \bibinfo{author}{\bibfnamefont{E.}~\bibnamefont{Litvinova}},
  \bibinfo{journal}{Eur. Phys. J. A} \textbf{\bibinfo{volume}{52}},
  \bibinfo{pages}{205} (\bibinfo{year}{2016}).

\bibitem[{\citenamefont{Robin and Litvinova}(2018)}]{RobinLitvinova2018}
\bibinfo{author}{\bibfnamefont{C.}~\bibnamefont{Robin}} \bibnamefont{and}
  \bibinfo{author}{\bibfnamefont{E.}~\bibnamefont{Litvinova}},
  \bibinfo{journal}{Phys. Rev. C} \textbf{\bibinfo{volume}{98}},
  \bibinfo{pages}{051301} (\bibinfo{year}{2018}).

\bibitem[{\citenamefont{Robin and Litvinova}(2019)}]{Robin2019}
\bibinfo{author}{\bibfnamefont{C.}~\bibnamefont{Robin}} \bibnamefont{and}
  \bibinfo{author}{\bibfnamefont{E.}~\bibnamefont{Litvinova}},
  \bibinfo{journal}{Phys. Rev. Lett.} \textbf{\bibinfo{volume}{123}},
  \bibinfo{pages}{202501} (\bibinfo{year}{2019}).

\bibitem{Bor19} P.F. Bortignon, E.E. Saperstein, and M. Baldo, Eur. Phys. J. A {\bf 246}, (2019).

\bibitem{Rom09} P. Romaniello, D. Sangalli, J. A. Berger, F. Sottile, L. G. Molinari, and L. Reining, 
J. Chem. Phys. {\bf 130}, 044108 (2009). 

\bibitem[{\citenamefont{Litvinova and
  Schuck}(2019{\natexlab{b}})}]{LitvinovaSchuck2019a}
\bibinfo{author}{\bibfnamefont{E.}~\bibnamefont{Litvinova}} \bibnamefont{and}
  \bibinfo{author}{\bibfnamefont{P.}~\bibnamefont{Schuck}},
  \bibinfo{journal}{arXiv:1912.12585}  (\bibinfo{year}{2019}{\natexlab{b}}).

\bibitem[{\citenamefont{Papakonstantinou and
  Roth}(2009)}]{PapakonstantinouRoth2009}
\bibinfo{author}{\bibfnamefont{P.}~\bibnamefont{Papakonstantinou}}
  \bibnamefont{and} \bibinfo{author}{\bibfnamefont{R.}~\bibnamefont{Roth}},
  \bibinfo{journal}{Phys. Lett. B} \textbf{\bibinfo{volume}{671}},
  \bibinfo{pages}{356} (\bibinfo{year}{2009}).

\bibitem[{\citenamefont{Litvinova et~al.}(2007)\citenamefont{Litvinova, Ring,
  and Tselyaev}}]{LitvinovaRingTselyaev2007}
\bibinfo{author}{\bibfnamefont{E.}~\bibnamefont{Litvinova}},
  \bibinfo{author}{\bibfnamefont{P.}~\bibnamefont{Ring}}, \bibnamefont{and}
  \bibinfo{author}{\bibfnamefont{V.}~\bibnamefont{Tselyaev}},
  \bibinfo{journal}{Phys. Rev. C} \textbf{\bibinfo{volume}{75}},
  \bibinfo{pages}{064308} (\bibinfo{year}{2007}).

\bibitem[{\citenamefont{Litvinova et~al.}(2008)\citenamefont{Litvinova, Ring,
  and Tselyaev}}]{LitvinovaRingTselyaev2008}
\bibinfo{author}{\bibfnamefont{E.}~\bibnamefont{Litvinova}},
  \bibinfo{author}{\bibfnamefont{P.}~\bibnamefont{Ring}}, \bibnamefont{and}
  \bibinfo{author}{\bibfnamefont{V.}~\bibnamefont{Tselyaev}},
  \bibinfo{journal}{Phys. Rev. C} \textbf{\bibinfo{volume}{78}},
  \bibinfo{pages}{014312} (\bibinfo{year}{2008}).

\bibitem[{\citenamefont{Litvinova et~al.}(2010)\citenamefont{Litvinova, Ring,
  and Tselyaev}}]{LitvinovaRingTselyaev2010}
\bibinfo{author}{\bibfnamefont{E.}~\bibnamefont{Litvinova}},
  \bibinfo{author}{\bibfnamefont{P.}~\bibnamefont{Ring}}, \bibnamefont{and}
  \bibinfo{author}{\bibfnamefont{V.}~\bibnamefont{Tselyaev}},
  \bibinfo{journal}{Phys. Rev. Lett.} \textbf{\bibinfo{volume}{105}},
  \bibinfo{pages}{022502} (\bibinfo{year}{2010}).

\bibitem[{\citenamefont{Litvinova et~al.}(2013)\citenamefont{Litvinova, Ring,
  and Tselyaev}}]{LitvinovaRingTselyaev2013}
\bibinfo{author}{\bibfnamefont{E.}~\bibnamefont{Litvinova}},
  \bibinfo{author}{\bibfnamefont{P.}~\bibnamefont{Ring}}, \bibnamefont{and}
  \bibinfo{author}{\bibfnamefont{V.}~\bibnamefont{Tselyaev}},
  \bibinfo{journal}{Phys. Rev. C} \textbf{\bibinfo{volume}{88}},
  \bibinfo{pages}{044320} (\bibinfo{year}{2013}).

\bibitem[{\citenamefont{Tselyaev et~al.}(2016)\citenamefont{Tselyaev,
  Lyutorovich, Speth, Krewald, and Reinhard}}]{Tselyaev2016}
\bibinfo{author}{\bibfnamefont{V.}~\bibnamefont{Tselyaev}},
  \bibinfo{author}{\bibfnamefont{N.}~\bibnamefont{Lyutorovich}},
  \bibinfo{author}{\bibfnamefont{J.}~\bibnamefont{Speth}},
  \bibinfo{author}{\bibfnamefont{S.}~\bibnamefont{Krewald}}, \bibnamefont{and}
  \bibinfo{author}{\bibfnamefont{P.~G.} \bibnamefont{Reinhard}},
  \bibinfo{journal}{Phys. Rev. C} \textbf{\bibinfo{volume}{94}},
  \bibinfo{pages}{034306} (\bibinfo{year}{2016}).

\bibitem[{\citenamefont{Tselyaev et~al.}(2018)\citenamefont{Tselyaev,
  Lyutorovich, Speth, and Reinhard}}]{Tselyaev2018}
\bibinfo{author}{\bibfnamefont{V.}~\bibnamefont{Tselyaev}},
  \bibinfo{author}{\bibfnamefont{N.}~\bibnamefont{Lyutorovich}},
  \bibinfo{author}{\bibfnamefont{J.}~\bibnamefont{Speth}}, \bibnamefont{and}
  \bibinfo{author}{\bibfnamefont{P.~G.} \bibnamefont{Reinhard}},
  \bibinfo{journal}{Phys. Rev. C} \textbf{\bibinfo{volume}{97}},
  \bibinfo{pages}{044308} (\bibinfo{year}{2018}).

\bibitem[{\citenamefont{Niu et~al.}(2015)\citenamefont{Niu, Niu, Col{\`o}, and
  Vigezzi}}]{NiuNiuColoEtAl2015}
\bibinfo{author}{\bibfnamefont{Y.}~\bibnamefont{Niu}},
  \bibinfo{author}{\bibfnamefont{Z.}~\bibnamefont{Niu}},
  \bibinfo{author}{\bibfnamefont{G.}~\bibnamefont{Col{\`o}}}, \bibnamefont{and}
  \bibinfo{author}{\bibfnamefont{E.}~\bibnamefont{Vigezzi}},
  \bibinfo{journal}{Phys. Rev. Lett.} \textbf{\bibinfo{volume}{114}},
  \bibinfo{pages}{142501} (\bibinfo{year}{2015}).

\bibitem[{\citenamefont{Niu et~al.}(2018)\citenamefont{Niu, Niu, Col{\`o}, and
  Vigezzi}}]{Niu2018}
\bibinfo{author}{\bibfnamefont{Y.}~\bibnamefont{Niu}},
  \bibinfo{author}{\bibfnamefont{Z.}~\bibnamefont{Niu}},
  \bibinfo{author}{\bibfnamefont{G.}~\bibnamefont{Col{\`o}}}, \bibnamefont{and}
  \bibinfo{author}{\bibfnamefont{E.}~\bibnamefont{Vigezzi}},
  \bibinfo{journal}{Phys. Lett. B} \textbf{\bibinfo{volume}{780}},
  \bibinfo{pages}{325} (\bibinfo{year}{2018}).

\bibitem[{\citenamefont{Shen et~al.}(2020)\citenamefont{Shen, Col{\`o},
  Roca-Maza et~al.}}]{Shen2020}
\bibinfo{author}{\bibfnamefont{S.}~\bibnamefont{Shen}},
  \bibinfo{author}{\bibfnamefont{G.}~\bibnamefont{Col{\`o}}},
  \bibinfo{author}{\bibfnamefont{X.}~\bibnamefont{Roca-Maza}},
  \bibnamefont{et~al.}, \bibinfo{journal}{Phys. Rev. C}
  \textbf{\bibinfo{volume}{101}}, \bibinfo{pages}{044316}
  (\bibinfo{year}{2020}).

\bibitem[{\citenamefont{Litvinova}(2015)}]{Litvinova2015}
\bibinfo{author}{\bibfnamefont{E.}~\bibnamefont{Litvinova}},
  \bibinfo{journal}{Phys. Rev. C} \textbf{\bibinfo{volume}{91}},
  \bibinfo{pages}{034332} (\bibinfo{year}{2015}).

\bibitem[{\citenamefont{Erokhova et~al.}(2003)\citenamefont{Erokhova, Yolkin,
  Izotova, Ishkhanov, Kapitonov, Lileeva, and Shirokov}}]{Erokhova2003}
\bibinfo{author}{\bibfnamefont{V.~A.} \bibnamefont{Erokhova}},
  \bibinfo{author}{\bibfnamefont{M.~A.} \bibnamefont{Yolkin}},
  \bibinfo{author}{\bibfnamefont{A.~V.} \bibnamefont{Izotova}},
  \bibinfo{author}{\bibfnamefont{B.~S.} \bibnamefont{Ishkhanov}},
  \bibinfo{author}{\bibfnamefont{I.~M.} \bibnamefont{Kapitonov}},
  \bibinfo{author}{\bibfnamefont{E.~I.} \bibnamefont{Lileeva}},
  \bibnamefont{and} \bibinfo{author}{\bibfnamefont{E.~V.}
  \bibnamefont{Shirokov}}, \bibinfo{journal}{Bulletin of the Russian Academy of
  Science, Physics} \textbf{\bibinfo{volume}{67}}, \bibinfo{pages}{1636}
  (\bibinfo{year}{2003}).

\bibitem[{\citenamefont{Tselyaev}(2013)}]{Tselyaev2013}
\bibinfo{author}{\bibfnamefont{V.~I.} \bibnamefont{Tselyaev}},
  \bibinfo{journal}{Phys. Rev. C} \textbf{\bibinfo{volume}{88}},
  \bibinfo{pages}{054301} (\bibinfo{year}{2013}).

\bibitem{Pap10} P. Papakonstantinou and R. Roth, Phys. Rev. C {\bf 81}, 024317 (2010). 
\bibitem{Rai19} F. Raimondi and C. Barbieri, Phys. Rev. C {\bf 99}, 054327 (2019). 
\bibitem{Gra20} M. Grasso and D. Gambacurta, Phys. Rev. C {\bf 101}, 064314 (2020). 
\bibitem{Dan94} D. Danielewicz and P. Schuck, Nucl. Phys. A {\bf 567}, 78 (1994).
\bibitem{Urb14} M. Urban and P. Schuck, Phys. Rev. C {\bf 90}, 023632 (2014).
\bibitem{Noz85} P. Nozi\`eres and S. Schmitt-Rink, J. Low Temp. Phys. {\bf 59}, 159 (1985).
\bibitem{Sto03} A. Storozhenko, P. Schuck, J. Dukelsky, G. R\"opke, and A. Vdovin, Ann. Phys. (NY) {\bf 307}, 308 (2003).
\bibitem{Sch73a} P. Schuck, F. Villars, and P. Ring, Nucl. Phys. A {\bf 208}, 302 (1973).
\bibitem{Rij96} G. A. Rijsdijk, W. J. W. Geurts, K. Allaart, Phys. Rev. C {\bf 53}, 201 (1996). 
\bibitem{Kho82} V.A. Khodel and E. E. Saperstein, Phys. Rep. {\bf 92}, 183 (1982).
\bibitem{Gne14}  N.V. Gnezdilov, I.N. Borzov, E.E. Saperstein, and S.V. Tolokonnikov, Phys. Rev. C {\bf 89}, 034404 (2014). 
\bibitem{Nam19} Nam Vu, Ion Mitxelena, and A. E. DePrince, J. Chem. Phys. {\bf 151}, 244121 (2019).
\bibitem{Jem20} M. Jemai, and P. Schuck (submitted to Eur. Phys. J. A). 
\bibitem{Lip65} H. J. Lipkin, N. Meshkov and A. J. Glick, Nucl. Phys. A {\bf 62}, 188 (1965).
\bibitem{Vid04} J. Vidal, G. Palacios, and C. Aslangul, Phys. Rev. A {\bf 70}, 062304 (2004). 
\bibitem{Rib07} P. Ribeiro, J. Vidal, and R. Mosseri, Phys. Rev. Lett. {\bf 99}, 050402 (2007). 
\bibitem{Col18} G. Col\'o and S. De Leo, Int. J. Mod. Phys. E {\bf 27},  1850039 (2018).
\bibitem{Cam15} S. Campbell, G. De Chiara, M. Paternostro, G. M. Palma, and R. Fazio, Phys. Rev. Lett.  {\bf 114}, 177206 (2015). 
\bibitem{PS20} P. Schuck is very greatful to V.V. Baran who has pointed this out to him. 
\bibitem{Rop20} G. R\"opke, in "Nuclear Particle Correlations and Cluster Physics", edited by W.-U Schr\"oder 
(World Scientific, Singapore, 2017). 
\bibitem{Sch90} M. Schmidt, G. R{\"o}pke, and H. Schulz, Ann. Phys. (NY) {\bf 202}, 57 (1990).
\bibitem{Hor06} C.J. Horowitz and A. Schwenk, Nucl. Phys. A {\bf 776}, 55 (2006).
\bibitem{Sch08} P. Schuck, Int. J. Mod. Phys. E {\bf 17}, 2136 (2008).
\bibitem{Rop18} G. R\"opke, D. N. Voskresensky, I. A. Kryokov, D. Blaschke, Nucl.Phys. A {\bf 970}, 224 (2018).
\bibitem{Rop79} G. R\"opke, R. Der, Phys. Stat. Sol. (b) {\bf 92}, 501 (1979).
\bibitem{Sch90} M. Schmidt, G. R{\"o}pke, and H. Schulz, Ann. Phys. (NY) {\bf 202}, 57 (1990).
\bibitem{Men10} Meng Jin, M. Urban, and P. Schuck, Phys. Rev. C {\bf 82}, 024911 (2010). 
\bibitem{Alm95} T. Alm, G. R\"opke, and M. Schmidt, Z. Phys. A {\bf 351}, 295 (1995).
\bibitem{Rop15} G. R\"opke, Phys. Rev. C {\bf 92}, 054001 (2015).  
\bibitem{Ste95} H. Stein, A. Schnell. T. Alm, and G. R\"opke, Z. Physik A {\bf 351}, 295 (1995).
\bibitem{Rop94} G. R\"opke,  Ann. Physik (Leipzig) {\bf 3}, 145 (1994); R. Haussmann,  Z. Physik B {\bf 91}, 291 (1993).
\bibitem{Boz99} P. Bozek, Nucl. Phys. A {\bf 657}, 187 (1999).
\bibitem{Hau09} R. Haussmann and W. Zwerger, Phs. Rev A {\bf 80}, 063612 (2009).
\bibitem{Lom01} U. Lombardo, P. Nozi\'eres, P. Schuck, H.-J. Schulze, and A. Sedrakian, Phys. Rev. {\bf C 64}, 064314 (2001). 
\bibitem{Sog10} T. Sogo, G. R\"opke, and P. Schuck, Pys. Rev. C {\bf 81}, 064310 (2010). 
\bibitem{Sed05} A. Sedrakian, J. Mur-Petit, A. Polls, and H. M\"uther, Phys. Rev. A {\bf 72}, 013613 (2005). 
\bibitem{Sch99a} A. Schnell,  G. R\"opke, and P. Schuck,  Phys. Rev. Lett. {\bf 83}, 1926 (1999).
\bibitem{Ran97} M. Randeria, Varenna Lectures 1997, cond-mat/9710223.
\bibitem{Dic04} W. H. Dickhoff and C. Barbieri, Prog. Part. Nucl. Phys. {\bf 52},  377 (2004); 
A. Schnell, Ph.D. Thesis, Rostock (1996). 
\bibitem{Cha16} Chang Xu, Z. Ren, G. R\"opke, P. Schuck, Y. Funaki, H. Horiuchi, A. Tohasaki, T. Yamada, and Bo Zhou,
Phys. Rev. C {\bf 93}, 011306 (2016).
\bibitem{Cha17} Chang Xu, G. R\"opke, P. Schuck, Z. Ren, Y. Funaki, H. Horiuchi, A. Tohsaki, T. Yamada, and Bo Zhou, 
Phys. Rev. C {\bf 95}, 061306 (2017).
\bibitem{Sog09} T. Sogo, R. Lazauskas, G. R\"opke, and P. Schuck, Phys. Rev. C {\bf 79}, 051301 (2009).
\bibitem{Bey00} M. Beyer, S.A. Sofianos, C. Kuhrts, G. R\"opke, and P. Schuck, Phys. Lett. B {\bf 488}, 247 (2000).
\bibitem{Rop98} G. R\"opke, A. Schnell, P. Schuck, and P. Nozi\`eres, Phys. Rev. Lett. {\bf 80}, 3177 (1998).
\bibitem{Mal69} R.A. Malfliet and J.A. Tjon, Nucl. Phys. A {\bf 127}, 161 (1969).
\bibitem{Kam05} H. Kamei and K. Miyake, J. Phys. Jpn. {\bf 74}, 1911 (2005).
\bibitem{Toh01} A. Tohsaki, H. Horiuchi, P. Schuck, and G. R\"opke, Phys. Rev. Lett. {\bf 87}, 192501 (2001).
\bibitem{Bli86} A.H. Blin, R.W. Hasse, B. Hiller, P. Schuck, and C. Yannouleas, Nucl. Phys. A {\bf 456}, 109 (1986).
\bibitem{Ebr20} J.-P. Ebran, M. Girod, R. Lasseri, and P. Schuck, Phys. Rev. C {\bf 102}, 014305 (2020).
\bibitem{Toh17} A. Tohsaki, H. Horiuchi, P. Schuck, and G. R\"opke, Rev. Mod. Phys. {\bf 89}, 011002 (2017).
\bibitem{Saw62} J. Sawicki, Phys. Rev. {\bf 126}, 2231 (1962).
\bibitem{Dro90} S. Dro$\dot{\rm z}$d$\dot{\rm z}$, S. Nishizaki, J. Speth and J. Wambach, Phys. Rep. {\bf 197}, 1 (1990).
\bibitem{Toh04} M. Tohyama and P. Schuck, Eur. Phys. J. A {\bf 19}, 203 (2004). 
\bibitem{Wan85} S. J. Wang and W. Cassing, Ann. Phys. {\bf 159}, 328 (1985);
W. Cassing and S. J. Wang, Z. Phys. A {\bf 328}, 423 (1987).
\bibitem{Toh04a} M. Tohyama, S. Takahara and P. Schuck, Eur. Phys. J. A {\bf 21}, 217 (2004).
\bibitem{Tak88} K. Takayanagi, K. Shimizu and A. Arima, Nucl. Phys. A {\bf 477}, 205 (1988).
\bibitem{Tho61-b} D. J. Thouless, Nucl. Phys. A {\bf 22}, 78 (1961).
\bibitem{Toh04b} M. Tohyama and P. Schuck, Eur. Phys. J. A {\bf 32}, 139 (2004). 
\bibitem{Toh89} M. Tohyama and M. Gong, Z. Phys. A {\bf 332}, 269 (1989).
\bibitem{Toh05} M. Tohyama, Prog. Theor. Phys. {\bf 114}, 1021 (2005).
\bibitem{Toh07} M. Tohyama, Phys. Rev. C {\bf 75}, 044310 (2007).
\bibitem{Toh08} M. Tohyama and P. Schuck, Eur. Phys. J. A {\bf 36}, 349 (2008).
\bibitem{sen-model} P.Schuck, Int. J. Mod. Phys. E {\bf 29}, 2050023 (2020).
\bibitem{ST16} P. Schuck and M. Tohyama, Eur. Phys. J. A {\bf 52}, 307 (2016).
\bibitem{MT17} M. Tohyama, Prog. Theor. Exp. Phys.  {\bf 2017}, 093D06 (2017).
\bibitem{Lui01} Y.-W. Lui, H. L. Clark, D. and H. Youngblood, Phys. Rev. C {\bf 64}, 064308 (2001).
\bibitem{usman} I. Usman {\it et al.}, Phys. Lett. B {\bf 698}, 191 (2011).
\bibitem{hartmann} T. Hartmann {\it et al.}, Phys. Rev. Lett. {\bf 85}, 274 (2000).
\bibitem{Hot74} A. Hotta, K. Itoh and T. Saito, Phys. Rev. Lett. {\bf 33}, 790 (1974).
\bibitem{Vas18} O. Vasseur, D. Gambacurta and M. Grasso, Phys. Rev. C {\bf 98}, 044313 (2018).
\bibitem{Hed65} L. Hedin, Phys. Rev. A {\bf 139}, 796 (1965).
\bibitem{Kad62}  G. Baym, L. P. Kadanoff, Phys. Rev. {\bf 124}, 287 (1961).
\bibitem{Bon96} G. Baym, Phys. Rev. {\bf 127}, 1391 (1962). 
\bibitem{Del16} D.S. Delion, P. Schuck, and  M. Tohyama, Eur. Phys. J. B {\bf 89}, 1 (2016).
\bibitem{Aou91} Z. Aouissat, G. Chanfray, P. Schuck, and G. Welke, Z. Phys. {\bf 340}, 347 (1991). 
\bibitem{Aou95} Z. Aouissat, P. Schuck, R. Rapp, J. Wambach, and G. Chanfray, Nucl. Phys. A {\bf 581}, 471 (1995). 
\bibitem{Bar08} X. Barillier-Pertuisel, S. Pittel, L. Pollet, and P. Schuck, Phys. Rev. A {\bf 77}, 012115 (2008).

\end{thebibliography}
\end{document}